\documentclass[preprint2]{aastex}

\usepackage{epsf}
\usepackage{graphics}
\usepackage{color}

\def\eps@scaling{1.6}%

\newcommand\plotthree[3]{{%
 \typeout{Plottwo included the files #1 #2 #3}
 \centering
 \leavevmode
 \columnwidth=.30\columnwidth
 \includegraphics[width={\eps@scaling\columnwidth}]{#1}%
 \hfil
 \includegraphics[width={\eps@scaling\columnwidth}]{#2}%
 \hfil
 \includegraphics[width={\eps@scaling\columnwidth}]{#3}%
}}%

\newcommand{\be}{\begin{equation}}
\newcommand{\ee}{\end{equation}}

\newcommand{\bx}{$\beta_{\rm X}$}
\newcommand{\axb}{$\alpha_{\rm X2}$}
\newcommand{\axc}{$\alpha_{\rm X3}$}

\newcommand{\plm}{$\pm$}

\newcommand{\swift}{\mbox{\it Swift}}	    % defines Chandra name
\newcommand{\meszaros}{M\'esz\'aros~}
\shorttitle{GRB Statistics}
\shortauthors{Grupe et al.}

\begin{document}

%\input DGrupe_clipfig.tex
%\useunitmm

\def\etal{{\it et\thinspace al.}\ }
\def\alp{{$\alpha$}\ }
\def\al2{{$\alpha^2$}\ }


\title{Evidence for New Relations between Gamma Ray Burst Prompt and X-ray 
Afterglow Emission from 9 Years of  \swift\ 
}


\author{Dirk Grupe\altaffilmark{1,2},\email{dgrupe007@gmail.com} 
John A. Nousek\altaffilmark{1,2}, P\'eter Veres\altaffilmark{2},
Bin-Bin Zhang\altaffilmark{2, 3},
Neil Gehrels\altaffilmark{4}
}

\altaffiltext{1}{\swift\ Mission Operation Center, 2582 Gateway Dr., State College, PA 16801}

\altaffiltext{2}{Department of Astronomy and Astrophysics, 
Pennsylvania State
University, 525 Davey Lab, University Park, PA 16802} 

\altaffiltext{3}{Center for Space Plasma and Aeronautic Research (CSPAR), University of Alabama in Huntsville, Huntsville, AL 35899}

\altaffiltext{4}{Astrophysics Science Division, Astroparticle Physics Laboratory,
Code 661, NASA Goddard Space Flight Center, Greenbelt, MD 20771 }


%\altaffiltext{1}{Visiting Astronomer, Cerro Tololo Inter-American Observat}


\begin{abstract}
When a massive star explodes  as a Gamma Ray Burst  information
about this explosion is retained in the properties of the prompt and
 afterglow emission.
We report on new relationships between the prompt and X-ray afterglow
emission of \swift-detected Gamma Ray Bursts found from BAT and XRT data
between 2004 December and  2013 August (754 GRBs). 
These relations suggest that the prompt and 
afterglow emission are closely linked. 
In particular,
we find very strong correlations between the BAT 15-150keV 
$T_{90}$ and the break
times before and after the plateau phase in the X-ray 0.3-10keV 
afterglow light curves. 
 We also find a strong anti-correlation between the photon
index of the GRB prompt emission and the X-ray spectral slope of the afterglow.
Further, anti-correlations exist between the rest frame peak energy 
in the prompt emission, $E_{\rm peak,z}$, 
and the X-ray afterglow decay slope during the plateau phase and the break 
times after the plateau phase.
The rest-frame break times before and after the plateau phase are also 
anti-correlated with the rest-frame 
15-150keV luminosity and isotropic energy during the prompt emission. 
A Principal Component Analysis suggests that GRB properties 
 are primarily driven by the luminosity/energy release in the 15-150 keV band. 
Luminosity functions derived at various redshifts from logN-logS analysis, indicate  
that the 
density of bright bursts 
is significantly lower in the local Universe compared with the Universe
 at z$\approx$3, where the 
  density of bright GRBs peaks. Using cluster analysis, 
 we find that the duration of \swift\ BAT-detected short-duration
 GRBs is less than 1s. 
 We also present the catalogue of all \swift-onboard detected bursts.
\end{abstract}

\keywords{GRBs: general
}

\section{Introduction}
The \swift\ mission \citep{gehrels04} has revolutionized the study of
Gamma-Ray Burst (GRB) afterglows. With \swift\ it was possible for the first time
to access the earliest phase 
of X-ray afterglows. In the pre-\swift\ era
observations of the X-ray afterglow were limited to the 'normal' decay phase, 
typically starting roughly a day after detecting the
burst. During the BATSE era only a relatively
small number of bursts were actually followed up by X-ray observatories such as
BeppoSAX\citep{costa99, depasquale06}.
 All this
changed when \swift\ was launched in November 2004. \swift\ is an autonomous
robot making it  possible to
observe the X-ray afterglow starting typically 1 or 2 minutes after the trigger
by the \swift\ Burst Alert telescope \citep[BAT,][]{barthelmy05}. 
Over almost 9 years in orbit (August 2013), \swift\ has discovered more than 800 GRBs,
providing the largest sample of GRBs with prompt and afterglow emission
observations. 
The prompt emission is likely the result of internal shocks produced in the
newly formed jet \citep[e.g.][]{meszaros06}. Further models include photospheric
emission including either baryon- \citep{Peer+06phot,Beloborodov10phot} or
magnetic field \citep{Drenkhahn+02,McKinney+11switch} dominated ejecta.
Additionally magnetic reconnection irrespective of the photospheric emission
can be responsible for the prompt emission \citep{Thompson94, Zhang+11icmart}.

For the afterglow, the external shock scenario \citep{Meszaros+97ag} is
observationally the most supported mechanism with modifications involving jet
opening angle effects \citep{Rhoads99jet}, the nature of the interstellar
medium (wind or constant density profile \citet{panaitescu00}) or possible late
energy injection \citep{Rees+98refresh}.

As shown by
\citet{nousek06} and \citet{zhang06}, X-ray afterglows typically display a
canonical light curve which is described by a very steep initial
decay slope  followed by a much shallower slope - usually referred to as
the 'plateau phase'. After this plateau phase we see the 'normal' decay phase.
The initial decay slope in the X-ray light curve always regards as 
the tail of the prompt emission phase 
\citep{zhangbb07, zhangbb09}.
 The plateau phase marks the beginning of
the afterglow phase which is generally caused by 
external shocks when the jet starts interacting with
the interstellar medium.  \swift's ability to observe GRBs in soft 
X-rays within minutes after the explosion has lead to the discovery of
flares in roughly 1/3 of all GRBs \citep[e.g.][]{falcone07, margutti11b}.
\swift\ not only revolutionized GRB studies, but the large
number of bursts covered with multi-wavelengths observation and redshift measurements
allows for the first time a very detailed statistical analysis of GRBs including
cosmological studies of GRBS \citep[e.g.][]{wanderman10}

Although a connection between the prompt and the afterglow emission of a GRB 
is
expected according to the standard fireball model \citep[e.g.,][]{meszaros06},
in the early days of \swift\ there was no evidence for such a connection
\citep{willingale07} except for the fluences in
the prompt and afterglow emission. As shown by \citet{gehrels08} based on the
observations of the GRBs detected during the first two and half years of \swift,
there is a strong correlation between the fluence in the 15-150 keV band and the
flux density of the X-ray afterglow emission. 
 \citet{obrien06} showed that the
\swift\ BAT and XRT light curves of the prompt and afterglow emission can be
typically  consistently connected.  
Recently \citet{margutti12} performed a statistical analysis of all \swift-detected 
GRBs until December 2010 and found clear correlations between the energetics of the prompt
and the afterglow emission which also seem to suggest a universal scaling relation between 
short and long duration 
GRBs \citep{bernardini12, nava12}. \citet{davanzo12} showed that X-ray afterglow
luminosities at various times after the trigger correlate strongly with prompt emission
properties, such as the isotropic energy $E_{\rm iso}$ and the peak energy in the high energy
spectrum $E_{\rm peak}$.
Previously, statistical analysis of the X-ray afterglow light
curves were also performed by \citet{evans09} and \citet{racusin09}. 
While \citet{evans09} focused on the automated analysis of \swift\ XRT light curves and found that about 40\% 
of \swift\ bursts with X-ray observations are GRBs with canonical light curves, \citet{racusin09} did an
analysis of the spectral and temporal parameters of the \swift\ X-ray data. 
They found that different phases of 
the light curves are all consistent with closure relations that explain the various states of GRB afterglow 
light curves due to the jet geometry and physics, 
the environment around the GRB, and the election
density and cooling \citep[e.g.][and references therein]{zhang04}. 

One of the 
best-known relations between  GRB properties is the relation of the peak energy in the 
$\gamma$-ray spectrum $E_{\rm peak}$ with the isotropic energy $E_{\rm iso}$
 or the 
collimation-corrected energy $E_{\gamma}$, the Amati and Ghirlanda relations, respectively \citep{amati02,
ghirlanda04}. Similar to these are the relations found by \citet{yonetoku04} and \citet{schaefer07}
between $E_{\rm peak}$ and the burst luminosity.

Although various efforts have been made to link the energetic properties of the
prompt and afterglow emission of GRBs \citep[e.g.,][]{margutti12}, what is still
missing is to find correlations between  prompt and afterglow properties,
such as $T_{90}$, break times, or spectral slopes. While this effort was
hampered at the beginning of the \swift\ mission due to the small number of
bursts \citep{willingale07}, with  more than 800 bursts
that \swift\ has detected in
almost  9 years in orbit (August 2013),
the GRB sample has grown allowing to find new relations among prompt and afterglow emission parameters.

One motivation for this study comes from the relation we found
between redshift and excess absorption seen in the observed X-ray
spectra of GRB afterglows \citep{grupe07}.
However, this method is
rather limited in the sense that it can only say if a burst has a large
amount of excess absorption then it is a low-redshift burst. Our goal has
been to find other means to distinguish between high and low redshift bursts more
precisely and finally to estimate the redshift based on early
available \swift\ data. Hydrogen column density derived from X-ray observations 
 was found to correlate using canonical correlation analysis \citep{Balazs+11multi} with the
  properties of the prompt emission. 
The most sophisticated statistical analysis on \swift\ GRBs to date has been
performed by \citet{morgan11} who applied a Random Forest algorithm in order to
predict high redshift bursts. 
By looking through the data of \swift-detected GRBs we noticed several relations
between prompt and afterglow properties which are not necessarily expected. In
particular we noticed a strong relation of the BAT $T_{90}$ and the 15-150 keV
luminosity with the break times
in the X-ray afterglow light curve \citep{grupe12}.

 Here we 
present a detailed analysis of the whole \swift-detected GRB dataset and will
show that indeed there are many correlations that establish a tight relation
between the prompt and afterglow emission in GRBs. 
This paper is organized as follows: In Section\,\ref{sample} we describe the sample
selection and the data analysis, in Section\,\ref{results} we show the distributions
and correlation analysis of the \swift-detected GRBs prompt and afterglow properties,
followed by a discussion in Section\,\ref{discuss}.
Throughout
the paper spectral indices $\beta$  and light curve decay indices $\alpha$ 
are defined as $F_{\nu}(\nu)\propto\nu^{-\beta}t^{-\alpha}$.  
All errors are 1$\sigma$ unless stated otherwise. Cosmological parameters like the
luminosity distance or the comoving volume were derived from the cosmology
calculator by \citet{wright06}. For all of these values we assumed the standard
cosmology with $H_0$=71 km s$^{-1}$ Mpc$^{-1}$, $\Omega_{\rm M}$=0.27, and $\Omega_{\Lambda}$=0.73.
 
\section{GRB Sample and Observations \label{sample}}

\subsection{Sample Selection} 
By the end of August 2013 \swift\ has discovered  804 bursts\footnote{
A full interactive table of all \swift\ bursts can be found at:
http://swift.gsfc.nasa.gov/docs/swift/archive/grb\_table/.}.
Of these,  40 GRBs were
discovered through ground processing and 9 burst through the \swift\ BAT slew survey. There is also one burst with no data available.
Excluding these 50 bursts from the initial
sample leaves a total of 754 \swift-discovered onboard triggered bursts. 
In Appendix\,\ref{grb_catalog} we present the whole catalog of these 754 GRBs as a machine readable file that is available
online. This
file contains the GRB name, \swift\ BAT trigger number, redshift, and BAT 
and XRT measured observed parameters. 
A full list of these parameters is given in Appendix\,\ref{grb_catalog} in
 Table\,\ref{grb_table}.

\begin{figure}
%\epsscale{0.75}
\epsscale{0.75}
\plotone{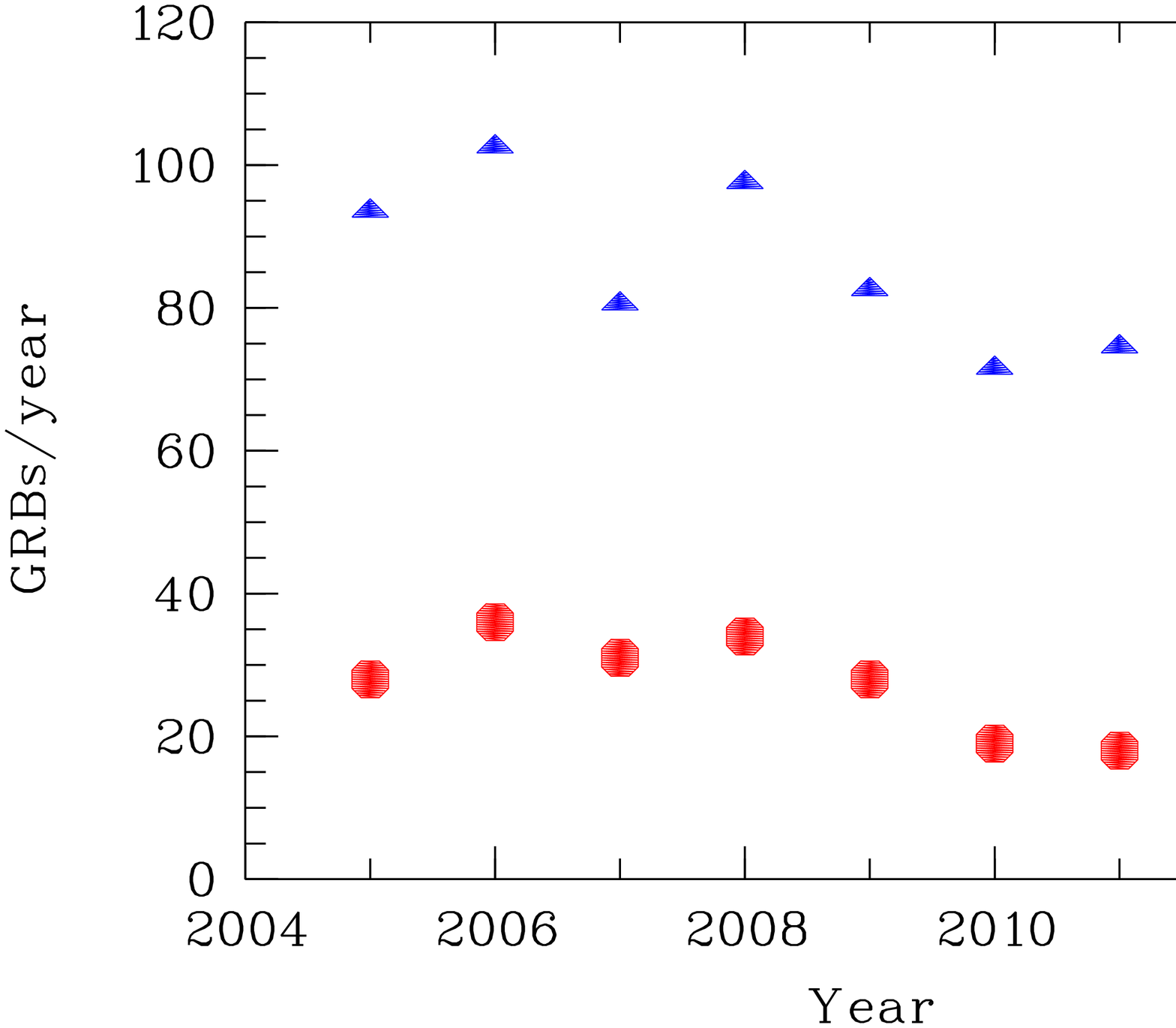}
\caption{\label{grbs_year} 
Onboard \swift-BAT-detected GRBs per year. The blue triangles displays all GRBs detected and the red circles those GRBs which had spectroscopic redshift
measurements. Note that the 2013 numbers were projected based on the GRBs 
detected in the first 8 months of 2013.  
}
\end{figure}

Figure\,\ref{grbs_year} displays the number of onboard detected GRBs per year (blue triangles). 
The red circles display the 
number of GRBs per year with spectroscopic redshift measurements.
In 663 cases 
XRT observed the field 
of the GRBs and 607 X-ray afterglows were detected (91\%)
In 30 cases no X-ray afterglow was found although \swift\ was on the target with
the XRT and UVOT within 300 seconds after the trigger.
XRT did not observe the remaining
97 bursts due to observing constraints by the Sun, the Moon, or the Earth. 
The \swift\ UVOT detected a total of 251 GRB afterglows out of 655 UVOT
observations performed.
In this paper,
 however, we focus on the high energy properties of the bursts and
leave the UVOT analysis for a later publication. Crucial to the analysis
of  physical parameters of GRBs are redshift measurements.
Up to 2013 August, 232 GRBs out for the 754 GRBs in our sample
or their host galaxies
have spectroscopic redshift measurements  (about 31\%).
Burst redshifts were taken from GCN circulars and the GRB redshift 
catalogues by \citet{fynbo09, jakobsson12} and \citet{kruehler12}.
Note that in the pre-\swift\ era there were redshift measurements for only 
43 bursts \citep{gehrels12}.

Of the 754 GRBs in our sample 63 are short duration and 689 are long duration
 GRBs (one early GRB did not had a $T_{\rm 90}$ measurement)
 following the standard 
division at $T_{90}$=2s as defined by \citet{kouveliotou93}.
This is a significantly lower percentage compared with the ~25\% found in the BATSE GRB
sample. The reason is purely a selection effect due to the lower energy range of the \swift\
BAT compared with BATSE. As we will see later, the 2s devision line is too long for
\swift\ BAT-discovered bursts \citep[see also][]{bromberg12}.

\subsection{\label{observe} Observations and data reduction}

The 0.3-10 keV X-ray count rate light curves and spectra were derived from the GRB repository website at
the University of Leicester
\citep[][]{evans07}\footnote{http://www.swift.ac.uk/xrt\_curves/}. 
The light curves were fitted using multi-segment power law models
\citep[e.g.][]{evans09}
in which period of flares, in particular at early times after the explosion, were excluded from the analysis.
In almost all cases (90-90\%) we made use of the Leicester online light curve fitting routine which is used by the \swift\
team to make predictions on the XRT count rate of an X-ray afterglow.  In the remaining cases we fitted the light curves in
XSPEC manually. For these light curves we converted the light curve file into pha and an rmf that can be read into
XSPEC by applying the FTOOL command {\it flu2xsp} to the light curve ASCII file. 

For most spectra, we  used the photon counting mode data
\citep[pc,][]{hill04} that primarily cover the afterglow phase. 
We extracted 
source and background
spectra and auxiliary response 
files provided by the GRB catalogue website at Leicester. 
These data typically cover the entire afterglow phase.
The Windowed Timing mode data at the beginning of the
light curves are often contaminated by emission from flares which tend to be
much harder than the afterglow emission \citep{falcone07,
margutti11a,margutti11b}.
We did not rely on the automated fitting results on the GRB catalogue website 
in Leicester. In the majority of 
cases the automated routine uses a free fit `intrinsic' absorber which results in
excess absorption with large uncertainties even though the spectrum is consistent 
with just the Galactic absorption column density. 
Therefore all spectra were
 analyzed manually. 
In September 2007 the substrate voltage on the 
\swift\ XRT detector was increased from 0 to 6V \citep{godet09}.   
Accordingly, for all spectra we used the most current response files. For spectra
 before August 2007 we used the 
response file {\it swxpc0to12s0\_20070901v011.rmf} and 
 after the end of August 2007 the response file 
{\it swxpc0to12s6\_20010101v013.rmf}. 
All data were fit to absorbed power law models in
XSPEC \citep{arnaud96}. First we fit the data with the absorption parameter
fixed to the Galactic value derived from the HI maps by \citet{kalberla05}. 
If this model does not result in an acceptable fit,
we thaw the absorption parameter to search for excess absorption above
the Galactic value as described in \citet{grupe07}. For afterglows with
spectroscopic redshifts we determine the intrinsic absorption column density at
the redshift of the GRB.

The parameters measured from the BAT data were derived from the BAT GRB analysis
pages\footnote{http://swift.gsfc.nasa.gov/docs/swift/results/BATbursts/} and
BAT refined circulars and GCN burst reports. 
Whenever possible we also made use of the peak energy $E_{\rm peak}$ in the high energy
spectrum of a burst given by primarily Konus-Wind \citep{aptekar95} and the 
{\it Fermi} Gamma-Ray Burst Monitor
\citep[GBM;][]{meegan09}. 
 Before the launch of {\it Fermi} \citep{atwood09}, we primarily used the 
$E_{\rm peak}$ measurements by Konus-Wind and the Suzaku WAM published in GCN Circulars.
After the {\it Fermi} launch, we used the {\it Fermi} GBM measurements when available, which now exist for roughly 
50\% of all bursts after the {\it Fermi} launch.

In the end we derive $T_{90}$, the 15-150 keV photon spectral slope $\Gamma$,
and the 15-150 keV fluence from the BAT, and the 0.3-10 keV X-ray spectral slope
\bx, the break times at the beginning and the end of the plateau phase $T_{\rm break
1}$ and $T_{\rm break 2}$, the slope during the plateau phase \axb\ and 
the 'normal' decay slope \axc. For bursts with spectroscopic redshift measurements
the fluence was k-corrected \citep[e.g.][]{humason56, oke68}
and $T_{90}$, $T_{\rm break 1}$, and $T_{\rm break 2}$
were transferred into the rest frame. The rest-frame 15-150 keV luminosity 
is the mean isotropic luminosity during the time $T_{90, z}$ = $T_{90}$/(1+z)
determined from the k-corrected fluence.
A complete description of each parameter is given in Appendix\,\ref{parameters}.

\section{\label{results} Results}
One of the goals of this paper is to establish  new
connections between GRB prompt and afterglow emission
properties. Therefore we need to examine the dataset by means of statistics.
In this section we will first present the distributions of the observed and
rest-frame parameters of \swift-detected GRBs. In particular we will look at differences between short and long-duration GRBs
We will then look at 
bi-variate corrections among these parameters with the gaol of finding evidence for connections between GRB prompt and afterglow phases.
This is then followed by a principal component analysis in order to search for underlying properties that drive the observed 
parameters in GRBs.
 The section closes with a discussion of the GRB
luminosity functions and looking into estimating the redshift of a burst including an 
 update on the redshift - excess absorption relation
\citep{grupe07}.

\subsection{Distributions \label{distribution}}

\begin{figure*}
%\epsscale{0.75}
\epsscale{1.5}
\plottwo{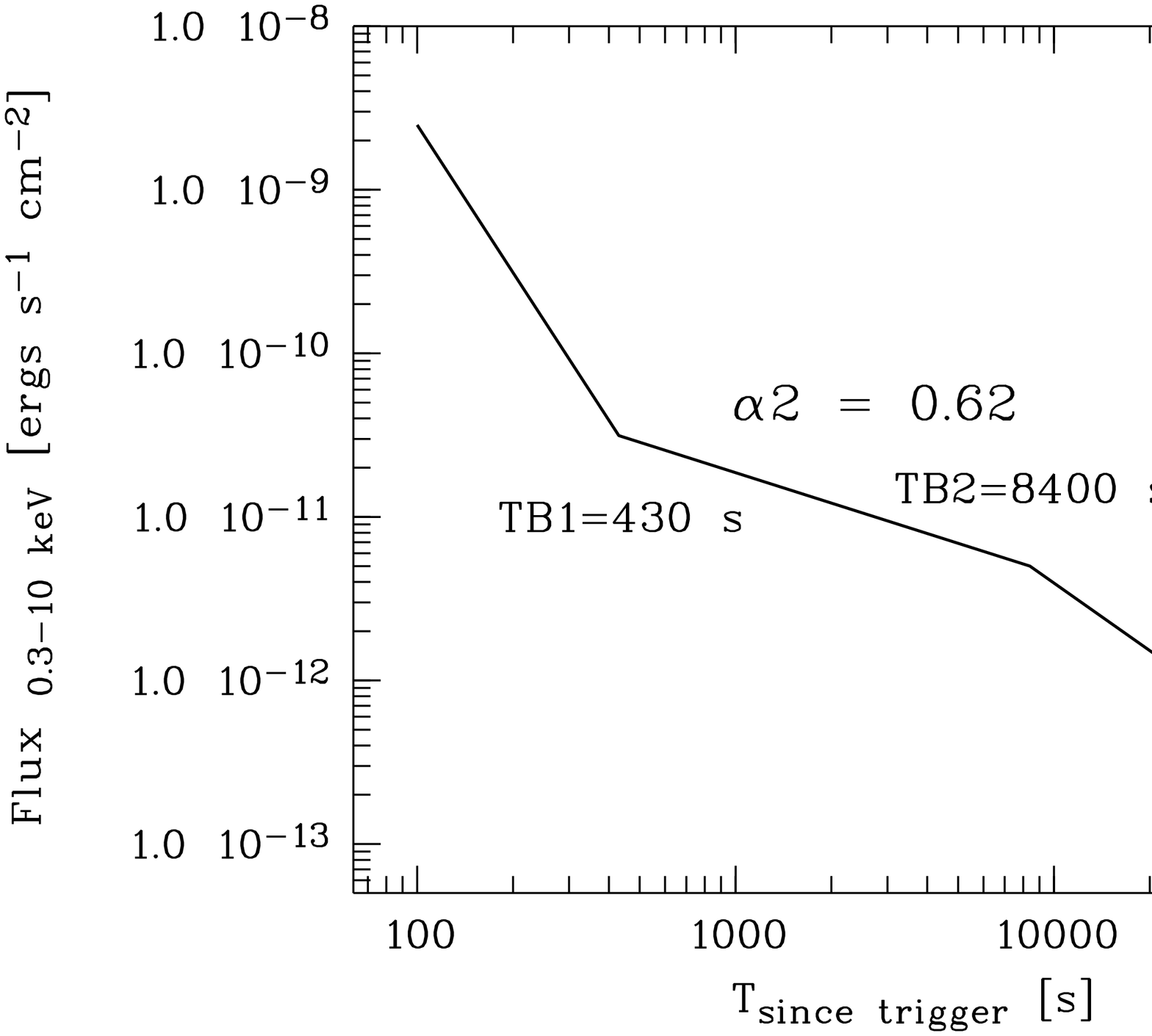}{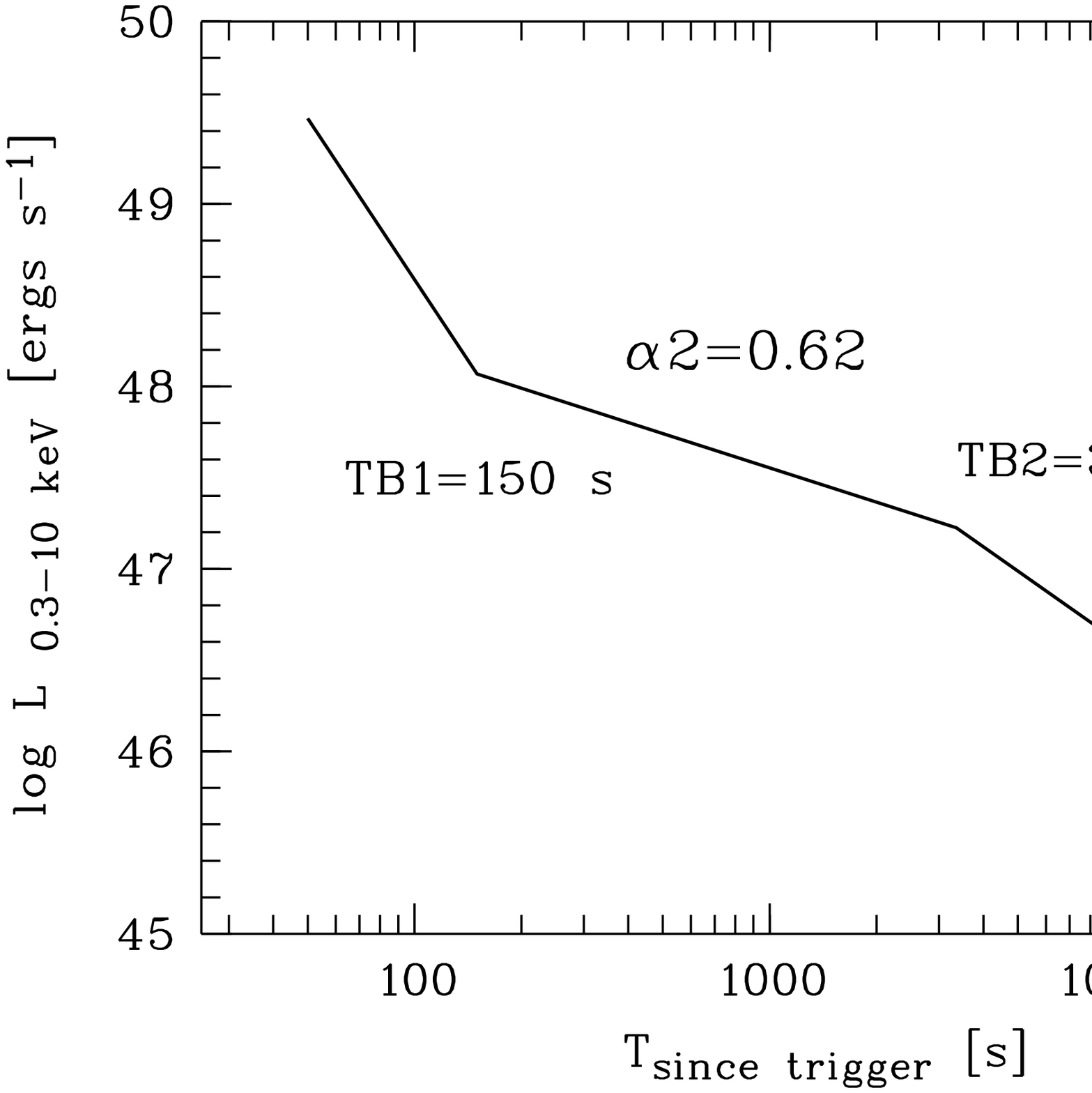}
\caption{\label{median_xrt_lc} 
Median observed XRT flux light curve and rest-frame luminosity light curve of \swift-detected GRBs
(left and right panel, respectively, see also Table\,\ref{grb_statistics}).
}
\end{figure*}

The mean, standard deviation and median of all \swift-discovered GRBs, as well as for long and
short-duration GRBs, and high redshift GRB (z$>$4.0)
are listed
in Table\,\ref{grb_statistics}. We included high-redshift bursts to this table in order to examine if these bursts
appear to be somewhat special and have different properties than bursts at lower redshifts 
or bursts without redshift measurements. In this section we show the distributions of $T_{\rm 90}$, the break times 
in the X-ray light curves, decay slopes, and redshifts. 
All other distribution are shown in Appendix\,\ref{distr_append}.
In Figures\,\ref{distr_t90_tb} through \ref{distr_z} the statistical analysis of the
property distributions are visualized through three standard tools in data mining
\citep[e.g.][]{feigelson12, torgo11, crawley07}: 
\begin{itemize}
\item Histogram and kernel density estimator
\item Box plot
\item q-q plot
\end{itemize}

We explain the purpose of these statistical tools in Appendix\,\ref{distr_append}.

Figures\,\ref{distr_t90_tb} and \ref{distr_t90z_tbz}
display the distributions of the BAT $T_{90}$ and the
break times $T_{\rm break 1}$ and $T_{\rm break 2}$ before and after the plateau
phase in the X-ray light curve. The  observed
times are displayed in Figure\,\ref{distr_t90_tb} and
the plots in Figure\,\ref{distr_t90z_tbz} show the values shifted into the
rest-frame. 
The median $T_{90}$ value for all \swift-discovered GRBs is 38s and for long GRBs only, 
46s. The density estimate of the $T_{90}$ distribution suggests that  for \swift-detected GRBs we see
a bimodal distribution, separating the bursts into short and long bursts, as had been suggested based
on the BATSE results \citep{kouveliotou93}. The division between the two groups, however, appears to be
at a shorter $T_{90}$
 than  with the BASTE detected bursts,in agreement with the results by
\citet{bromberg12}: the observed $T_{90}$ distribution suggests that the devision line between short and long bursts
detected by BAT is on the order of 1s. These short bursts also appear clearly as outliers in the $T_{90}$ box diagram
for all GRBs.  
 In the rest-frame (as shown in Figure\ref{distr_t90z_tbz})
 the median $T_{90}$ of all
bursts (with redshifts) is 13.4s, 16.0s for long, and 0.27s for short bursts.

Although break times before the plateau phase of short and long bursts are 
all of the order of roughly 500s, 
there are significant differences in the break times after the plateau phase. 
While long GRBs show breaks   after the plateau phase in the observed
frame  at about  8400s 
after the trigger, short GRBs show a median break time after the plateau of 300s. 
In the rest frame the break times are 150s and 3360s for the long bursts. 
 These break times before and after the plateau phase are comparable to the values
published by \citet{evans09} and \citet{margutti12}.
It is interesting to note that from the relation between redshift and the length of
the plateau phase as reported by \citet{stratta09} one would naively expect short GRBs
to have longer plateau phases than long GRBs as they occur (as we will show later) at
lower redshifts. 
Note, however, that for the 
short bursts we do not have enough bursts with redshift and break time 
measurements that will allow to
state reliable numbers. 

Figure\,\ref{distr_ax} displays distributions of the X-ray light curve decay
slopes during and after the plateau phase \axb\ and \axc\ (upper and lower panels, respectively).
The distribution of decay slope during the plateau phase
\axb\ is almost Gaussian with an extended wing towards the
very flat end of the distribution (see the Q-Q plot in Figure\,\ref{distr_ax})
Note, although this phase is commonly 
called the `plateau phase',  the median decay slope of all bursts is of the order of \axb=0.6
 and that there is a large scatter in the distribution of
 decay slopes ($\sigma$=0.44). 
 Some plateau decay slopes can be relatively steep with \axb$>$1.0.  
Our mean and median values of \axb\ are slightly steeper than what has been found be \citet{evans09} and
\citet{margutti12}, 0.27 and 0.32, respectively. 
 One reason why we found a
slightly steeper decay slope during the plateau phase can be that we accept even slopes with \axb$>$1.2
(see Figure\,\ref{distr_ax})
which is not the case in the \citet[e.g.][]{margutti12} sample.

The distribution of the `normal' decay slopes after the plateau is shown in the
lower panel of Figure\,\ref{distr_ax}. Note again that the distribution of \axc\
is quite broad and  a significant number of bursts have decay slopes that exceed
\axc$>$2.0 which is typically assumed to be a decay slope after a jet break,  particularly for
short-duration GRBs. 
 The mean and standard deviation of all GRBs
of \axc\ is 1.55 and 0.73 and the median value is 1.38, suggesting a non-Gaussian 
distribution. The median decay slope during the 'normal' decay phase for long GRBs is 
1.37. Note that there are three outliers in the \axc\ distribution, which are all short GRBs 
(GRBs 051210, 120305A, and 120521A). All these decay slopes are extreme of the order of 7 or 8.
With the median values as discussed above and
listed in Table\,\ref{grb_statistics} we can construct  a 
median GRB  X-ray
afterglow light curve. Figure\,\ref{median_xrt_lc} displays the observed
 median 0.3-10 keV X-ray
flux and the 
rest-frame luminosity light curves (left and right panel, respectively). The light curves 
were constructed by using the median 
values for the break times before and after the plateau phase and the decay slope \axb\ 
and \axc. We used the median values 
for the fluxes and luminosities at the break time after the plateau phase 
($F(T_{\rm break,2})=5\times 10^{-12}$ erg s$^{-1}$ cm$^{-2}$
and $L(T_{\rm break,2})=1.7\times 10^{47}$ erg s$^{-1}$) as the normalization points 
of the light curve.

The redshift distribution of the 232 \swift-detected bursts with spectroscopic 
redshift measurements is shown in Figure\,\ref{distr_z}. The mean redshift of 
all \swift\ bursts 
is z=2.03 and the median z=1.76
which is lower than for the bursts detected during the first two years of \swift\ 
operation \citep[z=2.30; ][]{grupe07}. Recently \citet{coward12} have studied the 
evolution  of the 
mean redshifts of \swift-detected GRBs since 2005 and shown that the mean redshift of
\swift-discovered bursts has decreased over the course of the mission.  
There is a significant difference in the redshift
distributions of short and long GRBs, with short GRBs being detected only at the lower end of
the redhshift distribution. This is a selection effect.
As shown by e.g. \citet{bernardini12} and \citet{margutti12}, short
bursts are less energetic than long bursts (see also below) which results
 in a lower fluence.
 The cause of the different energetics of short and long GRBs is most
likely the result of their different progenitors 
\citep[e.g.][]{salvaterra08, fong13, Rowlinson+13magnetar} which then results in different redshift
distributions. 
As a consequence, short bursts which have lower energy appear to be 
undetectable at high redshifts.

\subsection{Selection Effects \label{selection}}
\begin{figure}
\epsscale{0.75}
\plotone{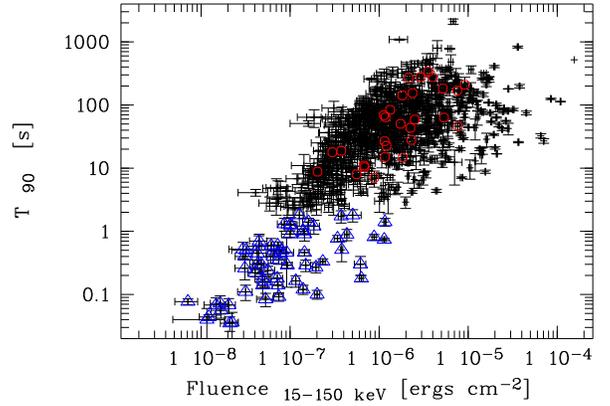}
\caption{\label{fluence_t90} Fluence in the 15-150 keV \swift\ BAT band and $T_{90}$
}
\end{figure}

Before we start any discussion on   results of the correlation analysis from our sample 
we need to be aware of any 
selection biases in the sample that may lead to non-physical results. 
One of these selection biases is displayed in Figure\,\ref{fluence_t90} which shows the
observed 15-150 keV fluence in the \swift\ BAT and $T_{90}$ which seem to be very strongly correlated
($r_s$=0.667, $T_s$=22.85). However, this is purely a selection effect driven by the detector
properties. We only detect low-fluence bursts when their energy is released in a relatively short 
amount of time. 
If the energy is spread over a longer time span
 the signal will be dominated by noise and the 
BAT will not be able to trigger on this event. This selection effect means that low fluence bursts 
have to be short in order to be detected,
explaining the upper left part of the diagram. 
On the other hand the void at the lower right part of
Figure\,\ref{fluence_t90} may be of physical nature. There are two effects here: 
\begin{enumerate}
\item a highly energetic burst 
(and high fluence burst) requires a certain amount of time to release the energy and 
\item related to this, short bursts
generally speaking have lower
luminosities/energies than long GRBs \citep[e.g][]{margutti12}. 
\end{enumerate}

1) The reason why high fluence bursts are seen with longer $T_{90}$ may be
that there is only some maximum amount of flux (or
luminosity) that can be generated by the burst. Therefore a high fluence burst would
 need a longer time span to release
its energy than a low fluence burst.  

2) In order to detect  a short burst
with a high fluence requires that this burst occurs at close distances. For example the short
GRB with the highest fluence ($1.16\times 10^{-6}$ erg cm$^{-2}$)
is GRB 051221A \citep{burrows06} with a redshift of z=0.547,
resulting in a luminosity distance of 3150 Mpc. To detect this GRB with a fluence of the
order of $10^{-4}$ erg cm$^{-2}$ it needs to occur within 390 Mpc  equivalent
to a redshift of z=0.085. Although this is not impossible and \swift\ has detected a few bursts
with redshifts lower than 0.085 (4 to be precise) the probability of detecting a relatively
luminous short burst like GRB 051212A at such a low distance is low. This is mostly due to the
the general space density of GRBs  and their luminosity functions (see also
Section\,\ref{lum_func})
and the much lower comoving volume at a redshift of 0.085
than at higher cosmological redshifts. Having a less luminous burst explode and detect it with
a fluence in the $10^{-4}$ range requires an even closer distance. 

How does the $T_{90}$ - fluence selection effect affect other correlations? We found a mild
correlation between $T_{90}$ and the isotropic energy in the 15-150 keV BAT band. However,
this correlation may mostly been driven by this selection effect. 

Another question that may raises concerns is whether $T_{90}$ really is a good parameter to
describe a burst. As pointed out by e.g. \citet{zhang12} and \citet{qin13}, $T_{90}$ is strongly
detector dependent. This is  similar to a hardness ratio which does not
allow a direct comparison between different missions. However, as long as
we use both parameters from samples using only data from the same detector
it may not be a problem. Still, for \swift-BAT GRBs we typically use the 2s
division line between short and long duration GRBs as it was defined by
\citet{kouveliotou93} for Gamma-Ray Observatory BATSE-detected GRBs which was
operating at much higher energies than the \swift\ BAT. 
However, as we have pointed out in Section\,\ref{distribution} the division between 
\swift\ BAT-detetced short and long duration GRBs occurs at earlier times.
It has been recently shown also by \citet{bromberg12} that the cutoff line at 2s 
between short and long duration GRBs detected by the \swift-BAT is not a good choice
and it is more appropriate for \swift\ discovered bursts to have the cutoff at around 0.8s.
We will discuss this in more detail in Section\,\ref{short-long}.

\begin{figure*}
%\epsscale{0.75}
\epsscale{1.5}
\plottwo{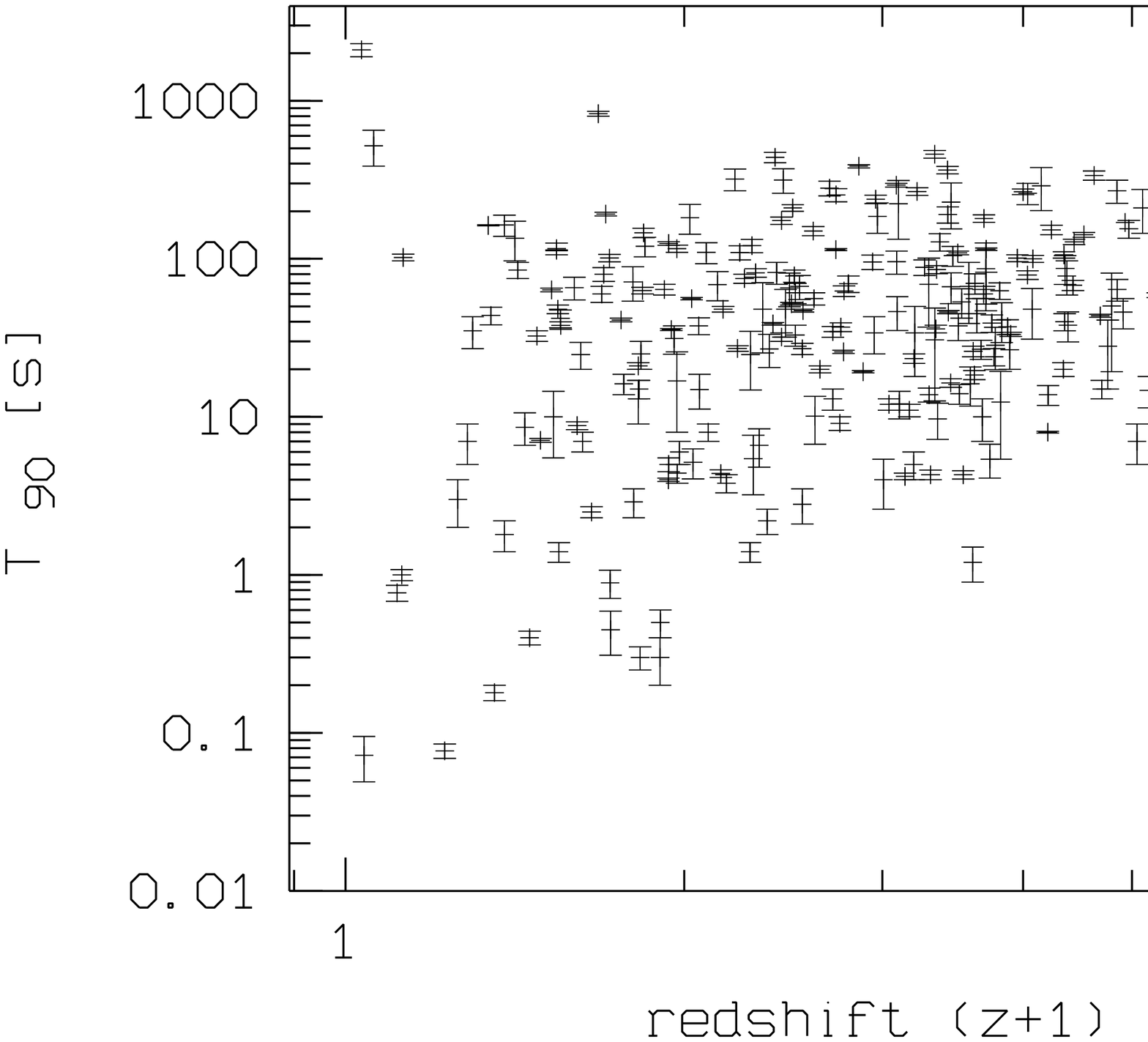}{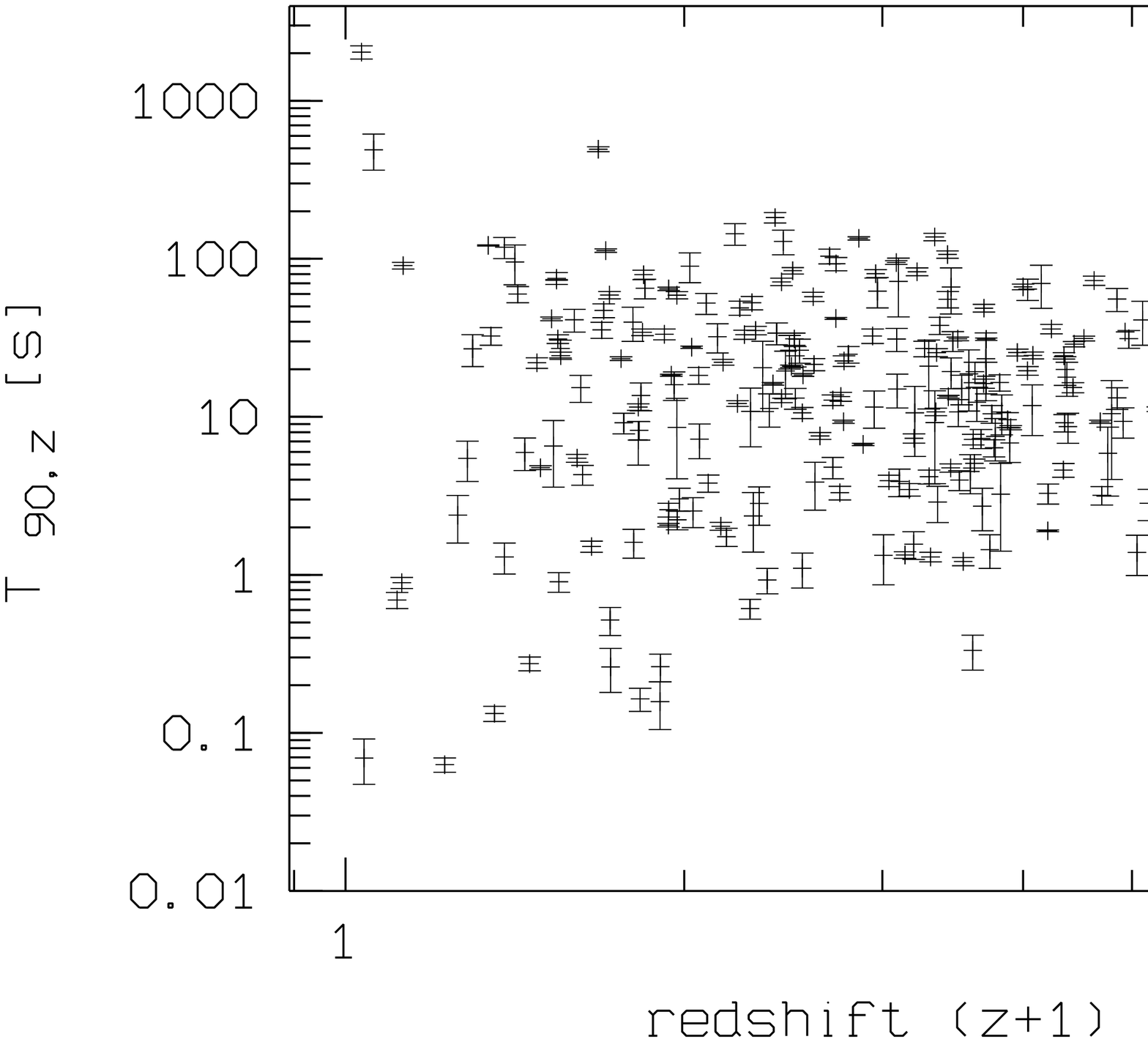}
\caption{\label{z_t90} 
Relation between redshift and the observed and rest-frame BAT 15-150 keV
$T_{90}$.
}
\end{figure*}

There
is another problem that has recently been pointed out by several authors:
high-redshift bursts tend to have rather short $T_{90}$ \citep{littlejohns13, tanvir12, kovecski12}.
Bursts like GRB 090423 or GRB 080913
\citep[e.g.][]{zhang09}
which are
the GRBs with the highest spectroscopically measured redshifts (8.2 and 6.7),
had $T_{90}$ of 8.0 and 10.3s, respectively, which in the rest-frame suggests
that these maybe short bursts. However, the reason for these short $T_{90}$ may
not be a physical property of the bursts, rather a consequence
of the detector threshold. Because the flux light curve of the prompt emission 
observed from a high-redshift burst appears to be fainter than that of a
low-redshift burst on average, a detector triggers on the prompt emission later
than  for a low-redshift burst because the main part of the prompt emission
light curve is below the detector threshold \citep{littlejohns13}.
We basically observe the tip of the iceberg of the prompt emission of high-redshift 
bursts where most of the prompt emission is lost in the noise. 
Nevertheless, the short duration of the observed prompt emission is not a
general property of high-redshift bursts (E.g. GRB 050904 which is a burst at
the redshift of z=6.2 had a $T_{90}$ =181.7s).
The concern is how much does this threshold effect influence the results we
found for our \swift\ GRB sample? If there is a significant effect on the sample
we would expect to see a decrease of $T_{90}$ with increasing redshift. This
relation is plotted for the observed and rest-frame $T_{90}$ in
Figure\,\ref{z_t90}. However, we do not observe  such an effect. Therefore we 
conclude that although some high-redshift bursts appear to have rather short
duration prompt emission due to the detector threshold this effect is not
significant for the whole sample.  
   
Nevertheless there is another concern regarding high-redshift bursts
\citep{coward12}  
As we mentioned at the beginning of this subsection, the fluence and $T_{90}$ 
are strongly correlated because of how the BAT triggers on bursts. 
Although the mean and median fluences and $T_{90}$ of all GRBs and high-redshift bursts 
are essentially the same (Table\ref{grb_statistics})
this picture changes when looking at the rest-frame parameters:
We do not find very long $T_{90}$ GRBs at high redshifts and high-redshift bursts are significantly  
more luminous and energetic than compared with the total GRB sample. The same selection bias
that there are long GRBs with low fluence also excludes long high-redshift GRBs from detection. 
The problem here is that due to time-dilation ($T^{'} = (1+z)\times T$) a rest-frame long burst 
becomes even dramatically longer in the observed frame.
This means that the fluence of these bursts will be smeared out over a longer time span with the consequence 
that due to the detector noise the burst remains undetected. 
 In other words, in order to detect a GRB with 
the \swift\ BAT at high redshifts, its rest-frame $T_{90}$ needs to be
relatively short and it has to be a highly energetic burst. Consequently this means that the majority of 
high-redshift bursts are missed by the BAT. As pointed out by
\citet{wanderman10} this prediction, however, also strongly depends on the star formation rate at high redshifts.

\subsection{Correlation Analysis \label{correlation}}

Throughout this subsection we look at correlations between observed and 
rest-frame GRB prompt and afterglow emission parameters. In particular we are interested in strong relations
between prompt and afterglow properties. One of the goals is to be able to make predictions of 
the behavior of the X-ray afterglow based on prompt emission properties. 
The correlations between these 
parameters are listed in Tables\,\ref{correlation_tab_all} and \ref{correlation_tab_z}
using Spearman rank order correlation
coefficients and Student's T-tests. 
While Table\,\ref{correlation_tab_all} lists the correlation results 
of the observed parameters 
of all 754 GRBs in our sample,
Table\,\ref{correlation_tab_z} lists the correlations between the rest-frame parameters of 232 GRBs 
with spectroscopic redshifts only.
 In this subsection we will 
only discuss those correlations which are statistically significant,
meaning that the probability of the correlation being just random is $P<10^{-3}$. 
 Correlations quoted in this subsection apply to the whole GRB sample if not noted otherwise.
We will primarily focus on 
those relations which clearly connect the prompt and the afterglow phase. Other relations are
shown in  Appendix\,\ref{appendix_corr}. In addition to the Spearman rank order correlation coefficient,
we also determined Kendall's $\tau$ value and the Pearson correlation coefficient. 
In Tables\,\ref{correlation_kendall_all} and \ref{correlation_tab_z_kendall} in Appendix\,\ref{appendix_corr}
the 
Kendall $\tau$ value and the
Pearson correlation coefficients for all GRBs and  those GRBs with spectroscopic redshift 
measurements are shown, respectively.

We have already mentioned in section\,\ref{distribution} that the photon spectral index in 
the 15-150 keV BAT energy band 
is flatter for short-duration GRBs than for long-duration GRBs (Figure\,\ref{distr_gamma_bx}).
Figure\,\ref{gamma_t90} 
shows the relation between 
$T_{90}$ in the observed and rest-frame and the 15-150 keV photon index $\Gamma$. 
In the observed frame clearly there are 
two groups of GRBs. This effect, however, becomes smeared out in the rest-frame. 
The reason that the percentage of
short-duration GRBs with spectroscopic redshift is significantly lower than in 
long-duration GRBs, is simply because they are much
more difficult to follow up due to their lower flux/fluence and faster decay slopes compared
with long-duration GRBs.

\begin{figure*}
%\epsscale{0.75}
\epsscale{1.5}
\plottwo{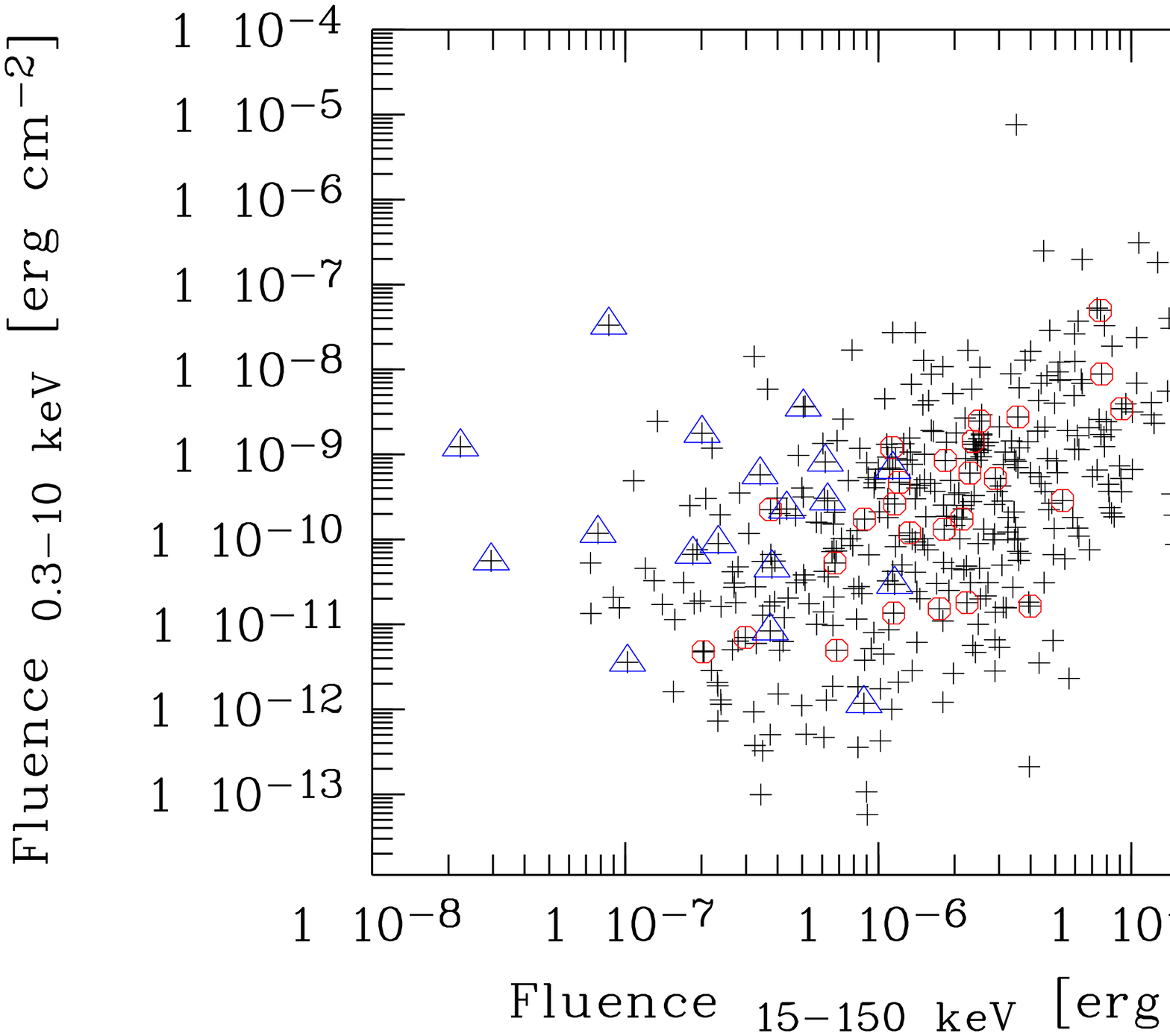}{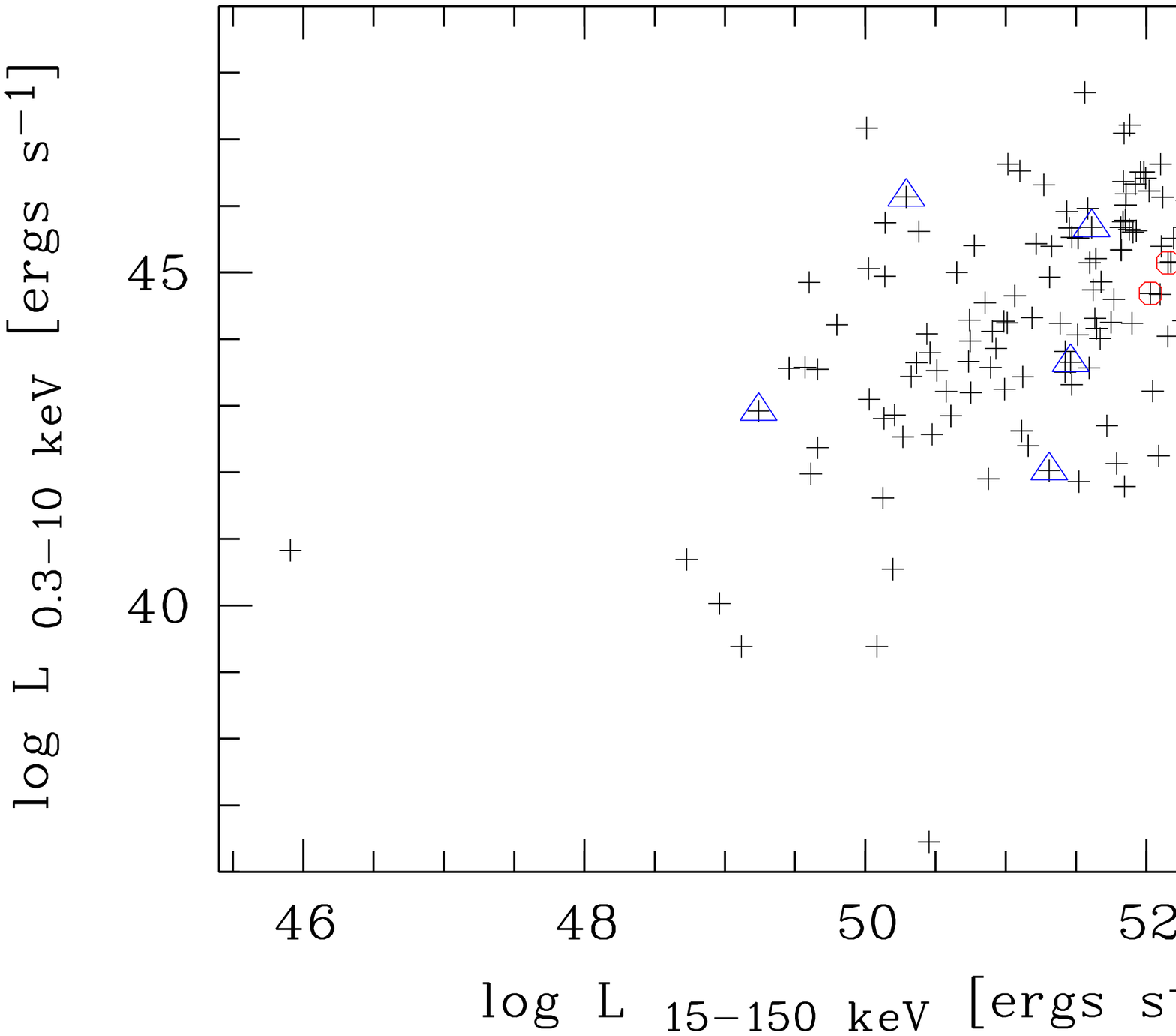}
\caption{\label{lum_lum} 
Correlations between the fluences and the luminosities in the 15-150 keV BAT band and the 0.3-10 keV XRT 
energy band (left and right panel, respectively). Short bursts are marked as blue triangles and high redshift bursts
(z$>3.5$) as red circles.
}
\end{figure*}

Early studies of \swift-detected GRBs  noticed a close connection of the 
energetics of the prompt and
the afterglow emission \citep[e.g. ][]{obrien06, willingale07, gehrels08} and more 
recent studies by e.g. \citet{margutti12}
confirm these findings. The left panel of
Figure\,\ref{lum_lum} shows the correlations between the fluence in the 15-150 eV  
band during the prompt emission and
the 0.3-10 keV fluence in the afterglow emission. The right panel displays the 
luminosities in the 15-150 keV band of the prompt and the
0.3-10 keV luminosity in the X-ray afterglow emission. Clearly, as expected, the 
energetics of the prompt and afterglow emission
are very strongly correlated: bursts with high 15-150 keV fluence in their prompt emission 
will have high X-ray 
luminosities/fluxes in the afterglows and vice versa.  

\begin{figure*}
%\epsscale{0.75}
\epsscale{1.5}
\plottwo{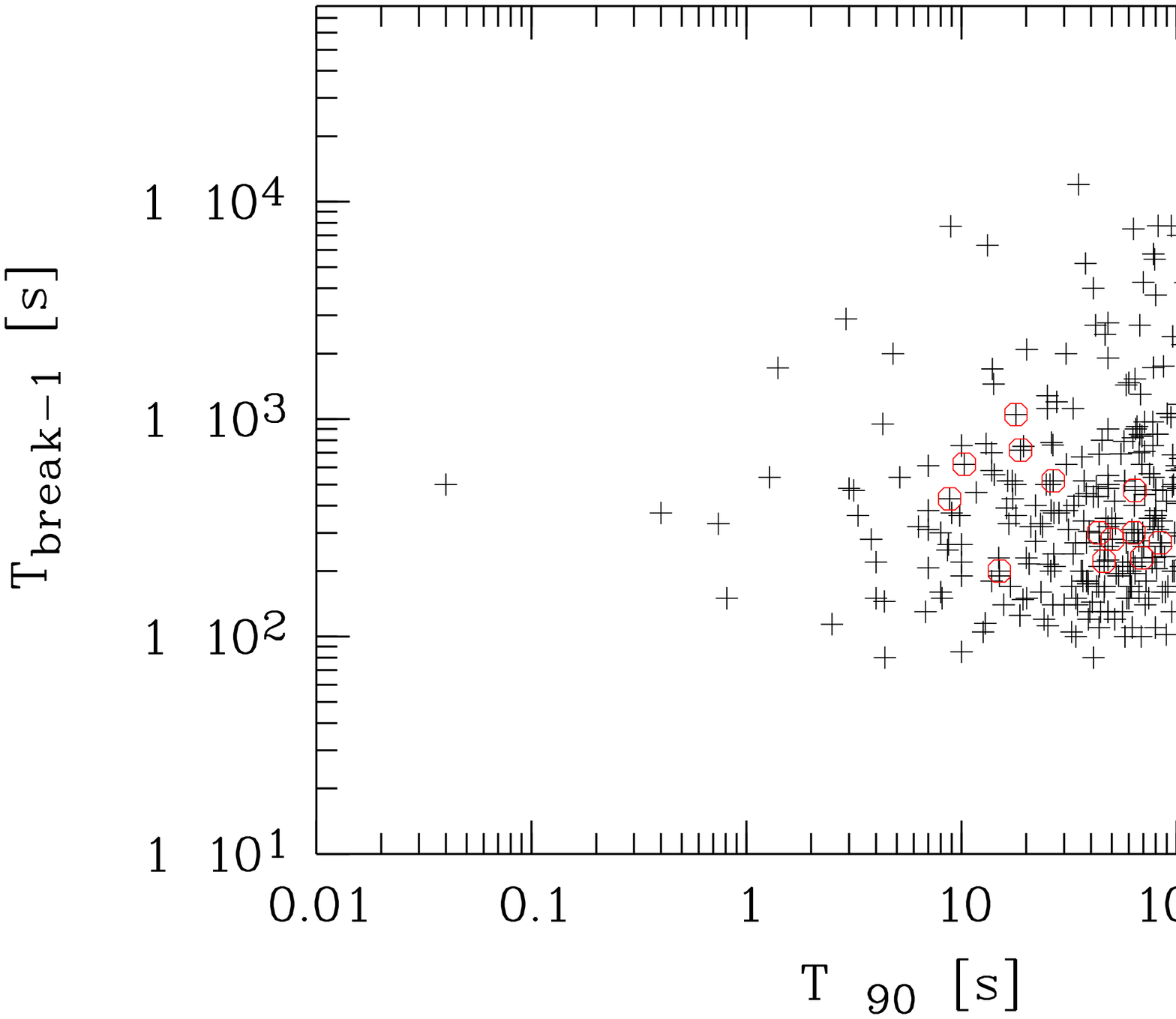}{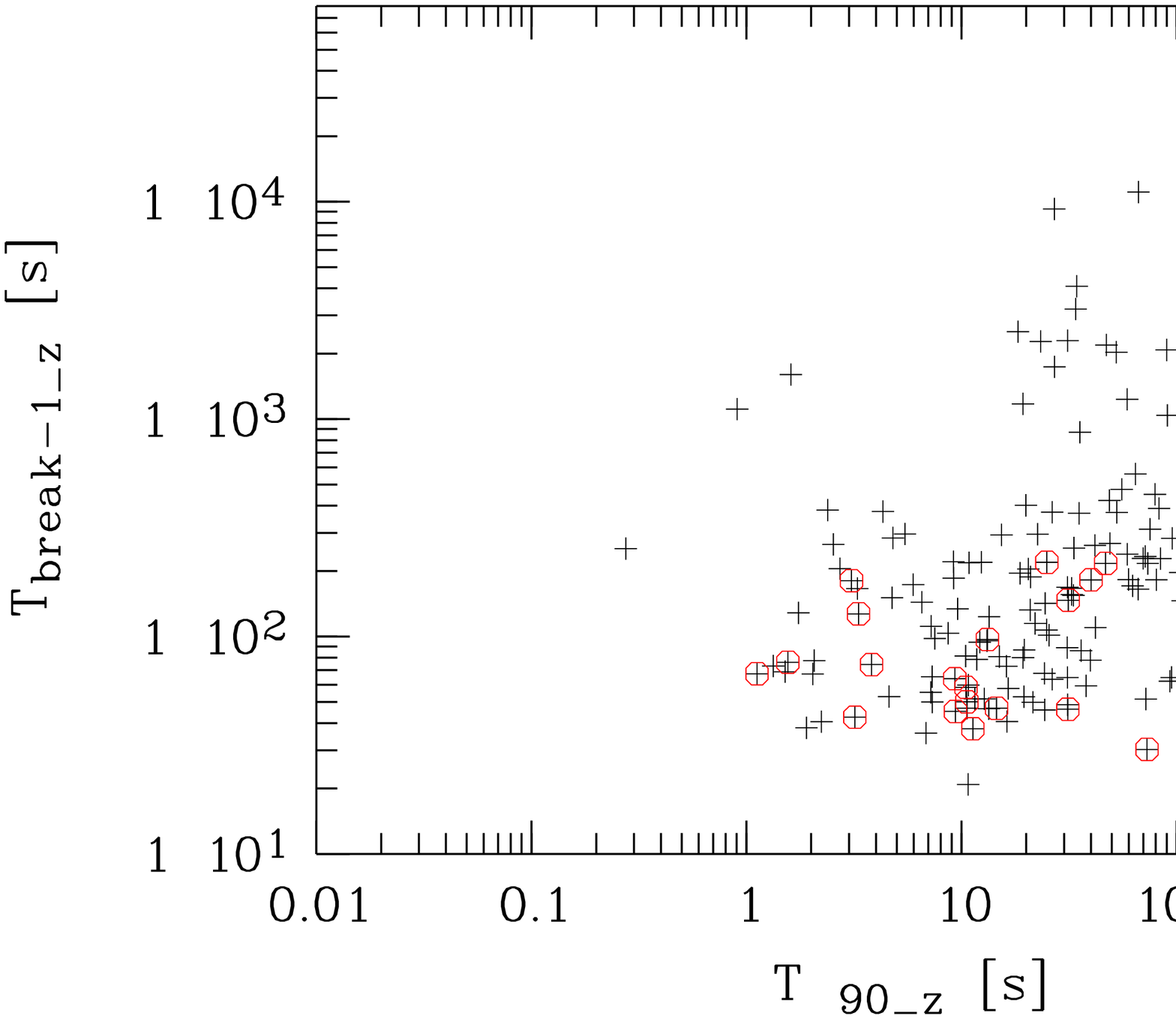}
\caption{\label{t90_tb1} 
Correlation between the observed BAT 15-150 keV $T_{90}$ and 
the break time $T_{\rm break 1}$ in the 
X-ray afterglow light curves before the plateau phase in the observed and
rest-frame (left and right panel, respectively). GRBs with a redshift z$>$3.5 are marked as
red circles.
}
\end{figure*}

\begin{figure*}
%\epsscale{0.75}
\epsscale{1.5}
\plottwo{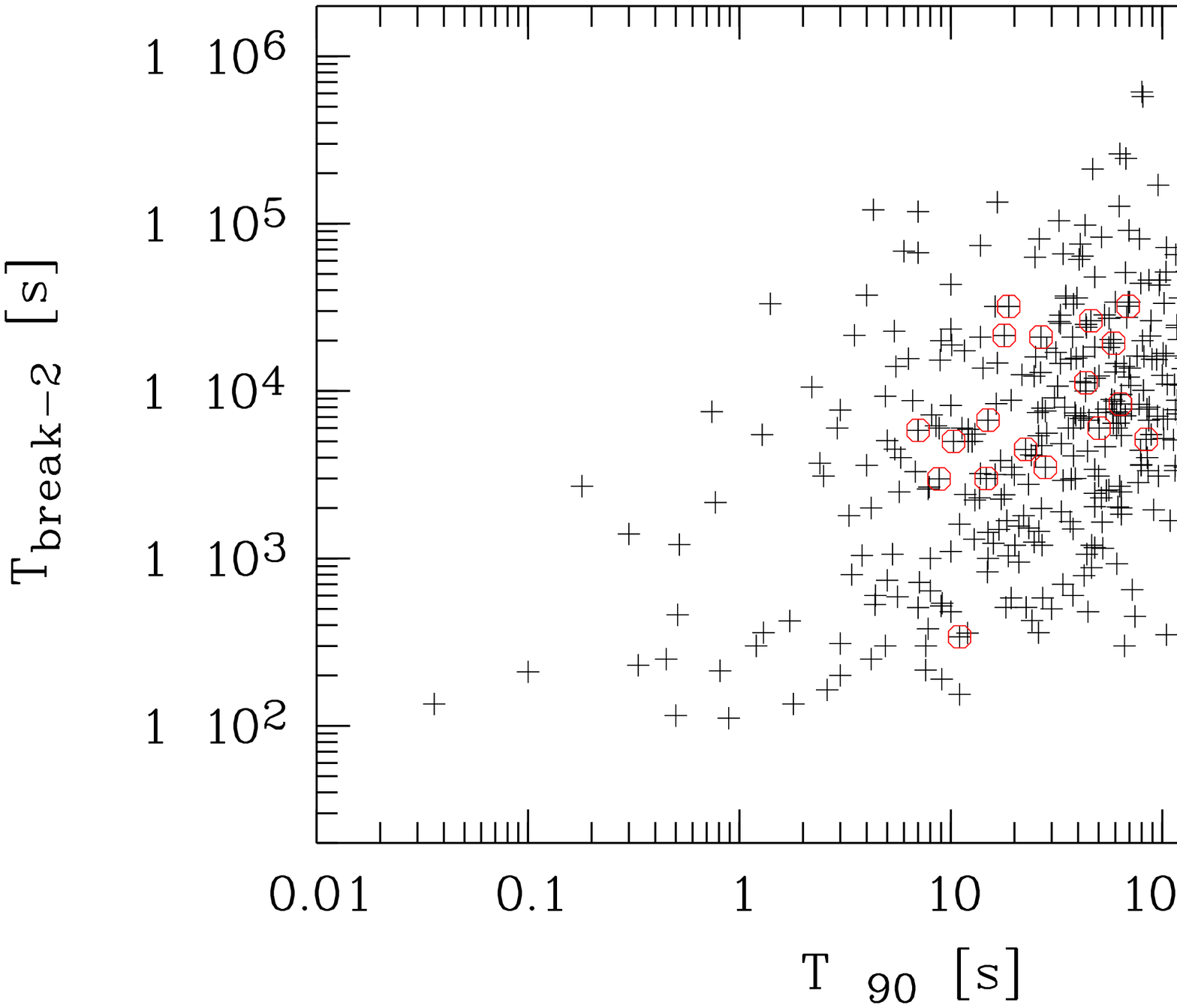}{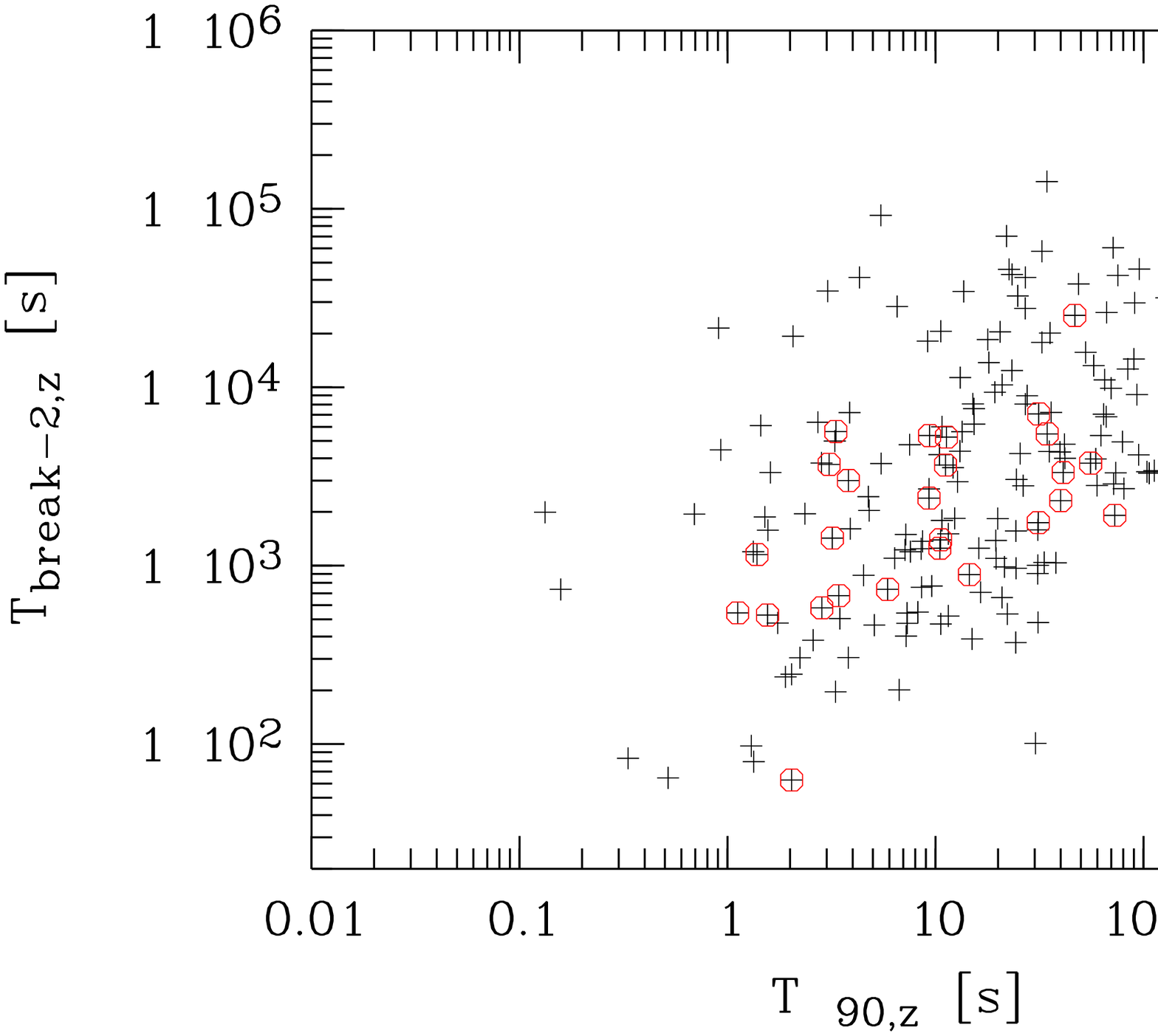}
\caption{\label{t90_tb2} 
Correlation between the observed BAT 15-150 keV $T_{90}$ and 
the break time $T_{\rm break 2}$ in the 
X-ray afterglow light curves  after the plateau phase in the observed and
rest-frame (left and right panel, respectively).
}
\end{figure*}

In our paper we go a step beyond the energetics
and ask what other prompt and afterglow properties are correlated.
One of our main results is that there are clear correlations between the 
BAT 15-150 keV $T_{90}$ and the break times in the X-ray afterglow
light curves before and after the plateau phase.
These findings apply to observed as well as to rest-frame parameters. 
 The left panel in
Figures\,\ref{t90_tb1} and \ref{t90_tb2}
show the observed values, while the right panel display the values
 in the rest-frame of the burst. Although there is a large scatter in all
these  relations, there  are clearly correlations:
GRBs with long $T_{90}$ start
their X-ray afterglow plateau phase at later times and end the plateau phase
 later than GRBs with shorter $T_{90}$. 
In correlations between $T_{90}$ and  break time before the plateau phase
$T_{\rm break, 1}$ we found a Spearman rank order correlation $r_s$=0.328,
$T_s$=6.776 (N=384 GRBs) which implies a probability of $P<10^{-8}$ of a
random result for the observed values, and $r_s$=0.371,
$T_s$=4.767 (N=144 GRBs) which  $P=4.5\times 10^{-6}$ for the values in the
rest-frame.
This relationship between the prompt and afterglow emission is especially strong
between $T_{90}$ and the break times after the plateau phase 
$T_{\rm break, 2}$.
For the observed values we found $r_s$=0.495, $T_s$=11.615 (N= 417 GRBs) with
$P<10^{-8}$, and $r_s$=0.439, $T_s$=6.421 (N= 175 GRBs) with
$P< 10^{-8}$ for the times in the rest-frame. What the relations between
$T_{90}$ and the break times in the X-ray light curves suggest is that there seems
to be a strong connection between the prompt and the afterglow emission. What these relations 
- long $T_{90}$ - later afterglow break times allow is in principle to make predictions 
of the behavior of the X-ray afterglow light curve based on prompt emission properties.
If this is true
then we would also expect to see a correlation between the spectral slopes in the
BAT 15-150 keV and the X-ray 0.3-10 keV band. 

\begin{figure}
%\epsscale{0.75}
\epsscale{0.75}
\plotone{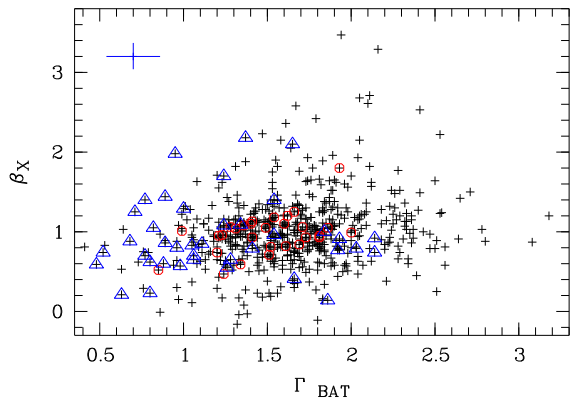}
\caption{\label{gamma_bx} 
Correlation between 15-150 keV photon spectra slope $\Gamma$ in 
the BAT and the 0.3-10 keV energy spectral index \bx. Of this plot is the
extremely soft GRB 060602B \citep{aharonian09} at $\Gamma=4.97\pm0.49$ and
\bx=3.30\plm0.91. 
The blue cross in the upper left corner displays the median uncertainty of each property.
}
\end{figure}

The correlation between the BAT 15-150 keV hard X-ray photon index $\Gamma$ 
and the X-ray energy spectral slope \bx\ is displayed in Figure\,\ref{gamma_bx}.
Again, although there is a large scatter in this relation, the two properties are
clearly correlated and is another hint that prompt and afterglow phases are clearly linked.
GRBs with steep spectral in the 15-150 keV band also show
steeper X-ray spectra. The Spearman rank order correlation coefficient is
$r_s$=0.179 with $T_s$=4.560 (632 GRBs) and a probability $P=6.32\times 10^{-6}$ of
a random result. We also checked if there is any (anti-)correlation between
spectral indices and the rest-frame break times before and after the plateau 
phase. We could not find any significant correlation between these properties 
(Table,\ref{correlation_tab_all}).
There is only a weak trend between the X-ray spectral slope and $T_{\rm break2,
z}$ that bursts with later break times have steeper X-ray spectra, but the
probability is 1.5\% that this is just a random result.

\begin{figure*}
%\epsscale{0.75}
\epsscale{1.5}
\plottwo{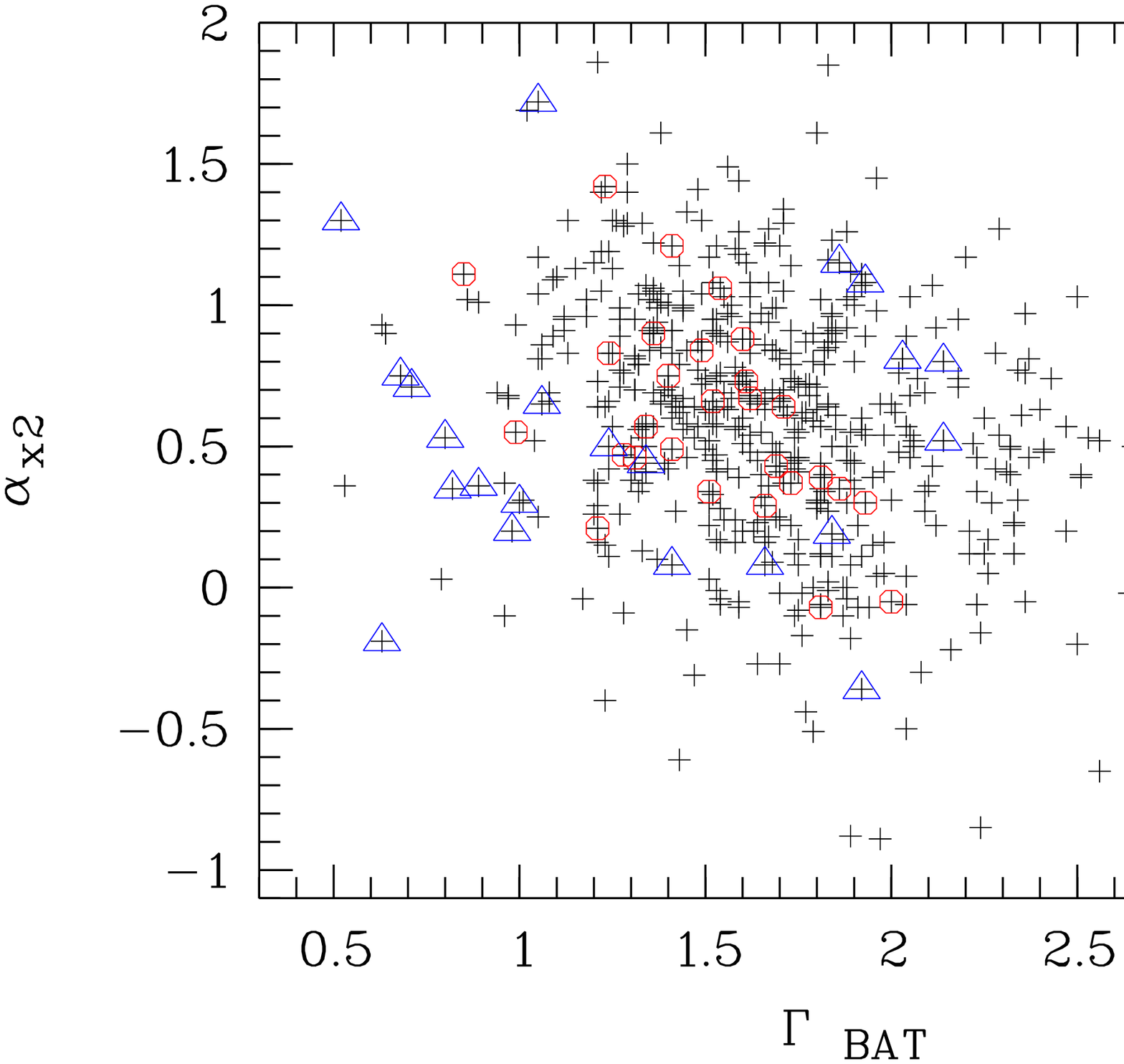}{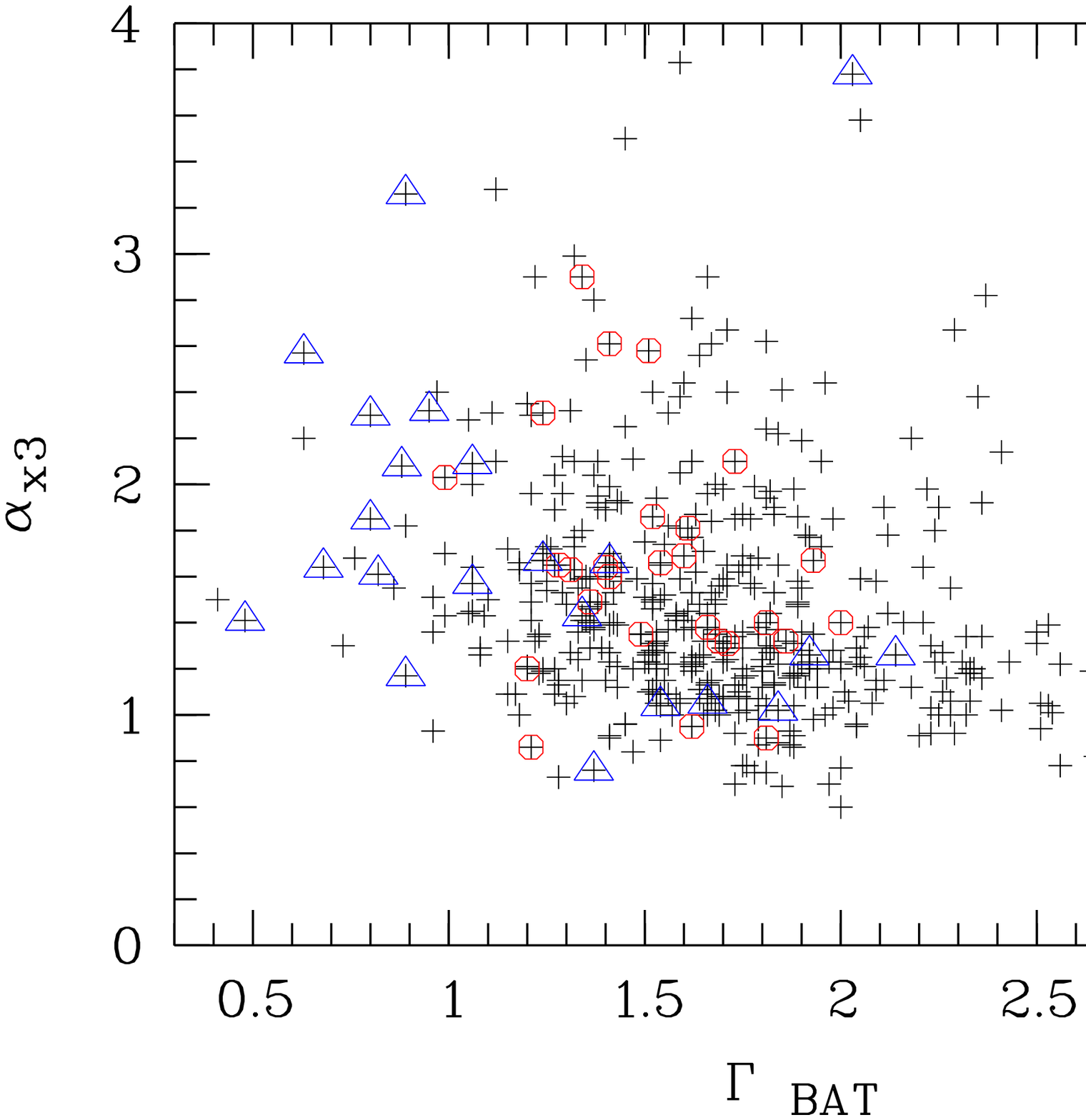}
\caption{\label{gamma_ax} 
Correlation between the  BAT 15-150 keV photon index $\Gamma_{\rm BAT}$ and the
decay slopes during the plateau phase and the `normal' decay phase.
 (left and right panel, respectively). Short duration GRBs are displayed as triangles.
The blue cross in the upper right corner displays the median uncertainty of each property.
}
\end{figure*}

The BAT 15-150 keV photon index $\Gamma$ also strongly 
anti-correlates with the decay slopes during the plateau and normal decay
phases \axb\ and \axc.
The decay slope during the plateau phase anti-correlates with $\Gamma$ with a
Spearman rank order coefficient $r_s =-0.260$ with a Student's T-test $T_s=-6.232$ and a 
probability of a random distribution of $P<10^{-8}$ (539 GRBs), 
and $r_s=-0.313$, $T_s=-7.154$ and $P<10^{-8}$ (475 GRBs) for the `normal decay slope \axc. 
These relations are displayed in Figure\,\ref{gamma_ax}. Again, these are relations clearly link the
prompt with the afterglow emission: bursts with steeper 15-150 keV spectra have flatter decay slopes during the 
plateau and normal decay phases. These relations can be used to 
 estimate the behavior of the X-ray light curve.

\begin{figure*}
%\epsscale{0.75}
\epsscale{1.5}
\plottwo{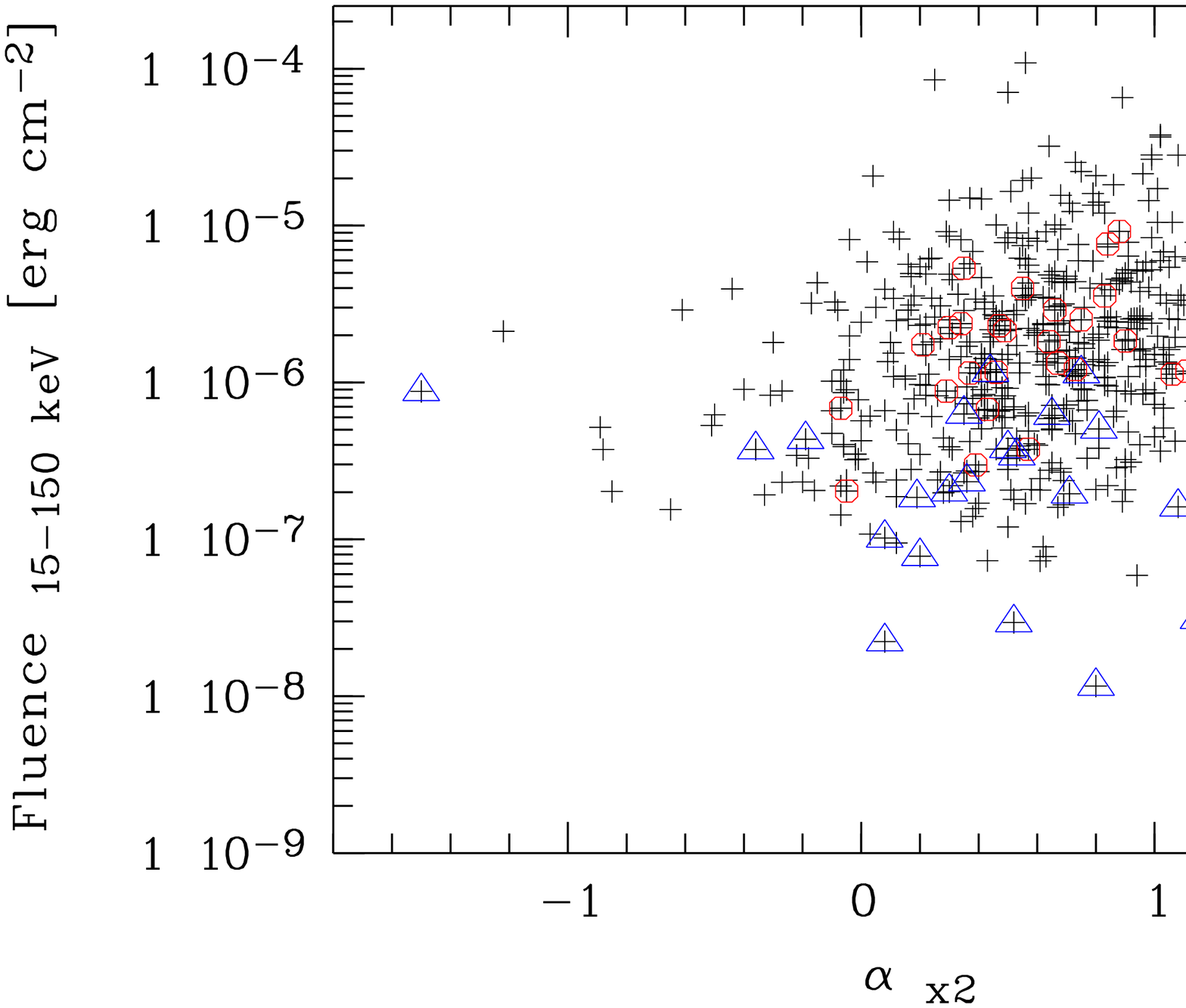}{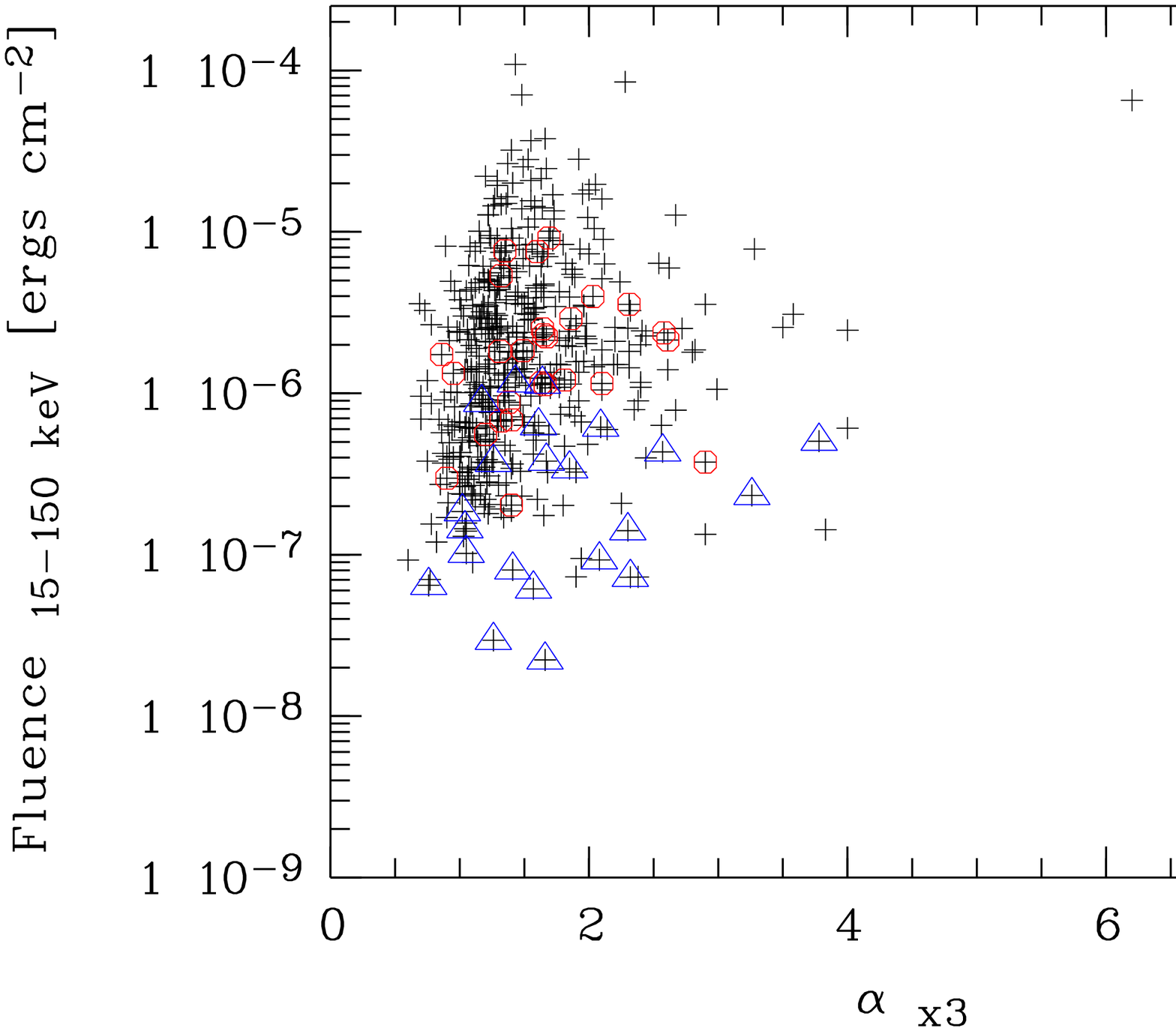}
\caption{\label{fluence_ax} 
Correlation between the  fluence in the 15-150 keV BAT and and the
decay slopes during the plateau phase and the `normal' decay phase.
 (left and right panel, respectively). Short duration GRBs are displayed as triangles.
}
\end{figure*}

We noticed strong correlations between the fluence in the 15-150 keV BAT
 energy band and the decay slopes in the
X-ray light curve \axb\, and \axc, as shown in Figure\,\ref{fluence_ax}. We find
Spearman rank order correlation coefficients of $r_s$=0.270 and 0.225,
 Student's T-test values $T_s$=6.482 (for 538 GRBs) and 
5.043 (475 GRBs) with probabilities of a random result of $P<10^{-8}$ and 
$5.6\times 10^{-7}$ for the fluence vs. \axb\ and \axc\
correlations, respectively. These relations become apparent in our \swift\ GRB sample due to the large number of bursts.
Although this is purely a phenomenological relation which is most likely due to selection effect,
 we can still take advantage of these relations
 to predict the behavior of the X-ray light curve.
Note that for the relations of luminosity and isotropic energy with \axb\ and \axc\ there are only  trends that bursts with
higher fluence in the BAT energy band decay  faster in X-rays 
(see Table\,\ref{correlation_tab_z})

\begin{figure*}
%\epsscale{0.75}
\epsscale{1.5}
\plottwo{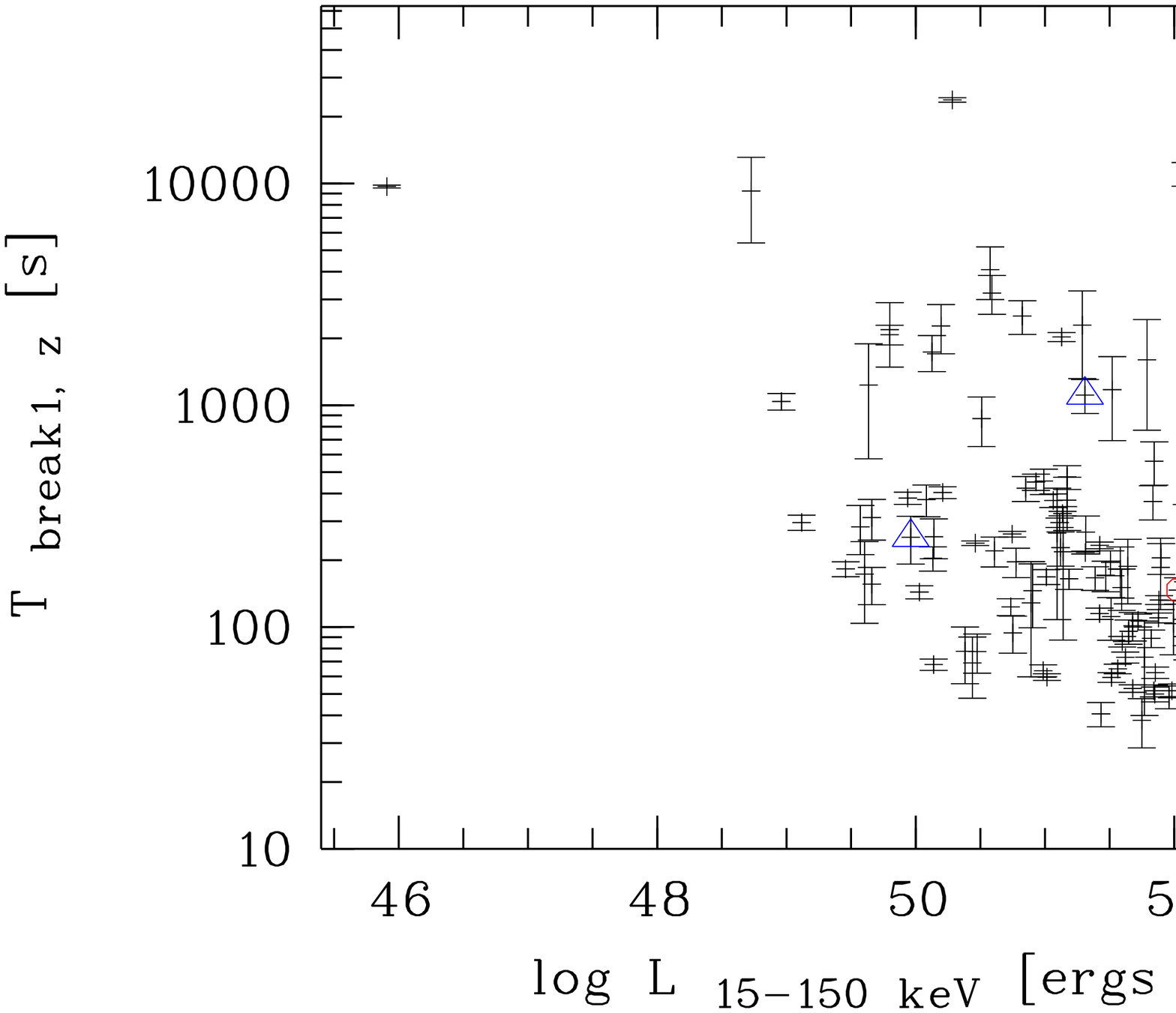}{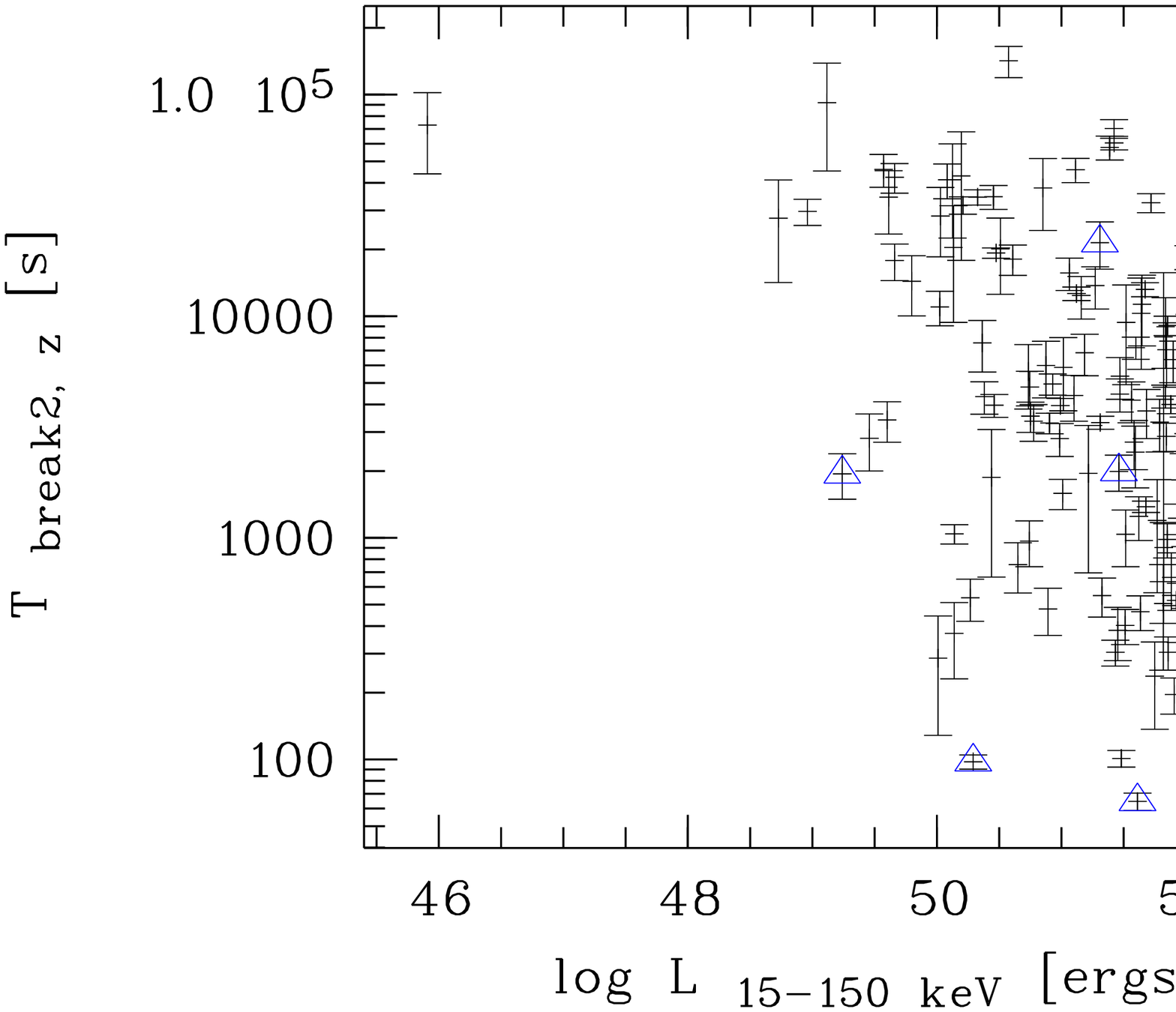}

\plottwo{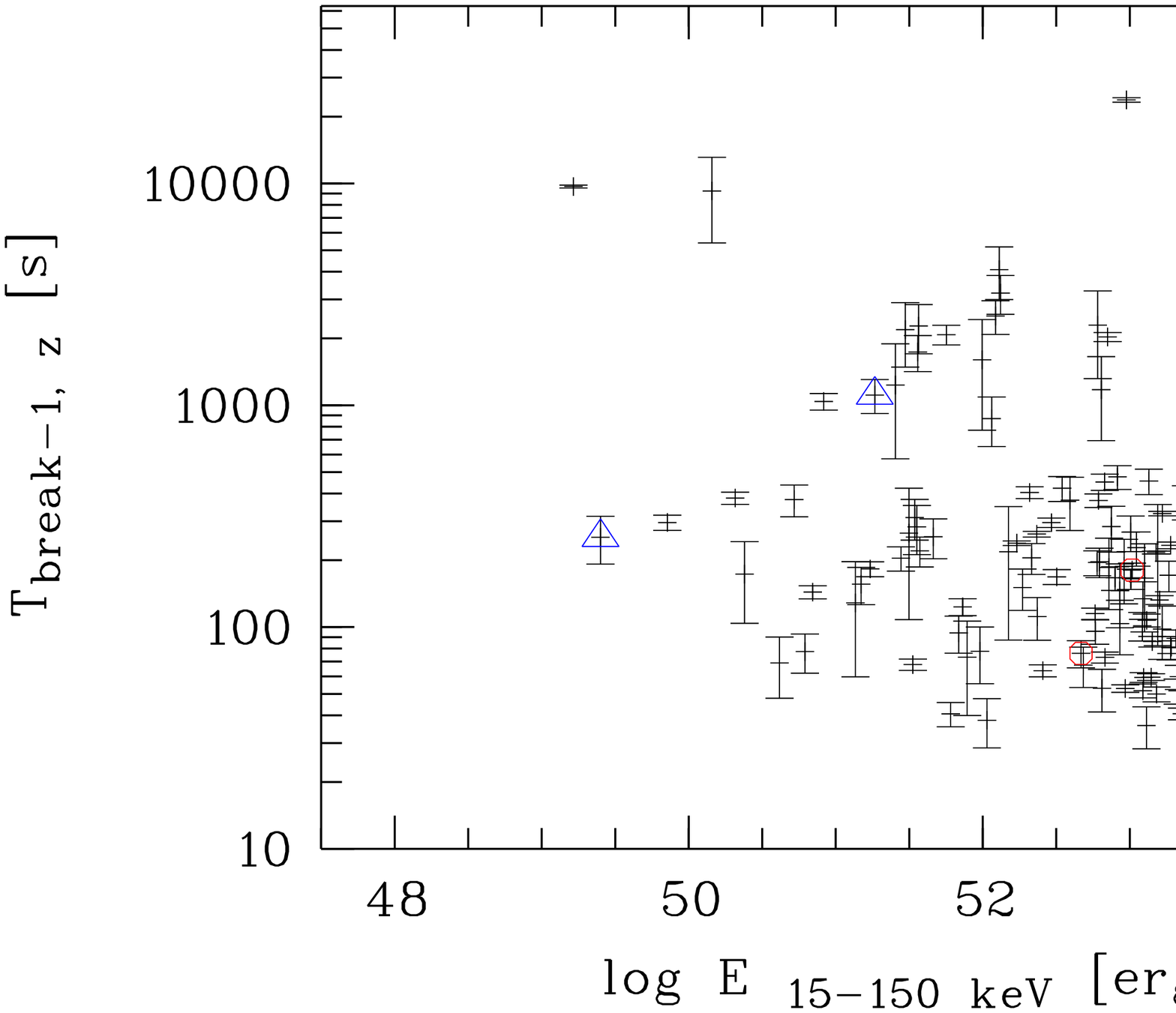}{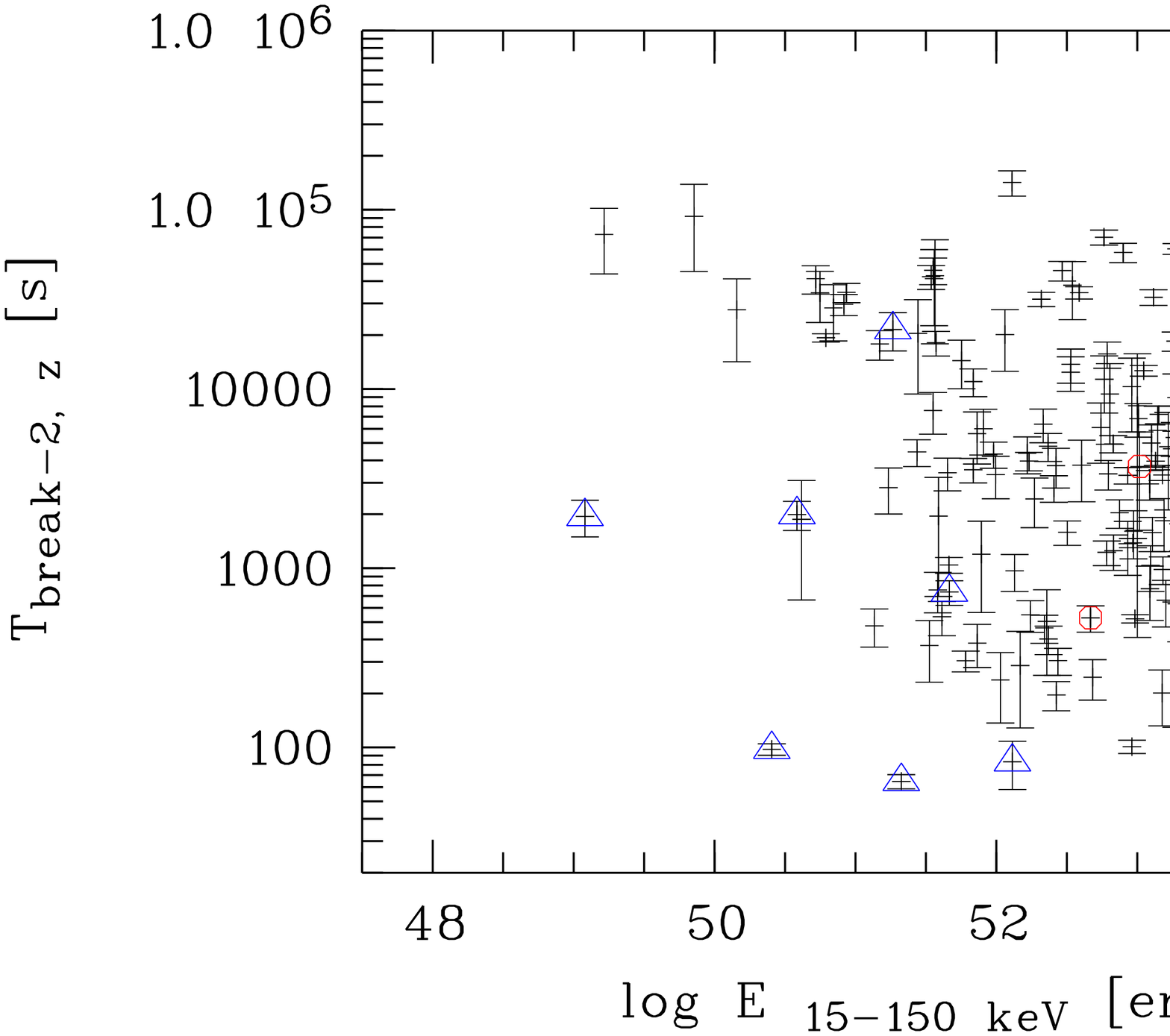}
\caption{\label{lum_tb1_tb2} 
Relation between the k-corrected BAT luminosity and the BAT photon index $\Gamma$
and the break times before and after the X-ray afterglow plateau phase
 (left and right panel, respectively). Short
bursts in these plots are marked as triangles.
}
\end{figure*}

The next step is to see if the luminosity in the prompt emission also
anti-correlates with the break times in the X-ray light curve. As shown in
Figure\,\ref{lum_tb1_tb2}, this is indeed the case for both the break times before
and after the plateau phase $T_{\rm break1, z}$ and $T_{\rm break2, z}$,
respectively. In the first case, the  Spearman correlation coefficient and Student
T-test values are $r_s=-0.587$, $T_s=-8.614$, with $P<10^{-8}$ (143 GRBs),
and for the break time after the plateau phase
$r_s=-0.413$, $T_s=-5.956$ with $P=1\times 10^{-8}$ for all GRBs (175 GRBs) and 
$r_s=-0.433$, $T_s=-6.187$ with $P< 10^{-8}$ for the long GRBs (168 GRBs)
for all long GRBs. 
 This
is a real anti-correlation between the prompt luminosity and the afterglow emission light curve break times.
 This is not necessarily 
true, however, for the connection found by e.g. \citet{dainotti08} between the break
time after the plateau phase $T_{\rm break2,z}$ and the luminosity at $T_{\rm
break2, z}$, which are strongly anti-correlated.  
The problem is that $T_{\rm break-2}$ and $L_{\rm 0.3-10 keV}(T_{\rm break-2})$ are not 
independent parameters. 
Because the decay slope of the `plateau' phase in not 0\footnote{In a matter of fact,
the median decay slope during the plateau phase is 0.6.}, a later break time will
automatically result in a lower flux/luminosity. 
 Although there
is a strong correlation between $T_{\rm break2,z}$ and the luminosity at $T\_{\rm break2, z}$ 
as reported by \citet{dainotti08} also in our sample ($r_{\rm s}=-0.559, T_{\rm s}=-8.785, P<10^{-8}$) 
there is also a strong
anti-correlation with the flux at that time. 
As a consequence, any possible physical cause for this relation can not be disentangled from the 
fact that at a later time a burst will automatically appear to be fainter than at earlier times.

Note that the anti-correlations
between prompt emission luminosity and isotropic energy and break times in the X-ray
afterglow light curve can be used as a diagnostic to determine if a redshift measured 
of a galaxy in the direction of the burst is actually associated with the burst or
just a random galaxy in the line of sight. A good example here is GRB 051109B \citep{tagliaferri05}.
This burst has an observed $T_{90}=15\pm1$s and a fluence of $(2.7\pm0.4)\times 10^{-7}$ erg cm$^{-2}$
\citep{hullinger05} and break times in the X-ray light curve at $T_{\rm break 1}$=200s 
and $T_{\rm break 2}$=1430s. \citet{perley05} reported of a galaxy in the direction of this burst and measured 
a redshift of z=0.080. Is this galaxy associated with the burst or not? 
The answer is that this association is most unlikely. 
At the redshift of the galaxy
the luminosity distance is $D_{\rm L}$=360 Mpc which results in a 15-150 keV luminosity $L_{\rm 15-150 keV}=3\times 10^{47}$ erg s$^{-1}$ cm$^{-2}$
and an isotropic energy $E_{\rm iso}=4\times 10^{48}$ erg. These values are far off the relations shown in Figures\,\ref{lum_t90}
and \ref{lum_tb1_tb2}. We can therefore conclude that the galaxy found by \citet{perley05} at a redshift of z=0.08
 is not associated with GRB 051109B and just a random foreground galaxy. 
As pointed out by \citet{campisi08} with increasing magnitude limits of newer
telescopes the chance coincident of a GRB with a foreground galaxy at redshifts z$<$1.5
is of the order of a several percent.

\subsection{Principal Component Analysis \label{pca}}

So far we have only looked at bivariate correlations and we have found strong correlations between GRB
prompt and afterglow properties.
One step further is the statistical
analysis in a multi-dimensional parameter space. The goal here is to search for any underlying
fundamental property that is driving these relation.
One of the standard tools in multivariate analysis that may answer this question
is the Principal Component Analysis 
\citep[PCA; ][]{pearson1901}.
The idea of a PCA is to reduce the number of
significant sample parameters to a small number of parameters that capture
most of the variance in the data. For example, in AGN, the measured parameters are primarily 
driven by the mass of the 
central black hole and the Eddington ratio $L/L_{\rm Edd}$ \citep[e.g. ][]{grupe04, boroson02}.
In a mathematical sense, the PCA  searches for the
eigenvalues and eigenvectors in a correlation coefficient matrix.
A good description for the application of a PCA in astronomy
can be found in \citet{francis99} and \citet{boroson92}.

We applied a PCA to the  bursts in
our sample for which the following input parameters were available:
log $T_{90, z}$, log $T_{\rm break2, z}$, \axb, \bx, $\Gamma$, and the rest-frame 15-150 keV 
luminosity (175 GRBs in total). The reason why we do not include all properties as
listed in Table\,\ref{correlation_tab_z} is because some of these have obvious
correlations, such as $T_{\rm break,1}$ and $T_{\rm break,2}$. We want the input
properties as independent as possible. 
We applied the PCA in the statistical package {\bf R} \citep[e.g.][]{crawley07, rteam09}.
All input parameters were normalized by the standard normalization $x_{\rm norm} = \frac{x-x_{\rm mean}}{sd(x)}$ where 
$sd(x)$ is the standard deviation of the parameter x.
The results of this PCA are summarized in Table\,\ref{pca_results}. 
 The first two Eigenvectors from the PCA 
account for 60\% of the variance of the sample. The most dominant of these, Eigenvector 1,
accounts for almost 40\% of the variance. It strongly correlates with the 15-150 keV rest-frame luminosity 
and anti-correlates with all other parameters, except for the decay slope during the plateau phase, \axb.
This may suggest that Eigenvector 1 in our sample presents the 15-150 keV luminosity. In order to test this
hypothesis we excluded the 15-150 keV luminosity from the input parameters and ran another PCA on the sample.
The results of this PCA are listed in Table\,\ref{pca_results2}. This analysis agrees with the first PCA that included 
the 15-150 keV luminosity.

\begin{figure*}
%\epsscale{0.75}
\epsscale{1.5}
\plotthree{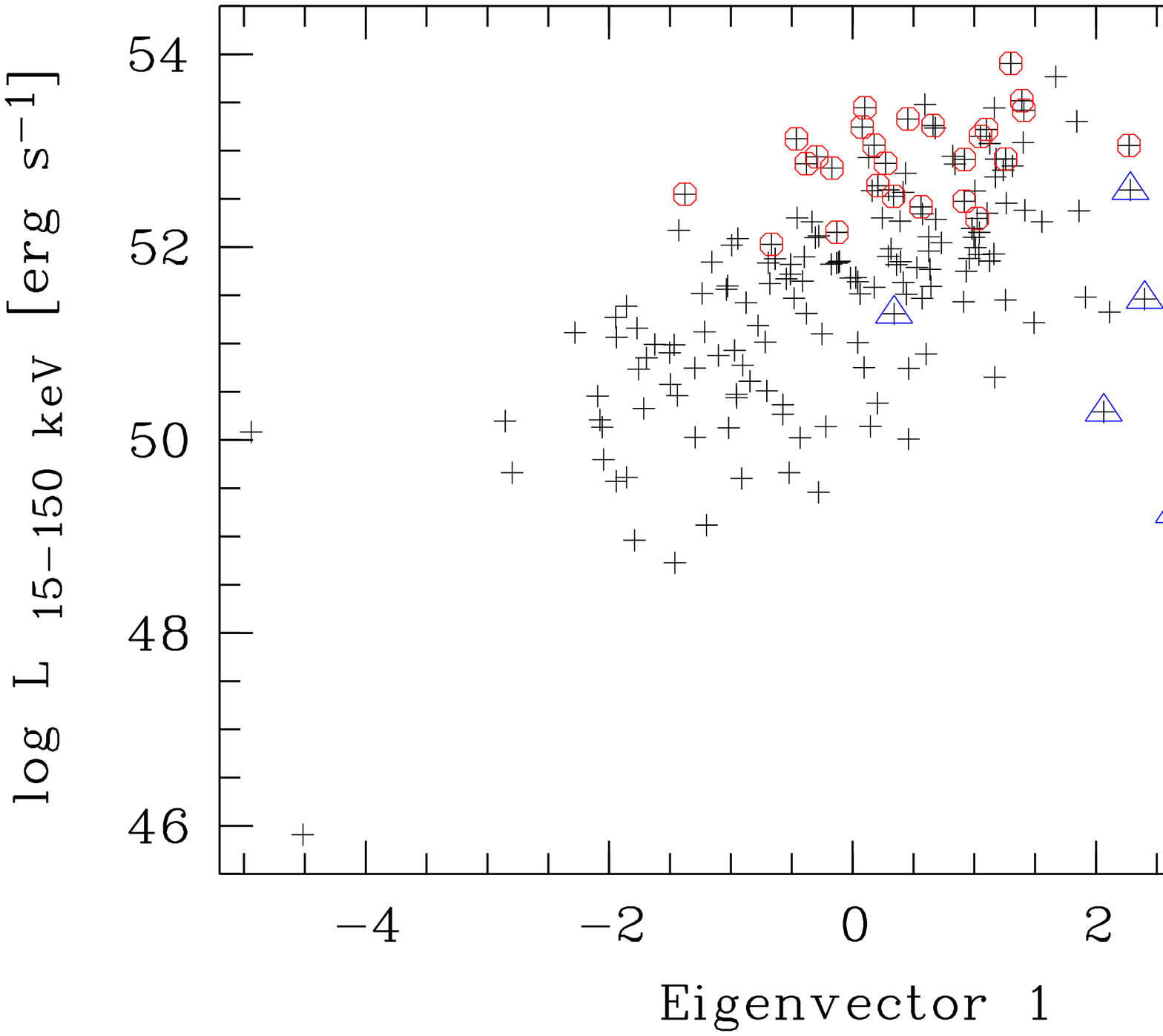}{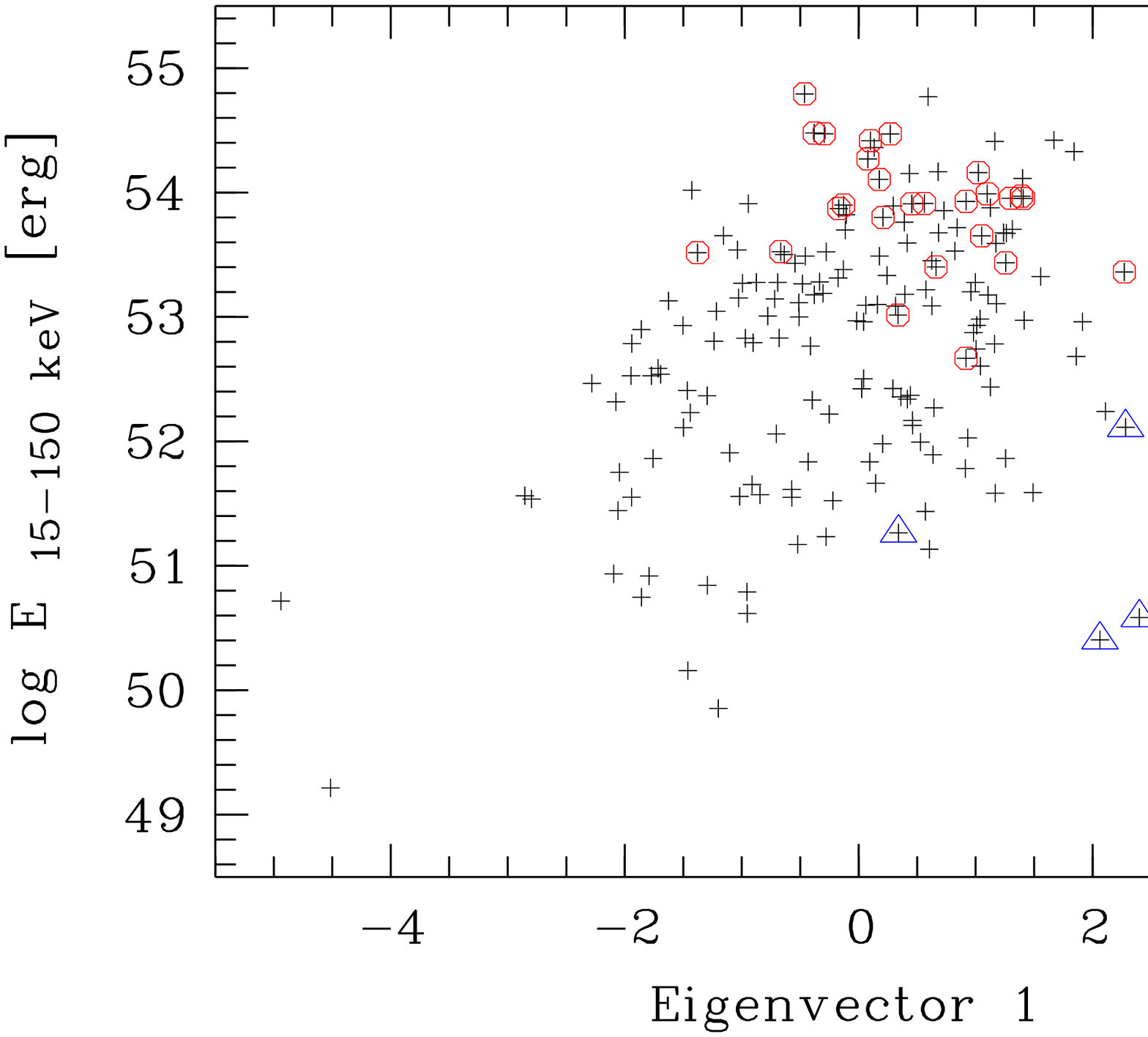}{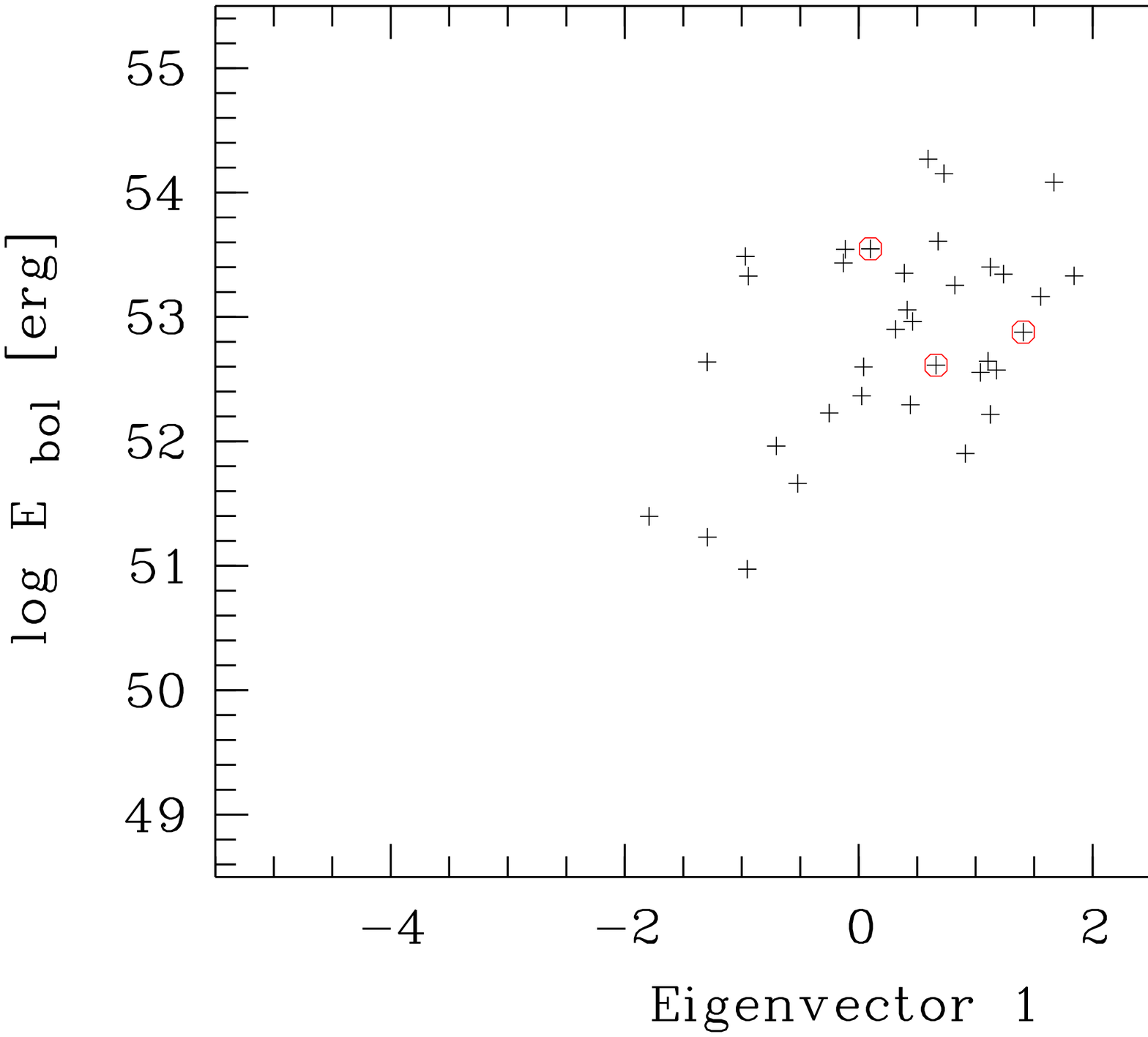}
\caption{\label{ev1_lum15_150} 
Eigenvector 1 vs. rest-frame 15-150 keV luminosity and isotropic energy in the 15-150
 keV BAT band in our GRB sample
(left and middle panel respectively). Eigenvector 1 was determined based on the 5 parameters used in the PCA listed in
Table\,\ref{pca_results2}. The right panel displays eigenvector 1 vs. the bolometric energy derived by \citet{nava12}
as a comparison.
}
\end{figure*}

Using the later analysis,
we calculated the first eigenvector for each GRB 
and plotted the eigenvector 1s vs. the 15-150 keV luminosity and isotropic energy 
as shown in Figure\,\ref{ev1_lum15_150}.
If the conclusion, that GRBs are primarily driven by their energetics is correct, then the 
eigenvector 1 - energy relation has to hold also for energies determined independently by other missions.
We used the bolometric energies listed in \citet{nava12} for about 30 GRBs. This relation is displayed in the right panel
of Figure\,\ref{ev1_lum15_150}. Although the number of GRBs is small, they still follow the trend expected from our PCA.
We can conclude that eigenvector 1 in our GRB sample represents the rest-frame 15-150 keV luminosity 
and/or the isotropic energy 
which seem to be the strongest drivers for the prompt and afterglow emission properties in our sample. 
Note, that \citet{margutti12} came to a similar conclusion using $E_{\rm \gamma, iso}$, $E_{\rm peak}$, $L_{\rm peak}$,
$T_{\rm 90,z}$, and the isotropic energy in the X-ray band as input parameters of their PCA of their GRB sample
(R. Margutti, priv. communications). 
Although one has to be very careful comparing PCAs from different samples with different input parameters, it seem to
suggest that the energetics is the most important underlying property of a GRB.

The second most relevant component, eigenvector 2,  strongly correlates with $T_{90}$ and the slope of the plateau phase \axb,
and anti-correlates with the spectral slope in the 15-150 keV and 0.3-10 keV energy bands $\Gamma$ and \bx.
There is, however, only a light correlation with the break time after the plateau phase $T_{\rm break 2}$.

\subsection{GRB in a cosmological context \label{lum_func}}

While Log N - Log S tests and luminosity functions are standard tools in quasar
cosmology studies \citep[e.g. ][]{richards06, ross12}
they have only been applied to GRBs in a few studies recently.
The luminosity function is usually defined as
\begin{equation}
\Phi(L, z) = \frac{\Phi^*(L_{\rm break}(z))}{(L/L_{\rm break}(z))^{a} + (L/L_{\rm break}(z))^{b}}
\end{equation}

where $\Phi^*$ is the number density at the break luminosity $L_{\rm break}(z)$, and $a$ and $b$ are the slopes of the 
luminosity function before and after the break.

The problem with GRBs in the pre-\swift\ era has been that only
for a handful of bursts redshift measurements existed. At the time it was only
possible to discuss GRB luminosity function in a theoretical context
\citep[e.g.][]{kumar00}. This has changed since \swift\ and luminosity functions 
have been derived from \swift\ bursts 
\citep[e.g. ][]{schmidt09, wanderman10, cao11, salvaterra08, salvaterra12}.

\begin{figure}
%\epsscale{0.75}
\epsscale{0.75}
\plotone{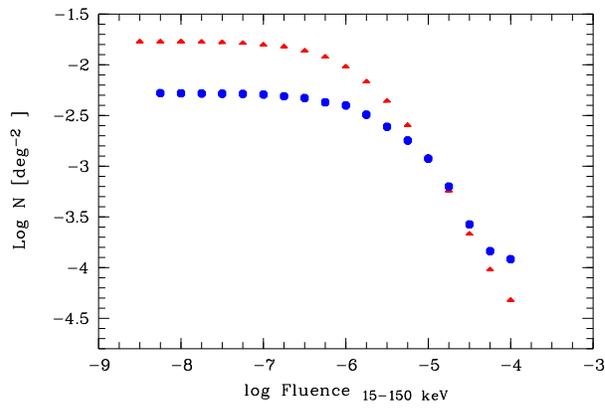}
\caption{\label{logn_logs} 
Log N - Log S Diagram of \swift-detected GRBs. The red triangles the observed
fluence in the 15-150 keV BAT band and the blue squares the k-corrected 15-150
keV fluence.
}
\end{figure}

The log N - log S diagram using the 15-150 keV fluence in the BAT band is shown 
in Figure\,\ref{logn_logs}.
This diagram contains the observed 15-150 keV fluence (754 bursts)
as well as the rest-frame k-corrected fluence (232 bursts). 
The parameters of the log N - log s function are $N^*=(1.733\pm0.018)\times 10^{-2}$, 
$a=(7.41\pm2.06\times 10^{-3}$, $b$=1.185\plm0.018,
 and the break fluence $S_{\rm break} = (1.339\pm-0.031)\times 10^{-6}$.

\begin{figure*}
%\epsscale{0.75}
%\epsscale{5.8}
\plotthree{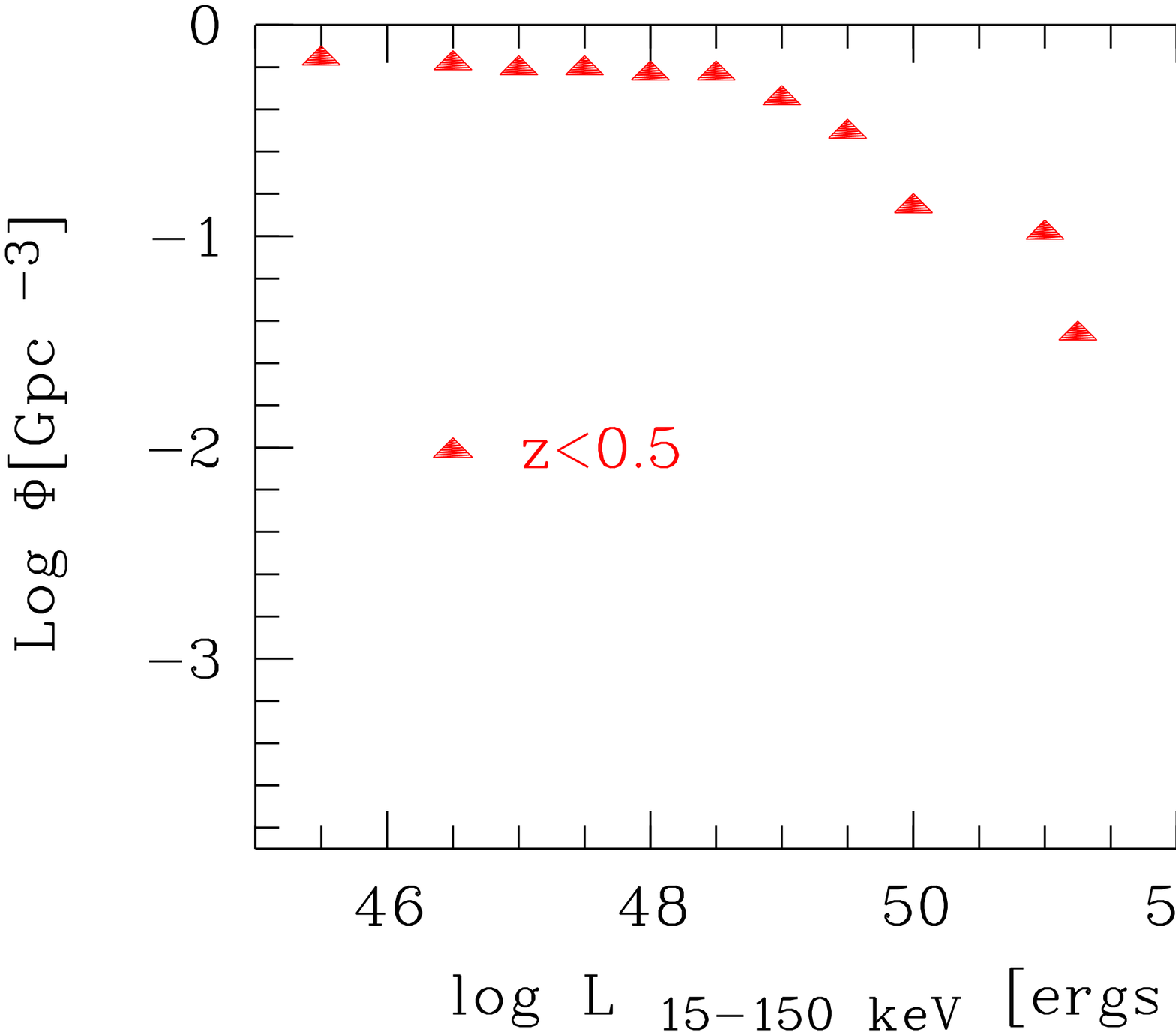}{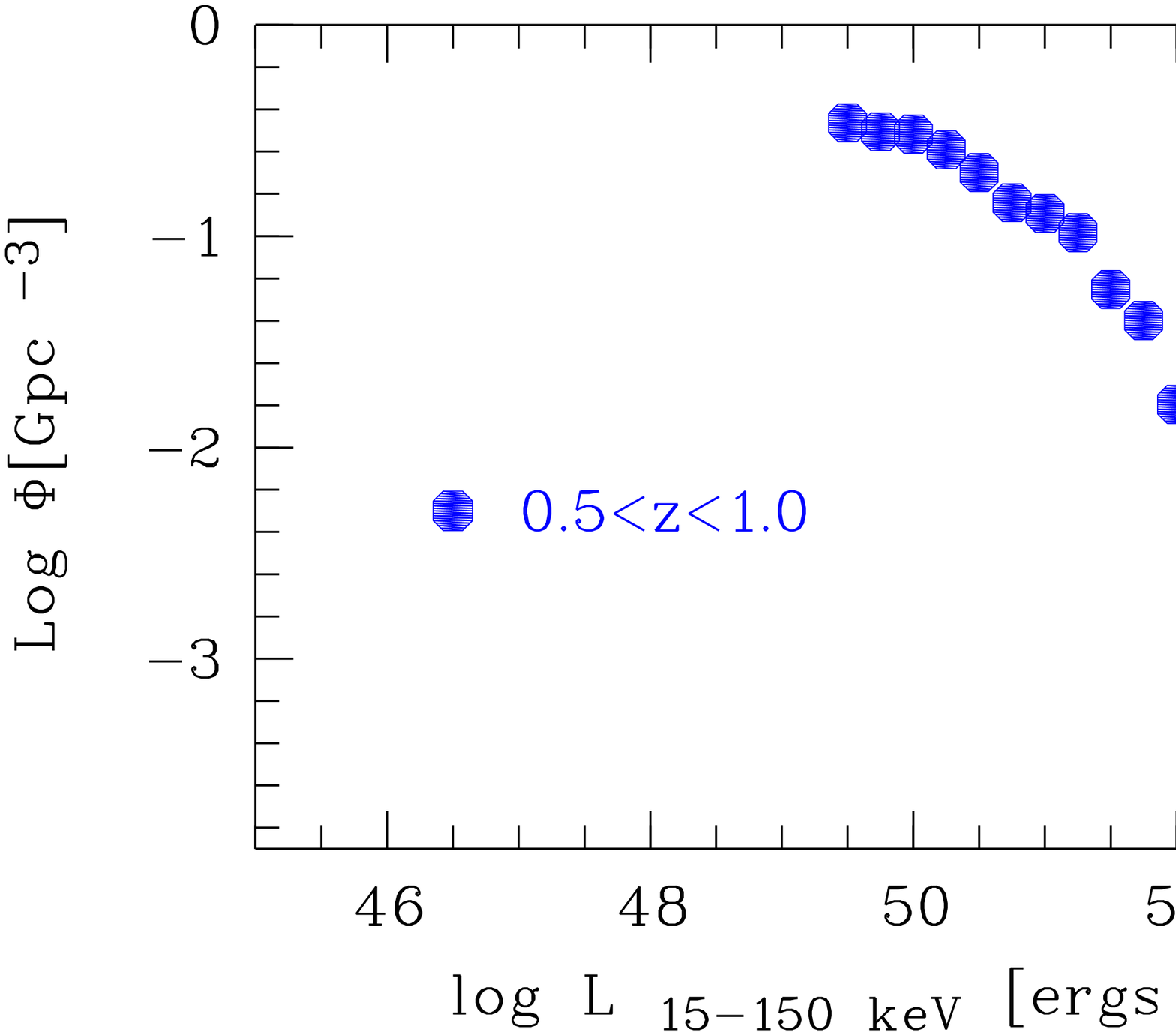}{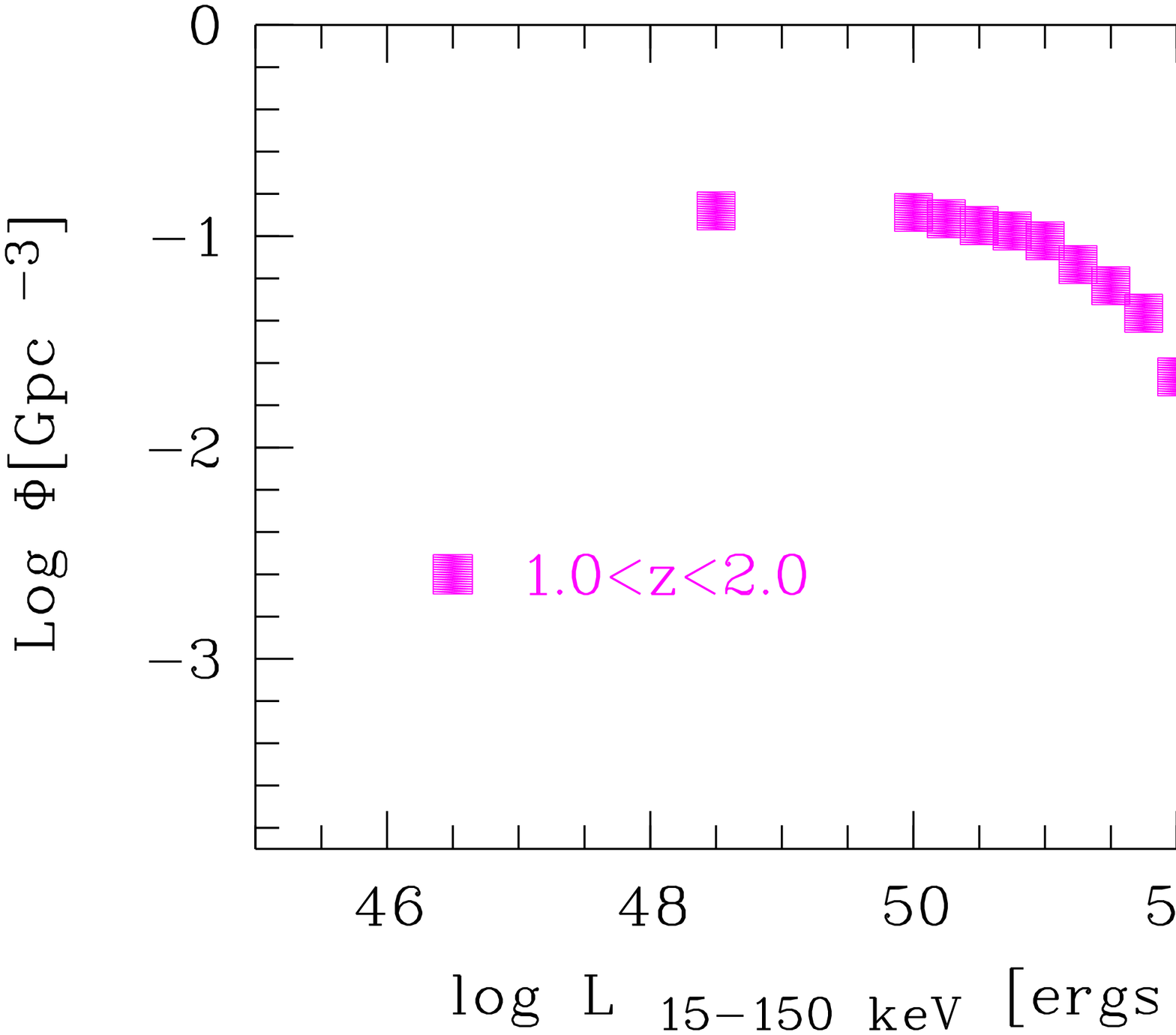}

\plotthree{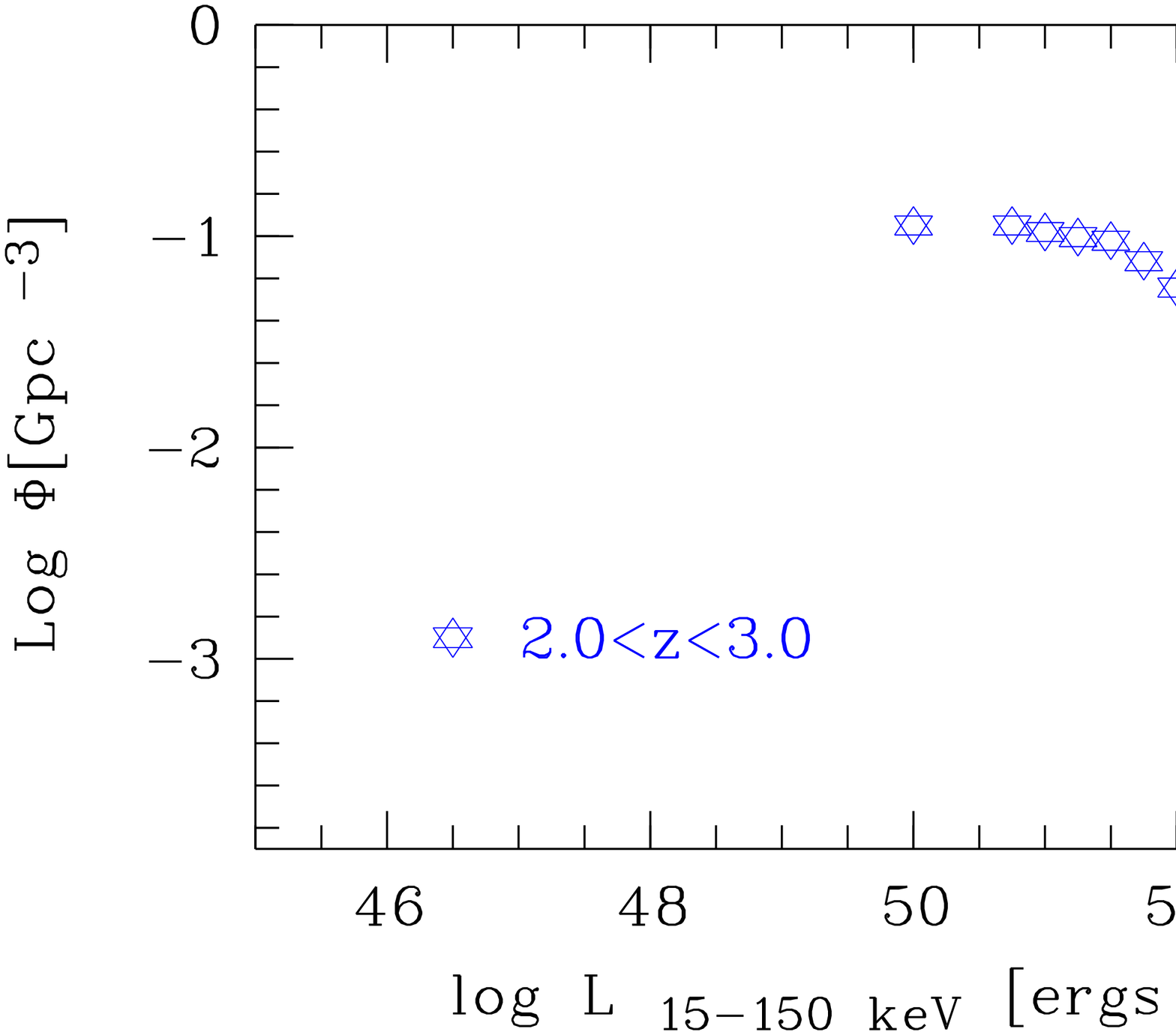}{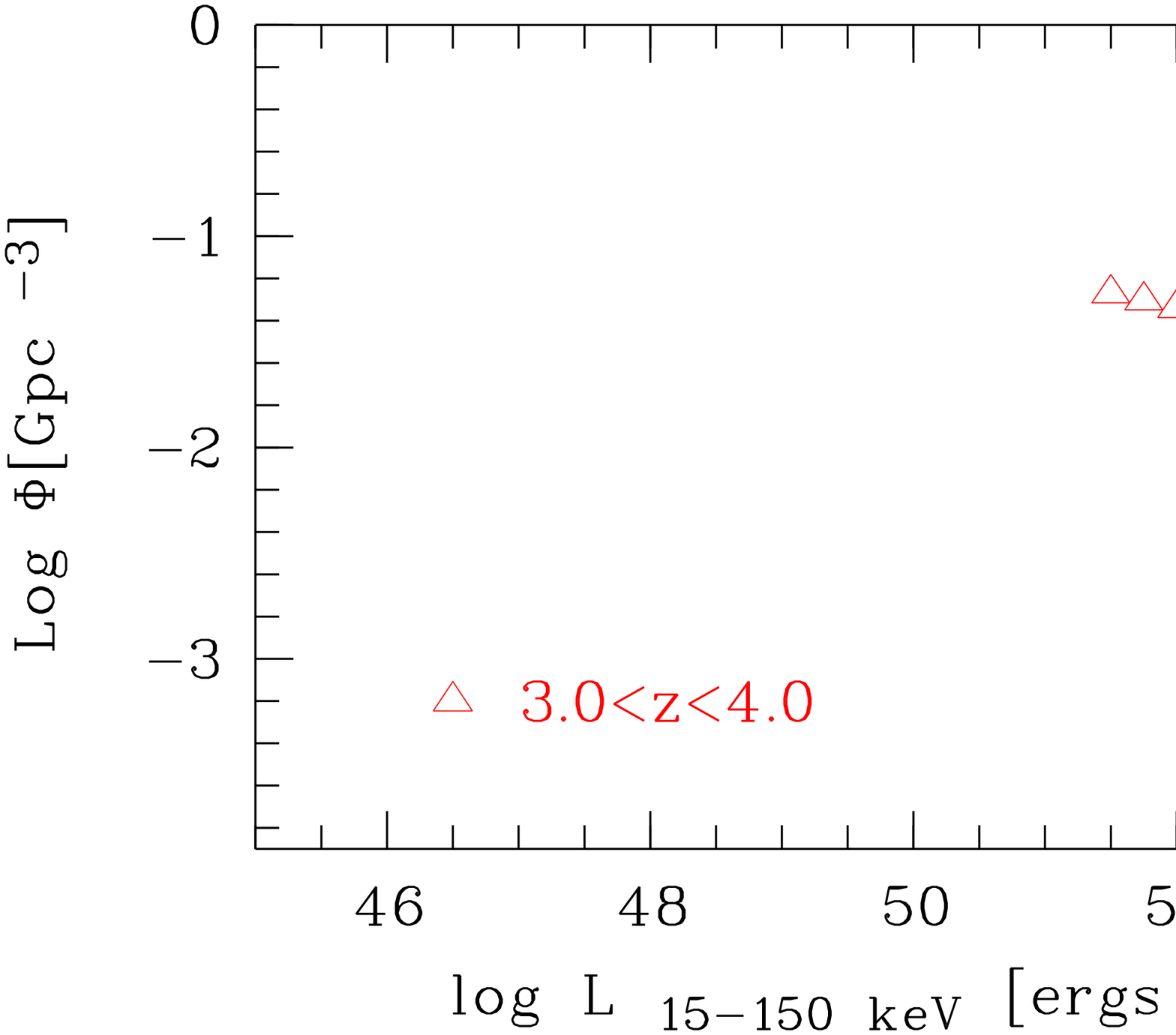}{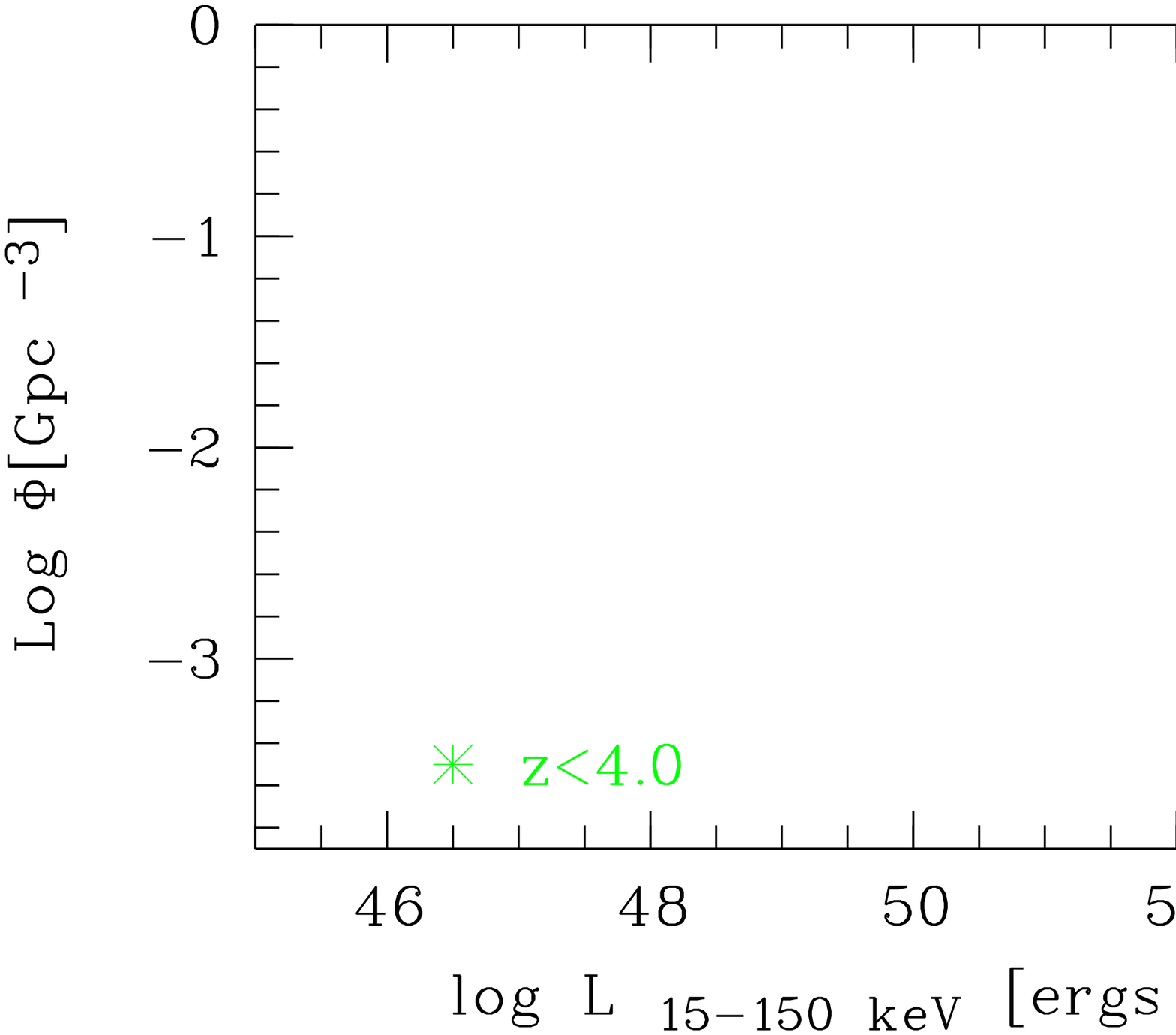}

\caption{\label{lum_function} 
Luminosity function of \swift-discovered GRBs in various redshift intervals. The fit parameters to these luminosity
functions and the number of GRBs in each function are listed in Table\,\ref{lum_func_res}.
}
\end{figure*}

We constructed luminosity functions in 6  redshift intervals as displayed in 
Figure\,\ref{lum_function}. The parameters for fits to the GRB luminosity functions as displayed in
that figure are listed in Table\,\ref{lum_func_res}.

\begin{figure}
%\epsscale{0.75}
\epsscale{0.75}
\plotone{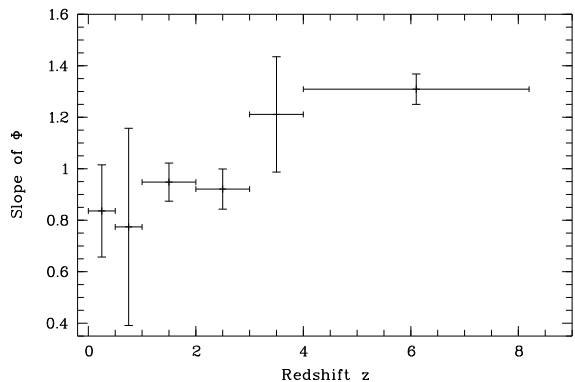}
\caption{\label{phi_slope} 
Slope b of the luminosity function (see Table\,\ref{lum_func_res}) for bursts with $L>L_{\rm break}$ vs. redshift.
}
\end{figure}

We noticed that the
slope of the high luminosity end of the luminosity functions with $L>L_{\rm break}$ 
becomes steeper with increasing redshift as displayed in Figure\,\ref{phi_slope}. 
This is the opposite to what has been 
reported by \citet{richards06} for quasars in the SDSS for which the high-luminosity end of the 
luminosity function becomes flatter with increasing redshift. As expected, the luminosity where the 
luminosity function breaks $L_{\rm break}$ shifts to higher luminosities with increasing redshift. The
GRB number per Gpc$^{-3}$ however, decreases with increasing redshift.

Now one step further is to look at the rates GRBs occur in a space volume per
year as shown in Figure\,\,\ref{grb_rate}. This rate contains long as well as
short bursts.
The way we estimated these numbers is to
take the number of \swift-discovered bursts with spectroscopic redshift
measurement per redshift interval and assume the same underlying redshift
distribution for the remaining \swift\ bursts. This number
was then multiplied by the
ratio of the whole sky in square degrees compared with the BAT sky coverage. This
plot suggests that the GRB rate is significantly higher in the current Universe
than at early times. This however, is a very na\"ive picture. The number of bursts
with spectroscopic redshifts measurements is biased against high redshift bursts.
Therefore we can not necessarily assume that the redshift distribution of the
\swift\ bursts without redshift measurements is the same as for those with
redshift measurement.

\begin{figure}
%\epsscale{0.75}
\epsscale{0.75}
\plotone{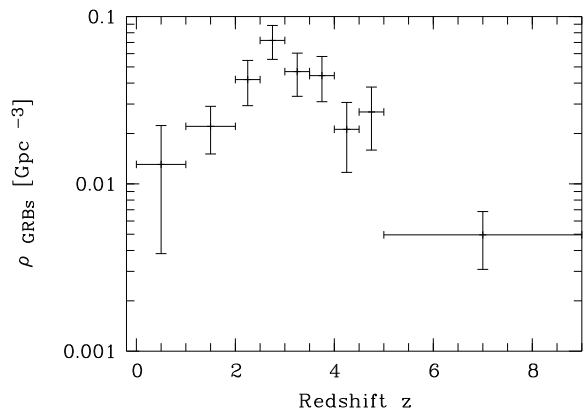}
\caption{\label{grb_density} 
Density of bright ($L>10^{52}$ erg s$^{-1}$) \swift-discovered bursts.
}
\end{figure}

\begin{figure}
%\epsscale{0.75}
\epsscale{0.75}
\plotone{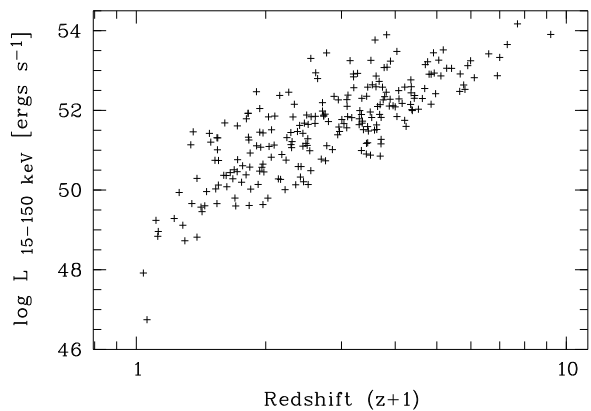}
\caption{\label{z_lum} 
15-150 keV luminosity vs. redshift
}
\end{figure}

Another attempt to tackle this problem is to have a look only at the space density
of the most luminous bursts. This is shown in Figure\,\ref{grb_density}. Here we
consider only those bursts who have a luminosity of $L_{\rm 15-150 keV}>10^{52}$ erg s$^{-1}$. 
 What we  see is that the number of \swift-detected bursts
per Gpc$^{3}$ decreases significantly in the local Universe. This is similar to
the ``cosmic downsizing'' that is known for bright quasars \citep[e.g.
][]{richards06}. If this is true, then we should see this effect of ``cosmic downsizing''
also in a luminosity - redshift plot. The 15-150 keV luminosity vs. redshift is shown in 
Figure\,\ref{z_lum} and indeed there are no bursts with $L_{\rm 15-150 keV}>3\times 10^{52}$ erg s$^{-1}$
at redshifts less than 1.5.
 Note, that the peak the quasar density is roughly at a redshift
of z=2.75 \citep[e.g.][]{richards06},
similar to that of our GRB sample. 
 This result is consistent with what has been found by
\citet{wanderman10}. 
The space density of GRBs, thus
 decreases with increasing redshift.
However, keep in mind that this may also be a selection effect. Because high
redshift bursts are those which are not detected in the UVOT, often no one wants
to take the risk of sacrificing valuable observing time on a burst that is simply
too faint and not necessarily highly redshifted. As we see later in this discussion, 
bursts with redshift measurements tend to be brighter than those without redshift measurements.
Nevertheless, the result shown in Figure\,\ref{grb_density} agrees with the cosmic star formation 
history \citep[e.g.][]{hopkins06}.

\subsection{Redshift Predictions \label{redshift}}

Spectroscopic redshift measurements exist for  232 \swift-onboard-discovered
bursts. The problem with obtaining GRB redshifts is that a) the afterglows are
faint to begin with, often fainter than 20th mag in R and b) they decay fast. So an
optical observer has to decide quickly if it is worth spending valuable telescope
time on a newly discovered burst. The capacity of detecting GRB afterglows in the
optical and near infrared either by spectroscopic of photoscopic measurements has
significantly increased in recent years especially with the arrival of X-shooter
at the ESO VLT and GROND at the ESO/MPI 2.2m telescope in La Silla \citep{greiner08, kruehler11}. 

In order to predict GRB redshifts  ideally one would look for a relation between a
redshift dependent (or distant-dependent) parameter and a redshift-independent parameter. 
This can be based purely of selection effects. 
In the past it has been suggested 
\citep{grupe07} that bursts with significant excess absorption column densities
above the Galactic value are low redshifts bursts. \citet{ukwatta10} found an
anti-correlation between the spectral lag times in the BAT data and the isotropic
luminosity. Recently \citet{morgan11} suggested a statistical method that applied a
Random Forest technique to predict high redshift GRBs. To date this is the 
most advanced method to predict if a burst is at high redshift based on \swift\ data. 

\begin{figure}
%\epsscale{0.75}
\epsscale{0.75}
\plotone{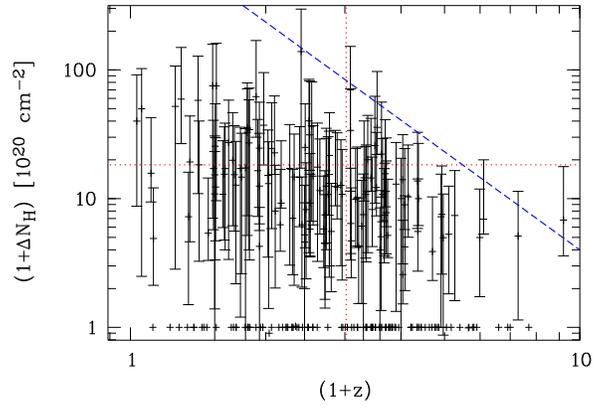}
\caption{\label{z_deltanh} 
Excess absorption column density ($\Delta_{\rm NH}$ + 1) vs. redshift (z+1). The red lines mark the mean values as listed in 
Table\,\ref{grb_statistics} and the blue line the $N_{\rm H}$ - redshift relation from \citet{grupe07}.
}
\end{figure}

Let us start with an update on the redshift - excess absorption relation presented
in \citet{grupe07}. At that time the \swift-GRB sample
only contained about 50 GRBs. Our new sample presented here contains more than 4
times as many bursts with spectroscopic redshifts measurements.
Figure\,\ref{z_deltanh} displays the relation between log (1+z) and log 
(1+$\Delta N_{H}$), showing that  in principle this method still holds. 
The only exception is the burst with
the highest redshift, GRB 090423 for which we obtained an excess absorption column
density of $5.8\times10^{20}$ cm$^{-2}$. Equation 1 in
\citet{grupe07} would have predicted a redshift of $z<$6.6, which, however,
would have been still a high redshift burst. Note also that the errors on the
absorption column density for this burst are rather large and within the errors,
GRB 090423 is still within the prediction. 

In order to obtain a better discriminator for high redshift bursts we have to
extend our simple redshift - $\Delta N_{\rm H}$ relation. One discriminator here
is a detection in the \swift\ UVOT. If the excess absorption is consistent with zero
this means it is either a high redshift burst or a low redshift burst with no
significant X-ray absorption. In the latter case this means that the burst will
most-likely not be reddened significantly on the Optical/UV so it should be
detectible in the UVOT. If UVOT does not detect this burst this makes it a high
redshift candidate. 

\begin{figure*}
%\epsscale{0.75}
\epsscale{1.5}
\plottwo{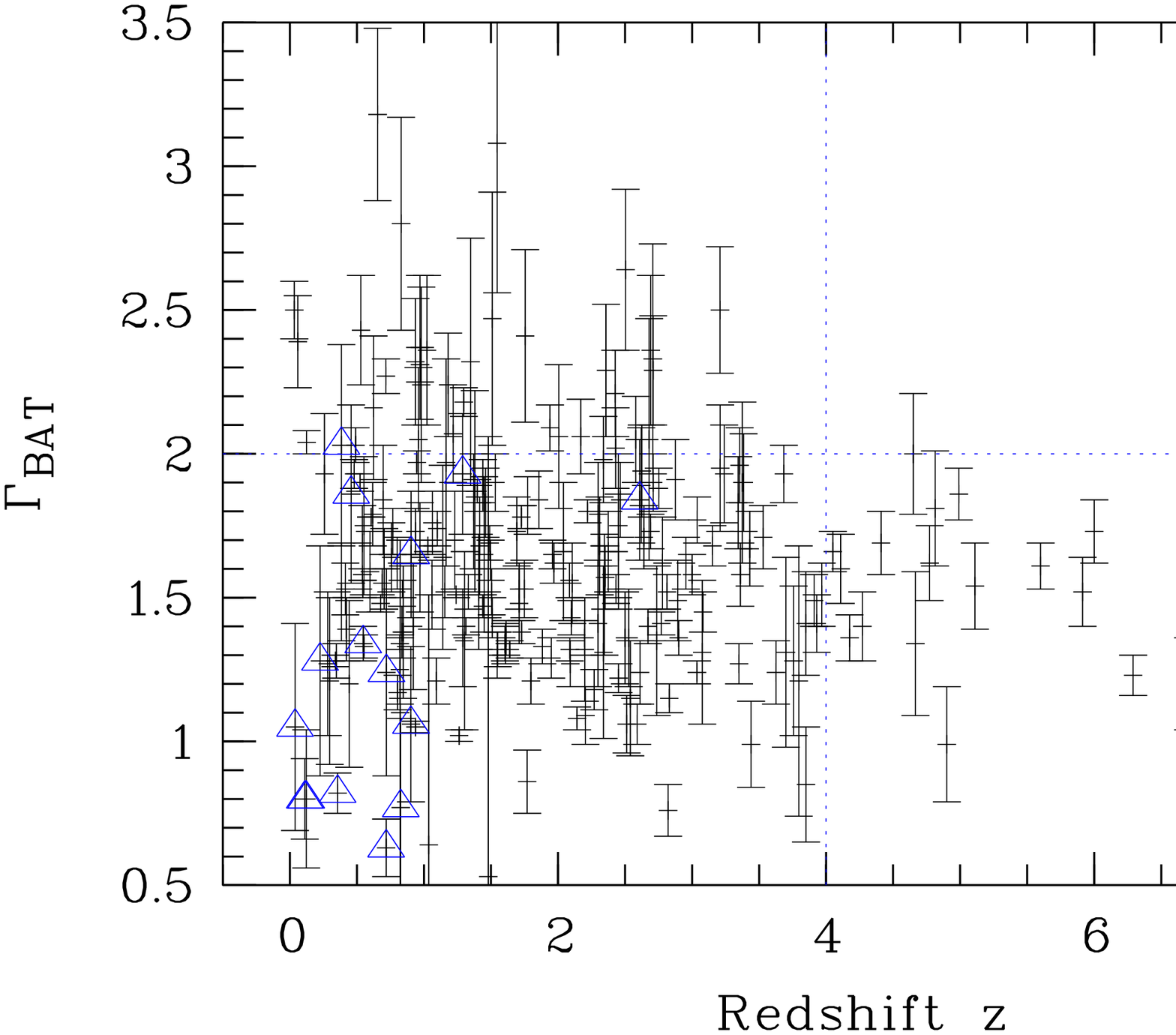}{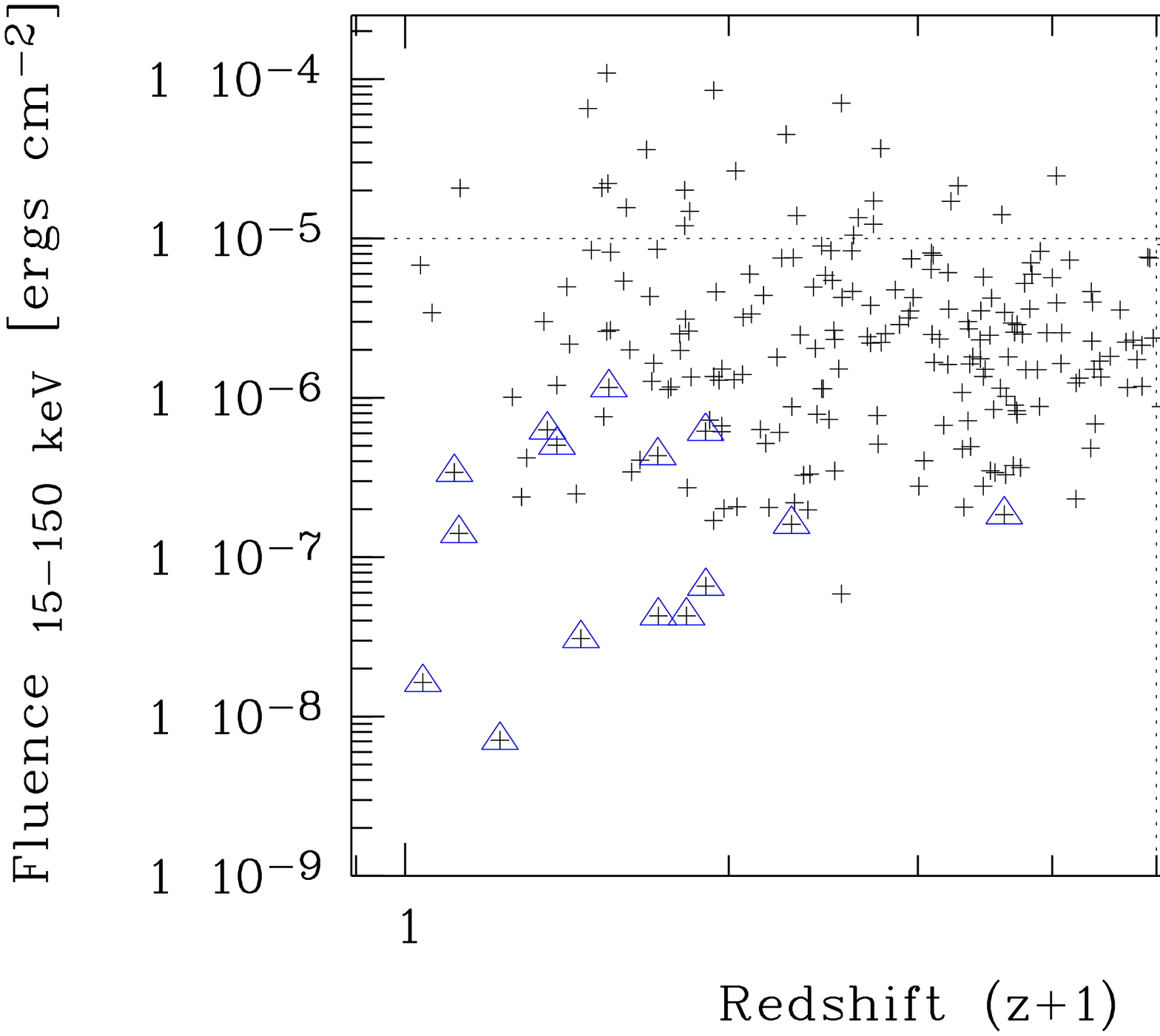}
\caption{\label{z_gamma_fluence} 
Relation between redshift and the photon index $\Gamma$ in the BAT band and the 
observed 15-150 keV fluence. Again, short duration GRBs are marked as triangles.
}
\end{figure*}

\begin{figure}
%\epsscale{0.75}
\epsscale{0.75}
\plotone{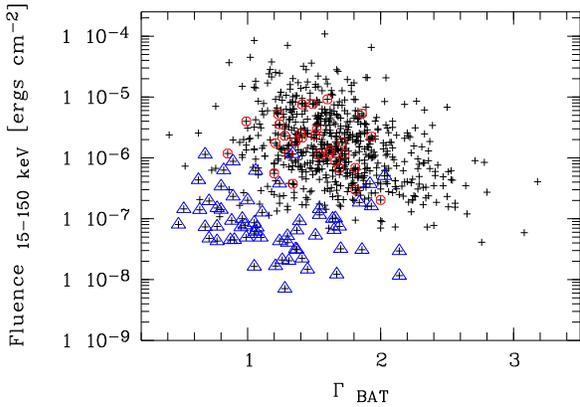}
\caption{\label{gamma_fluence} 
BAT photon index $\Gamma_{\rm BAT}$ vs. observed fluence in the 15-150 keV band. Short bursts are displayed as blue triangles.
}
\end{figure}

The next discriminators come from the BAT data.
Figure\,\ref{z_gamma_fluence} displays the relation between redshift and the
photon index $\Gamma$ in the BAT band and the observed 15-150 keV fluence. As we
can see from this plot, steep BAT spectra only occur in bursts with a redshift of
less than 4. This is a simple selection effect. As shown in Figure\,\ref{gamma_fluence}
there is as strong anti-correlation between the BAT photon index $\Gamma_{\rm BAT}$ and 
the fluence in the  15-150 eV band. 
As shown in the right panel of Figure\,\ref{z_gamma_fluence}, 
bursts with a fluence 
larger than $10^{-5}$ erg cm$^{-2}$ are only seen in GRBs with redshifts lower
than 4. Because of this and the anti-correlation between $\Gamma_{\rm BAT}$ and the 15-150 keV fluence,
we can observe GRBs with steep 15-150 keV spectra only in bursts with relatively low redshifts.

The question is can we find a combination of observed parameters that can be correlated with a 
redshift (distance) dependent property. To explore this question, we performed a PCA on the three
prompt emission parameters - $T_{90}$, fluence, and $\Gamma$. We then determined eigenvector 1 for each 
long-duration GRB and plotted this against the isotropic energy $E_{\rm iso}$. 
This relation is shown in Figure\,\ref{ev1_eiso}.
For the 216 long GRBs with spectroscopic redshifts in the sample,
this is a strong anti-correlation with a Spearman rank order correlation coefficient $r_{\rm s} =-0.572$ and
 $T_{\rm s}=-10.15$ with a probability of $P<10^{-8}$ that this is just a random distribution.
  This means that if we calculate eigenvector 1 for a newly discovered burst based on the BAT input parameters, we should
 be able to predict the isotropic energy in the BAT band. 
  Nevertheless there is a lot of scatter
 in this relation which prevents
  a clear statement of the redshift of the burst. 
  We examined this plot a bit further on 
 selected high and low redshift bursts. These are shown as red circles and blue triangles, respectively. 
 Clearly they form distinct groups 
 in this diagram. 
  For a given eigenvector 1, high redshift burst show larger isotropic energies than low-redshift bursts.
 
 If we can find another discriminator between low and high redshift bursts 
 then we  know in which group the newly discovered GRB belongs and
we can use the eigenvector 1 of this burst in order to get 
  an estimate of the isotropic energy.
 With this estimate  and the knowledge of the 15-150 keV fluence,
 we can then  estimate  the 
 distance to the burst and its redshift. Discriminators for high and low redshift bursts are: 
 a) does the X-ray spectrum show strong excess
 absorption? and b) is the afterglow detected in the UVOT or not?. 
 Table\,\ref{grb_statistics_redshift} lists the mean, standard deviation, and median of
 observed prompt and afterglow properties of low (z$<$1.0), intermediate (1.0$<$z$<$3.5) 
 and high redshift bursts (z$>$3.5). Besides that, low redshift bursts
 show enhanced excess absorption column densities, we also find that high redshift 
 bursts tend to have flatter 15-150 keV photon indexes, steeper later time
 afterglow decay slopes, and lower fluence than low redshift bursts. 
 
 As an example, we can estimate the redshift of GRB 051109B, the GRB that was
 suggested to be associated with a nearby galaxy \citep{perley05}, however, as we
 showed in Section\,\ref{correlation}, this is most likely not the case and the
 redshift is probably much larger. Applying the method described above, we calculated
 an eigenvector 1 of $+1.9$\footnote{See Appendix\,\ref{z_predict}
  for how to calculate 
 eigenvector 1.}
 for this burst.
  From
 Figure\,\ref{ev1_eiso} we can place the burst at a vicinity of $E_{\rm iso}\approx
 3 \times 10^{51}$ erg. The fluence of this burst was 
 2.7$\times 10^{-7}$ erg cm$^{-2}$. Therefore the luminosity distance is roughly
 $D_{\rm L}$=10 Gpc, which using the cosmology
calculator by \citet{wright06} corresponds to a redshift of z$\approx$1.5.

\section{\label{discuss} Discussion}

The main motivation for our project is to search for relationships between GRB prompt and afterglow
properties that suggest a close connection between these two  phases. 
With \swift\ it is now possible for the first time to perform statistics 
on the rest-frame physical parameters of GRBs.
Only since the launch of \swift\ have optical/NIR observatories been able to obtain
spectroscopic redshifts for a significant number of GRBs. About 30\% of all
\swift-discovered bursts have spectroscopic redshifts. 
As recent papers by \citet{margutti12, bernardini12, nava12}, and \citet{lu10} show 
there is a close link between the
energetics of the prompt and afterglow emission which seem to be driven by
energetics. Our study presented here found new relations that support the connection
between the prompt and afterglow phases.

\subsection{Correlation analysis}

As shown in the analysis section, there are several correlations between the prompt and afterglow properties of GRBs.
We found that:
\begin{itemize}
\item The length of the prompt emission (a.k.a $T_{90}$) correlates strongly 
with the break time after the plateau phase, which is essentially 
the length of the X-ray afterglow plateau phase.
\item The spectral  slopes
 of the 15-150 keV prompt emission and the 0.3-10 keV X-ray afterglow emission are closely correlated.
\item The X-ray afterglow decay slopes anti-correlate with the 15-150 keV spectral slope of the prompt emission. 
\item The X-ray afterglow decay slopes correlate with the rest-frame peak energy $E_{\rm peak}$ at high energies in the prompt
emission. 
\item The X-ray afterglow decay slopes also depend strongly on the energy/luminosity of the burst.
\end{itemize}

The underlying mechanism of the shallow decay phase is poorly understood.  
In X-ray
afterglow observations, there is no change in the spectral slopes from the
plateau phase to the normal decay phase, and in most of the cases the plateau
flux joins smoothly with the normal afterglow \citep{liang07b}.  On the other
hand the correlation between $T_{90}$ and $T_{\rm break,2}$ points to the fact
that the plateau is of internal origin, related to the central engine of the
bursts.  The fact that the spectrum does not change during  $T_{\rm break,2}$,
points to it either being determined by a geometric or a hydrodynamic effect.
A Similar correlation for a much smaller number of GRBs was found by \citet{liang07b}.
The correlation suggests an approximate relation  $T_{90} \propto T_{\rm break,2}$.
 Since there is significant scatter in the data, we only resort to
qualitative interpretation.

There are many theories in the literature addressing the origin of the  plateau
phase \citep[for a review, see ][]{zhangcjaa07}. The timescale of the plateau can be constrained by these. However, in most of these models it is difficult
to link the prompt duration $T_{90}$ to the plateau duration. 
One of the most discussed models is the refreshed shocks
\citep{Rees+98refresh,zhang06}: at the forward shock either by a smoothly
decaying central engine luminosity ($L\propto t^{-q}$) or by a distribution of
instantaneously ejected masses with a distribution of Lorentz factors down to
few tens \citep{granot06,Ghisellini:07a-late}.  Models for the plateau phase
include change in  microphysical parameters with time \citep{ioka06}, emission
by a distribution of ejected Lorentz factors \citep{granot06}, a strong reverse
shock contribution \citep{Genet+07shallow,Uhm+07shallow} or anisotropic jet
structure \citep{eichler06,toma06},

A class of models explaining the plateau phase invokes energy injection from
magnetars \citep{dailu98b,zhangmeszaros01a} into the forward shock, similar to
the refreshed shocks. These can reproduce the flattening in the X-ray light curve
\citep{DallOsso+11injection,Bernardini+12xrayag}. The supply of energy is
provided by the spin down energy of a newly born magnetar. In these scenarios the
collapse to a black hole is delayed by the magnetar phase
\citep{Lyons+10magnetar}.
In relation to short GRBs, the injection of spin down energy by a magnetar and
thus the plateau phase, can be linked to gravitational wave emission
\citep{Rowlinson+13magnetar,Corsi+09mag}.
In these models the prompt emission is tentatively produced by local processes
to the outflow (e.g.  magnetic dissipation or shocks close to the magnetar wind
photosphere \citet{Metzger+11grbmag}), which cannot be readily related to the
length of the plateau phase in the afterglow.

\begin{itemize}
\item  The most straightforward way of interpreting the $T_{90} - T_{\rm break}$
correlation is the model by \citet{Kumar+08acc} which addresses both the
prompt emission and the plateau phase. In this model the X-ray luminosity is
driven by the mass accretion rate. They envisage a massive star with a core and
an envelope. The part of the core collapses to form a black hole. Accreting the
remainder of the core produces the prompt phase. From the outermost parts of
the core until the inner part of the envelope the density drops suddenly,
producing the steep decay part. The envelope is accreted subsequently. The
duration of accretion is roughly the fallback time, $t_{fb}\propto M(<r)^{-1/2}\times
r^{3/2}$, $M(<r)$ is the mass within radius r, generally dominated by the black
hole mass.  Thus  $T_{90}$ corresponds to the core accretion time and  $T_{\rm
break,2}$ to the envelope accretion in this model.  The correlation results in
$r_{\rm core}\propto r_{\rm envelope}$.
The drawback of this scenario is that it predicts a steep decay at the end of
the shallow decay phase corresponding to the outer radius of the star. This is
only observed in a few cases.

\item A natural explanation can be given to the correlation if the reverse
shock is at the origin of the plateau. The details of such an interpretation
are described, for example in \citet{Genet+07shallow,Uhm+07shallow}. The
prompt duration ($T_{90}$) is defined as the active phase of the central engine
(e.g. in a photospheric model) and this also defines the width of the ejecta.
The plateau duration in this scenario corresponds to the ejecta crossing time
by the reverse shock which is then proportional to $T_{90}$.
\end{itemize}

The gamma-ray luminosity L anti-correlates with the plateau break time $T_{\rm
break,2}$. This can be understood in terms of an approximate energy
conservation. For a universal ratio between the prompt and plateau luminosity,
the increase in energy emission in the prompt phase will result in a smaller
emitted energy in the plateau, similar if we had a constant energy reservoir. 

There is a similar observed anti-correlation between the 
luminosity at the end time of the plateau 
\citep{dainotti08,  Bernardini+12xrayag} 
which can be explained in the magnetar model. 
 We can tentatively understand this anti-correlation within the framework of the
magnetar model as follows: both the prompt and afterglow luminosity  depend on 
some
positive power of the magnetic field (e.g. $L_{\rm plateau}\propto B^2$) while
the spin-down timescale, which is related to $T_{break,2}$ is proportional to
$B^{-2}$. This would qualitatively explain the anti-correlation. Other
properties such as the neutron star radius or the rotation period would act to
strengthen the anti-correlation.

\begin{figure*}
%\epsscale{0.75}
\epsscale{1.5}
\plottwo{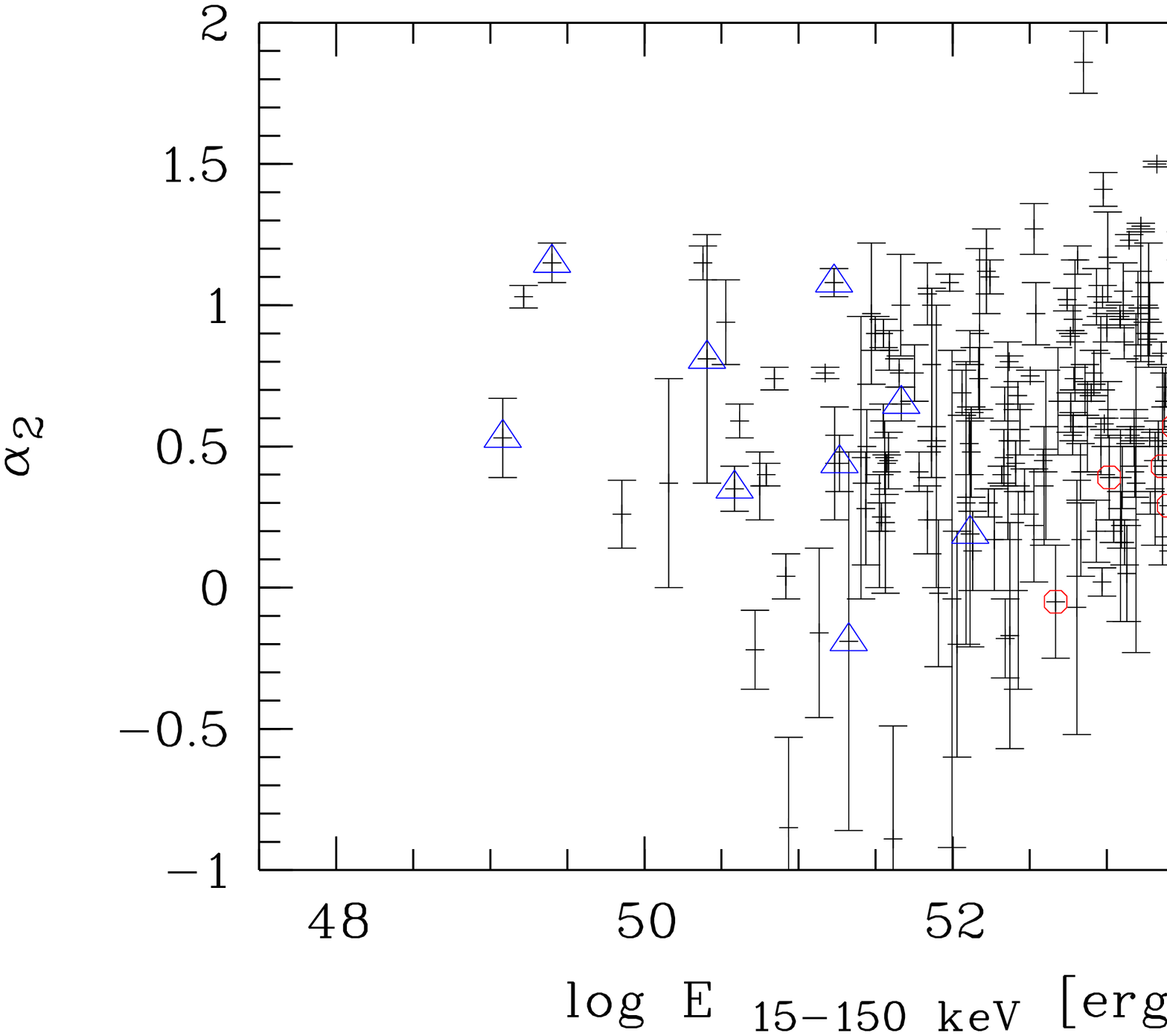}{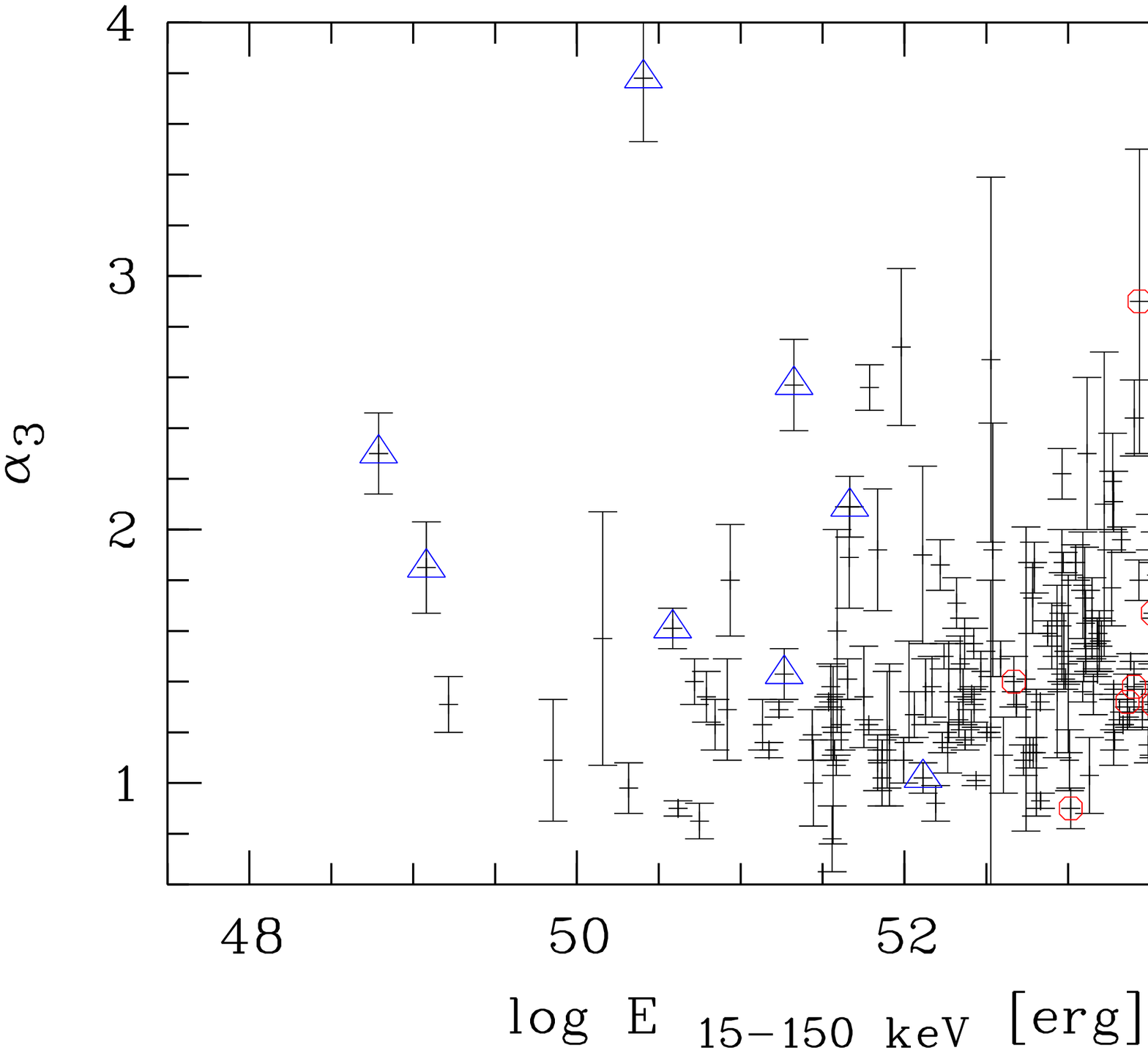}
\caption{\label{eiso_ax} 
Correlation between the 15-150 keV isotropic energy and 
X-ray afterglow decay slopes during and after the plateau phase \axb\ and\axc.
 Short bursts in these plots are marked as triangles.}
\end{figure*}

As we will see below (Section\,\ref{pca_discuss}),
it seems that these (anti)-correlations are primarily driven by the luminosity/energy of the burst.
What is interesting among these relations are those between the decay slopes of the X-ray afterglow emission with the rest-frame
$E_{\rm peak}$ and the energy and luminosity of the burst. While the energy, luminosity and rest-frame $E_{\rm peak}$ are all
redshift  
(or distance) dependent, 
the decay slopes are not. In principle we do a have a redshift indicator. However, at this point
we can not use the slopes \axb\ and \axc\ as redshift indicators due to the large scatter in these relations (see
Figure\,\ref{eiso_ax}). As we will discuss later, machine learning techniques may be the solution to take advantage of these
redshift indication relations.

\subsection{Multi-variate Analysis \label{pca_discuss}}

Using the log $T_{\rm 90, z}$, log $T_{\rm break2, z}$, \axb, $\Gamma$, and \bx\ as 
representatives of the prompt and afterglow emission of a subsample
of 175 GRBs with canonical X-ray light curves, 
the results from the PCA show that the first two eigenvectors account for more than 60\% of 
the variance in the sample. We can see directly that Eigenvector 1
represents the luminosity and/or the isotropic energy
of the GRB in the rest frame 15-150 keV band (Figure\,\ref{ev1_lum15_150}). 
This is somewhat expected. 
What is interesting to note is that although using different input parameters (see Section\,\ref{pca}),
\citet{margutti12} came to a similar conclusion.
The question that remains is what actually determines the luminosity 
and energy release of a burst?

Are  GRBs with spectroscopic redshifts, in particular those used in the PCA,
representative of the whole sample? 
In other words can conclusions that we draw from the 
statistical analysis on this sub-sample of  bursts, including 
 PCA, be applied to the 
rest of the sample which does not have redshift measurements?
This is an important question, because as we have discussed earlier, the entire sample 
is driven by selection biases.  
In order to test this, we need to look at the statistics between the GRB samples with
and without spectroscopic measurements. 

We have mentioned already in section\,\ref{selection},  \swift\
GRBs are affected by selection effects, primarily by the BAT properties. As pointed out by 
\citet{coward12} selection biases become even more important when dealing with high-redshift bursts. 
In order to answer this question we looked at the observed parameters of the
 samples for GRBs with and without spectroscopic redshift measurements and those used in the PCA.
The mean, standard deviations, and median of each sample are 
listed in Table\,\ref{grbs_z_noz}. 
All parameters like $T_{90}$ or \axb\ have very similar means, 
standard deviations,
and medians and suggest that these are all drawn from the same underlying distribution. However,
this is not true for the observed 15-150 keV fluence and the peak energy $E_{\rm peak}$. Here the 
distributions seem to be different: 
GRBs with spectroscopic redshifts have on average a fluence twice 
as high when compared with GRBs with no
redshift measurements. 
We performed Kolmogorov-Smirnov tests on the samples 
with GRBs with and without spectroscopic redshift measurements
and  found that indeed the distributions in the fluence and $E_{\rm peak}$ of these two samples are drawn 
from different populations.
 For the fluence we found 
D=0.2166 with a probability $P<10^{-4}$ of a random result and for $E_{\rm peak}$ D=0.30 and $P=0.002$.
Clearly, the bursts with redshift measurements are biased towards brighter GRBs. 
 As shown in Table\,\ref{grb_statistics_redshift}, this is really a selection
effect between burst with and without redshift and not driven by different redshift
intervals.
What this means is that these
GRBs also more likely to have optical counterpart and as a result will be observed preferentially by ground-based
observers. This becomes apparent when looking at the UVOT detections in both samples: while 2/3 of all GRBs with
spectroscopic redshift measurements have UVOT detections, this is only true for 
17\% of the GRBs without spectroscopic redshift measurements.
Nevertheless, the fact that otherwise the observed parameters of the bursts 
are very similar suggests that they are drawn from the same population. If this is true 
then the
correlations found among the rest-frame parameters in bursts with redshifts also apply to
 those bursts without
redshift measurements. 
In particular, the conclusion of our PCA that the energetics is the main driver in 
GRB properties does therefore apply to all GRBs.

\subsection{What is driving the prompt and afterglow properties? \label{pca_discuss}}

As we have seen from the PCA, 
 properties of the prompt and afterglow phases seem to be driven primarily by
energetics. 
This means the largest relative scatter in the multidimensional data cloud can
be accounted for by luminosity. In other words, if we know the luminosity,
we have accounted for a high percentage of the information otherwise
attainable. 
Also the luminosity changes by 5 orders of magnitude, more than any other
parameter. There are two viable mechanisms for extracting luminosity from the
black hole - accretion disk system

If the luminosity is extracted from the central engine by the Blandford Znajek
\citep{Blandford+77Znajek}  mechanism, this means $L\propto a^2 B^2 M^2$, where
$a$ is the dimensionless spin parameter, $B$ is the magnetic field threading
the central engine, and $M$ is its mass. In current models the mass can have a spread of
one order of magnitude, the rotational parameter is ideally around $1/2$ for
maximum efficiency, and it cannot introduce a large amount of variance.  Thus the
magnetic field has to account to the bulk of the variance.
If the jet is launched by neutrino emission from the disk, the luminosity
scales as: $L\propto \dot{M}^{9/4} M^{-3/2}$ \citep{Zalamea+11nu}. In this
scenario the large variance in luminosity can be accounted by variation in the
accretion rate by two orders of magnitude.

However, what is really behind all those properties, or in other words, what determines the
energetics of a burst? Part of the problem is that for all correlations mentioned here we measured
isotropic energies and luminosities. We know, however, that in a GRB the outflow is collimated. This may
explain the anti-correlation we have found between the luminosity and energy of the prompt emission with
(rest-frame) $T_{90}$ and break times before and after the plateau phase. Na\"ively one would assume that the
more energetic a burst is the longer its $T_{90}$ and the plateau phase start and end times. However, as
we have shown this seem to be not the case. We do, however, for all of our correlations assume isotropic
energies/luminosities. What may be the case is that bursts which short $T_{90}$ and break times before and
after the plateau phase are those bursts which outflows are highly collimated. Correcting for collimation
means that the total energy release of these bursts is significantly lower than in burst which are less
collimated. This means that if calibrated $T_{90}$ and the break times can be used as a measure of the jet
opening angle. This relation then would be somewhat similar to the jet opening angle - jet break times
relation found by \citet{frail01} for optical and radio afterglows. What this means is that bursts which
large opening angles are the bursts which are intrinsically less energetic and show the shorter $T_{90}$
and break times. This is exactly what we see in the second eigenvector in our PCA. The second Eigenvector
may be the jet opening angle. 

A connection between the X-ray spectral slope and the decay slopes in the light curve is 
expected from  for example the
closure relations \citep{racusin09, zhang04, zhang06}. 
The big question remaining is, can we predict the behavior of the 
X-ray afterglow, assuming a canonical behavior, based on properties
measured from the prompt emission. If this is possible then it
 gives us a handle on how to plan future observations
of the X-ray afterglow. We have already seen that $T_{90}$ and the break times before and after the plateau
in the X-ray afterglow light curves are strongly correlated. So  one prediction we can do is that bursts
with a short $T_{90}$ will show early breaks in the X-ray light curves. Although this correlation is
statistically very strong ($P<10^{-8}$), the scatter in the relation is large, making it impossible to
derive a precise prediction of the break times based on $T_{90}$. We have also found that the photon
spectrum measured in the prompt emission in the BAT 15-150 keV band in strongly anti correlated with the
X-ray afterglow decay slopes. So, if for example the 15-150 KeV hard X-ray photon index $\Gamma$ is flat, 
the the X-ray afterglow light curve decay slopes are most likely to be steep and vice versa. Again, this seem to be
linked to the energetics of the burst. More luminous/energetic bursts have flatter hard X-ray spectra (see
Figure\,\ref{lum_gamma_bx}). Nevertheless, we do not find a clear correlation between the luminosity and energy and
the decay slopes in the X-ray light curves. The correlation between $E_{\rm iso}$ and the decay slope in the
plateau phase \axb\ is mild ($r_{\rm s}$=0.240,  $P=5\times 10^{-4}$and the `normal' decay slope
 \axc\ with $r_{\rm s}$=0.306 and $P=3.1\times 10^{-3}$ (see
Table\,\ref{correlation_tab_z}). So, the X-ray afterglows of more energetic bursts decay faster than those of less
energetic bursts. This is somewhat the opposite than what one would normally expect. The reason here again may be
the opening angle of the burst. because we do all our correlations based on $E_{\rm iso}$
 we do not know, however,
 what the opening angle is and what the real energy release is.  

The 
question is: do the properties of the prompt emission of a burst determine the 
fate of the afterglow emission (assuming that the afterglow follows a canonical
light curve).
 Or in other words, can we use the properties measured from the prompt emission to make predictions 
of the behavior of the X-ray afterglow. How will the light curve evolve?

One of the strongest correlations has been found between the 15-150 keV $T_{90}$ and the break times at the beginning
 and end of the 
plateau phase in the X-ray light curve. This correlation gives us a rough estimate when these breaks happen 
based on $T_{90}$.
A burst with a short $T_{90}$ will have breaks in the X-ray light curve earlier than a burst with the 
long $T_{90}$. We have also found an 
anti-correlation between the 15-150 keV spectral slope $\Gamma$ and the decay slopes in the X-ray 
light curve: a burst with a soft 
15-150 keV spectrum will show flatter decay slopes than a burst with a harder 15-150 keV spectrum.
Last but not least there are strong correlations between the fluence in the 15-150 keV BAT energy band and the
decay slopes in the X-ray light curves. This results in that GRBs with high fluence decay faster in X-rays than 
GRBs with lower fluence.

\subsection{\swift\ BAT short and long duration GRBs \label{short-long}}

\begin{figure*}
%\epsscale{0.75}
\epsscale{1.5}
\plottwo{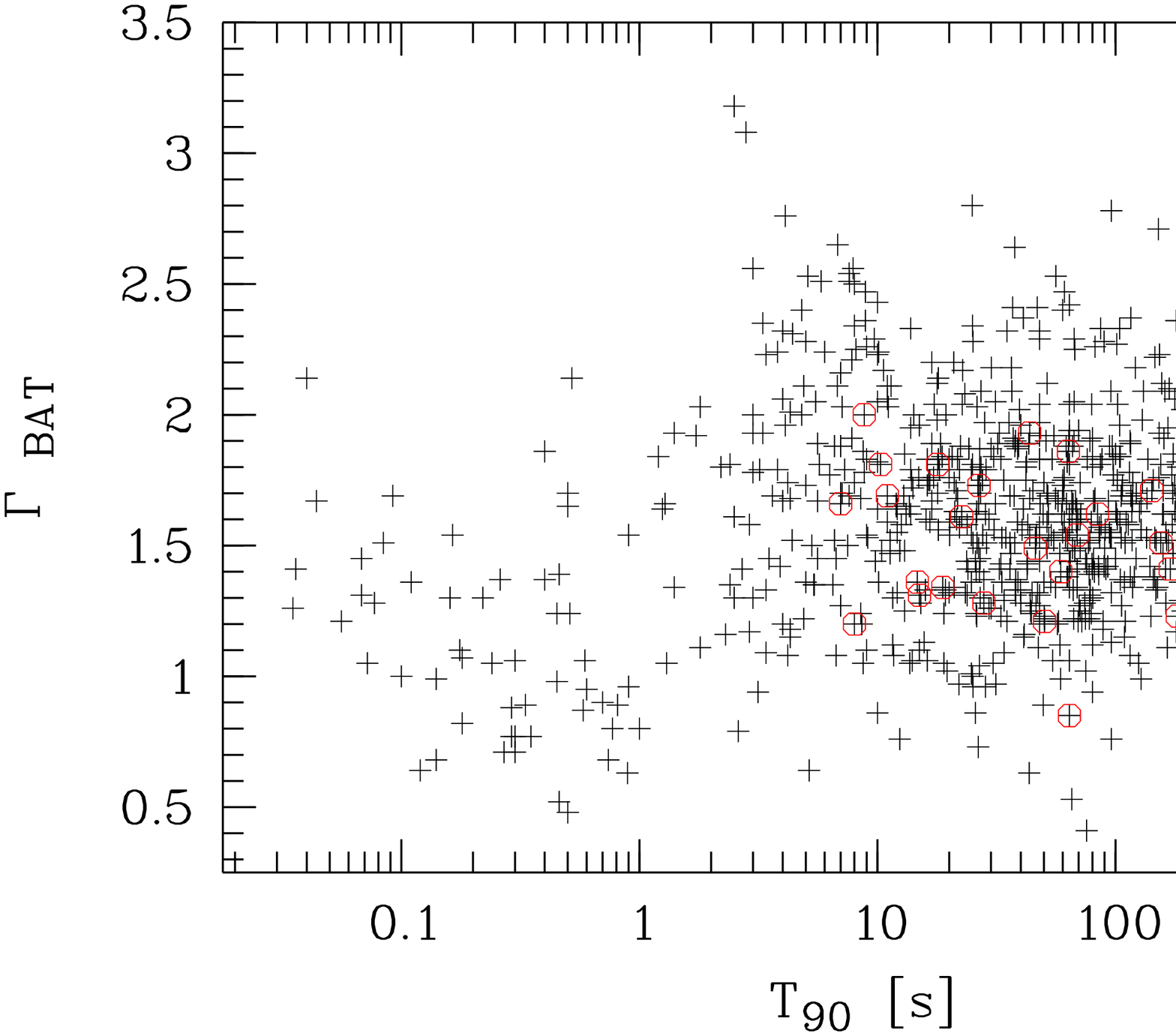}{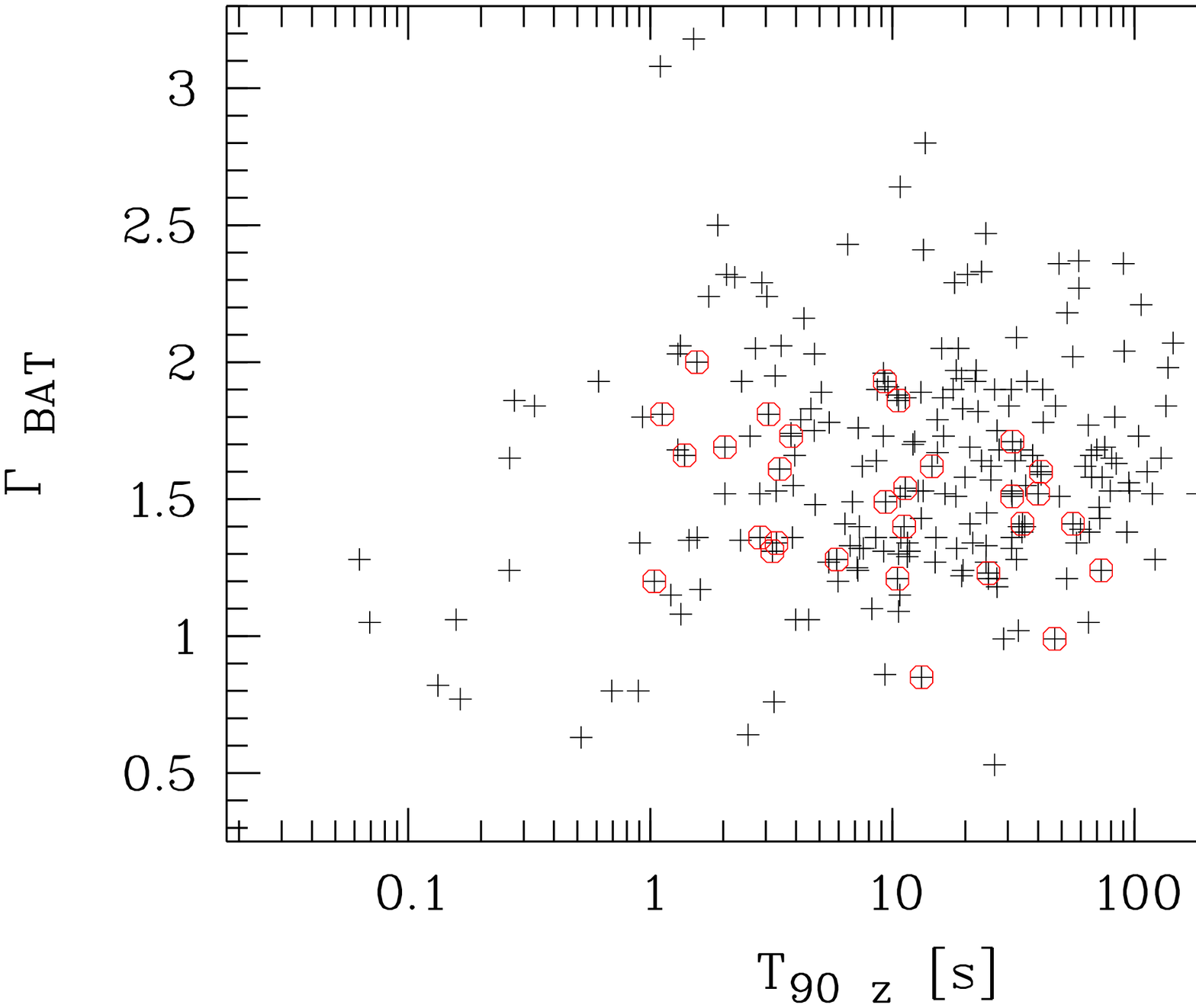}
\caption{\label{gamma_t90} 
Correlations between the BAT 15-150 keV photon index $\Gamma$ and the observed and rest-frame $T_{90}$.
High redshift bursts (z$>$3.5) are displayed as red circles.
}
\end{figure*}

The distribution of $T_{90}$ shown in Figure\,\ref{distr_t90_tb} from BAT discovered GRBs suggests a bimodal
distribution such has been found from BATSE bursts \citep{kouveliotou93}. However, due to the lower energy window of
the \swift\ BAT, the detection rate of short GRBs is significantly lower and the devision line between short and
long-duration GRBs appears to be at shorter times compared with BATSE bursts. As pointed out by \citet{bromberg12}
the devision between collapsar (long GRBs) and non-collapsar (short GRBs) is driven by the different physical
processes involved in these types. \citet{bromberg12} suggested that a devision between short and long bursts at
about 0.8s is more suitable for \swift-detected bursts. The observed $T_{90}$ distribution
(Figure\,\ref{distr_t90_tb}) seem to support this time. As already found from BASTE bursts, short GRB tend to show
harder spectra than long GRBs allowing another parameter to distinguish between the two classes. As shown in the
distributions of the 15-150 keV photon indices $\Gamma$ shown for long and short GRBs in
Figure\,\ref{distr_gamma_bx} we do see the same effect in \swift-detected short and long GRBs as well.
Figure\,\ref{gamma_t90} displays the relations between observed $T_{90}$ and $\Gamma$, suggesting that there are two
distinct groups in that diagram. As pointed out by \citet{margutti12}, short GRBs appear to be less energetic compared
with long GRBs. The consequence is that the fluence of short GRBs is significantly lower than that of long GRBs
(Figure\,\ref{distr_fluence}). 

\begin{figure*}
%\epsscale{0.75}
%\epsscale{5.8}
\plotthree{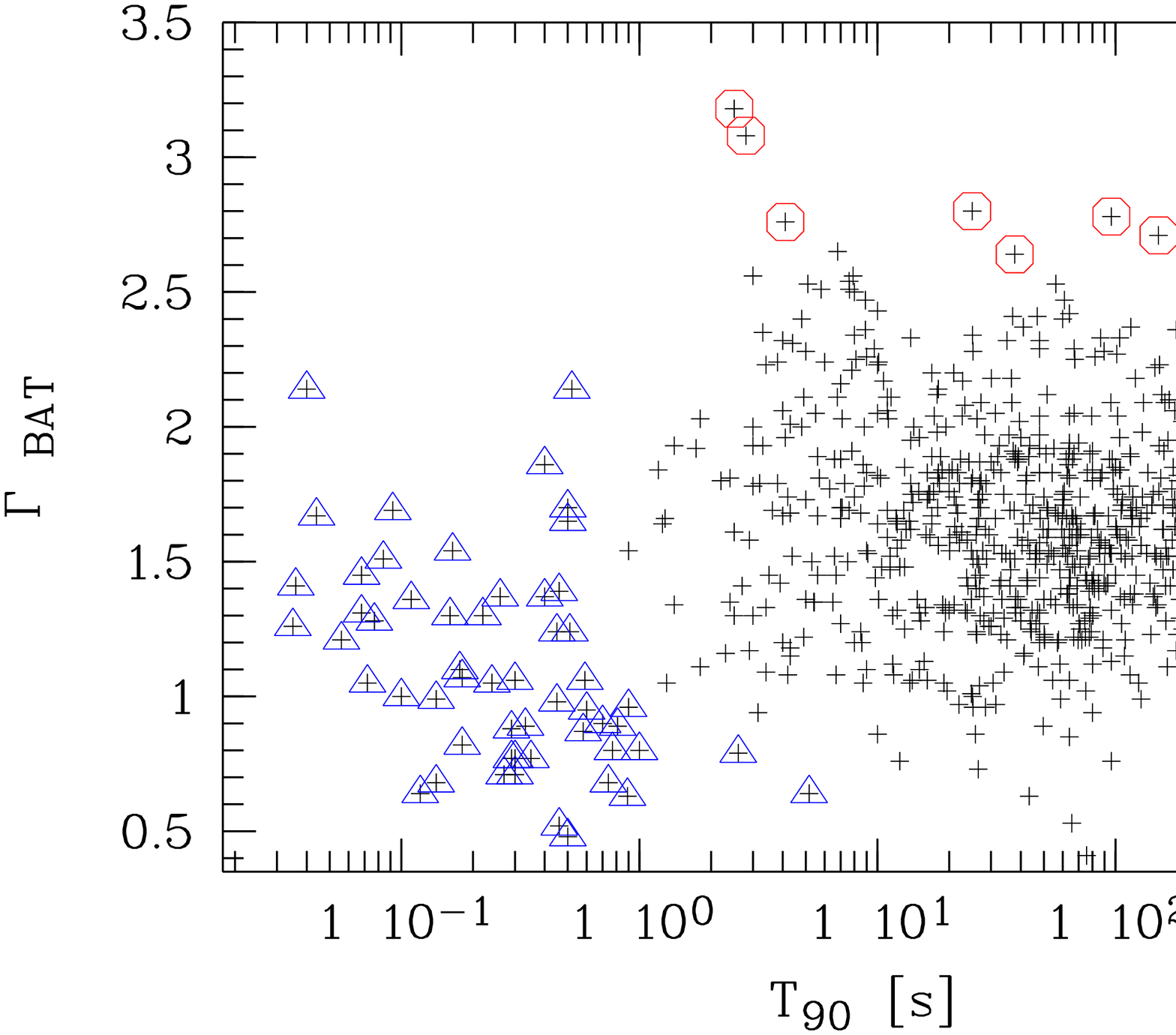}{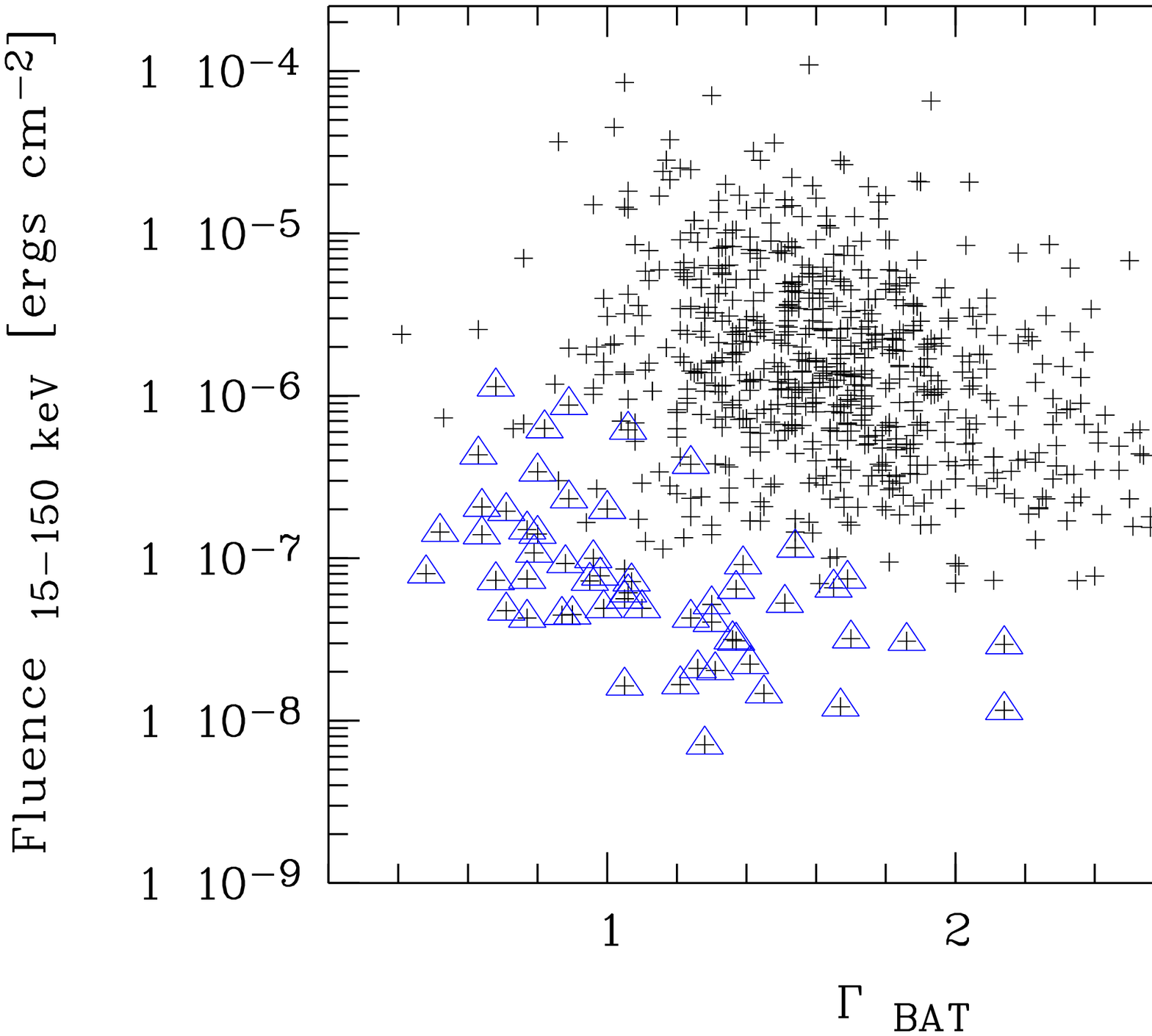}{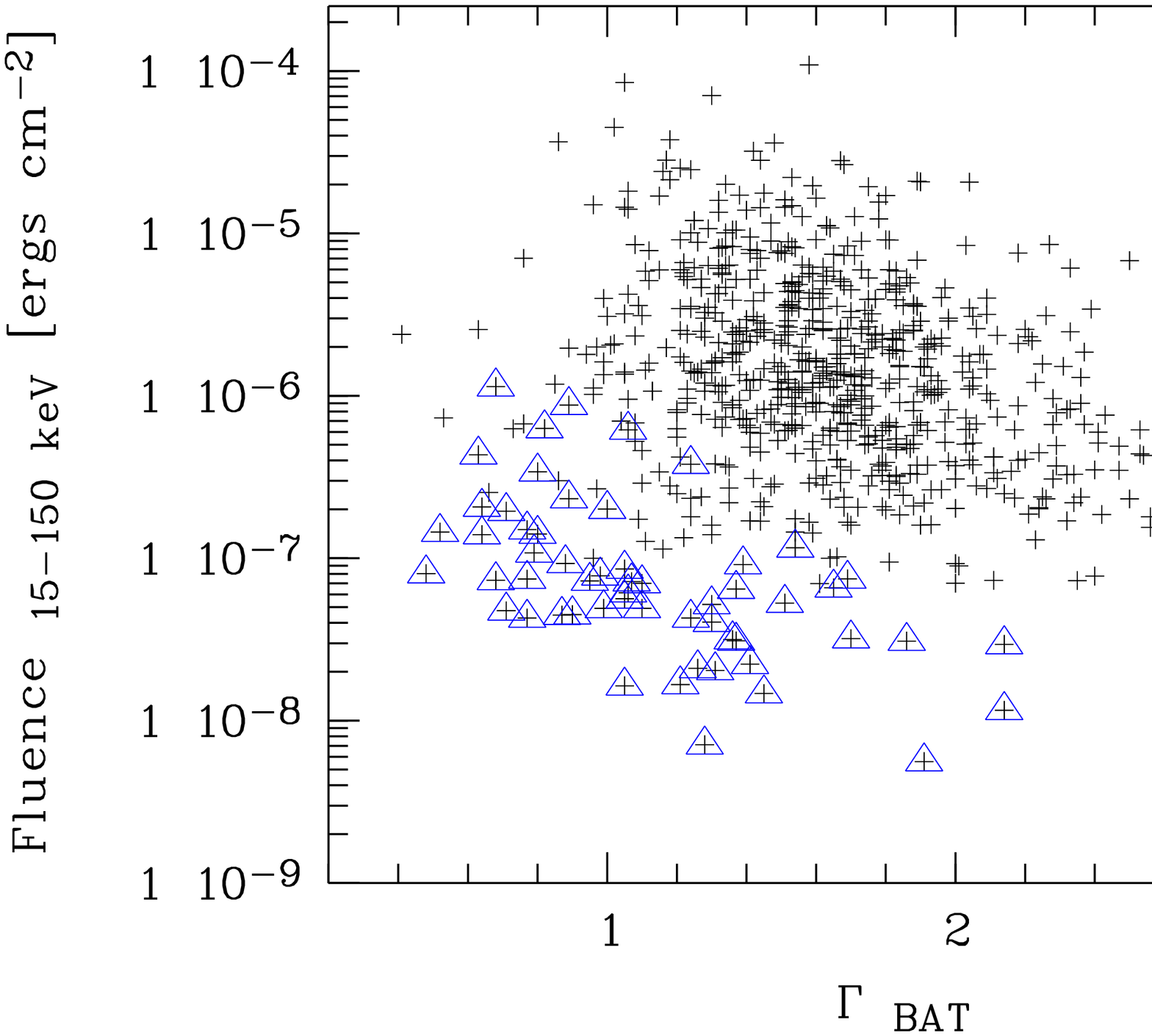}

\caption{\label{cluster_corr} 
The correlations between the three parameters used in the cluster analysis as discussed in
Section\,\ref{short-long}
 The
members of group 1 (long GRBs) from the cluster analysis are displayed as black crosses,
group 2 (short GRBs)  are displayed as (blue) triangles, and group 3 (XRFs) as red circles.
}
\end{figure*}

To find out if a burst belongs to one class of the other, one method that has been suggested in spectral
lag analysis \cite[[e.g.][]{ukwatta12, gehrels06, norris00}.
Another method is to apply statistical tools to the n-dimensional dataset and 
 classify a GRB as a short of a long-duration GRB using
cluster analysis \citep[e.g.][]{everitt11}. We use the observed $T_{90}$, 15-150 keV fluence, and $\gamma$ (all
normalized) to span a three-dimensional space. We then run a hierarchical cluster analysis with centroid linkage on
this dataset.  Besides outliers which consist of
GRB 060202B which has an extremely soft 15-150 keV X-ray spectrum \citep{aharonian09}, and
GRB 060218 which is the
low-luminosity GRB associated with supernova SN 2006aj \citep{soderberg06}, we see two main groups: Group 1 which
consists of 685 members, and Group 2 consisting of 55 members - all short GRBs. 
There is a small third group which consists of GRBs 050416A, 050819, 050824, 
060428B 061218, 080520, and 130608A. All these bursts are X-ray flashes
\citep[e.g.][]{sakamoto08}. This group has previously being suggested to be an intermediate duration GRB group by 
e.g \citet{mukherjee98} and \citet{veres10}.
Figure\,\ref{cluster_corr}
displays where these GRB groups appear in $T_{90}$ - $\Gamma$, $T_{90}$ - fluence, and $\Gamma$ - fluence diagrams. 
In these diagrams we displays group 1 as black crosses, group 2 as blue triangles and group 3 as red circles. 
These
diagrams suggest that there is still some overlap between the groups, although the XRFs (group 3) are clearly distinct in the 
$T_{90} - \Gamma$ and $\Gamma$-Fluence diagrams. 
However, these diagrams are just
two-dimensional projection of the three-dimensional space that is spanned by $T_{90}$, $\Gamma$ and the 15-150 keV
fluence. In order to really see if the three groups are disjoint we need to perform an axis transformation in the same
way as it was done for the PCA. Therefore we performed a PCA in the three parameter space and calculated the first
two Eigenvectors for each burst. This diagram is shown in Figure\,\ref{cluster_ev1_ev2}. Clearly short and long GRBs
and XRFs
occupy different areas in this diagram: short GRBs have low eigenvector 1 and high eigenvector 2. XRFs on the other hand have very high
eigenvector 2s.
 We can use this
diagram in order to determine if a burst with a border line $T_{90}$ is a short or a long duration GRB.
In Appendix\,\ref{long_short} we give the equations to determine where on
the eigenvector 1 - eigenvector 2 diagram (Figure\,\ref{cluster_ev1_ev2}) a burst lies based on its
 observed BAT properties.

What this analysis also shows is that the previously considered short bursts GRB 050724 and GRB 051221A 
\citep[e.g.][respectively]{grupe06, burrows06} are long GRBs. Other GRBs that were considered short burst in 
our previous analysis using the 2s cutoff line are: 
GRBs 070809, 071227, 080426, 080905, 081024, 
090426, 100724A, 120403A, and 121226A. 
These bursts however, need to be classified as long GRBs. 
On the other hand, what our analysis also found were short GRBs with an observed  $T_{90}$ of 2.6 and 5.2s, GRBs 081016B and 110726A, respectively.
Although their $T_{90}$ is significantly longer than the 0.8s that we suggest from our analysis consistent with the results by \citet{bromberg12}.
These burst can be classified as short bursts due to their low fluence of the order of $10^{-7}$ erg cm$^{-2}$ and hard X-ray 
spectra with $\Gamma$=0.79 and 0.64, respectively.

\subsection{Luminosity Function \label{lum_func_discuss}}

As for the luminosity functions we noticed a steepening of the slope for the higher 
luminosity part of the luminosity function with increasing redshift. However, one has 
to be very careful with drawing conclusions from these findings. The redshift distribution
of GRBs is strongly biased against high redshift bursts, due to selection effects. As mentioned above,
bursts without redshift measurements are drawn from a fainter population than bursts with redshift measurements.
As also pointed out by \citet{salvaterra12},  the \swift\ redshift sample is far from being complete.
In order to obtain a complete sample, \citet{salvaterra12} proposed to use only bursts with high 
peak photon flux in the BAT band. This is not true for all \swift\ bursts with redshift measurements. Because we do
have a bias against GRBs with low fluence, the sample of bursts with redshift measurements is biased against high
redshift bursts due to their typically lower fluence.

\begin{figure}
%\epsscale{0.75}
\epsscale{0.75}
\plotone{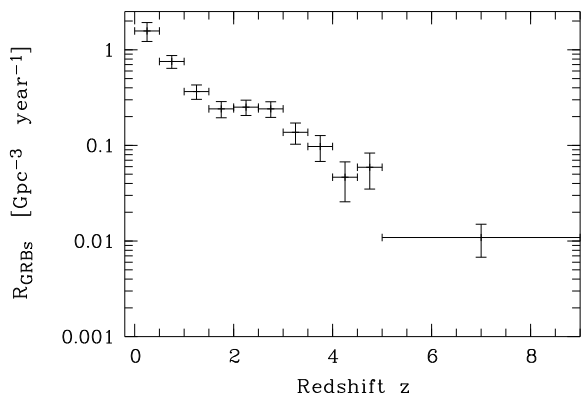}
\caption{\label{grb_rate} 
Total rate of GRBs per year and Gpc derived from the number of \swift-detected
GRBs
}
\end{figure}

While the selection effect of GRBs with redshift measurements can explain the steep decay in the GRB rate at higher
redshifts as shown in Figure\,\ref{grb_rate}, it does not explain the lower rate of bright GRBs in the low redshift
Universe. While the smaller comoving volume for the Universe with z$<$1 may explain part of this effect, it does not
explain the whole picture. The comoving volume of the Universe with z$<$1 is 153 Gpc$^{3}$, it is 453 Gpc$^{3}$ in the
redshift interval between z=1 and 2, and about 250 Gpc$^3$ in the redshift intervals between z=2 and 4 in z=0.5
intervals. Another effect maybe metallicity and the evolution of the star formation rate. As shown by \citet{grieco12},
the star formation rate in spiral galaxies
peaks at a redshift of about 3, which is consistent with the GRB rate shown in
Figure\,\ref{grb_rate}. The conclusion here is that in the local universe only a smaller number of very massive
stars is formed that will then end as a GRB, while at earlier epochs the rate of very massive stars has been higher.

\subsection{Redshift predictions for \swift-detected GRBs \label{redshift_discuss}}

Last but not least we raise the question whether we
 can make rough predictions on the X-ray light curve behavior based on
 prompt emission properties?
Can we use any relation between observed burst properties with redshift
to make predictions on the possible redshift of the burst?
We initially did this with the relation we found 
between the excess absorption in the X-ray afterglow X-ray spectrum 
\citep{grupe07}: GRBs with high excess absorption 
column density above the Galactic value are low redshift bursts.
 We found  similar relations between 
the BAT photon index $\Gamma_{\rm BAT}$ and the observed 15-150 keV 
fluence and redshift: GRBs with steep
 $\Gamma_{\rm BAT}$ ($>$2.0)
and high fluence ($>10^{-5}$ erg s$^{-1}$ cm$^{-2}$)
are only seen in bursts with redshifts $z<4$.  Again, we can 
only discriminate low-redshift GRBs by their 
observed properties. 
 One step further is the method we described at the end of
Section\,\ref{redshift}, where we use a PCA to determine an eigenvector1 
of a burst (see
also Appendix\,\ref{z_predict}) based on the observed BAT parameters $T_{\rm 90}$,
fluence and $\Gamma$. There is still strong scatter in this relation which still needs
independent redshift discriminators.

As we have shown in the discussion about the correlation analysis, 
most promising are the relations between the
energy/luminosity of the GRB's prompt emission and the decay slopes 
in the X-ray afterglow light curves. This, however, 
requires that the burst has to be observed until the decay slopes can be 
measured, which typically means at least several hours
after the burst for plateau phase as a minimum. 
So far no secure discriminator has been found for high-redshift bursts. 
Our plan is to apply machine learning techniques to the data set in order to
obtain better redshift predictions based on observed GRB properties measured from \swift\ data, similar to what has been 
shown by \citet{morgan11}. 
What is really needed is a relation between a redshift-independent parameter such as
$\Gamma$ and
redshift-dependent parameters such as energy or luminosity.

\section{Conclusions}
The main result of our statistical study of \swift-discovered GRBs is that we found
 new evidence that the GRB prompt and afterglow emission are linked.
This result is supported by the following findings:

\begin{itemize}
\item The BAT $T_{90}$ and the break times in the X-ray light curve before and 
after the plateau phase are strongly correlated (Sections,\ref{correlation}).
Bursts with longer $T_{90}$ tend to show later breaks in their X-ray afterglow
light curves. This is a statistically highly significant result, suggesting a
physical relationship between the high energy prompt and soft X-ray afterglow
emission. These correlations appear to be strong in the observed as well as in the
rest-frame.
\item The hard X-ray photon index $\Gamma$ of the prompt emission strongly 
anti-correlates with the decay slopes in the X-ray afterglow light curves. 
The observed fluence in the 15-150 keV BAT energy range anti-correlates with the decay
slopes in X-rays. Together with the $T_{90}$ - $T_{\rm break}$ correlations these
relations can be used to predict the behavior of the X-ray afterglow 
with a canonical light curve based on prompt
emission parameter measurements.
\item The Principal Component Analysis shows that the prompt and afterglow 
emission are driven 
primarily by the luminosity and/or isotropic energy release of the burst
(sections\,\ref{pca}, \ref{pca_discuss}).
\item A cluster analysis on the observed \swift-BAT parameters shows that short and 
long GRBs can be well-separated and
that the division between short and long-duration GRBs detected by the BAT 
is  less than 1s.
\item In addition we identified a third group in the cluster analysis: XRFs
\item The large number of \swift-detected bursts with redshift measurements allows to do
preform GRB cosmology (sections\,\ref{lum_func}, \ref{lum_func_discuss}).
The analysis of the GRB luminosity functions at different redshift intervals shows 
that the slope of the high-luminosity end of the luminosity function becomes steeper 
with increasing redshift.  
 The density of high luminous GRBs $L>10^{52}$ erg s$^{-1}$ rate peaks at a redshift of about z=3
and the results suggest a 'cosmic downsizing' of GRBs at lower redshifts, similar to what has been observed
for quasars. The peak of this GRB density is at a redshift of $~$2.7 which agrees with
the peak of the cosmic quasar density and the star formations rate history, suggesting a
close connection GRBs - star formation and quasar evolution.
\item The detection of a GRB is strongly depends on the BAT detection characteristics
(Section\,\ref{selection}). 
This is primarily a mixture
of the length of the  burst and the fluence. We can not detect very long bursts 
with low fluence.
\item Observed BAT parameters can be used to get a rough estimate of a burst redshift.
\item 
The statistical analysis presented in the paper can only be the beginning of
data mining the rich \swift\ GRB data set. In the future we need to look into
the analysis of the datasets including survival statistics to take upper and
lower limits, e.g. for the break times in the X-ray light curves into account.
Another important task in the future will be to develop a support vector machine
analysis that will allow to make predictions on the X-ray afterglow light curve
and its behavior based on prompt emission properties. Last but not least we
need to examine the possibility of estimating redshift predictions similar to
what has been done by \citet{morgan11}. Ultimately it is important that
ground-based observes increase the number of afterglows for which they obtain
spectroscopic data to measure redshifts. Only with redshifts and therefore
distance measurements we are able to derive the intrinsic physical properties of
the bursts. 
\end{itemize}

\acknowledgments
We would like to thank all observers at ground-based 
optical telescopes for their effort to obtain redshifts of the 
\swift\ afterglows.  We would also like to thank the anonymous referee 
for detailed and supportive reports.
We thank Raffaella Margutti for discussions on the PCA in her 2013 paper.
This research has made use of the
  XRT Data Analysis Software (XRTDAS) developed under the responsibility
  of the ASI Science Data Center (ASDC), Italy.
This research has made use of data obtained through the High Energy 
Astrophysics Science Archive Research Center Online Service, provided by the 
NASA/Goddard Space Flight Center. This work made use of data supplied
 by the UK Swift Science Data Centre at the University of Leicester.
PV thanks NASA NNX13AH50G and OTKA K077795, and Peter \meszaros for discussions.
At Penn State we acknowledge support from the NASA Swift program
through contract NAS5-00136.



\clearpage

\newcommand\plottwor[2]{{%
\typeout{Plottwo included the files from R #1 #2}
\centering
\leavevmode
\columnwidth=.65\columnwidth
\includegraphics*[angle=270, width={\eps@scaling\columnwidth}]{#1}%
\hfil
\includegraphics*[angle=270, width={\eps@scaling\columnwidth}]{#2}%
}}

\begin{figure*}
%\epsscale{0.75}
\epsscale{2.0}
\plottwor{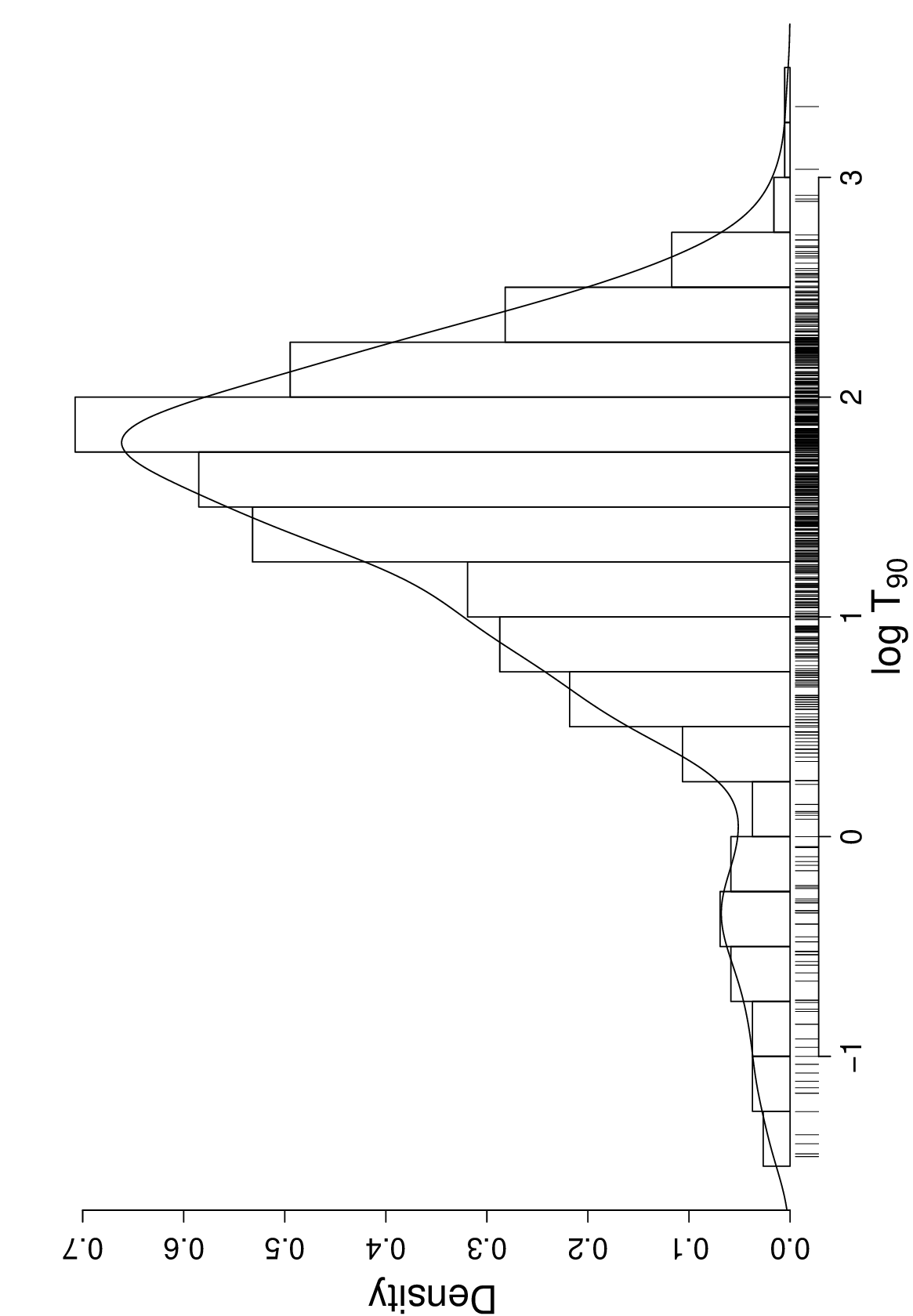}{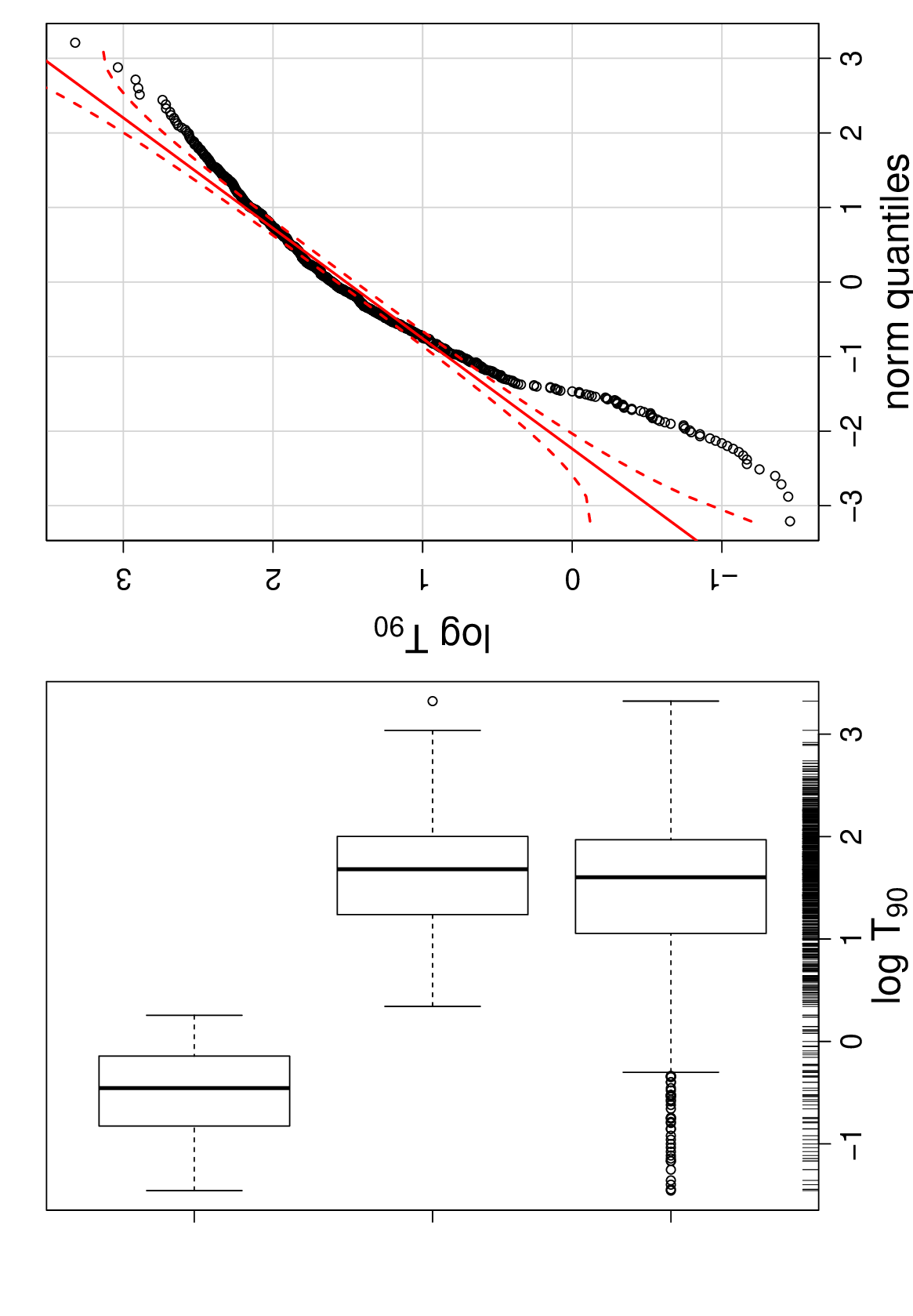}

\plottwor{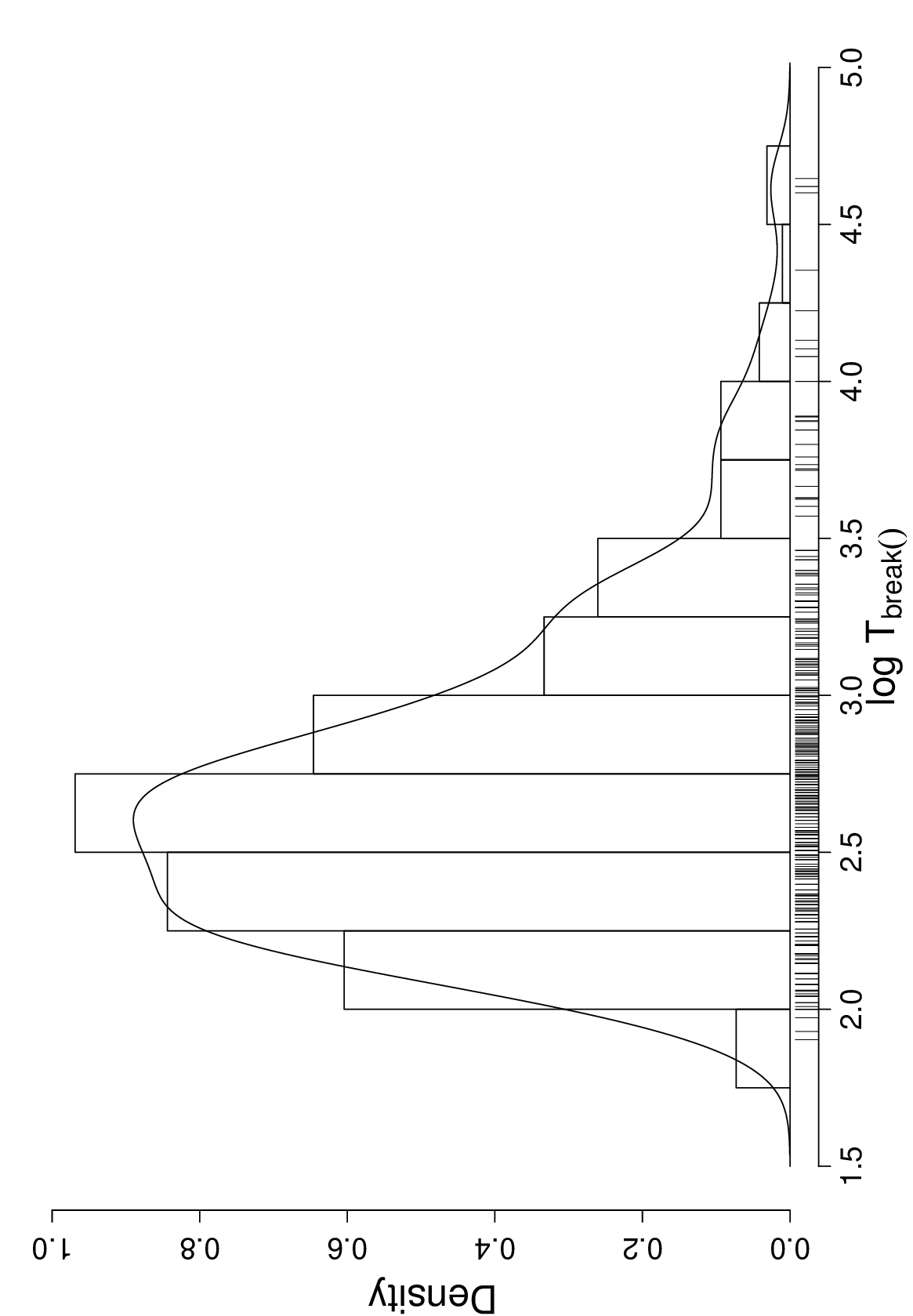}{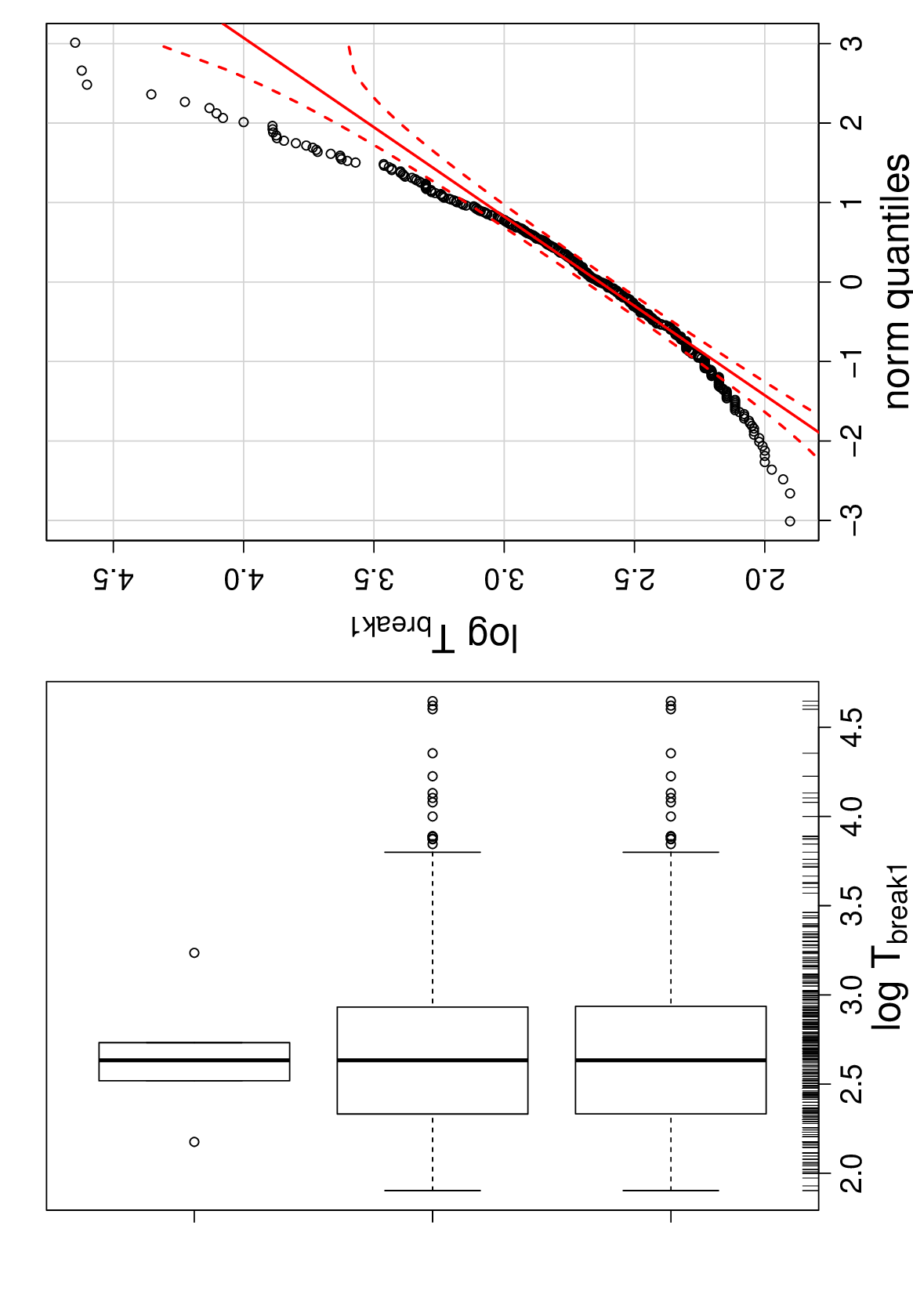}

\plottwor{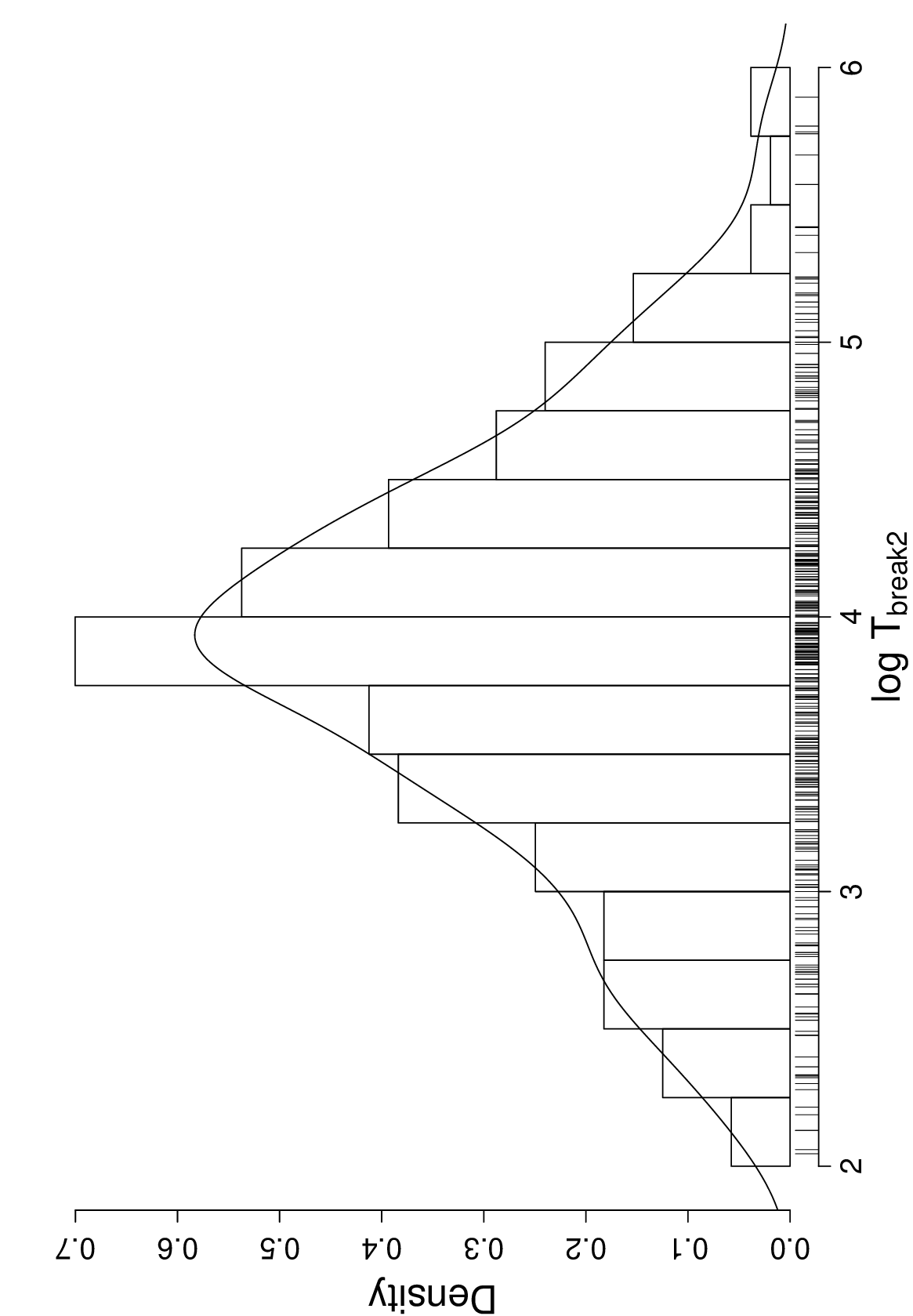}{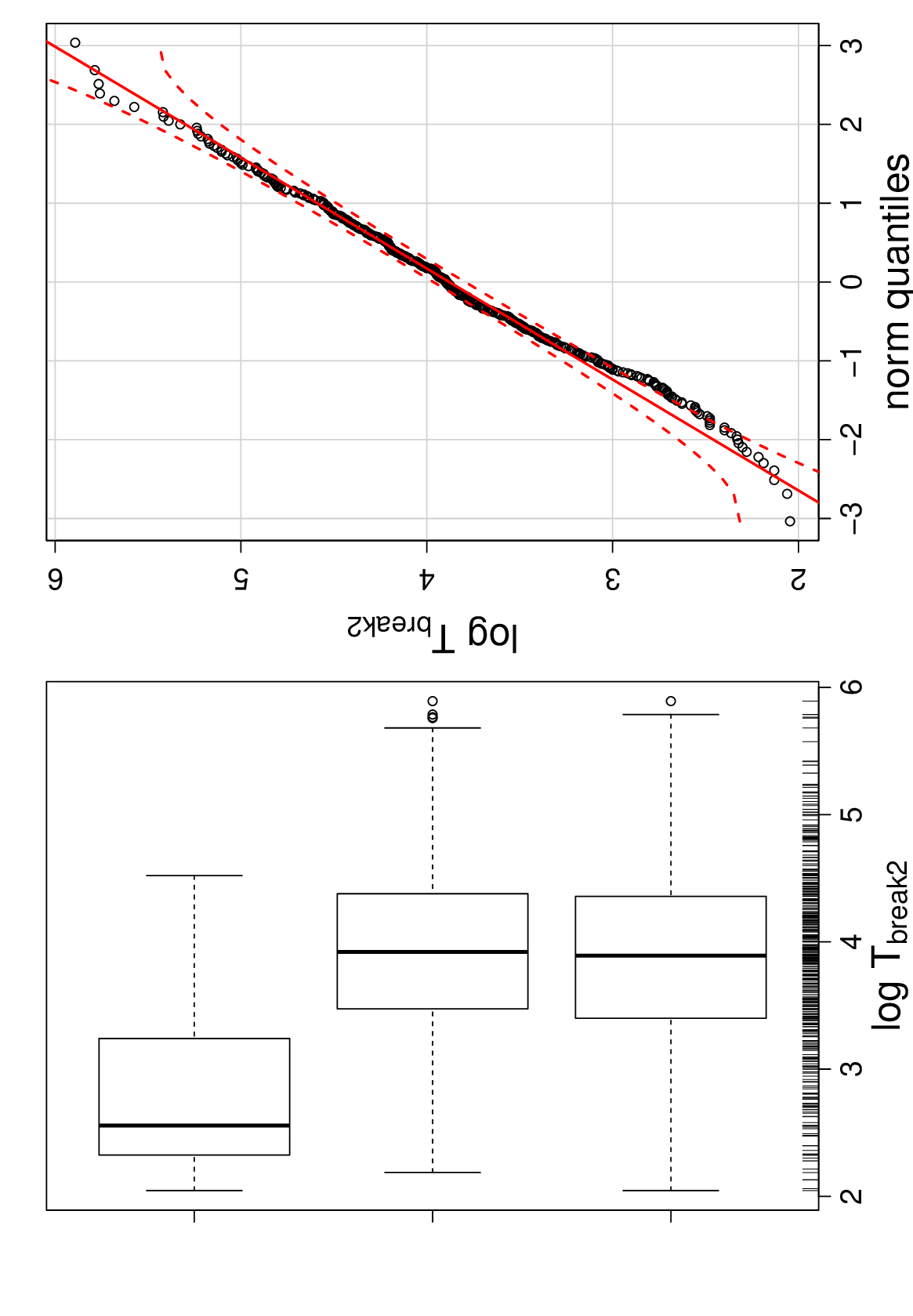}
\caption{\label{distr_t90_tb} 
Histograms, box plots and qq plots of the 
distributions of the 15-150 keV BAT $T_{90}$  (upper panels), the X-ray afterglow 
light curve break times $T_{\rm break 1}$ (middle panel) and $T_{\rm break 2}$
(lower panels)
before and after the plateau phase, respectively in the observed frame. 
The solid line in the histogram displays the kernel density estimator
\citep[e.g.][]{everitt10, feigelson12}.
In the box plots,
short bursts are displayed on top, long bursts in the middle and all bursts on the
bottom.
}
\end{figure*}

\begin{figure*}
%\epsscale{0.75}
\epsscale{1.5}
\plottwor{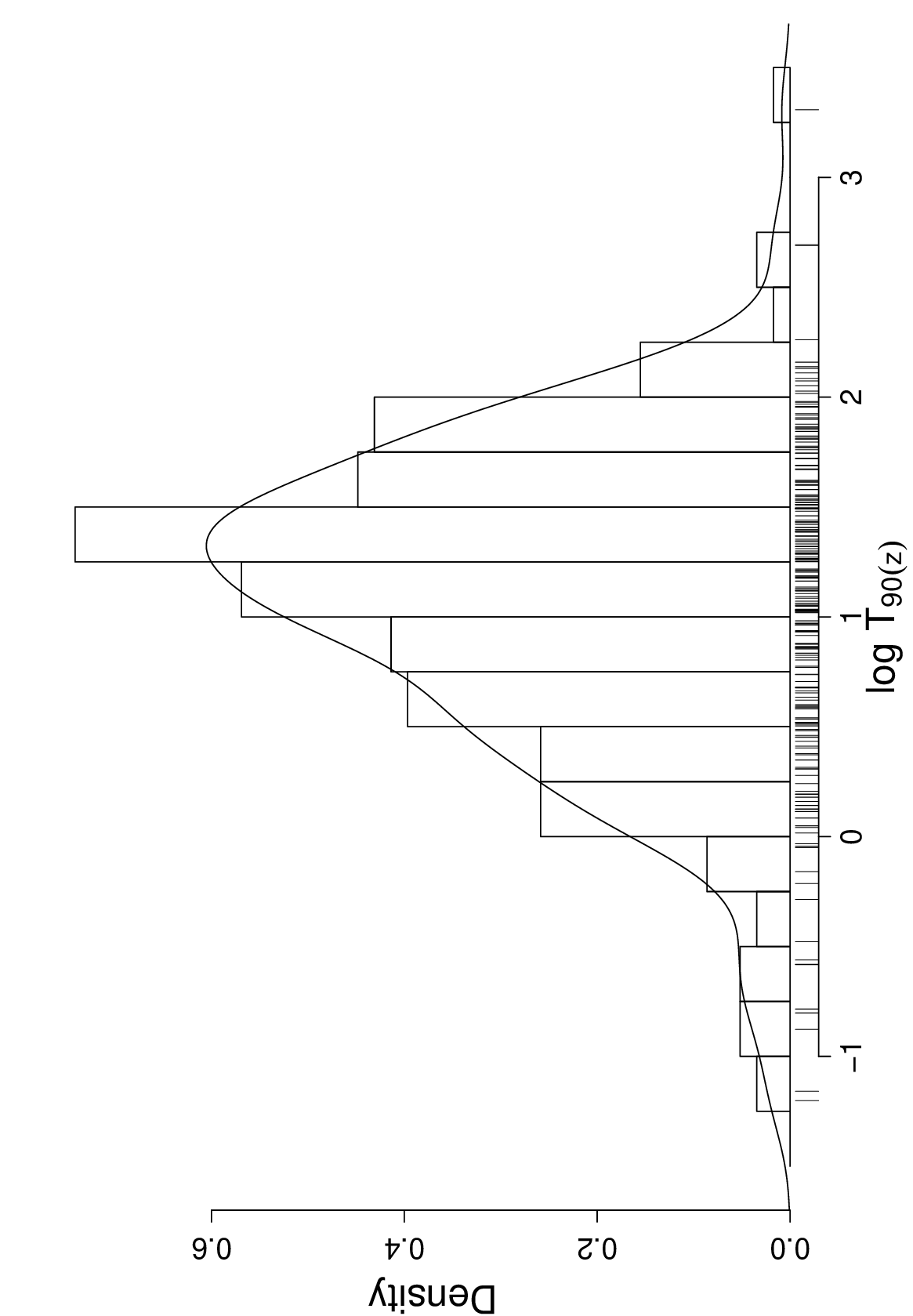}{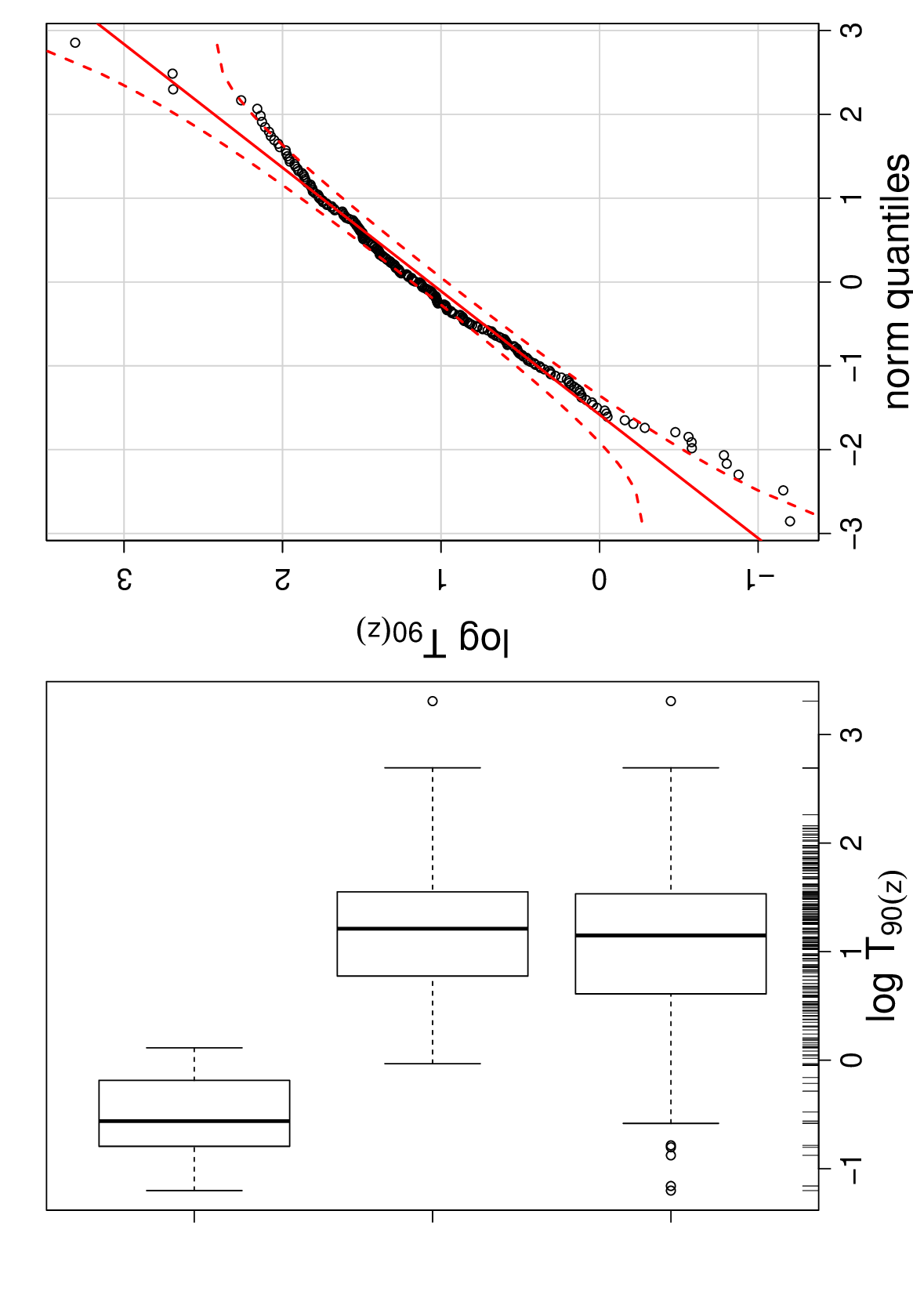}

\plottwor{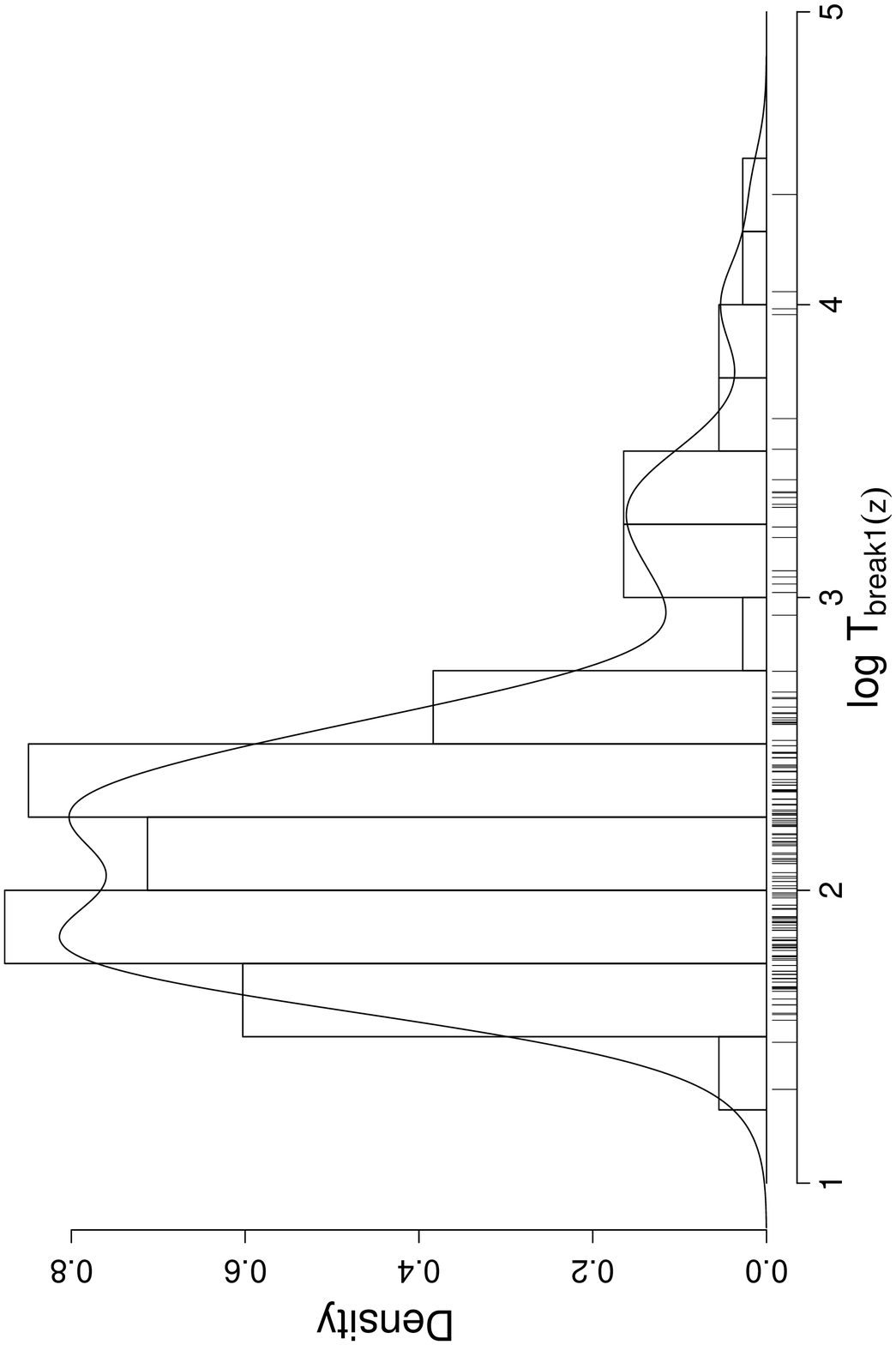}{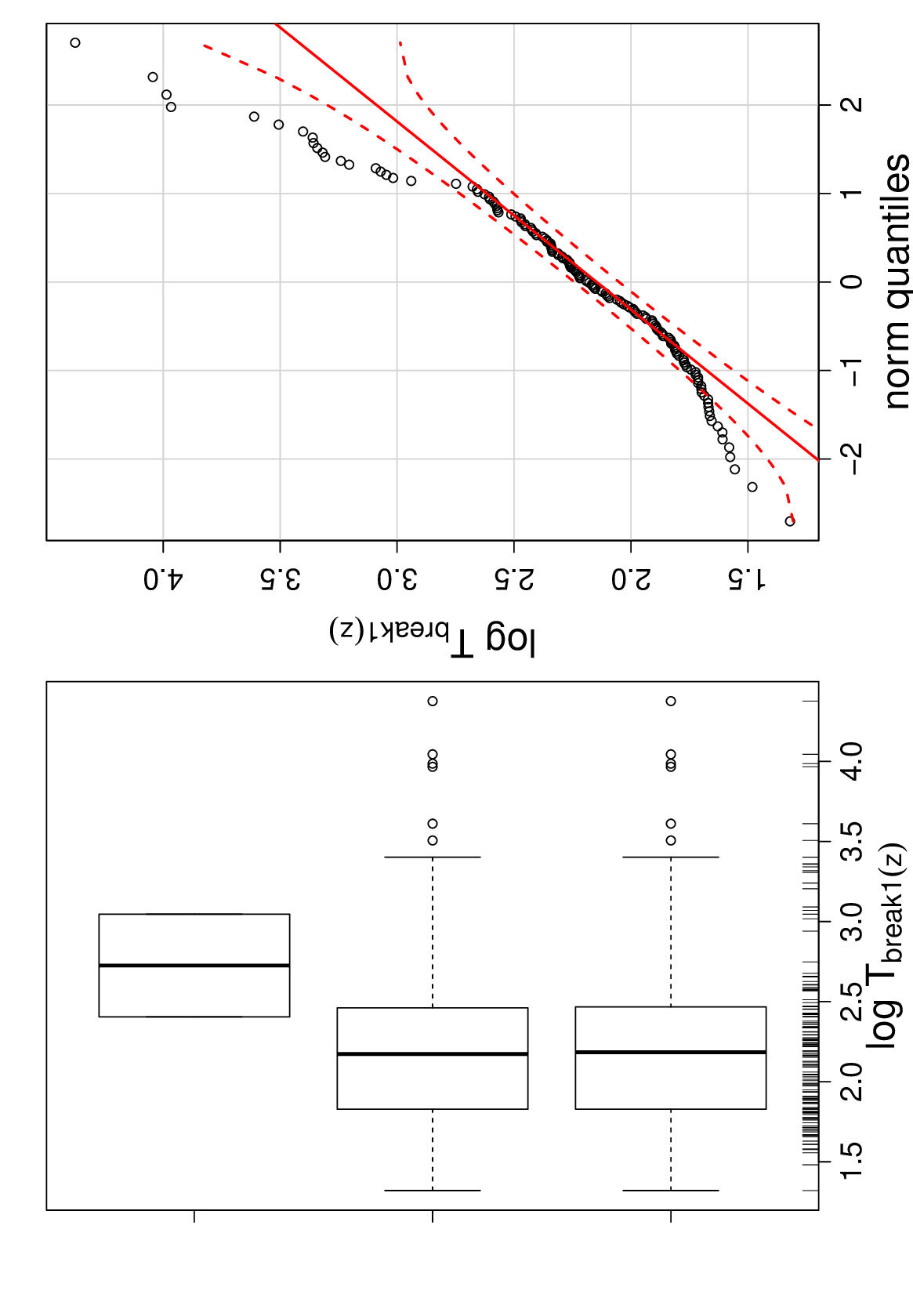}

\plottwor{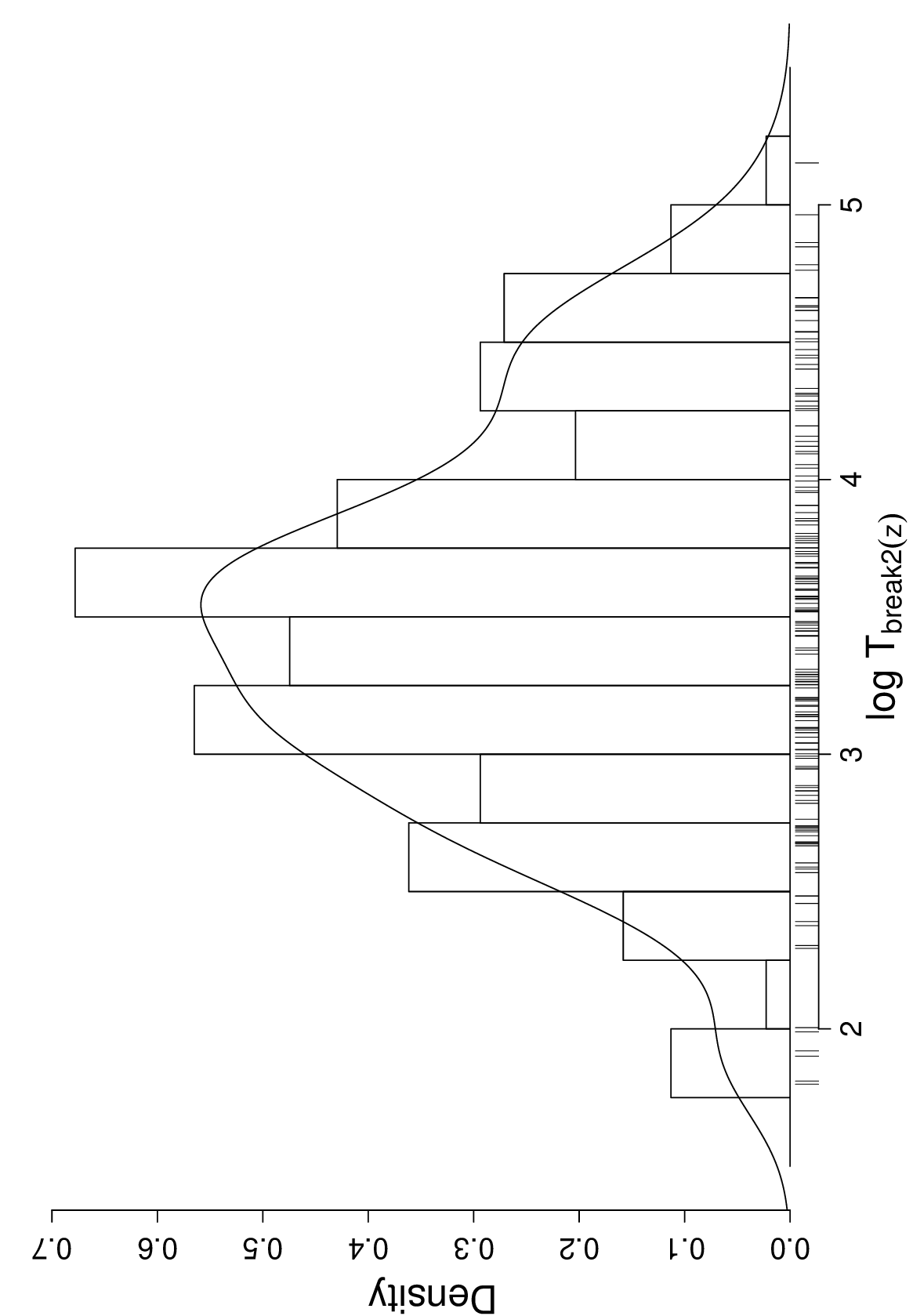}{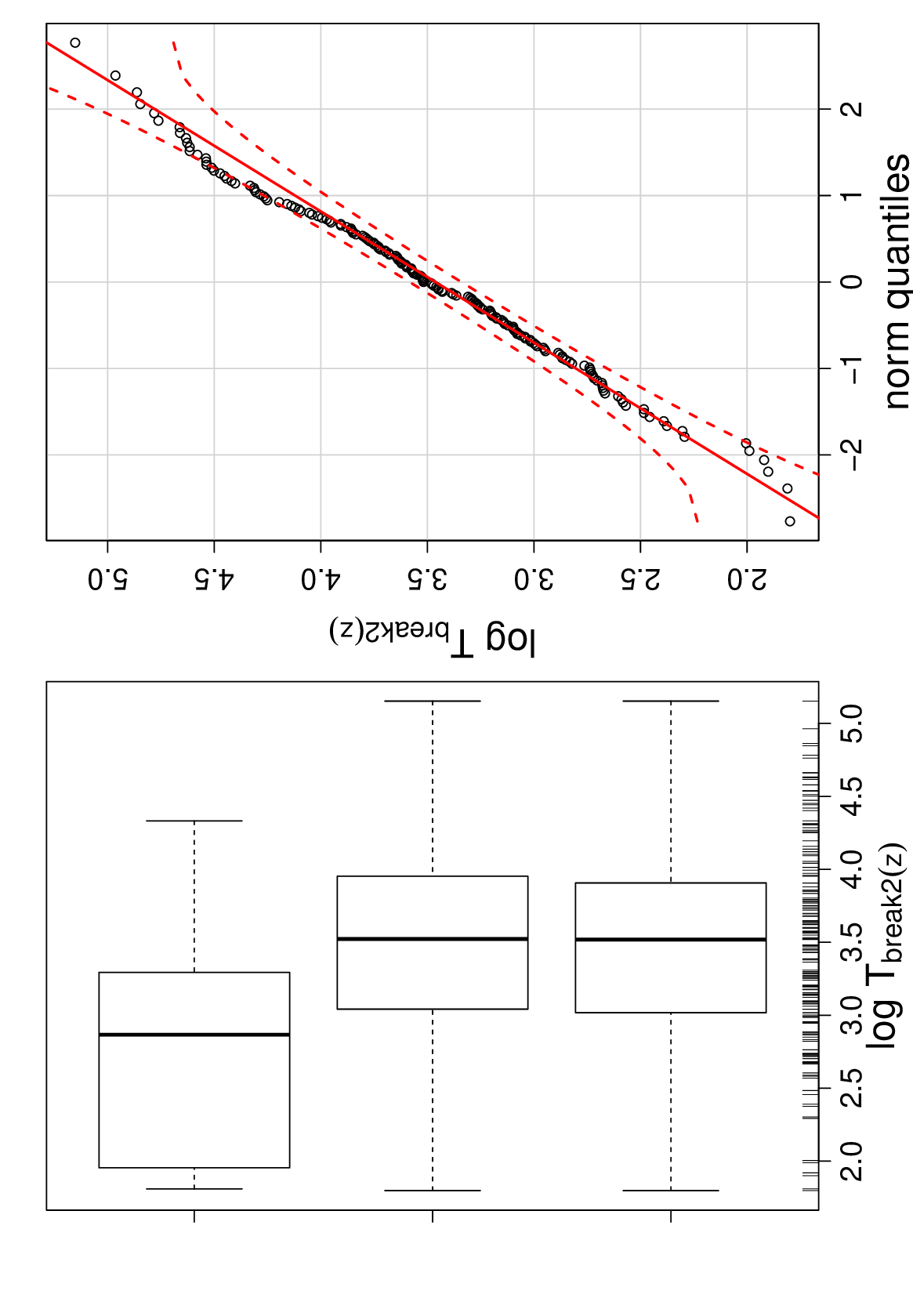}
\caption{\label{distr_t90z_tbz} 
Same as Figure\,\ref{distr_t90_tb}, but all times in the rest-frame.
}
\end{figure*}

\begin{figure*}
%\epsscale{0.75}
\epsscale{1.5}
\plottwor{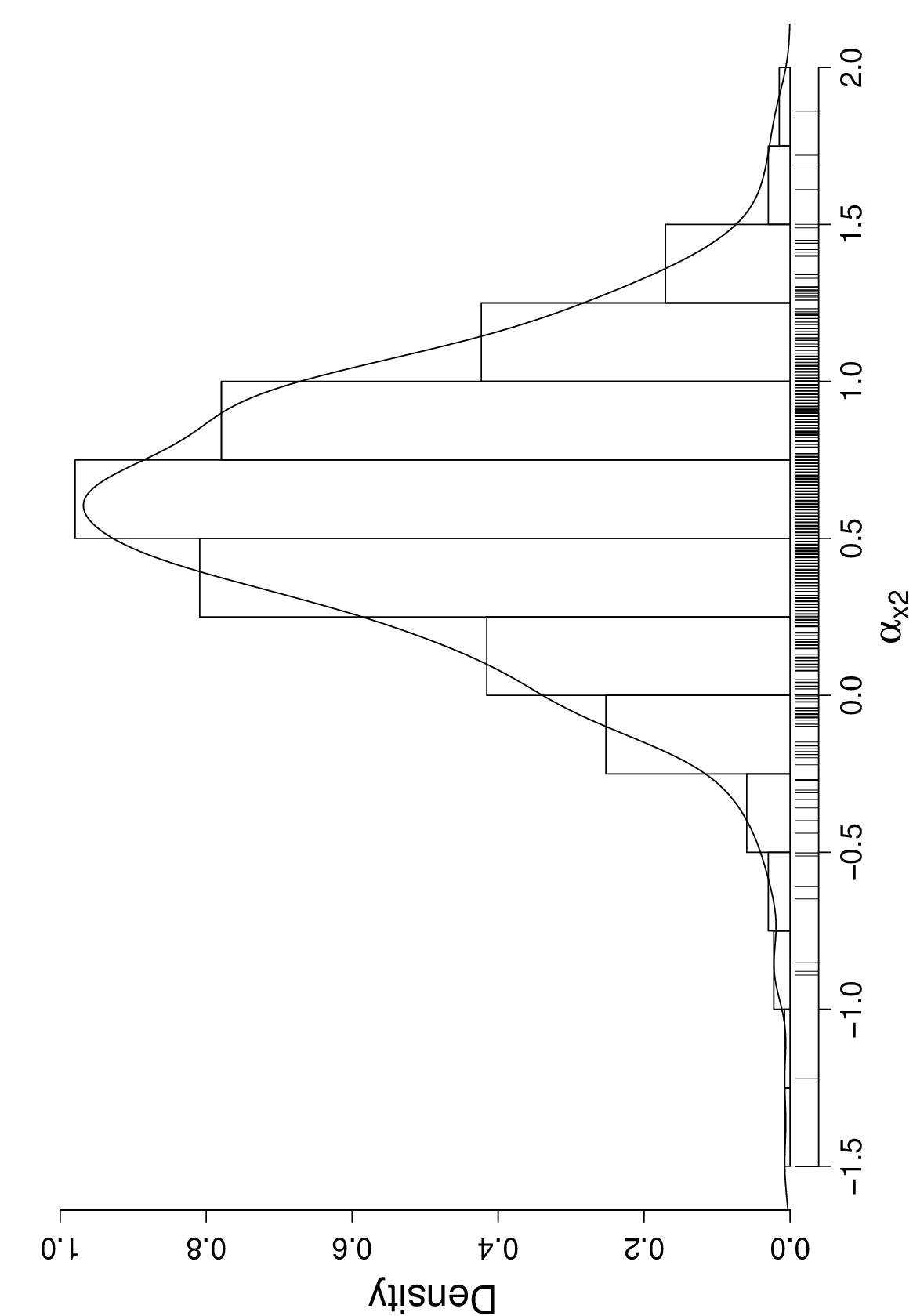}{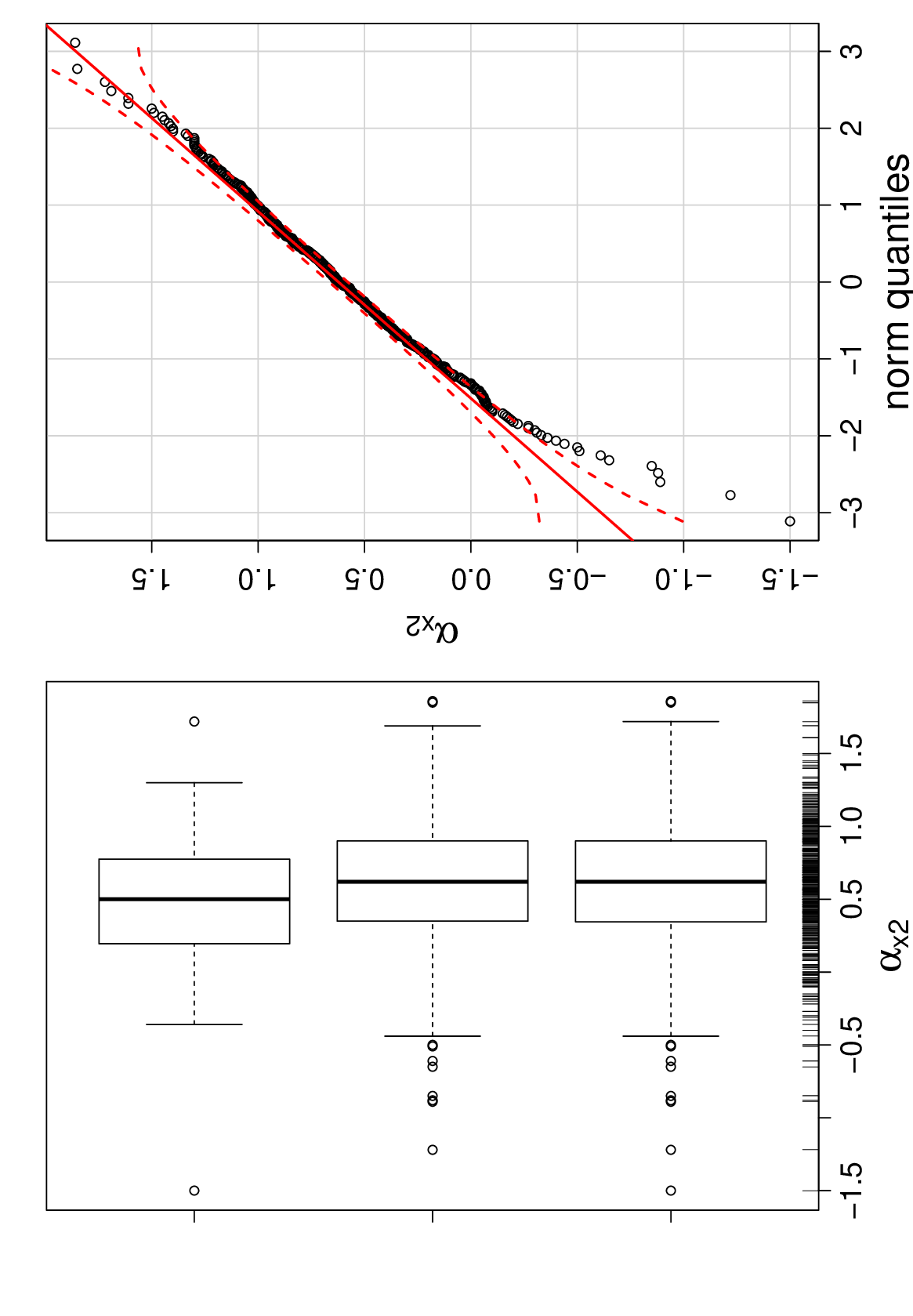}

\plottwor{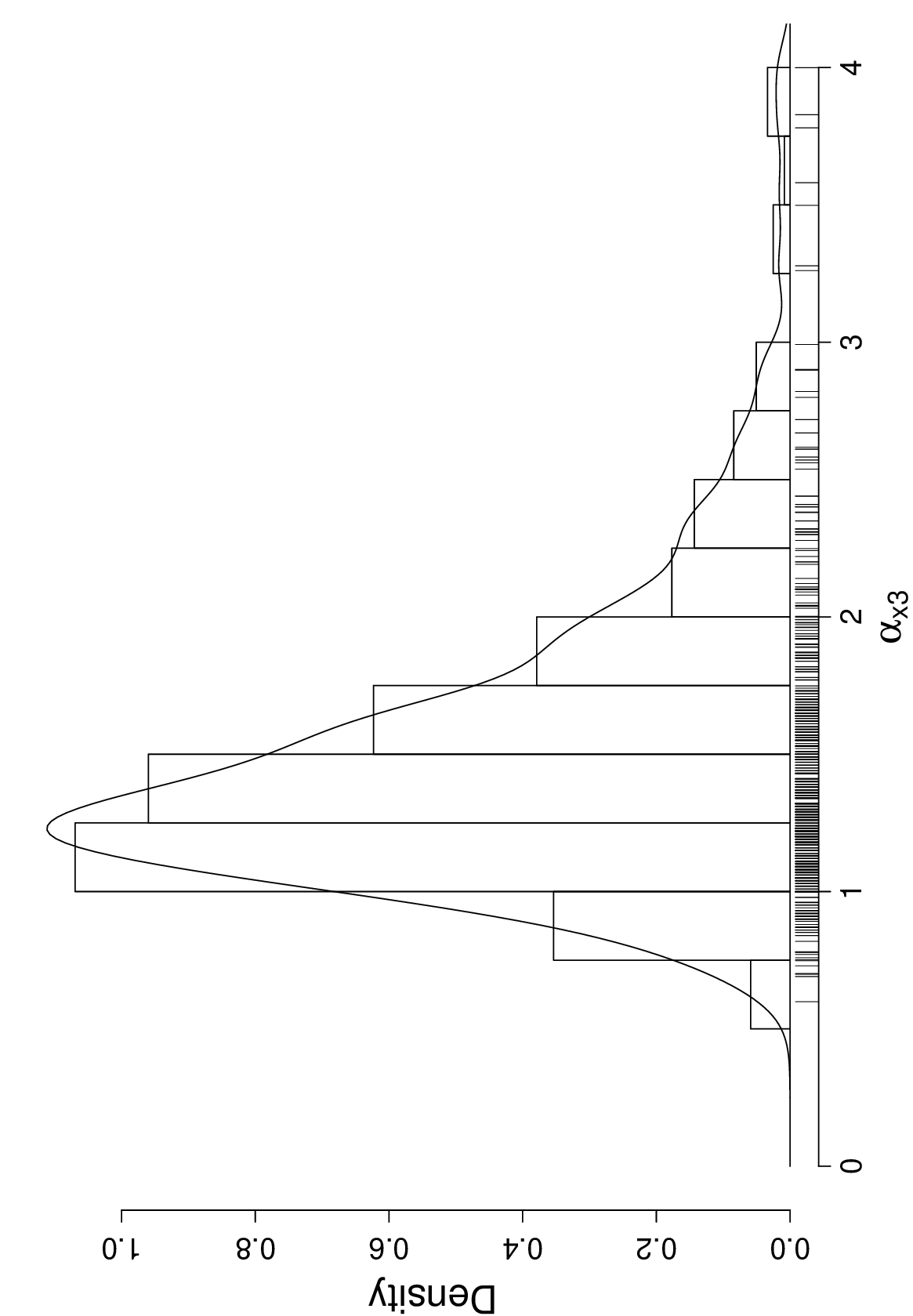}{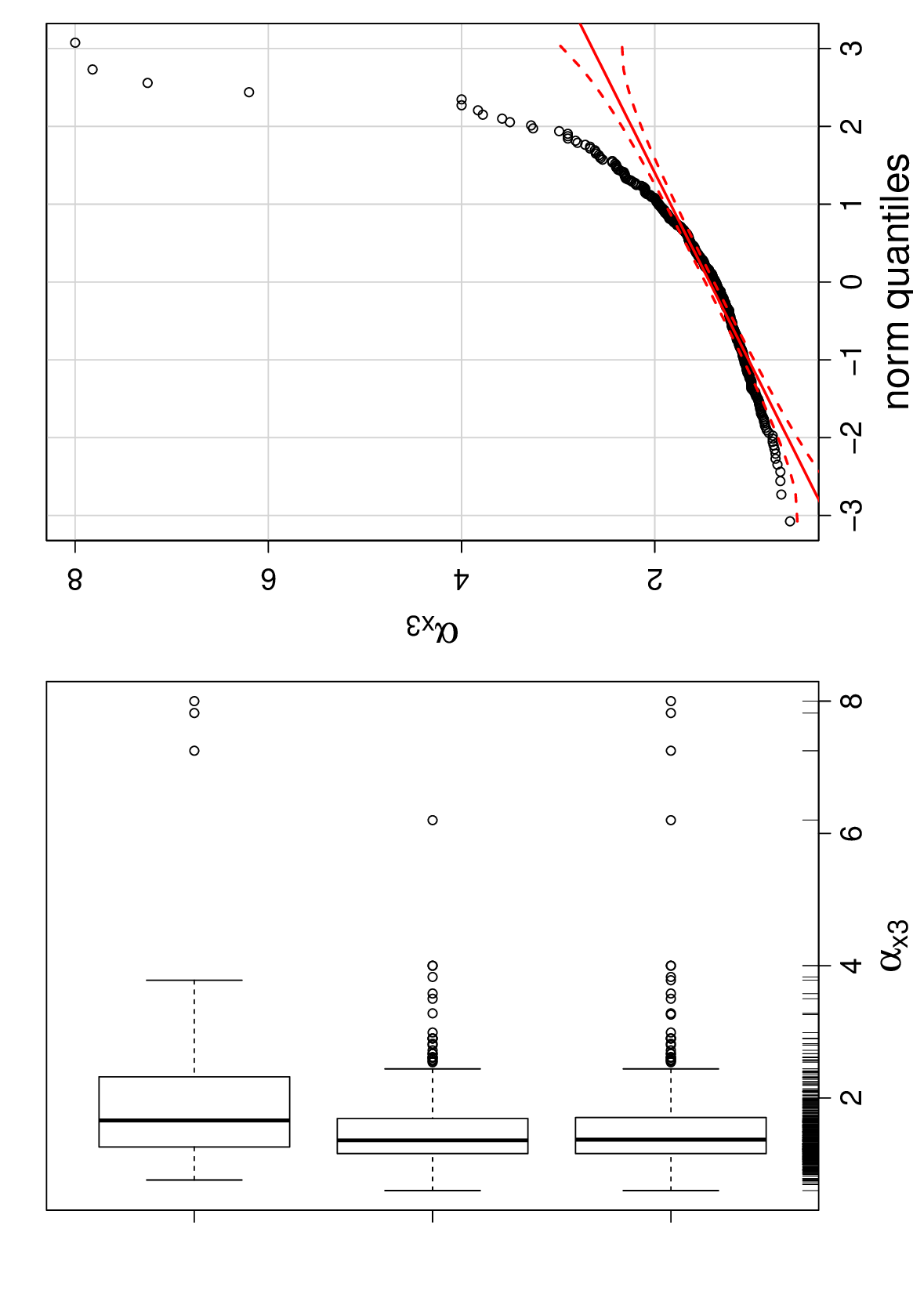}
\caption{\label{distr_ax} 
Distributions of the X-ray light curve decay slopes during the plateau and the
normal decay phase (upper and panel panel, respectively). As in
Figure\,\ref{distr_t90_tb}, the histogram, box plot, and q-q plot are shown. 
In the box plots,
short bursts are displayed on top, long bursts in the middle and all bursts on the
bottom.
}
\end{figure*}

\clearpage

\begin{figure*}
%\epsscale{0.75}
\epsscale{1.5}
\plottwor{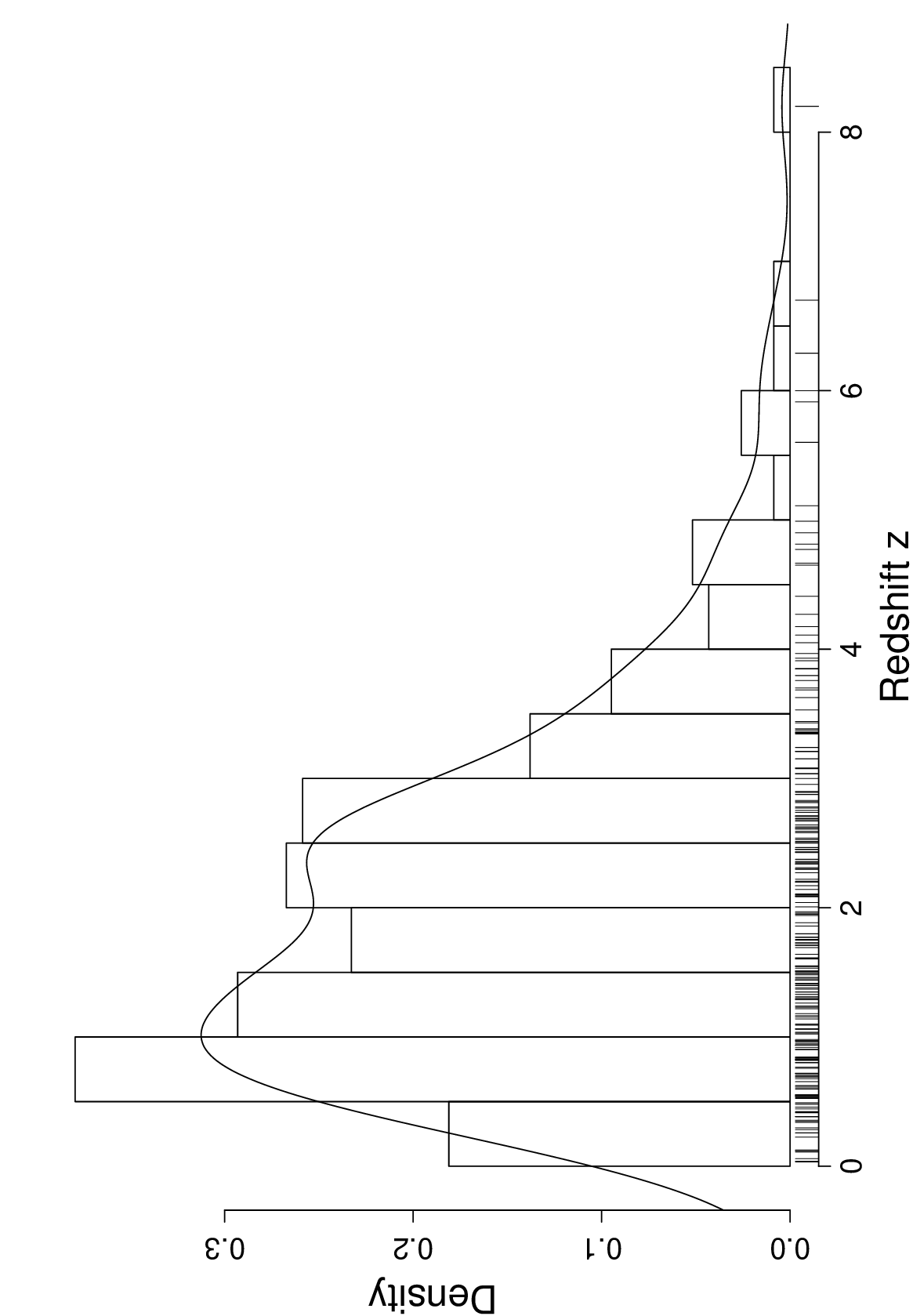}{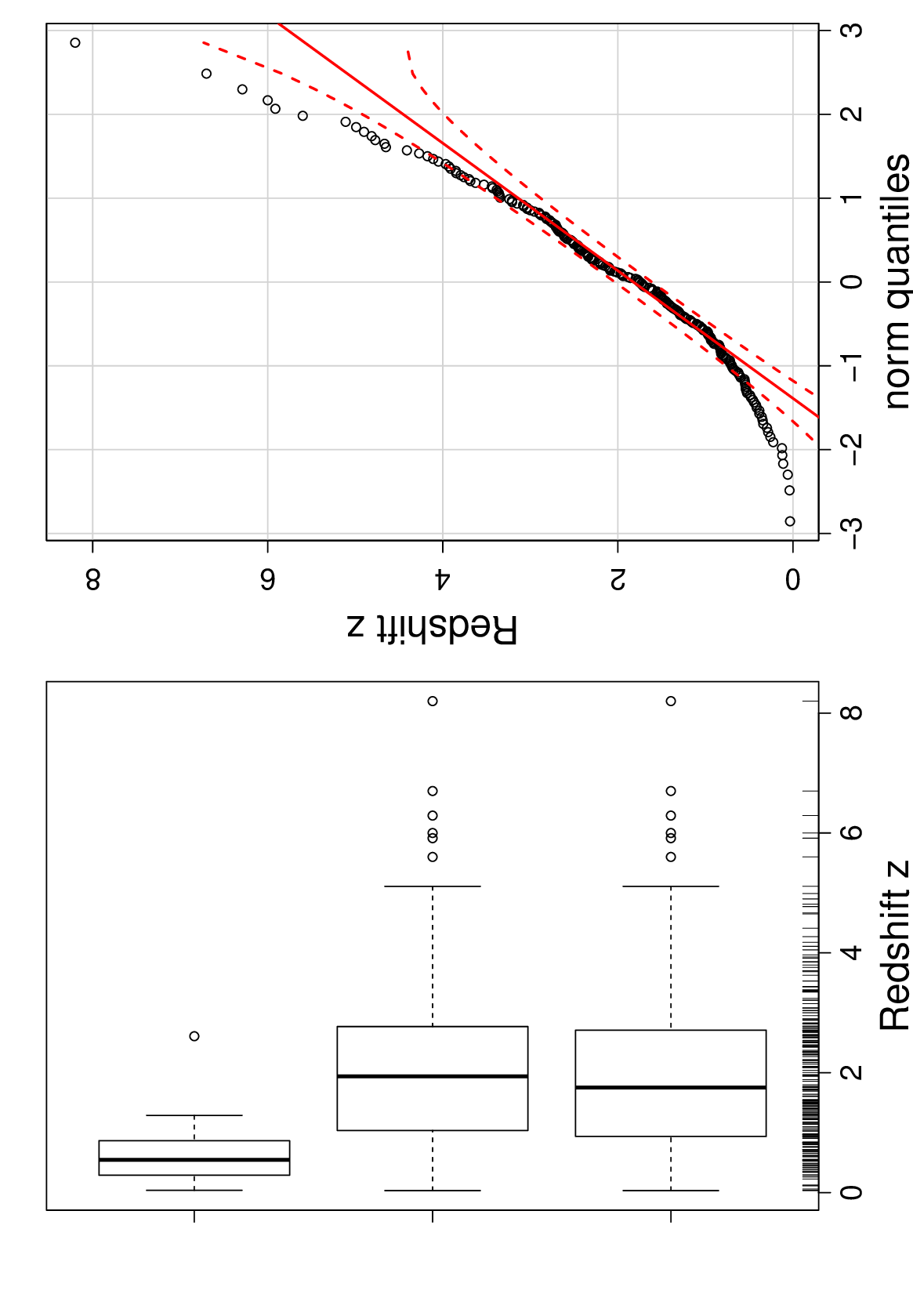}
\caption{\label{distr_z} 
Redshift distribution of all \swift-detected GRBs with spectroscopic 
redshifts. As in
Figure\,\ref{distr_t90_tb}, the histogram, box plot, and q-q plot are shown. 
In the box plots,
short bursts are displayed on top, long bursts in the middle and all bursts on the
bottom.
}
\end{figure*}

\clearpage

\clearpage

\clearpage

\clearpage

\clearpage

\begin{figure}
%\epsscale{0.75}
\epsscale{0.75}
\plotone{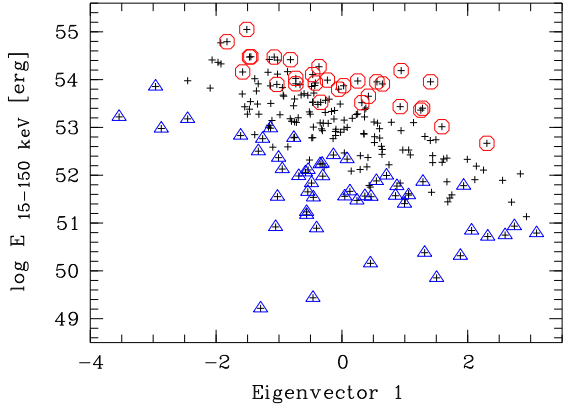}
\caption{\label{ev1_eiso} 
Eigenvector 1 from a PCA on long GRBs using $T_{90}$, Fluence, and $\Gamma$ as input properties vs. isotropic energy $E_{\rm iso}$ in the 15-150 keV band. 
Bursts with redshifts $z>3.5$ are marked as red circles and bursts with a redshift $z<1.0$ are marked as blue triangles.
}
\end{figure}

\begin{figure}
\epsscale{0.75}
\plotone{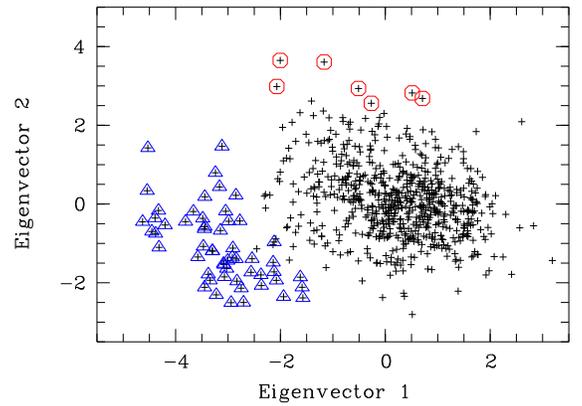}
\caption{\label{cluster_ev1_ev2} Eigenvector 1 vs. Eigenvector 2 plot of the GRBs of 
the cluster analysis using 
log $T_{90}$, $\Gamma$, and log 15-150 keV fluence.  The different groups are marked as
defined in Figure\,\ref{cluster_corr}. 
}
\end{figure}

\begin{deluxetable}{lcccccccccccccccc}
\rotate
\tabletypesize{\tiny}
\tablecaption{Mean, standard deviation, and median of GRB properties in the observed and rest-frame for all GRBS, long
and short duration GRBs, and high-redshift bursts (z$>$4.0).
\label{grb_statistics}}
\tablewidth{0pt}
\tablehead{
& \multicolumn{4}{c}{All GRBS} & \multicolumn{4}{c}{Long GRBs} & \multicolumn{4}{c}{Short GRBs} & \multicolumn{4}{c}{z$>4$ bursts }\\
\colhead{Property} 
& \colhead{Mean}
& \colhead{SD\tablenotemark{1}}
& \colhead{Median} 
& \colhead{\# of GRBs}
& \colhead{Mean}
& \colhead{SD\tablenotemark{1}}
& \colhead{Median} 
& \colhead{\# of GRBs}
& \colhead{Mean}
& \colhead{SD\tablenotemark{1}}
& \colhead{Median} 
& \colhead{\# of GRBs}
} 
\startdata
$\Gamma_{\rm BAT}$ & 1.614 & 0.420 & 1.600 & 751 & 1.651 & 0.400 & 1.620 & 687 & 1.205 & 0.418 & 1.100 & 63 & 1.554 & 0.260 & 1.600 & 18 \\
\bx & 1.002 & 0.432 & 0.970 & 632  & 1.007 & 0.429 & 0.970 & 592 & 0.934 & 0.478 & 0.830 & 39 & 0.974 & 0.171 & 0.960 & 18 \\
log $T_{90}$ & 1.435 & 0.798 & 1.601 & 753 & 1.612 & 0.550 & 1.681 & 690 & -0.501 & 0.465 & -0.456 & 63 & 1.557 & 0.554 & 1.354 & 18 \\
log $T_{90-z}$ & 1.074 & 0.706 & 1.129 & 232 & 1.184 & 0.580 & 1.208 & 217 & -0.511 & 0.404 & -0.561 & 15 & 0.770 & 0.561 & 0.535 & 18 \\
log $T_{\rm break-1}$ & 2.706 & 0.493 & 2.638 & 384 & 2.705 & 0.494 & 2.633 & 377 & 2.655 & 0.347 & 2.568 & 6 & 2.792 & 0.294 & 2.000 & 11 \\
log $T_{\rm break-1-z}$ & 2.258 & 0.564 & 2.190 & 146 & 2.252 & 0.564 & 2.178 & 144 & 2.725 & 0.453 & 2.725 & 2 & 1.984 & 0.285 & 1.881 & 11 \\
log $T_{\rm break-2}$ & 3.866 & 0.748 & 3.897 & 417 & 3.917 & 0.712 & 3.924 & 389 & 2.802 & 0.704 & 2.556 & 19 & 3.987 & 0.597 & 4.080 & 16 \\
log $T_{\rm break-2-z}$ & 3.484 & 0.697 & 3.365 & 177 & 3.512 & 0.674 & 3.522 & 170 & 2.786 & 0.935 & 2.867 & 7 & 3.201 & 0.615 & 3.255 & 16 \\
\axb & 0.592 & 0.439 & 0.620 & 539 & 0.597 & 0.427 & 0.630 & 515 & 0.455 & 0.639 & 0.500 & 23 & 0.582 & 0.375 & 0.660 & 17 \\
\axc & 1.550 & 0.748 & 1.380 & 475 & 1.499 & 0.555 & 1.370 & 450 &  2.475 & 2.088 & 1.660 & 25 & 1.591 & 0.476 & 1.490 & 17 \\
log fluence\tablenotemark{2} & -5.919 & 0.676 & -5.907 & 752 & -5.813 & 0.585 & -5.848 & 689 & -7.080 & 0.499 & -7.134 & 63 & -5.880 & 0.460 & -5.930 & 18 \\
log fluence-k\tablenotemark{3} &  -5.560 & 0.749 & -5.536 & 232 & -5.477 & 0.679 & -5.491 & 217 & -6.761 & 0.687 & -6.768 & 15 & -5.528 & 0.568 & -5.586 & 18 \\
log $L_{15-150 keV}$\tablenotemark{4} & 51.540 & 1.240 & 51.678 & 231 & 51.603 & 1.210 & 51.759 & 216 & 50.637 & 1.350 & 51.111 & 15 & 53.158 & 0.459 & 53.090 & 18 
\\
log $E_{\rm iso}$\tablenotemark{5} & 52.610 & 1.272 & 52.856 & 231 & 52.782 & 1.064 & 52.965 & 216 & 50.126 & 1.458 & 50.472 & 15 & 53.928 & 0.600 & 53.953 & 18 \\
$E_{\rm peak}$\tablenotemark{6} & 254.1 & 376.0 & 148.0 & 206 & 225.3 & 246.4 & 143.0 & 193 & 686,4 & 1108 & 370 & 13 & 154.4 & 96.5 & 122.5 & 8 \\
$E_{\rm peak, z}$\tablenotemark{6} & 750.6 & 5 & 893.2 & 104 & 669.2 & 532.5 & 537.4 & 98 & 2008 & 2947 & 933 & 6 & 966.4 & 570.2 & 824.0 & 8 \\
log $(\Delta_{\rm NH}+1)$\tablenotemark{7} & 0.606 & 0.715 & 0.000 & 618 & 0.626 & 0.715 & 0.000 & 578 & 0.377 & 0.698 & 0.000 & 39 & 0.259 & 0.380 & 0.000 & 18 \\
z & 2.018 & 1.373 & 1.763 & 232 & 2.111 & 1.363 & 1.950 & 217 & 0.681 & 0.640 & 0.547 & 15 & 5.201 & 1.080 & 4.860 & 18 \\
\enddata

\tablenotetext{1}{Standard deviation SD}
\tablenotetext{2}{Observer Fluence in the 15-150 keV BAT energy band
 in units of   erg cm$^{-2}$}
\tablenotetext{3}{K-corrected 15-150 keV Fluence
 in units of   erg  cm$^{-2}$}
\tablenotetext{4}{rest frame 15-150 keV luminosity given in unites of erg s$^{-1}$}
\tablenotetext{5}{k-corrected isotropic energy $E_{\rm iso}$ in the 15 - 150 keV BAT band}
\tablenotetext{6}{Peak energy $E_{\rm peak}$ in units of keV}
\tablenotetext{7}{The excess absorption column density $\Delta_{\rm NH}$ above the galactic value \citep[see ][]{grupe07} 
is given in units of 10$^{20}$ cm$^{-2}$}

\end{deluxetable}

\hspace{-5cm}
\begin{deluxetable}{lccccccccc}
\rotate
\tabletypesize{\tiny}
\tablecaption{Spearman rank order correlation 
and Student's t-test for the observed parameters of all \swift-detected GRBs.
\tablenotemark{1}
 \label{correlation_tab_all}}
\tablewidth{0pt}
\tablehead{
& \colhead{$\Gamma_{\rm BAT}$} 
& \colhead{\bx} 
& \colhead{\axb} 
& \colhead{\axc} 
& \colhead{$T_{\rm 90}$} 
& \colhead{$T_{\rm break 1}$} 
& \colhead{$T_{\rm break 2}$} 
& \colhead{15-150 keV fluence}
}
\startdata
$\Gamma_{\rm BAT}$   & ---   & 632, 6.23$\times 10^{-6}$ & 539, $< 10^{-8}$ & 475, $<10^{-8}$ & 750, 0.0231 & 384, 0.1638 & 417, 0.0162
& 750, 1.85$\times 10^{-4}$ \\
\bx    & $+$0.186, $+$4.557 & --- & 539, $4.55\times 10^{-4}$ & 473, $2.71\times 10^{-4}$ & 631, 0.4148 & 384, 0.1459 & 417, $9.33\times 10^{-4}$ 
& 631, 0.7742\\
\axb   & $-$0.259, $-$6.215 & $-$0.150, $-$3.528 & --- & 413, $<10^{-8}$ & 538, $.7.2\times 10^{-7}$ & 383, 0.4634 & 414, $1.7\times 10^{-7}$ 
& 538, $<10^{-8}$ \\
\axc   & $-$0.313, $-$7.181 & $-$0.167, $-$3.669 & $+$0.405, $+$8.980 & --- & 475, $1\times 10^{-8}$ & 278, 0.0387  & 414, $1.97\times 10^{-6}$
& 475, 5.6$\times 10^{-7}$ \\
$T_{\rm 90}$       &  $+$0.083, $+$2.279 & $+$0.033, $+$0.816 & $+$0.212, $+$5.015 & $+$0.262, $+$5.903 & --- & 383, $<10^{-8}$ & 417, $<10^{-8}$ 
& 752, $<10^{-8}$ \\
$T_{\rm break 1}$  &  $+$0.071, $+$1.395 & $-$0.074, $-$1.457 & $-$0.038, $-$0.734 & $+$0.124, $+$2.077 & $+$0.328, $+$6.776 & --- & 277, $<10^{-8}$ 
& 383, 0.2121 \\
$T_{\rm break 2}$  &  $+$0.118, $+$2.414 & $+$0.161, $+$3.334 & $+$0.254, $+$5.321 & $+$0.231, $+$4.825 & $+$0.495, 11.615 & $+$0.516, $+$9.994 & ---  
& 417, 7.49$\times 10^{-5}$ \\
15-150 keV fluence & $-$0.136, $-$3.758 & $+$0.011, $+$0.286 & $+$0.268, $+$6.450 & $+$0.227, $+$5.075 & $+$0.659, $+$23.99 & $-$0.064, $-$1.253
& $+$0.193, $+$3.999 & --- \\
\enddata

\tablenotetext{1}{The values below the diagonal list the Spearman rank order correlation coefficient $r_s$ and the
Student's T-test value $T_s$. The values above the diagonal list the number of GRBs in this correlation and the
probability $P$ that the result is drawn from a random distribution.}

\end{deluxetable}

\begin{deluxetable}{lccccccccccc}
\rotate
\tabletypesize{\tiny}
\tablecaption{Spearman rank order correlation 
and Student's t-test for the rest-frame parameters of  \swift-detected GRBs with spectroscopic redshifts.
\tablenotemark{1}
 \label{correlation_tab_z}}
\tablewidth{0pt}
\tablehead{
& \colhead{$\Gamma_{\rm BAT}$}
& \colhead{\bx} 
& \colhead{\axb} 
& \colhead{\axc} 
& \colhead{$T_{\rm 90,z}$}
& \colhead{$T_{\rm break 1,z}$}
& \colhead{$T_{\rm break 2,z}$}
& \colhead{$L_{\rm 15-150 keV}$}
& \colhead{$E_{\rm iso}$}
& \colhead{$E_{\rm peak, z}$} 
}
\startdata
$\Gamma_{\rm BAT}$   & ---    & 230, $1.68\times 10^{-3}$ & 208, $6.95\times 10^{-5}$ & 188, $1.68\times 10^{-4}$ & 232, 0.1360 & 144, 0.1991 & 
  175, 0.047 & 231, $<10^{-8}$ &  231, $4\times 10^{-8}$ & 104, $1.12\times 10{-5}$ \\
\bx                  & $+$0.206, $+$3.179 & --- & 208, 0.011 & 188, 2.75$\times 10^{-4}$ & 230, 0.2813 & 144, 0.2133 & 175, $4.61\times 10^{-4}$ &
  229, 0.0500 & 229, 0.0701 & 103, 0.1557 \\
\axb                 & $-$0.272, $-$4.064 & $-$0.175, $-$2.556 & --- & 175, 1.96$\times 10^{-5}$ & 208, 0.0197 & 143, 0.4256 & 
   175, 0.1557, & 207, 0.0702 & 207, 1.22$\times 10^{-3}$ & 98, 4.12$\times 10^{-3}$ \\
\axc                 & $-$0.271, $-$3.839 & $-$0.262, $-$3.708 & $+$0.317, $+$4.393 & --- & 188, 9.06$\times 10^{-3}$ & 115, 0.2404 & 174, 0.5166 &
   188, 0.0225 & 188, 2.98$\times 10^{-5}$ & 87, 0.0122 \\
$T_{\rm 90,z}$       & $+$0.098, $+$1.496 & $+$0.072, $+$1.082 & $+$0.162, $+$2.351 & $+$0.190, $+$2.637 & --- & 144, 4.53$\times 10^{-6}$ & 
   175, $<10^{-8}$ & 231, 7.64$\times 10^{-6}$ & 231, 2.64$\times 10^{-3}$ & 104, 0.7572 \\
$T_{\rm break 1,z}$  & $+$0.108, $+$1.291 & $+$0.104, $+$1.246 & $-$0.067, $-$0.799 & $-$0.111, $-$1.183 & $+$0.371, $+$4.767 & -- & 
    115, $<10^{-8}$ & 143, $<10^{-8}$ & 143, $<10^{-8}$ & 65, 0.2841 \\
$T_{\rm break 2,z}$  & $+$0.151, $+$2.004 & $+$0.262, $+$3.573 & $+$0.108, $+$1.425 & $+$0.050, $+$0.651 & $+$0.439, $+$6.421 & $+$0.628, $+$8.568 &
    --- & 175, $1\times 10^{-8}$ & 175, 0.0110 & 82, 0.1623 \\
$L_{\rm 15-150 keV}$ & $-$0.370, $-$6.032 & $-$0.130, $-$1.972 & $+$0.126, $+$1.816 & $+$0.166, $+$2.295 & $-$0.290, $-$4.579 & $-$0.587, $-$8.614 &
    $-$0.413, $-$5.956 &  --- & 231, $<10^{-8}$ & 103, 6.88$\times 10^{-8}$ \\
$E_{\rm iso}$        & $-$0.351, $-$5.678 & $-$0.120, $-$1.815 & $+$0.222, $+$3.264 & $+$0.300, $+$4.284 & $+$0.197, $+$3.048 & $-$0.472, $-$6.357 &
    $-$0.192, $=$2.570 & $+$0.850, $+$24.415 & --- & 103, 1.06$\times 10^{-6}$ \\
$E_{\rm peak, z}$    & $-$0.416, $-$4.616 & $-$0.141, $-$1.433 & $+$0.287, $+$2.938 & $+$0.268, $+$2.560 & $-$0.031, $-$0.309 & $-$0.136, $-$1.086 &
    $-$0.155, $-$1.405 & $+$0.426, $+$4.735 & $+$0.459, $+$5.188 & --- \\

\enddata

\tablenotetext{1}{The values below the diagonal list the Spearman rank order correlation coefficient $r_s$ and the
Student's T-test value $T_s$. The values above the diagonal list the number of GRBs in this correlation and the
probability $P$ that the result is drawn from a random distribution.}

\end{deluxetable}

\begin{deluxetable}{lcccccc}
\tabletypesize{\scriptsize}
\tablecaption{Results from the Principal Component Analysis for 175 GRBs 
of the  \swift\ sample\tablenotemark{1}.
\label{pca_results}}
\tablewidth{0pt}
\tablehead{
\colhead{Property} 
& \colhead{EV 1}
& \colhead{EV 2} 
& \colhead{EV 3}
& \colhead{EV 4}
& \colhead{EV 5} 
& \colhead{EV 6}
} 
\startdata
Proportion of Variance & 0.3742 & 0.2366 & 0.1316 & 0.1007 & 0.08319 & 0.07377  \\
Cumulative Proportion  & 0.3742 & 0.6108 & 0.7423 & 0.8430 & 0.92623 & 1.00000  \\
\\
log $T_{90, z}$          &  -0.3959 &  0.4782 & -0.2098 &  0.2080 &  0.7193 & -0.0997   \\
log $T_{\rm break2, z}$  &  -0.4859 &  0.2924 &  0.2970 &  0.1528 & -0.4977 & -0.5650   \\
\axb                     &   0.1043 &  0.6931 & -0.0393 & -0.6379 & -0.1961 &  0.2486   \\
$\Gamma_{\rm BAT}$       &  -0.3830 & -0.3865 & -0.5170 & -0.5799 &  0.0193 & -0.3163   \\
\bx                      &  -0.4105 & -0.2349 &  0.7128 & -0.3458 &  0.2755 &  0.2698   \\
log $L_{\rm 15-150 keV}$ &   0.5303 &  0.0277 &  0.3016 & -0.2658 &  0.3468 & -0.6605    \\
\enddata

\tablenotetext{1}{These are GRBs with canonical light curves and spectroscopic
redshift measurements.}

\end{deluxetable}

\begin{deluxetable}{lccccc}
\tabletypesize{\scriptsize}
\tablecaption{Same as Table\,\ref{pca_results}, but without the 15-150 keV luminosity.
\label{pca_results2}}
\tablewidth{0pt}
\tablehead{
\colhead{Property} 
& \colhead{EV 1}
& \colhead{EV 2} 
& \colhead{EV 3}
& \colhead{EV 4}
& \colhead{EV 5} 
} 
\startdata
Proportion of Variance &   0.3525 & 0.2836 & 0.1497 & 0.1172 & 0.09717 \\
Cumulative Proportion  &   0.3525 & 0.6360 & 0.7857 & 0.9028 & 1.00000 \\
\\
log $T_{90, z}$          & -0.4680  &  0.4521  & -0.3117 &   0.4303  & -0.5424 \\
log $T_{\rm break2, z}$  & -0.5854  &  0.2565  &  0.2216 &   0.1604  &  0.7188 \\
\axb                     &  0.0790  &  0.6985  & -0.0738 &  -0.7073  & -0.0044 \\
$\Gamma_{\rm BAT}$       & -0.4048  & -0.4096  & -0.7159 &  -0.3758  &  0.1210 \\
\bx                      & -0.5179  & -0.2719  &  0.5795 &  -0.3842  & -0.4176 \\
\enddata

\end{deluxetable}

\begin{deluxetable}{cccccc}
\tabletypesize{\scriptsize}
\tablecaption{Results of the fits to the GRB luminosity functions shown in Figure\,\ref{lum_function}
\label{lum_func_res}}
\tablewidth{0pt}
\tablehead{
\colhead{redshift interval} 
& \colhead{\# of GRBs}
& \colhead{a\tablenotemark{1}}
& \colhead{b\tablenotemark{1}} 
& \colhead{$\Phi^*$\tablenotemark{1,2}}
& \colhead{$L_{\rm break}$\tablenotemark{1,3}}
} 
\startdata
$<$0.5    & 21 &  +0.117\plm0.016 & 0.836\plm0.179 & 0.630\plm0.074 & $3.36\pm1.27\times 10^{49}$ \\
0.5 - 1.0 & 44 & -0.036\plm0.166 & 0.774\plm0.183 & 0.433\plm0.195 & $(2.81\pm349) \times 10^{50}$ \\
1.0 - 2.0 & 61 & -0.001\plm0.011 & 0.948\plm0.074 & 0.133\plm0.008 & $(2.35\pm0.36) \times 10^{51}$ \\
2.0 - 3.0 & 59 & -0.024\plm0.023 & 0.921\plm0.078 & 0.127\plm0.011 & $(9.21\pm2.13)\times 10^{51}$ \\
3.0 - 4.0 & 27 & -0.090\plm0.149 & 1.211\plm0.224 & 0.068\plm0.019 & $(1.85\pm0.95)\times 10^{52}$ \\
$>$4.0    & 18 & -0.107\plm0.040 & 1.309\plm0.059 & 0.014\plm0.001 & $(9.11\pm1.07) \times 10^{52}$ \\
\enddata

\tablenotetext{1}{The fit parameters are defined as $\Phi(L, z) = 
\frac{\Phi^*(L_{\rm break})}{(L/L_{\rm break})^{a} + (L/L_{\rm break})^{b}}$}
\tablenotetext{2}{The GRB density $\Phi^*$ at the break luminosity $L_{\rm break}$ is given in units if GRBs per Gpc$^{3}$.}
\tablenotetext{3}{The break luminosity is given in units of erg s$^{-1}$}

\end{deluxetable}

\begin{deluxetable}{lcccccccccccc}
\tabletypesize{\scriptsize}
\tablecaption{Mean, standard deviation, and median of observed GRB properties of GRBs with spectroscopic redshifts and
those without
\label{grbs_z_noz}}
\tablewidth{0pt}
\tablehead{
& \multicolumn{4}{c}{GRBs with redshift} & \multicolumn{4}{c}{GRBs without redshift} & \multicolumn{4}{c}{GRBs in PCA}\\
\colhead{Property} 
& \colhead{Mean}
& \colhead{SD\tablenotemark{1}}
& \colhead{Median} 
& \colhead{\# of GRBs}
& \colhead{Mean}
& \colhead{SD\tablenotemark{1}}
& \colhead{Median} 
& \colhead{\# of GRBs}
} 
\startdata
$\Gamma_{\rm BAT}$    & 1.618 & 0.421 & 1.590 & 232 & 1.612 & 0.420 & 1.600 & 519 & 1.647 & 0.397 & 1.620 & 175 \\
\bx                   & 1.032 & 0.366 & 1.020 & 230 & 0.985 & 0.465 & 0.930 & 402 & 1.040 & 0.319 & 1.040 & 175  \\
log $T_{90}$          & 1.511 & 0.730 & 1.630 & 232 & 1.401 & 0.825 & 1.595 & 521 & 1.577 & 0.652 & 1.653 & 175 \\
log $T_{\rm break-1}$ & 2.710 & 0.507 & 2.607 & 145 & 2.703 & 0.485 & 2.645 & 239 & 2.658 & 0.471 & 2.568 & 115 \\
log $T_{\rm break-2}$ & 3.941 & 0.678 & 3.924 & 176 & 3.820 & 0.793 & 3.887 & 241 & 3.932 & 0.675 & 3.924 & 175 \\
\axb                  & 0.620 & 0.419 & 0.590 & 208 & 0.575 & 0.451 & 0.620 & 331 & 0.556 & 0.392 & 0.550 & 175 \\ 
\axc                            & 1.559  & 0.567 & 1.430  & 188 & 1.545  & 0.847 & 1.320 & 287 & 1.527 & 0.445 & 1.430 & 174 \\
log fluence\tablenotemark{2}    & -5.729 & 0.671 & -5.716 & 232 & -6.004 & 0.662 & -5.975 & 520 &  -5.656 & 0.568 & -5.636 & 175 \\
$E_{\rm peak}$\tablenotemark{3} & 285.1 & 442.8 & 195.0 & 104 & 223.1 & 291.8 & 110.5 & 102 & 296.0 & 488.6 & 187.5 & 82 \\
z                     & 2.019 & 1.373 & 1.763 & 232 & \nodata & \nodata & \nodata & 0 & 2.115 & 1.369 & 1.884 & 175 \\
\# of UVOT detections\tablenotemark{4} & \multicolumn{4}{c}{155} & \multicolumn{4}{c}{91} & \multicolumn{4}{c}{123} \\
\enddata

\tablenotetext{1}{Standard deviation SD}
\tablenotetext{2}{Observer Fluence in the 15-150 keV BAT energy band
 in units of   erg s$^{-1}$ cm$^{-2}$}
\tablenotetext{3}{Peak energy $E_{\rm peak}$ in units of keV}
\tablenotetext{4}{Of all GRBs, 218 had UVOT detections}

\end{deluxetable}

\begin{deluxetable}{lcccccccccccc}
\rotate
\tabletypesize{\tiny}
\tablecaption{Mean, standard deviation, and median of GRB properties in the observed for low (z$<$1.0), intermediate (1.0$<$z$<$3.5),
 and high (z$>$3.5) long-duration  GRBs (group 1 in the cluster analysis)
\label{grb_statistics_redshift}}
\tablewidth{0pt}
\tablehead{
& \multicolumn{4}{c}{z$<$1.0} & \multicolumn{4}{c}{1.0 $<$ z $<$ 3.5}  & \multicolumn{4}{c}{z$>3.5$ bursts }\\
\colhead{Property} 
& \colhead{Mean}
& \colhead{SD\tablenotemark{1}}
& \colhead{Median} 
& \colhead{\# of GRBs}
& \colhead{Mean}
& \colhead{SD\tablenotemark{1}}
& \colhead{Median} 
& \colhead{\# of GRBs}
& \colhead{Mean}
& \colhead{SD\tablenotemark{1}}
& \colhead{Median} 
& \colhead{\# of GRBs}
} 
\startdata
$\Gamma_{\rm BAT}$                         & 1.693 & 0.425 & 1.630 & 50 & 1.635 & 0.401 & 1.620 & 136 & 1.494 & 0.274 & 1.510 & 29 \\
\bx                                        & 1.105 & 0.483 & 1.035 & 50 & 1.023 & 0.307 & 1.020 & 136 & 0.974 & 0.250 & 0.990 & 29 \\
log $T_{90}$                               & 1.455 & 0.662 & 1.552 & 50 & 1.652 & 0.530 & 1.681 & 136 & 1.689 & 0.531 & 1.702 & 29 \\
log $T_{\rm break-1}$                      & 2.867 & 0.638 & 2.681 & 31 & 2.667 & 0.471 & 2.565 & 90  & 2.646 & 0.310 & 2.633 & 19 \\
log $T_{\rm break-2}$                      & 4.017 & 0.761 & 3.959 & 39 & 3.923 & 0.645 & 3.924 & 103 & 4.018 & 0.513 & 4.002 & 26 \\
\axb                                       & 0.579 & 0.421 & 0.570 & 44 & 0.639 & 0.431 & 0.630 & 125 & 0.600 & 0.352 & 0.560 & 28 \\
\axc                                       & 1.469 & 0.528 & 1.340 & 41 & 1.517 & 0.394 & 1.440 & 112 & 1.652 & 0.509 & 1.630 & 27 \\
log fluence\tablenotemark{2}             & -5.632 & 0.695 & -5.703 & 50 & -5.674 & 0.556 & -5.632 & 136 & -5.775 & 0.418 & -5.740 & 29 \\
$E_{\rm peak}$\tablenotemark{6}            & 2.214 & 0.430 & 2.210 & 23 & 2.275 & 0.328 & 2.301 & 61  & 2.078 & 0.250 & 2.081 & 12 \\
log $(\Delta_{\rm NH}+1)$\tablenotemark{7} & 0.931 & 0.711 & 1.167 & 49 & 0.585 & 0.574 & 0.709 & 136 & 0.265 & 0.377 & 0.000 & 29 \\
\enddata

\tablenotetext{1}{Standard deviation SD}
\tablenotetext{2}{Observer Fluence in the 15-150 keV BAT energy band
 in units of   erg cm$^{-2}$}
\tablenotetext{3}{K-corrected 15-150 keV Fluence
 in units of   erg  cm$^{-2}$}
\tablenotetext{4}{rest frame 15-150 keV luminosity given in unites of erg s$^{-1}$}
\tablenotetext{5}{k-corrected isotropic energy $E_{\rm iso}$ in the 15 - 150 keV BAT band}
\tablenotetext{6}{Peak energy $E_{\rm peak}$ in units of keV}
\tablenotetext{7}{The excess absorption column density $\Delta_{\rm NH}$ above the galactic value \citep[see ][]{grupe07} 
is given in units of 10$^{20}$ cm$^{-2}$}

\end{deluxetable}

\clearpage

%%%% **********************************************************
%% Figures for Appendix

%%%   Distributions

\begin{figure*}
%\epsscale{0.75}
\epsscale{1.5}
\plottwor{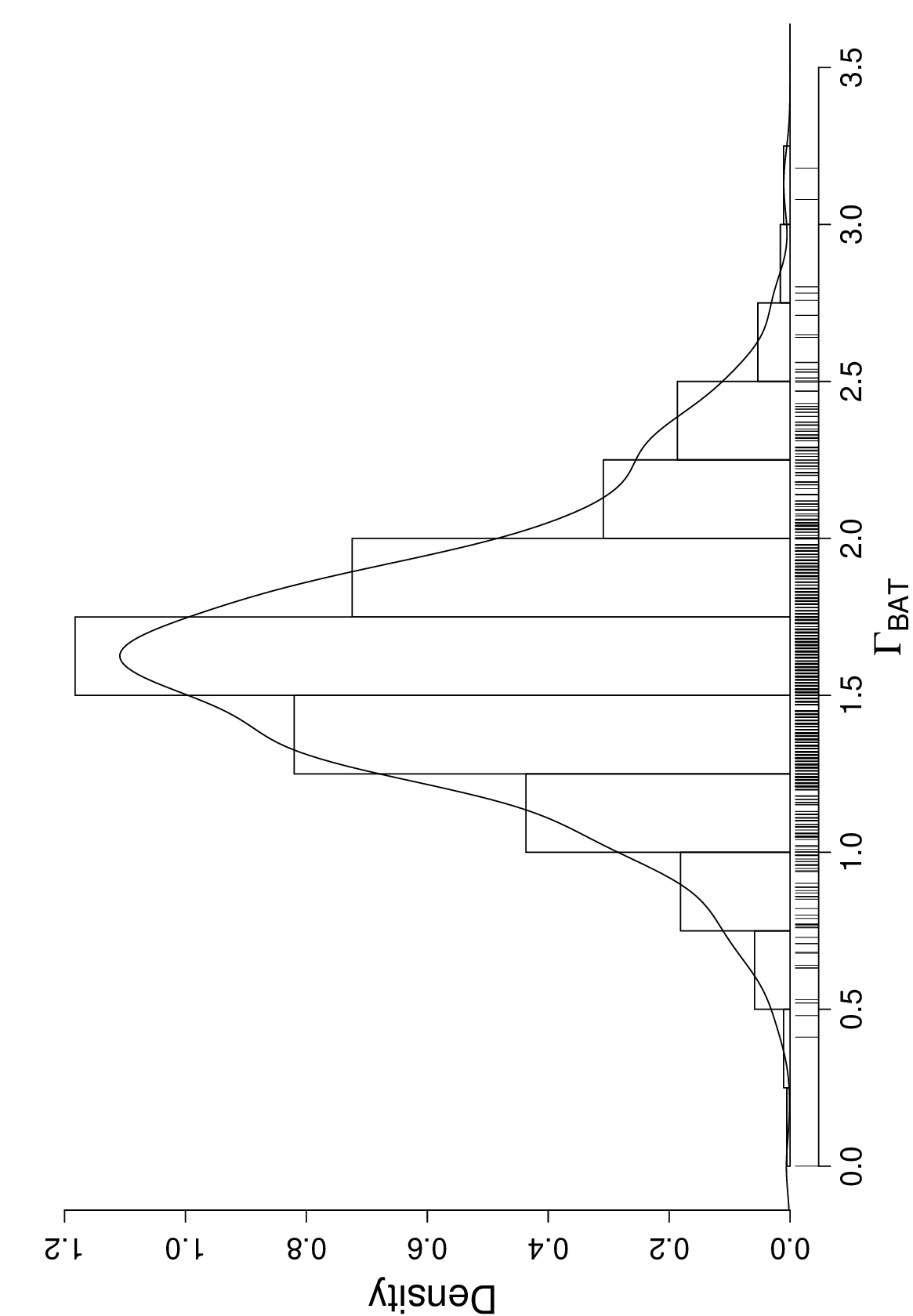}{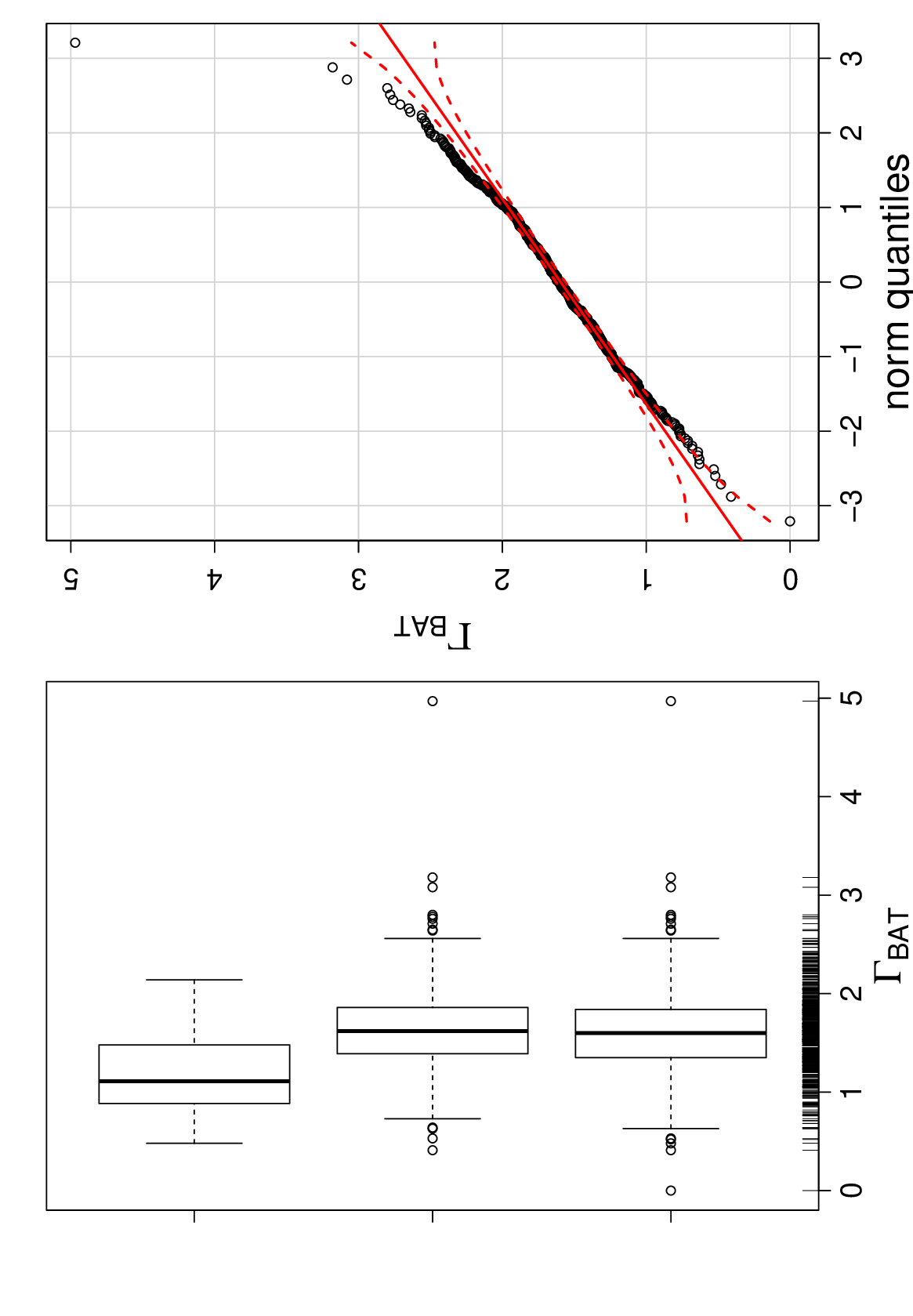}

\plottwor{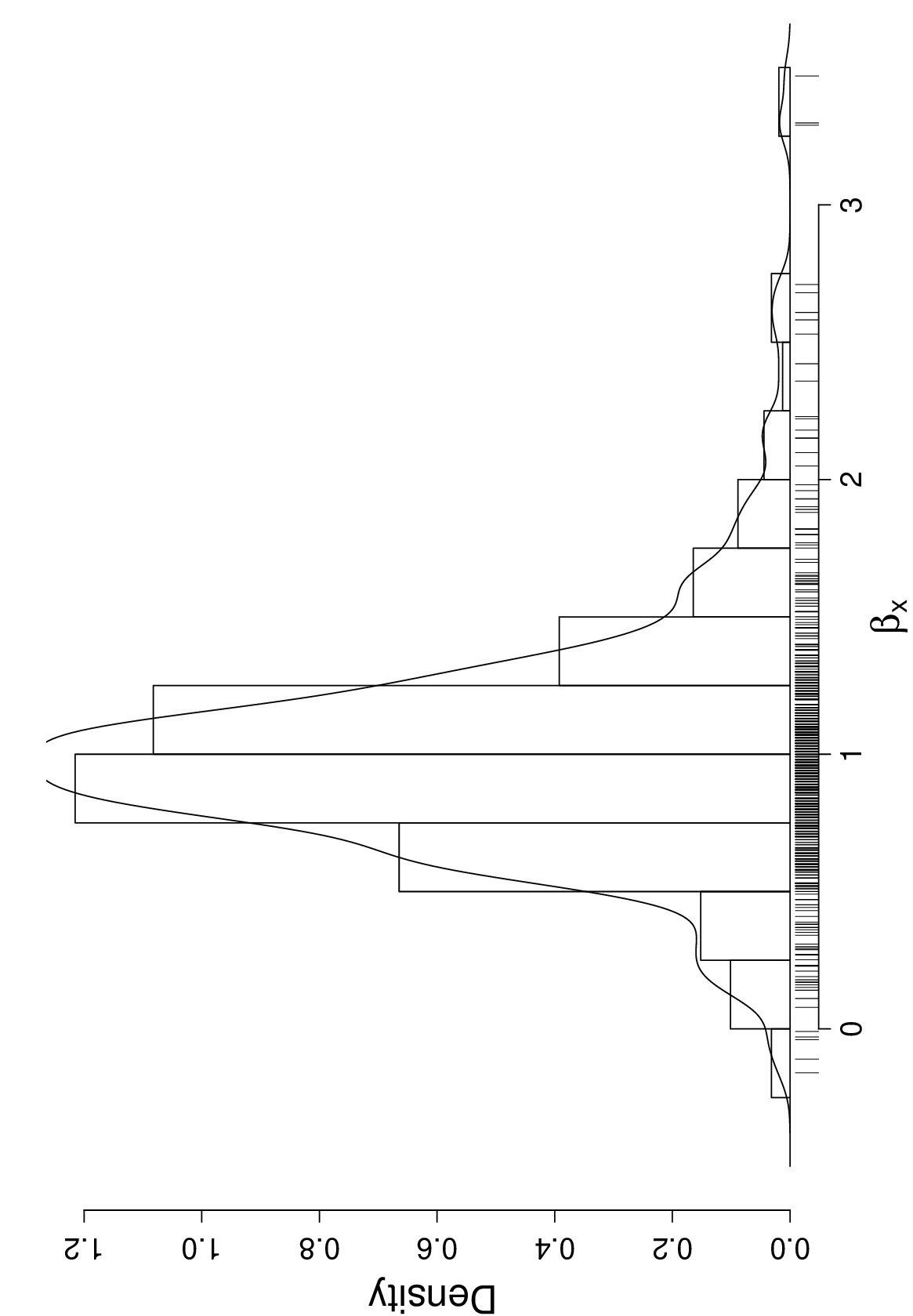}{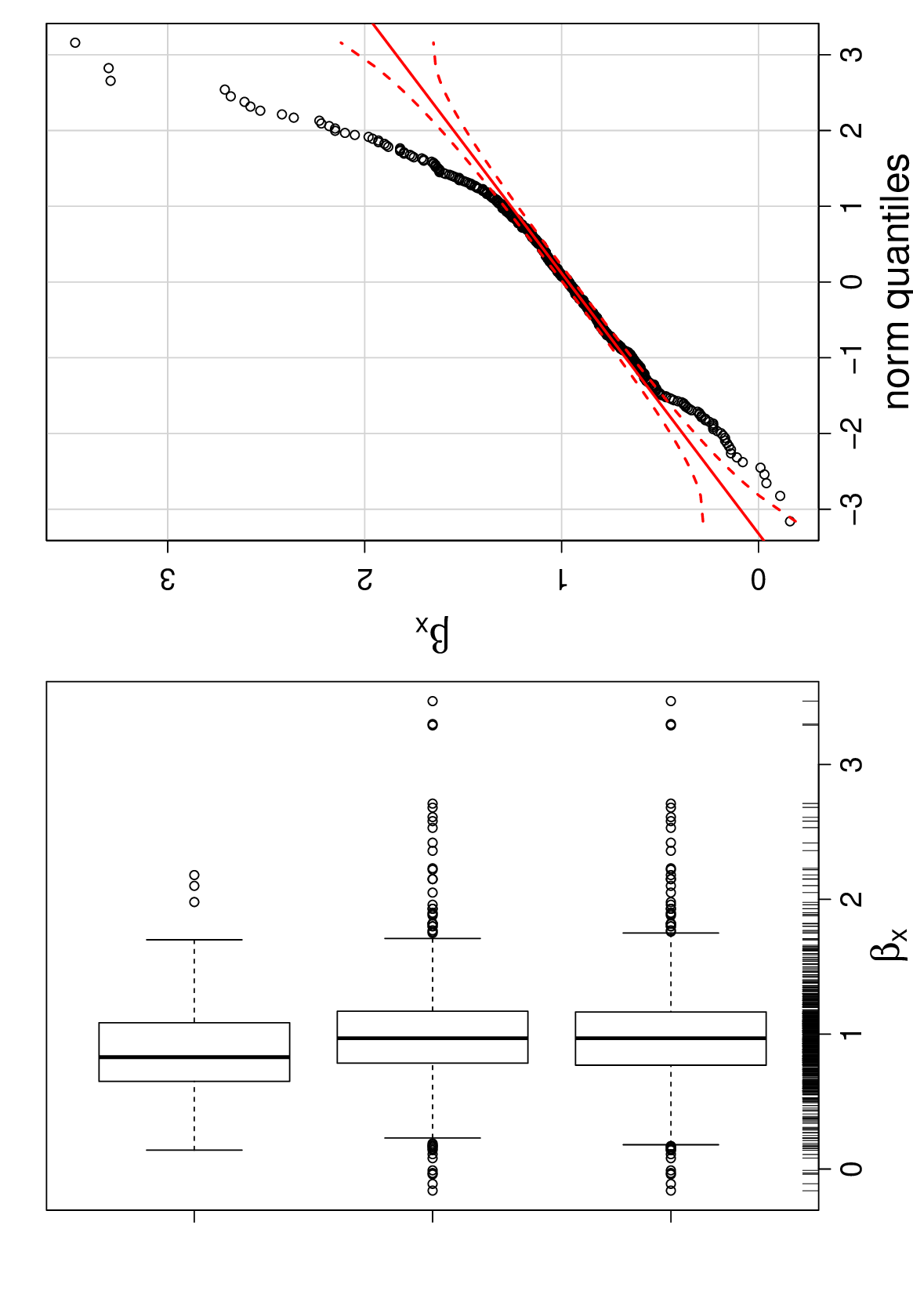}
\caption{\label{distr_gamma_bx} 
Distributions of the 15-150 keV BAT photon index and the 
0.3-10 keV XRT energy spectral slope \bx, shown in the upper and lower panels,
respectively.
As in
Figure\,\ref{distr_t90_tb}, the histogram, box plot, and q-q plot are shown. 
In the box plots,
short bursts are displayed on top, long bursts in the middle and all bursts on the
bottom.
}
\end{figure*}

\begin{figure*}
%\epsscale{0.75}
\epsscale{1.5}
\plottwor{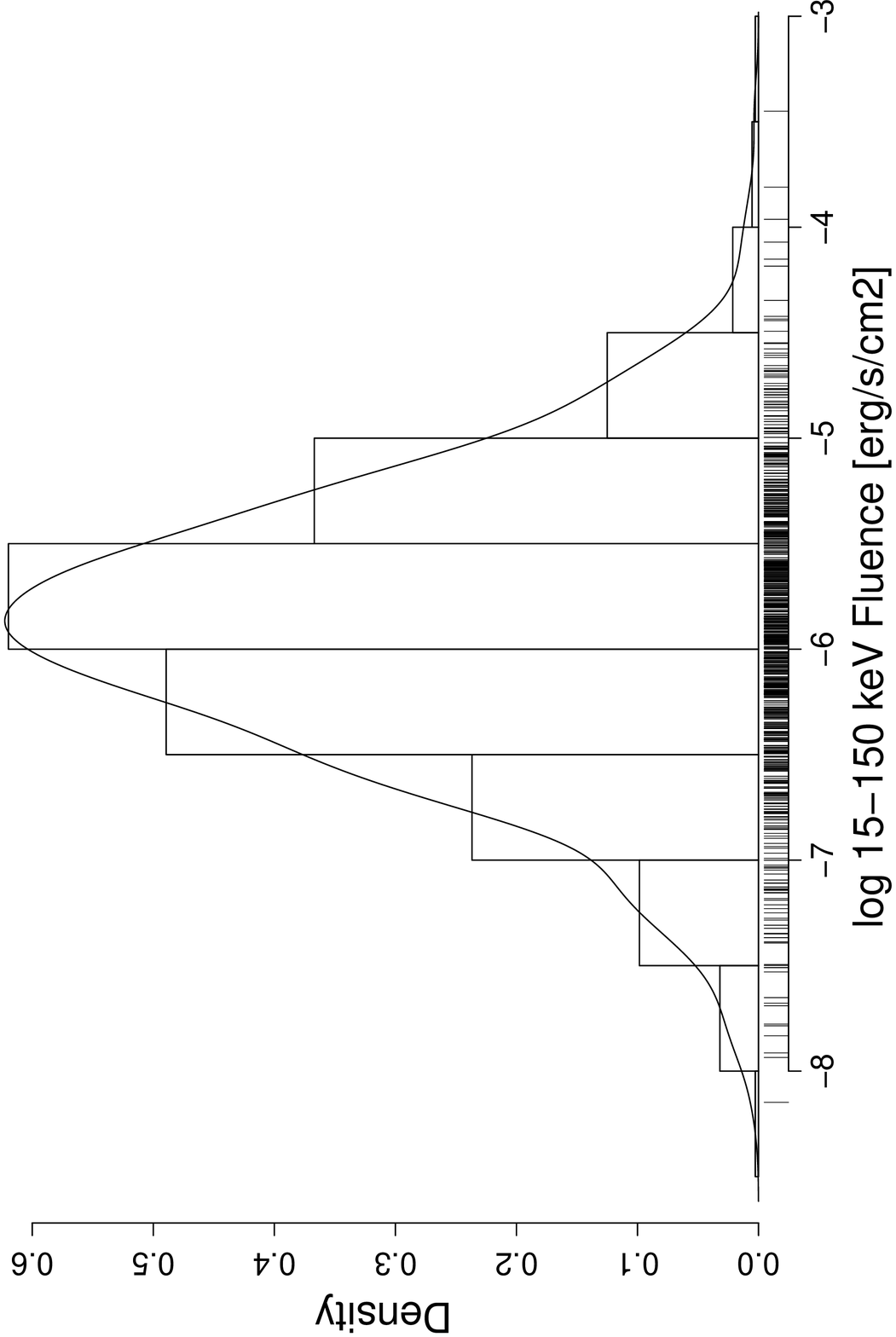}{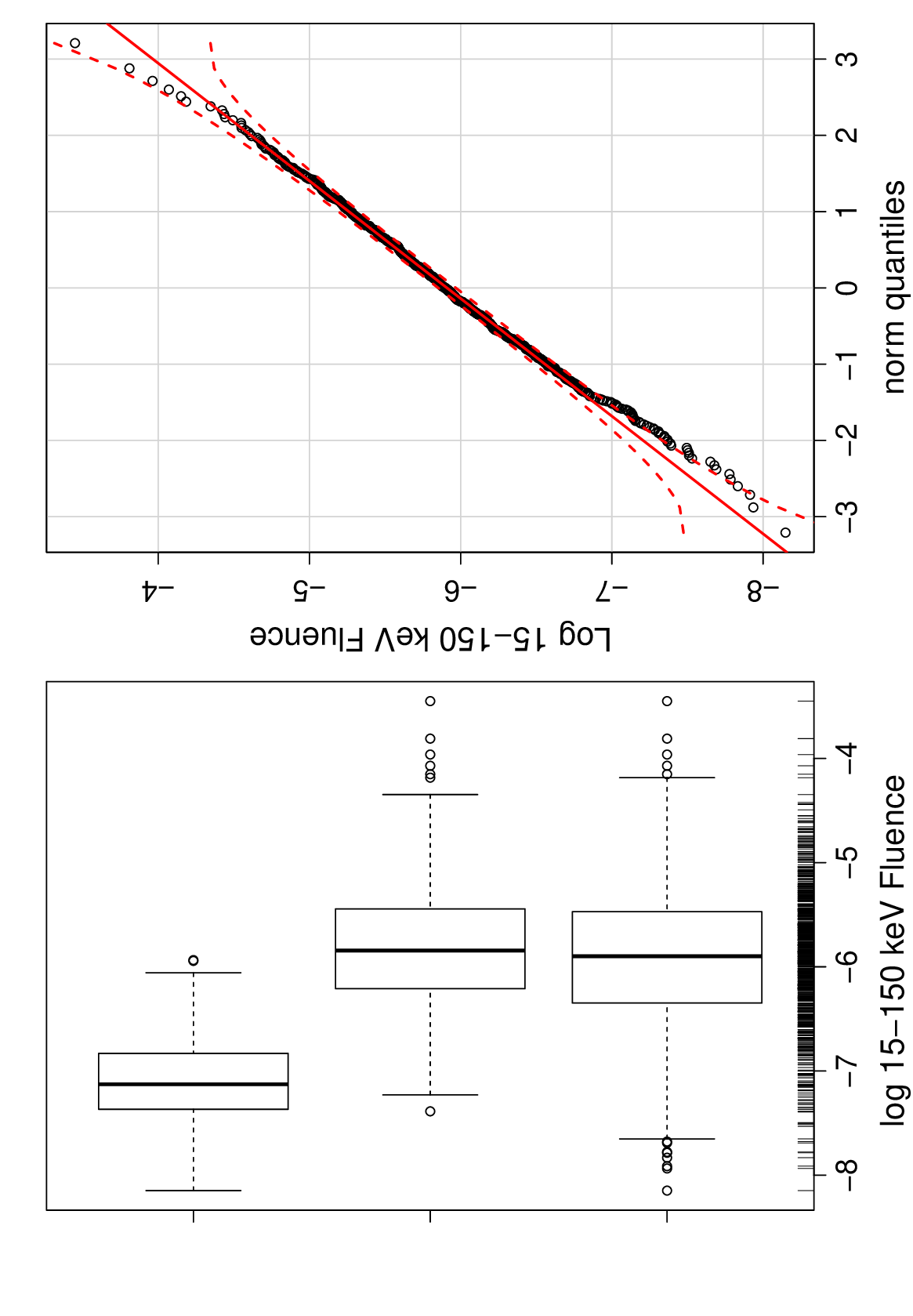}

\plottwor{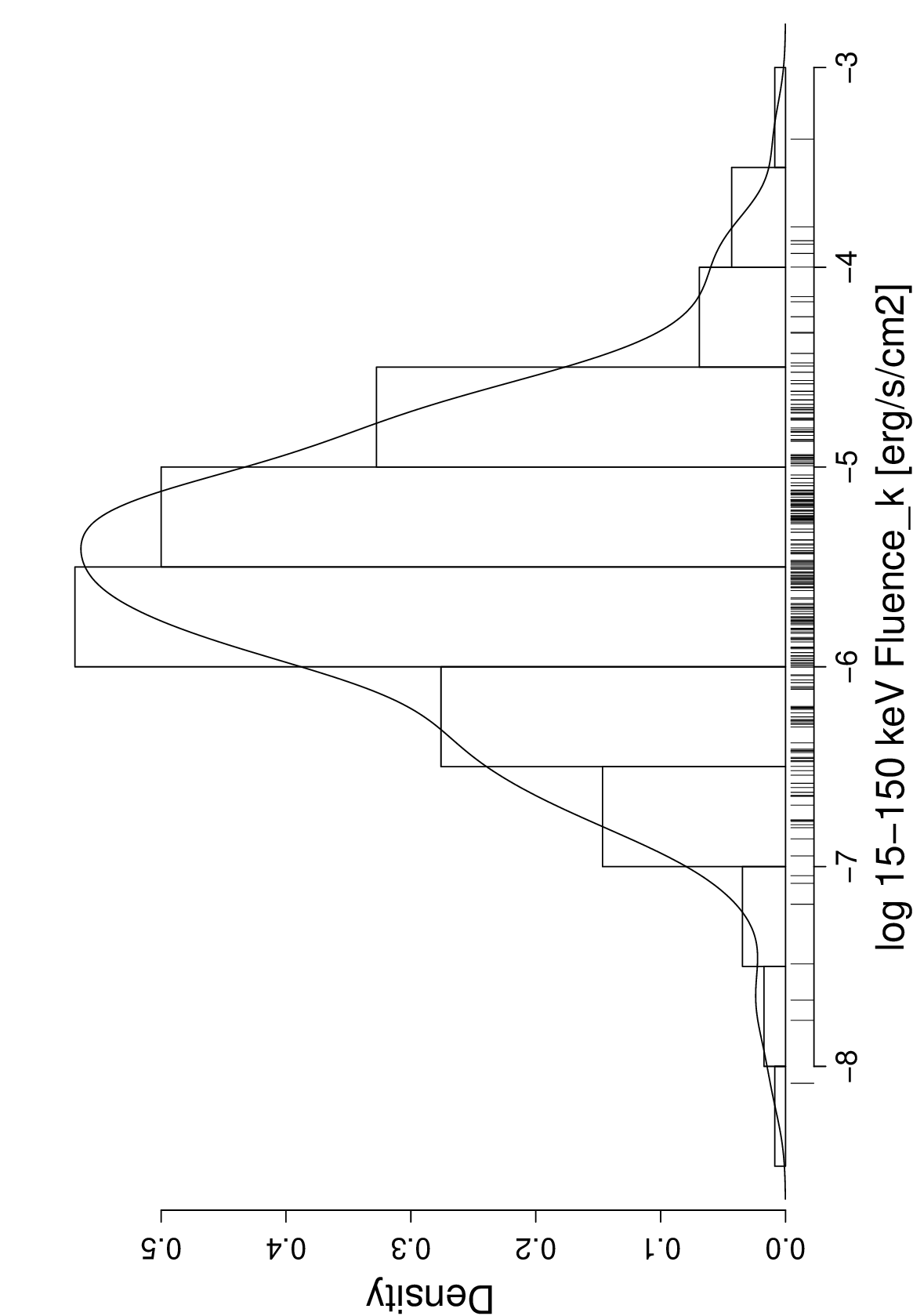}{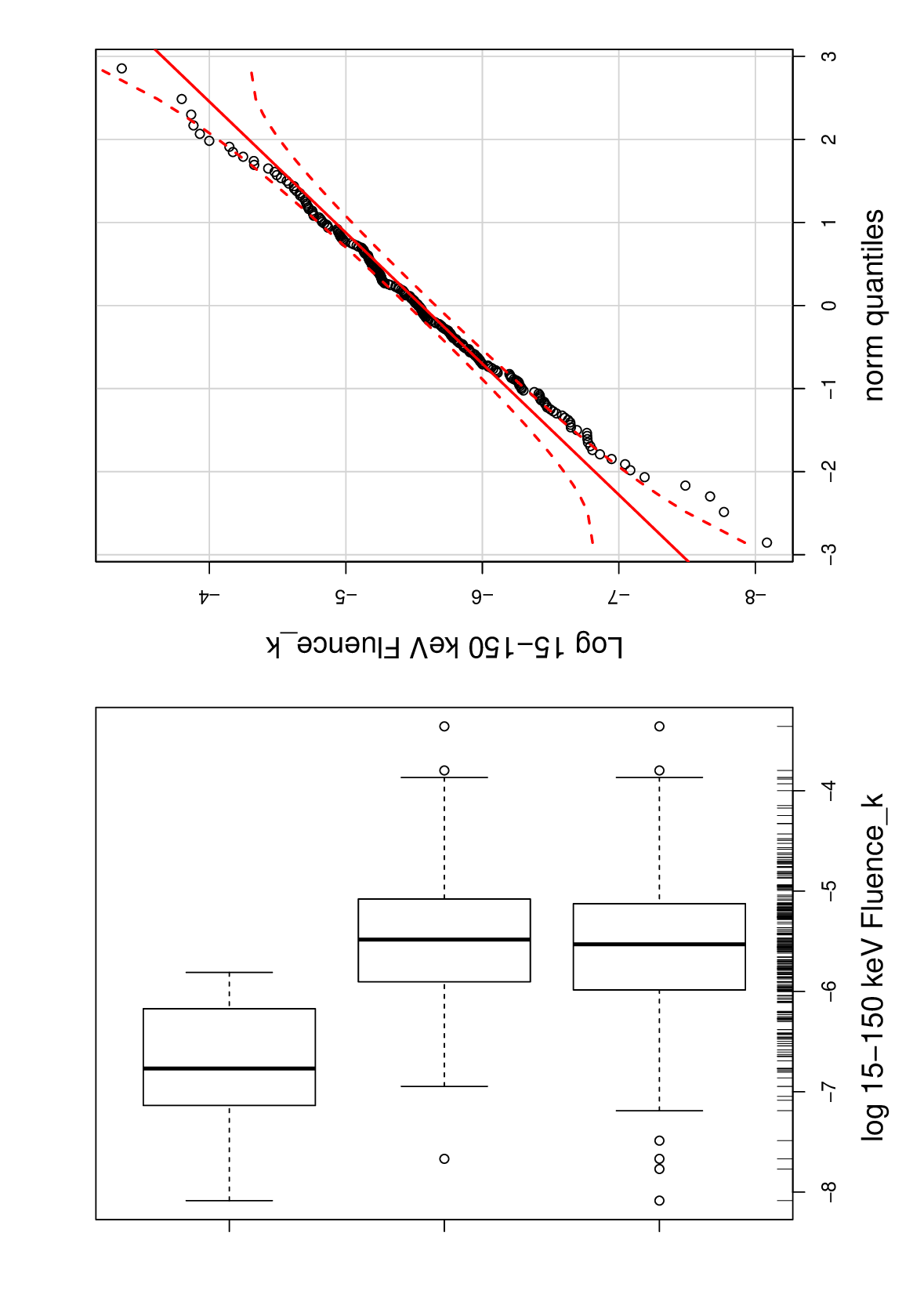}
\caption{\label{distr_fluence} 
Distribution of the observed and k-corrected 
15-150 keV fluence in the BAT band in units of 
erg cm$^{-2}$ (upper and lower panel, respectively). As in
Figure\,\ref{distr_t90_tb}, the histogram, box plot, and q-q plot are shown. 
In the box plots,
short bursts are displayed on top, long bursts in the middle and all bursts on the
bottom.
}
\end{figure*}

\begin{figure*}
%\epsscale{0.75}
\epsscale{1.5}
\plottwor{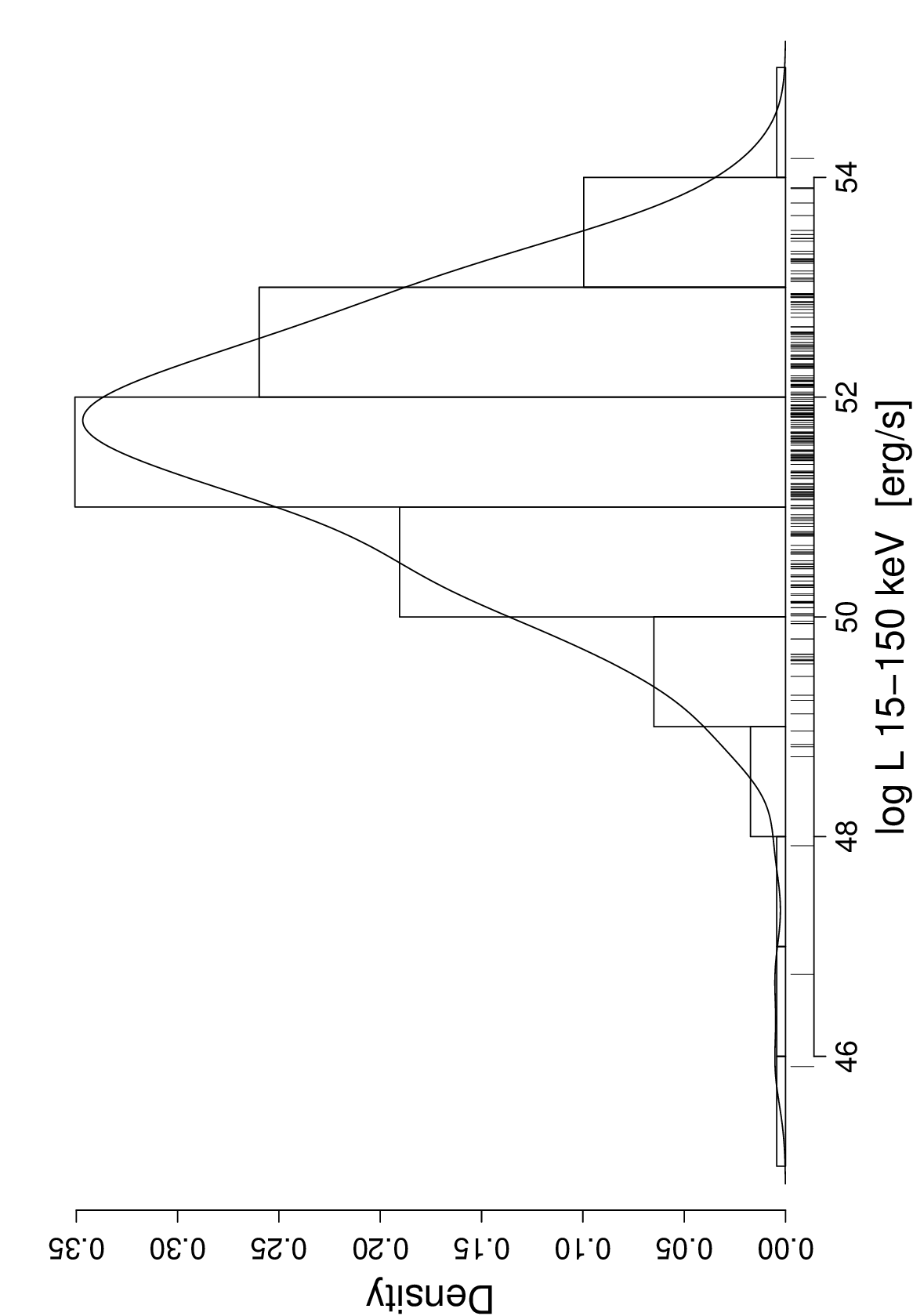}{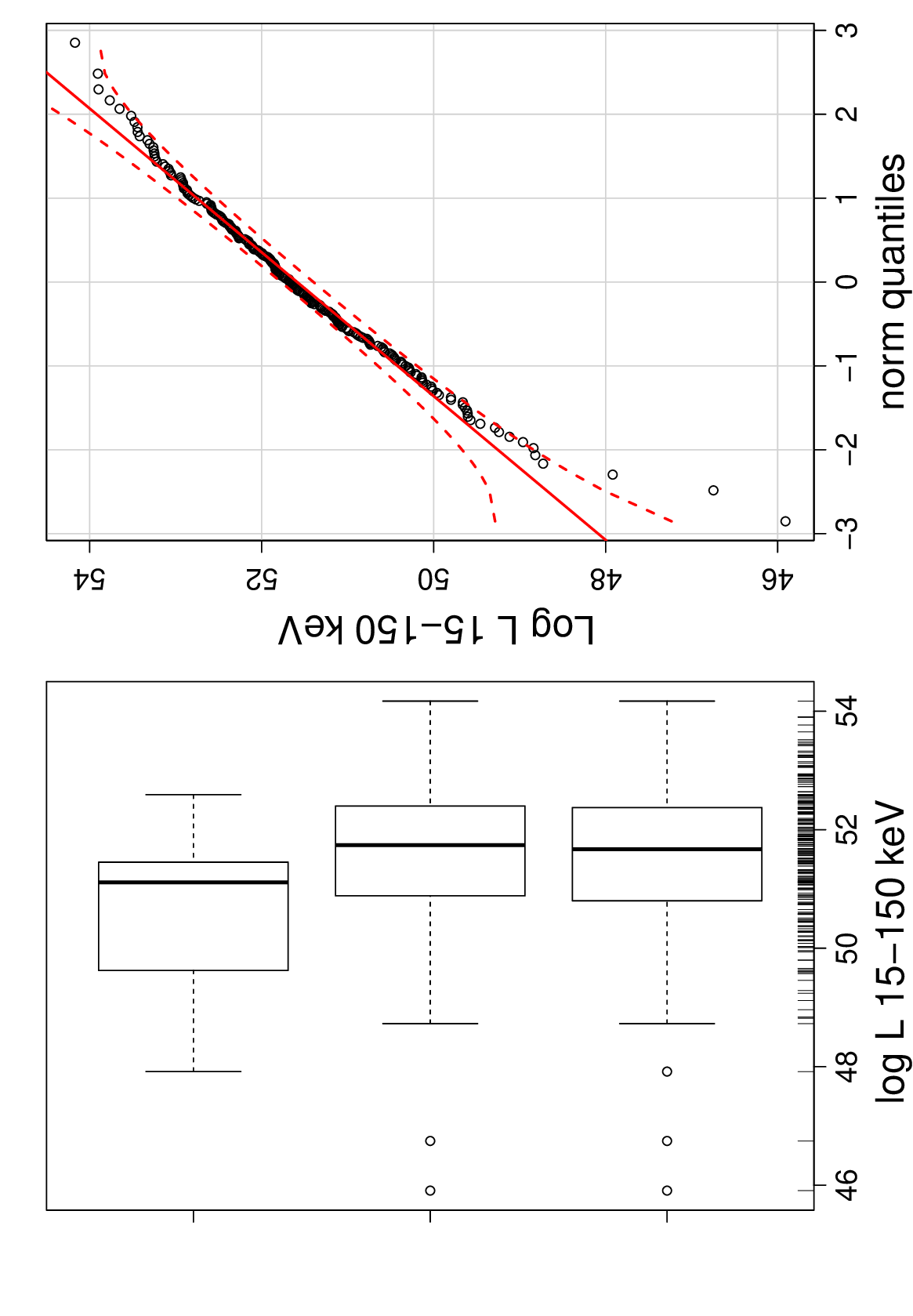}

\plottwor{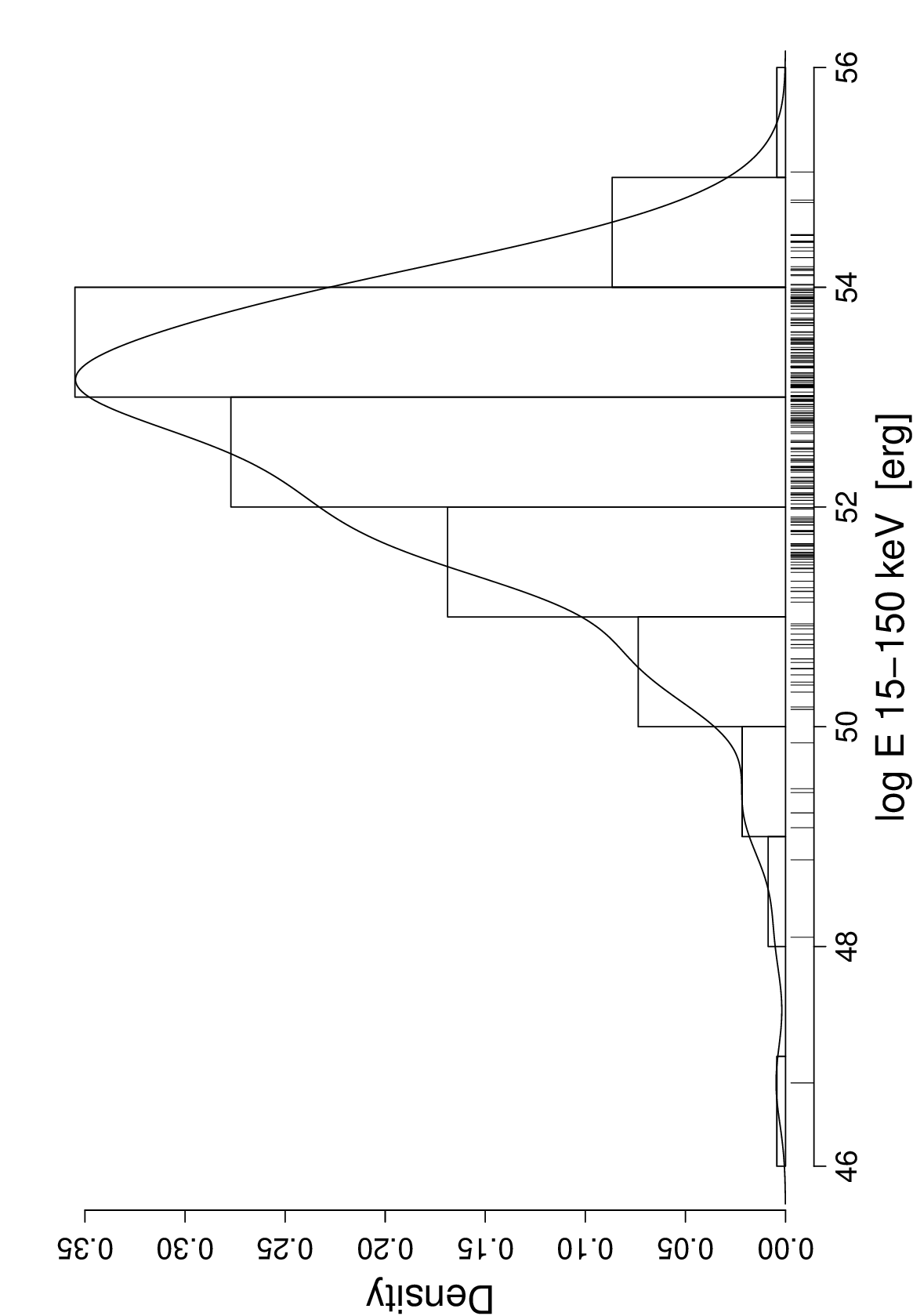}{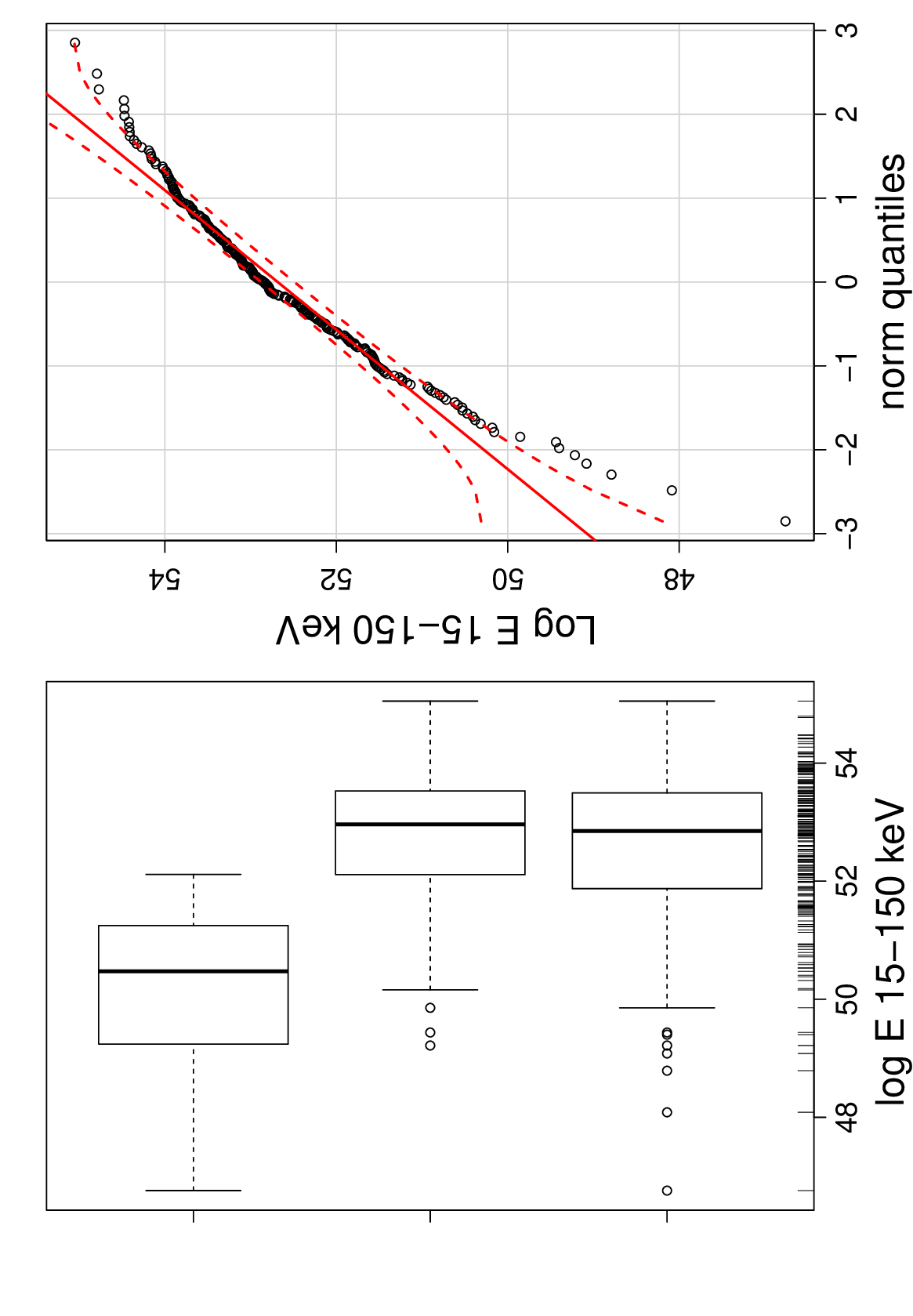}
\caption{\label{distr_l15_150} 
Distribution of the rest-frame 15-150 keV luminosity and isotropic energy $E_{\rm iso}$
of  \swift-detected GRBs with spectroscopic 
redshifts (upper and lower panel, respectively). As in
Figure\,\ref{distr_t90_tb}, the histogram, box plot, and q-q plot are shown. 
In the box plots,
short bursts are displayed on top, long bursts in the middle and all bursts on the
bottom.
}
\end{figure*}

\begin{figure*}
%\epsscale{0.75}
\epsscale{1.5}
\plottwor{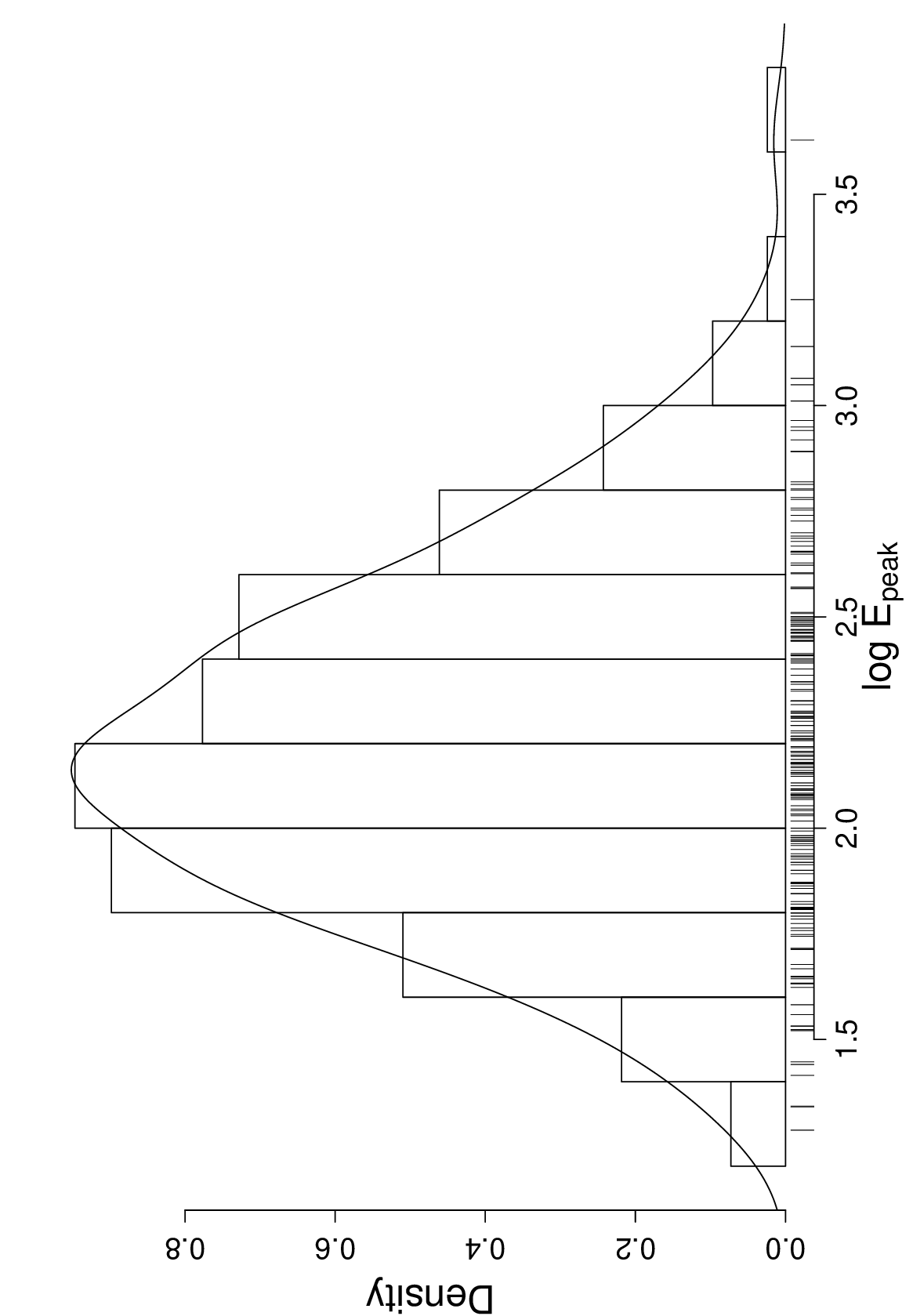}{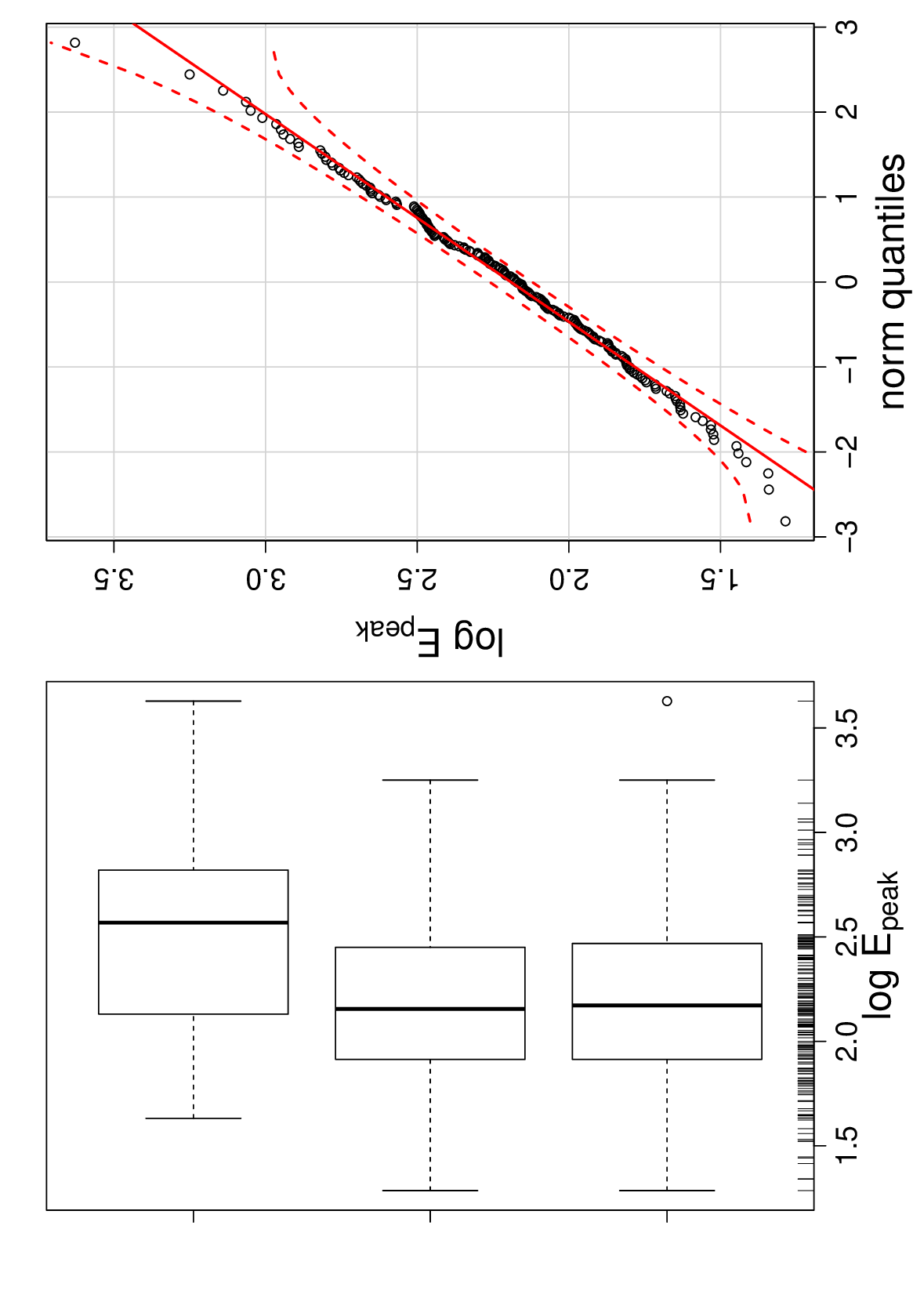}

\plottwor{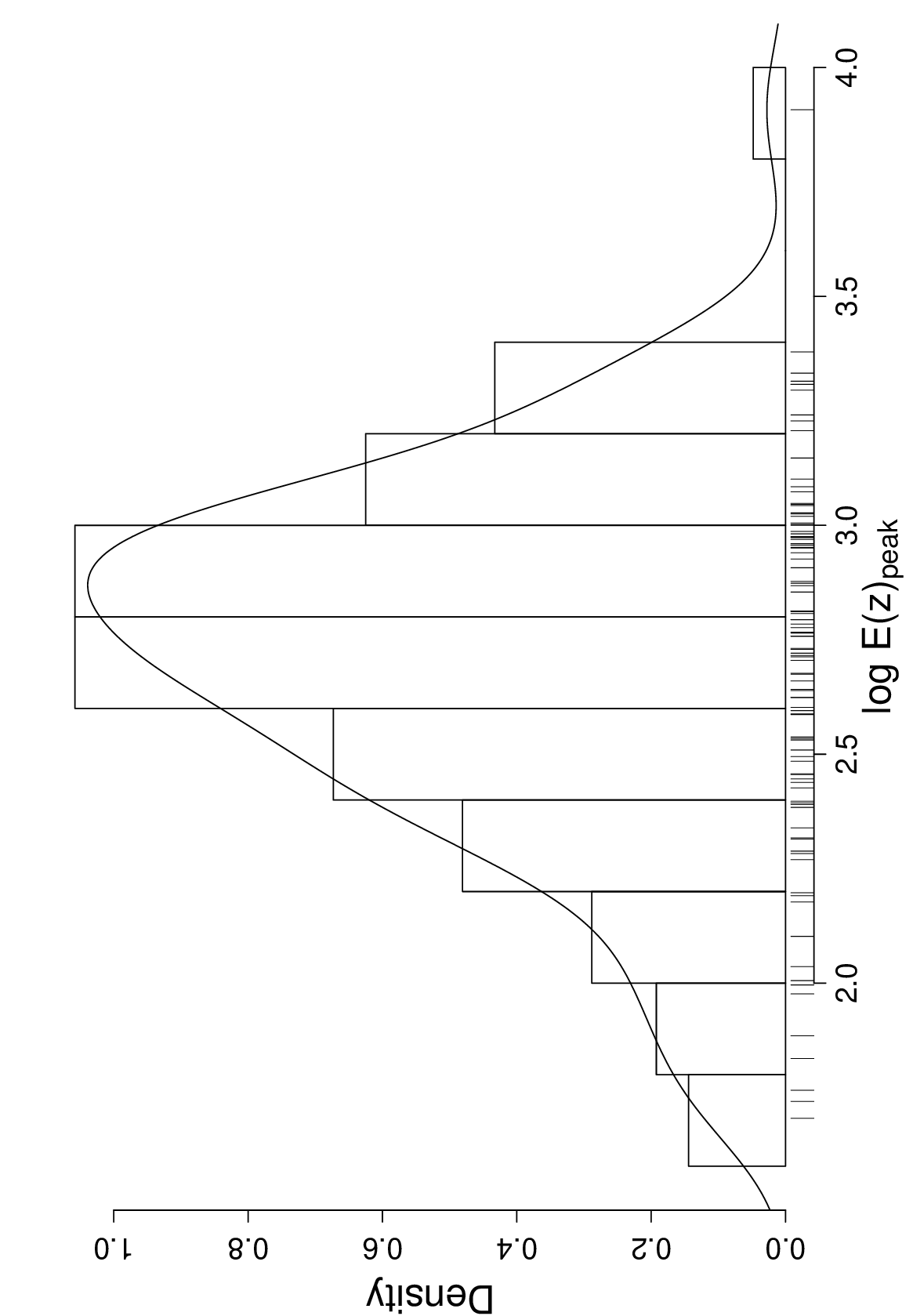}{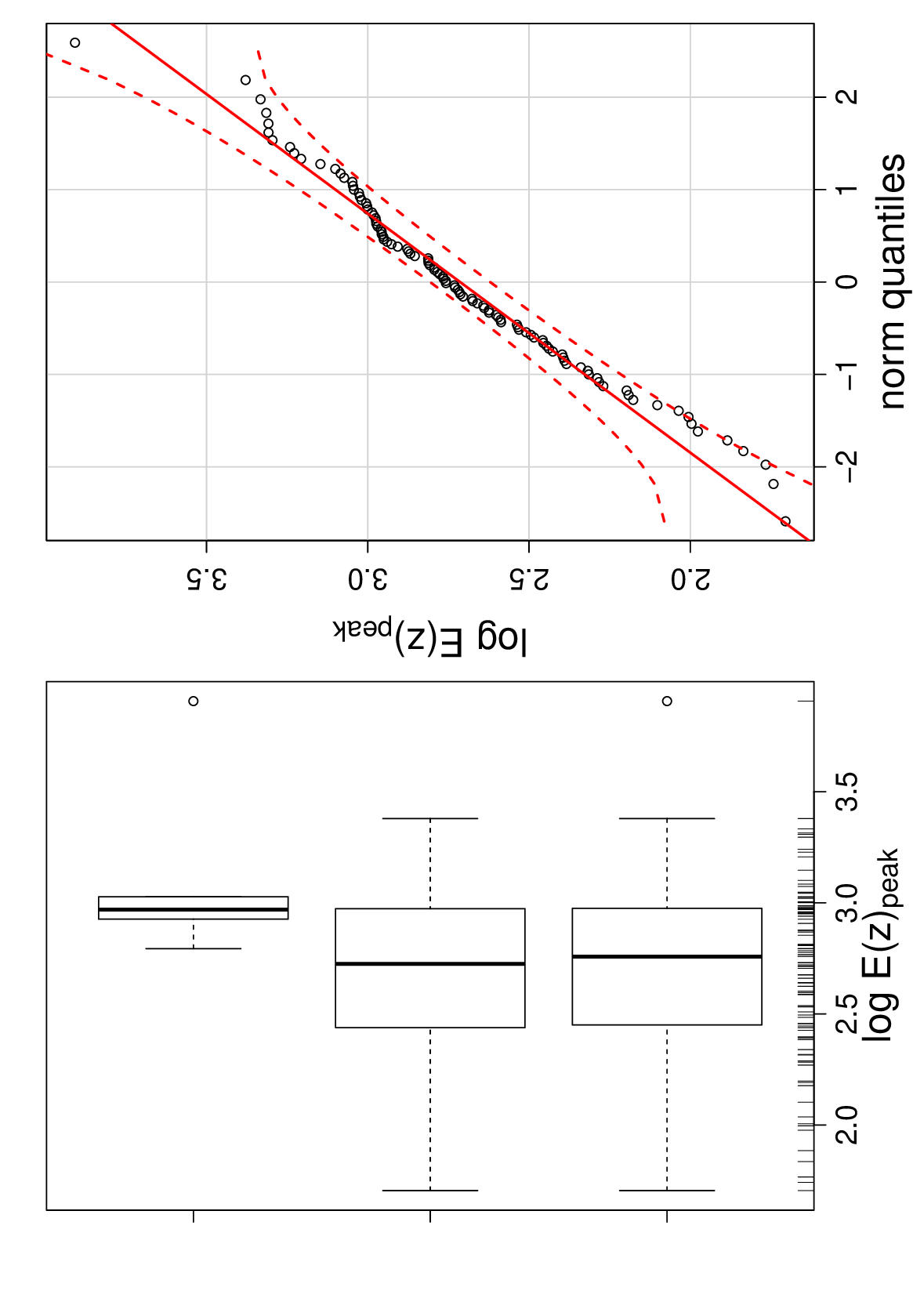}
\caption{\label{distr_epeak} 
Distributions of the observed and rest-frame Peak energy $E_{\rm peak, z}$ of 
\swift-detected GRBs with spectroscopic 
redshifts (upper and lower panels, respectively).  As in
Figure\,\ref{distr_t90_tb}, the histogram, box plot, and q-q plot are shown. 
In the box plots,
short bursts are displayed on top, long bursts in the middle and all bursts on the
bottom.
}
\end{figure*}

%% ************************************
% %  Correlations

\begin{figure*}
%\epsscale{0.75}
\epsscale{1.5}
\plottwo{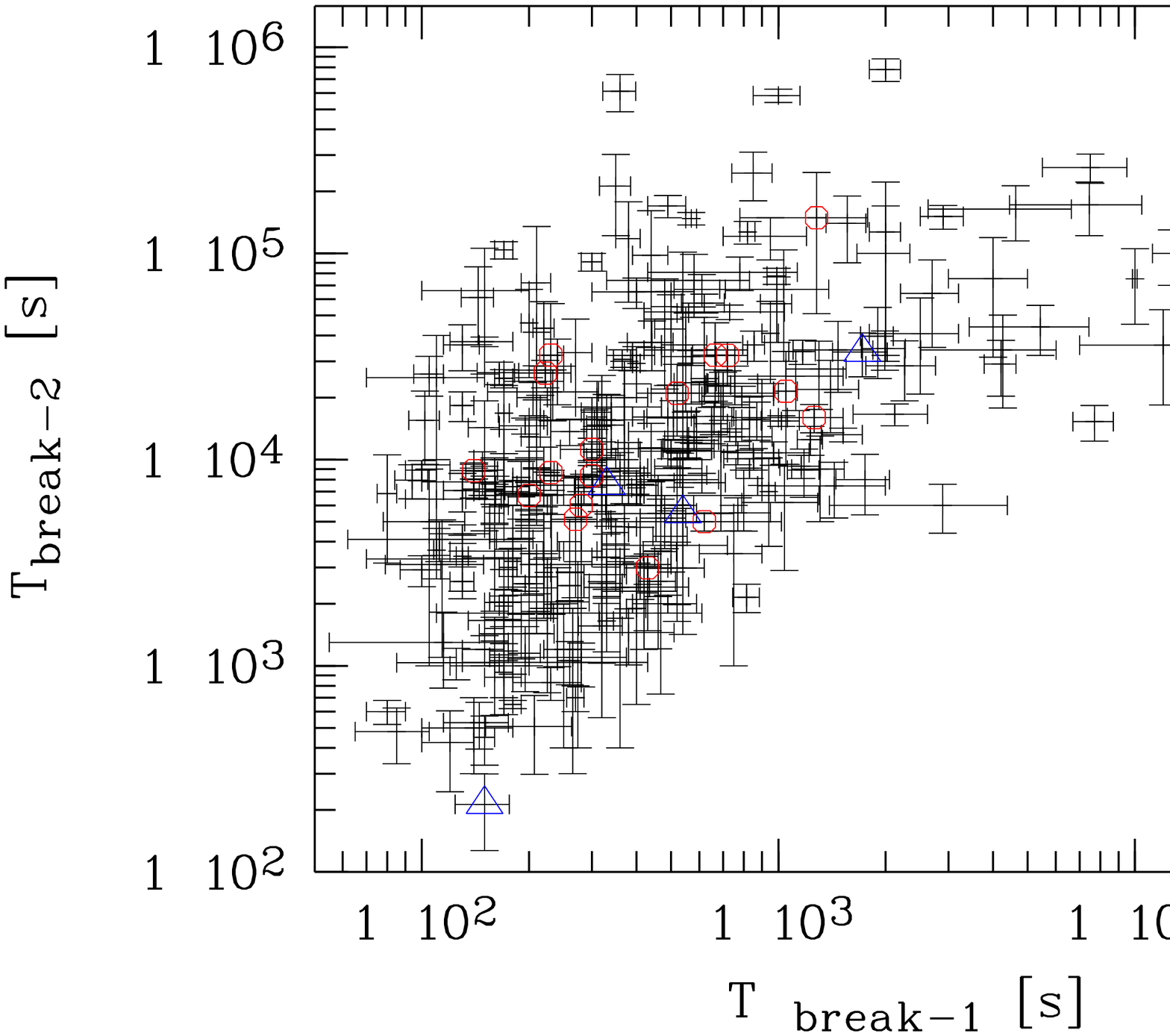}{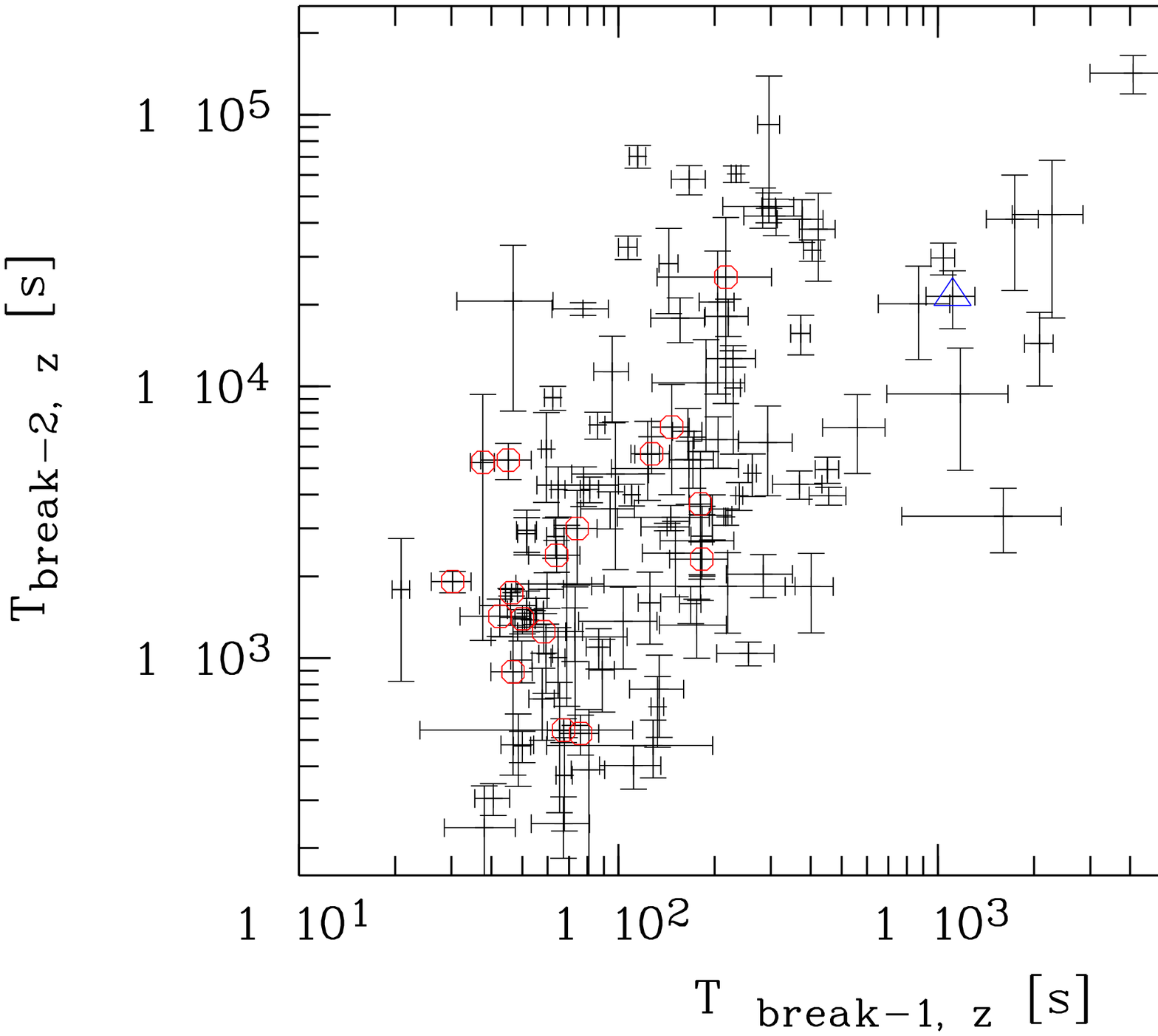}
\caption{\label{tb1_tb2} 
Correlation between the between the break times before and after the plateau phase $T_{\rm break1}$ and
$T_{\rm break2}$ in the observed and rest-frame (left and right panel, respectively)..
}
\end{figure*}

\begin{figure}
%\epsscale{0.75}
\epsscale{0.75}
\plotone{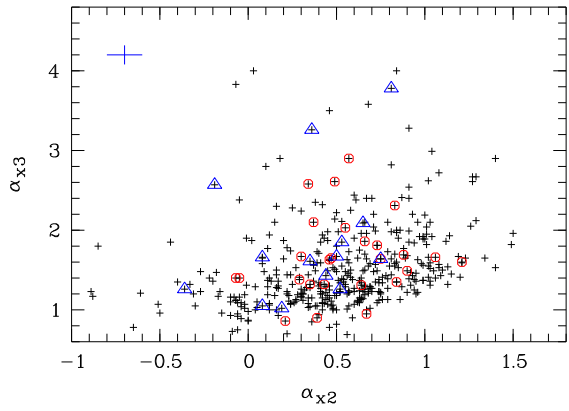}
\caption{\label{ax2_ax3} 
Correlation between decay slopes during the plateau phase \axb\ and the `normal' decay phase \axc.
The blue cross in the upper left corner displays the median uncertainty of each property.
}
\end{figure}

\begin{figure*}
%\epsscale{0.75}
\epsscale{1.5}
\plottwo{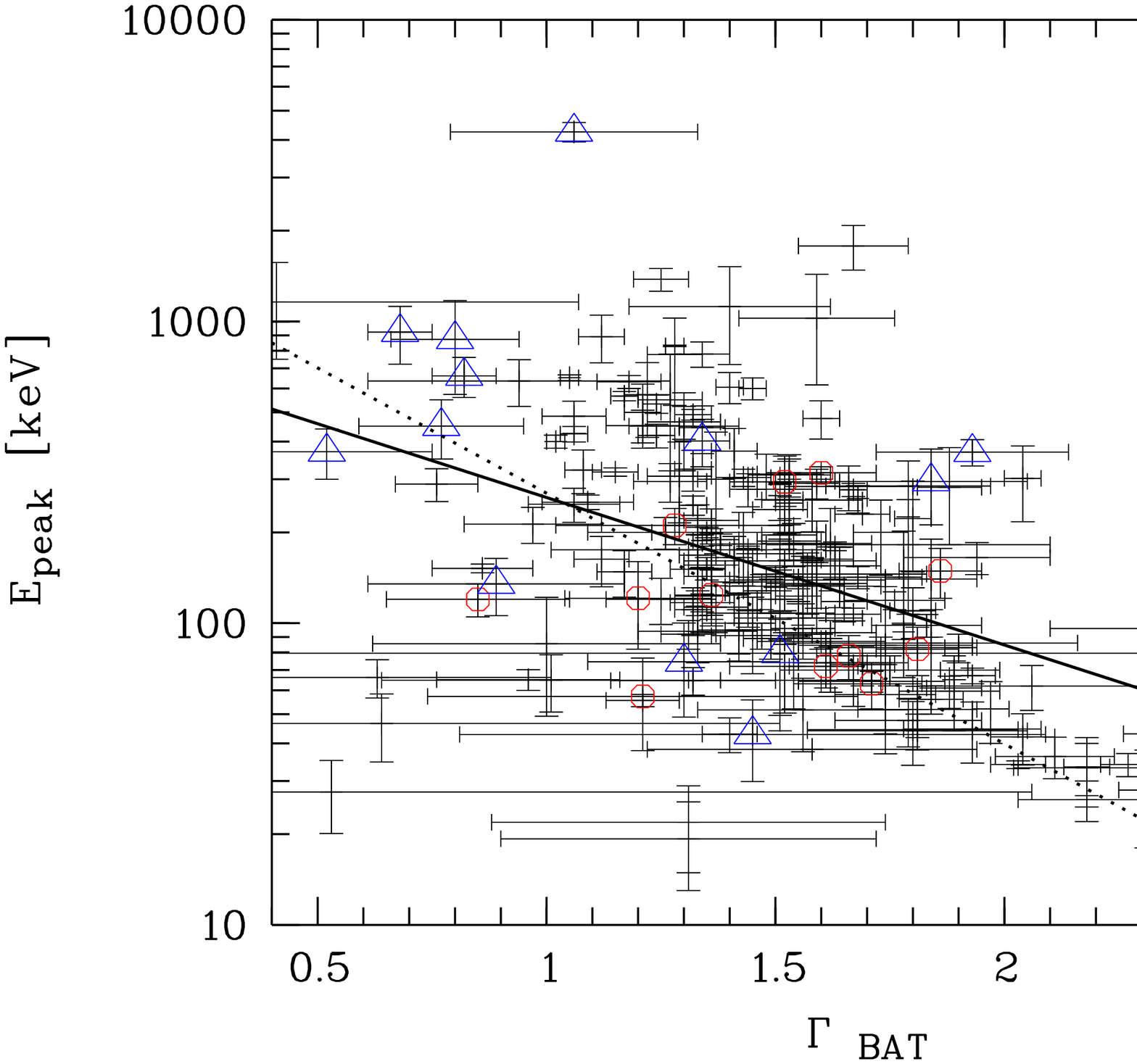}{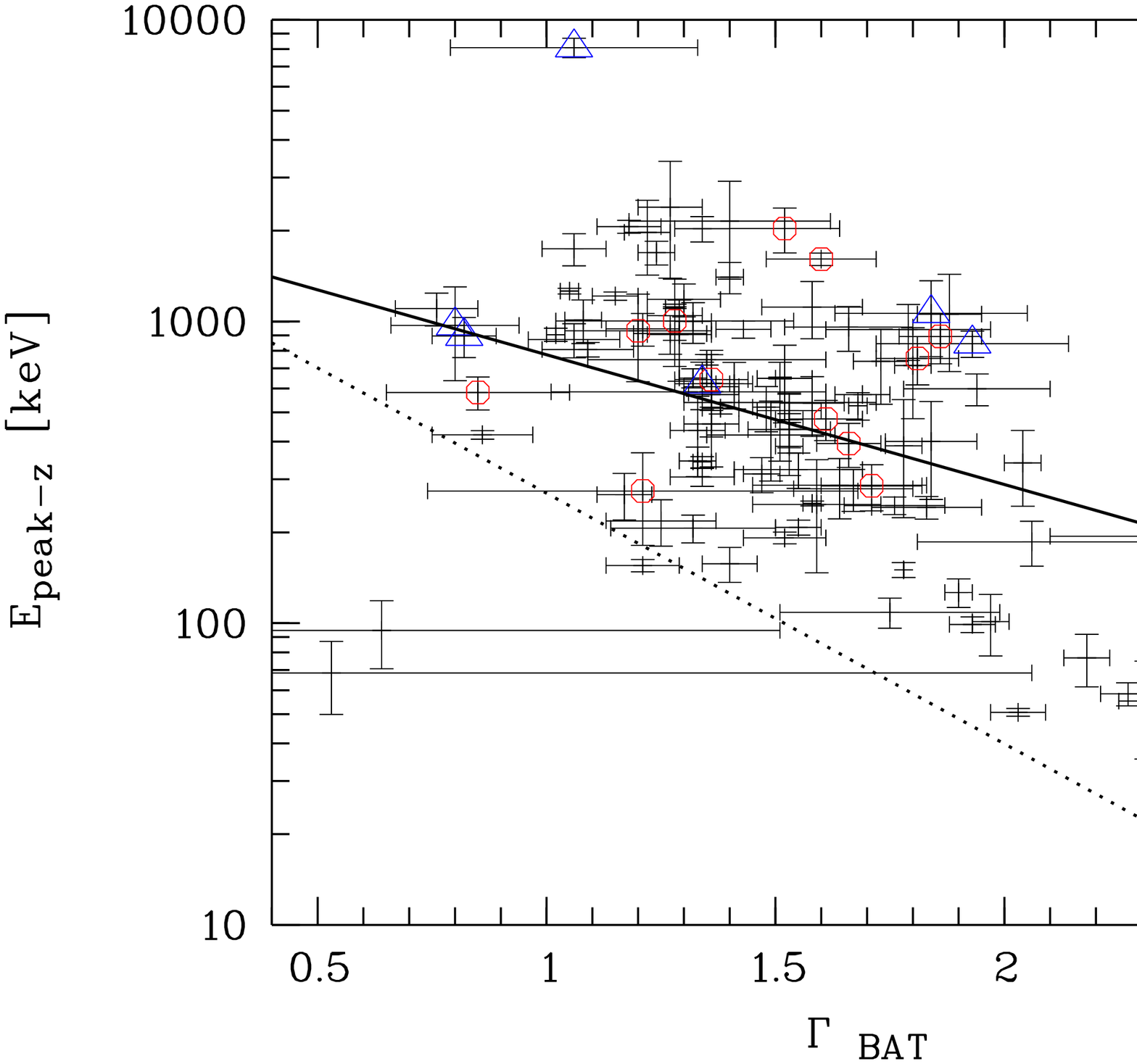}
\caption{\label{gamma_epeak} 
Correlation between the BAT 15-150 keV $\Gamma$ and 
the peak energy $E_{\rm peak}$  in the observed and
rest-frame (left and right panel, respectively).
The dashed lines  display the relation found by
\citet{sakamoto09} and the solid line displays the
relations found by us (see Appendix\,\ref{appendix_corr})
}
\end{figure*}

\begin{figure*}
%\epsscale{0.75}
\epsscale{1.5}
\plottwo{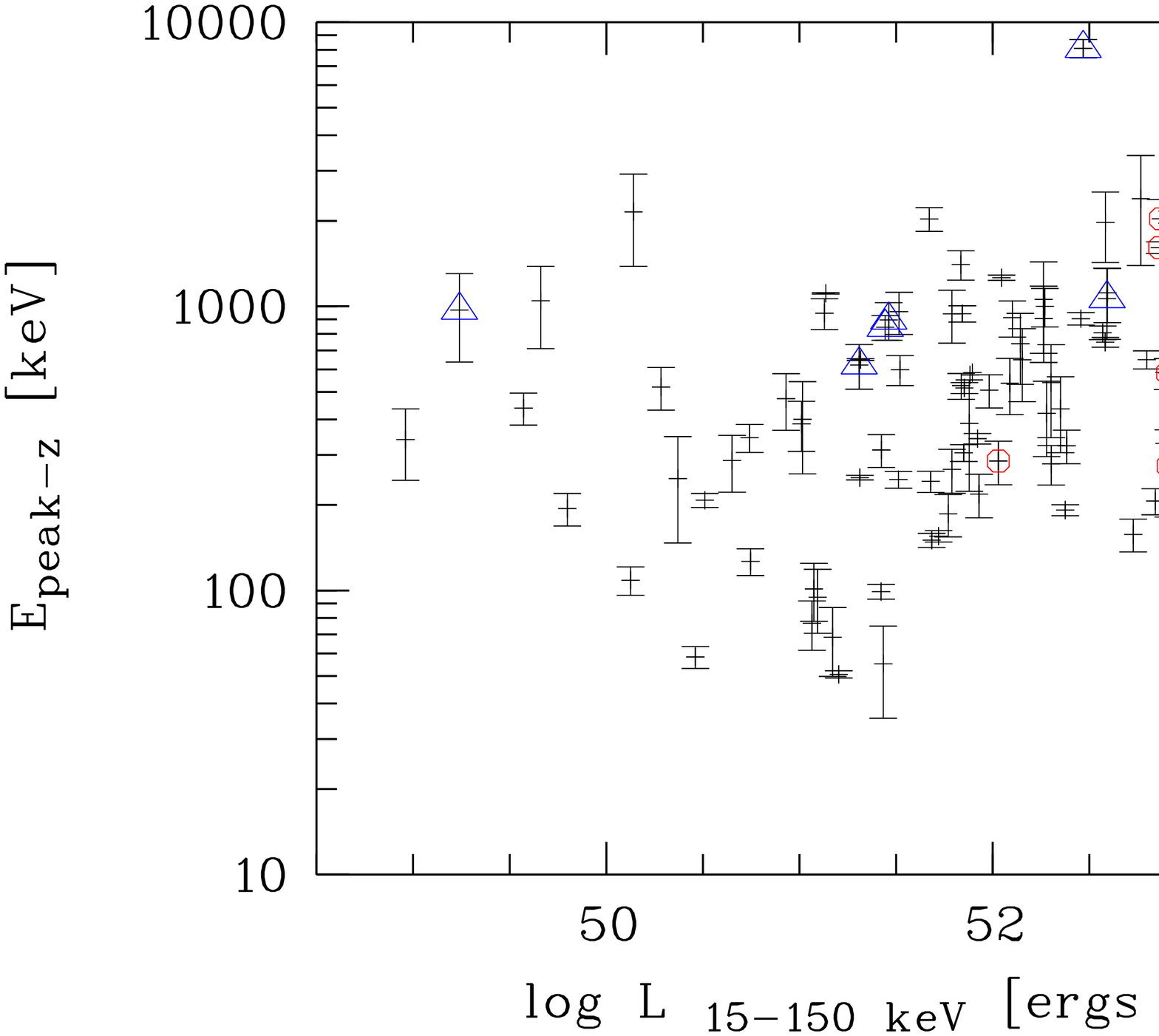}{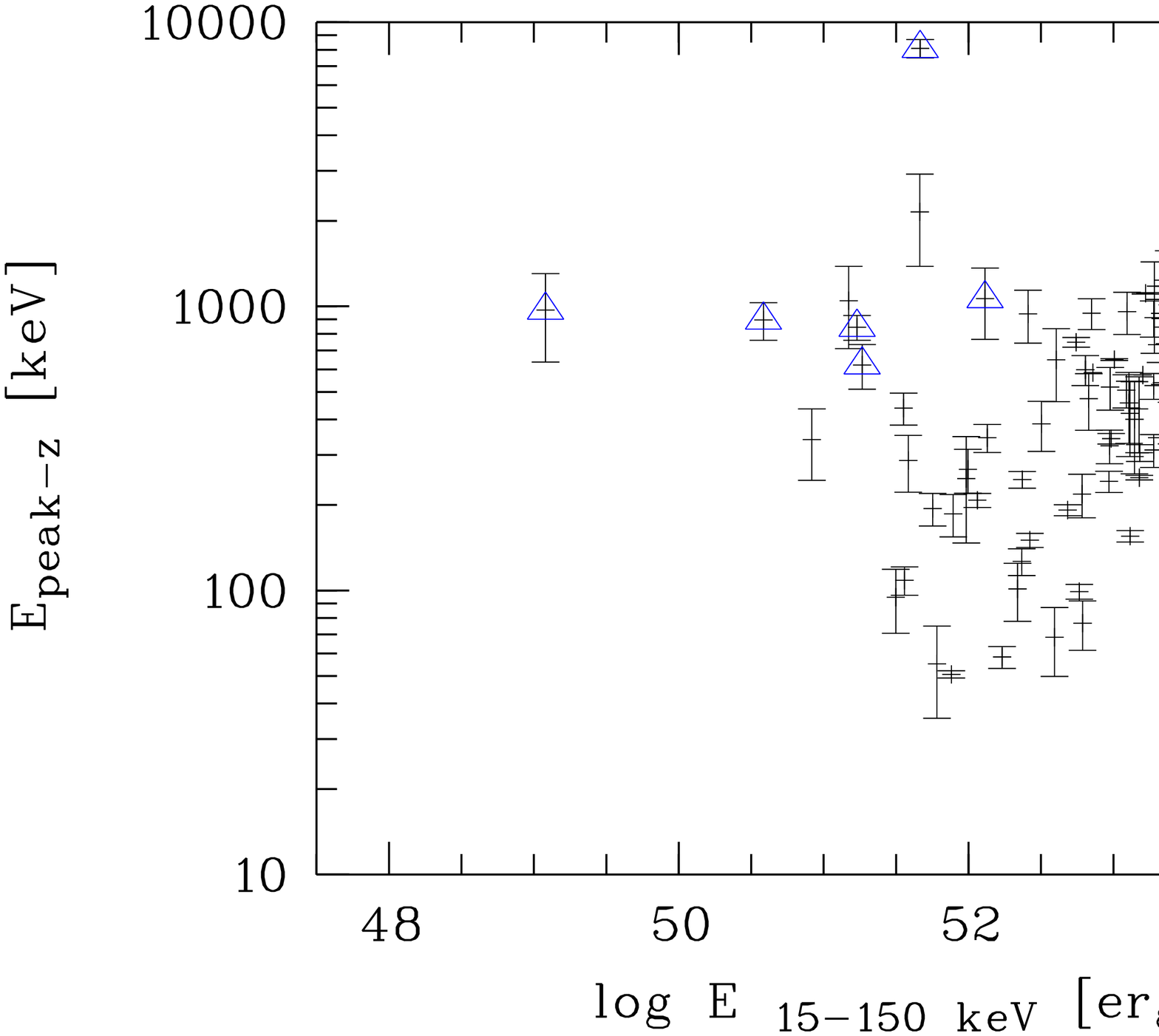}
\caption{\label{epeak_lum} 
Correlation between the peak energy in the rest-frame $E_{\rm peak, z}$ and the
k-corrected luminosity in the 15-150 keV BAT band. 
}
\end{figure*}

\begin{figure*}
%\epsscale{0.75}
\epsscale{1.5}
\plottwo{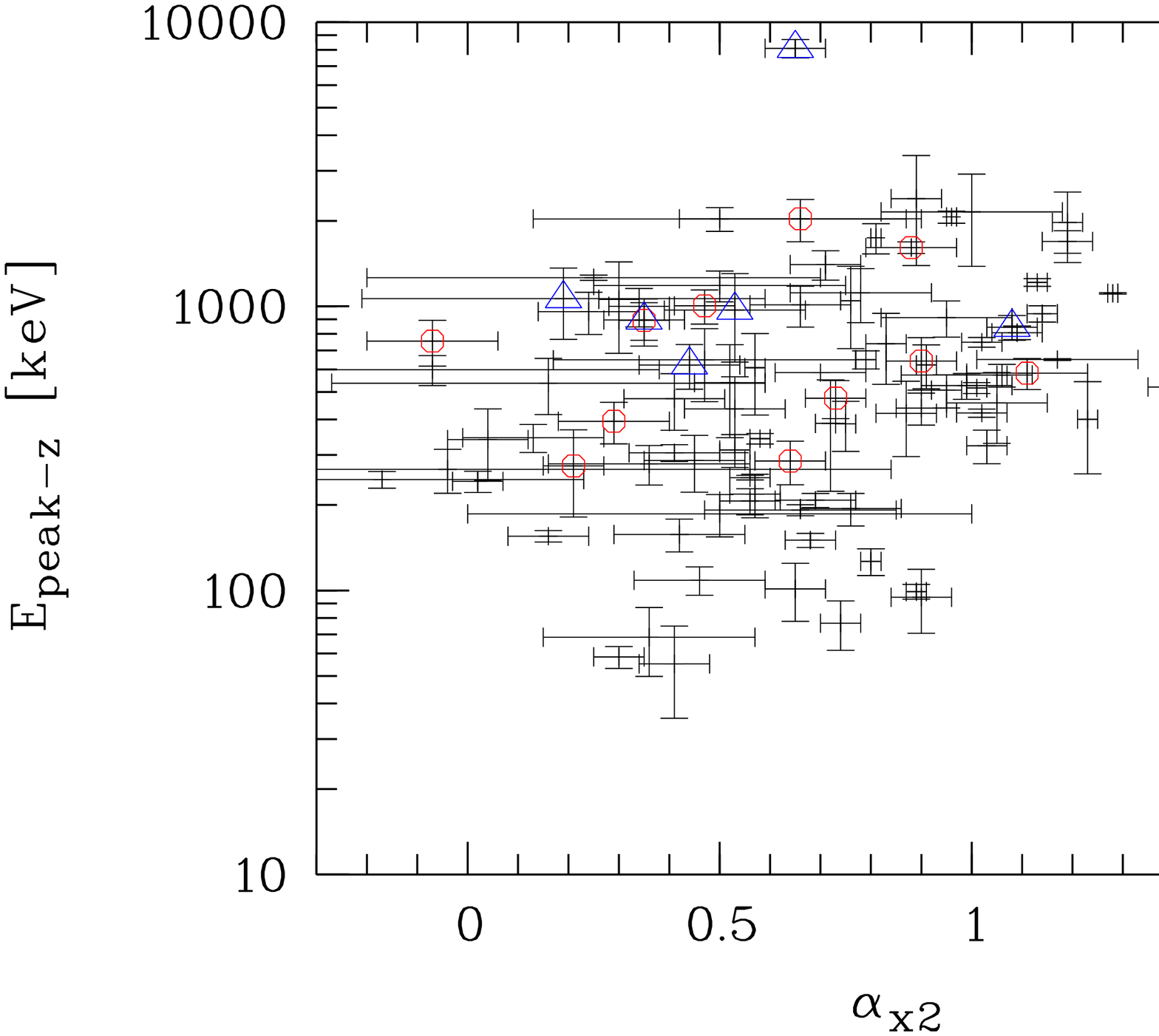}{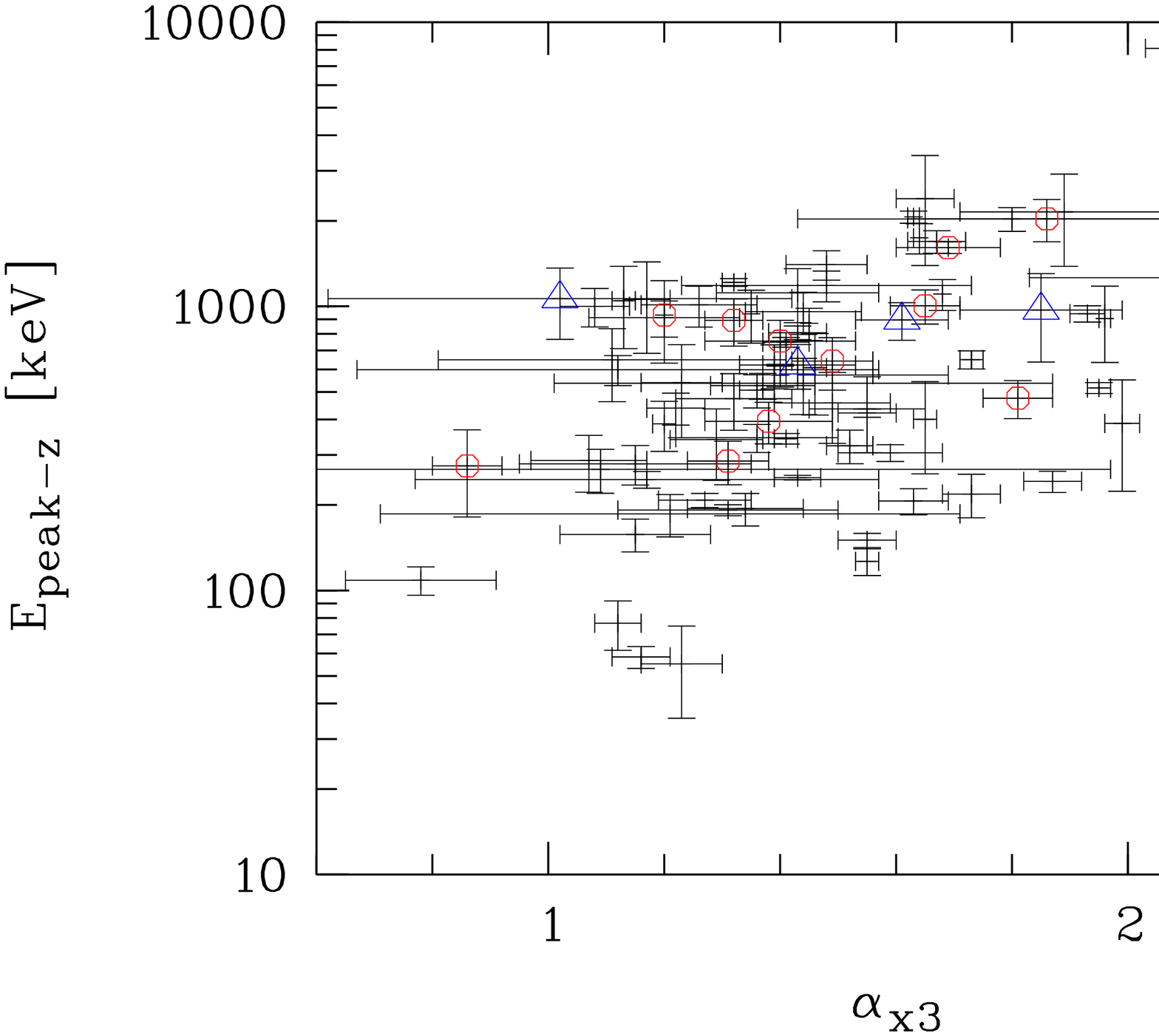}

\plottwo{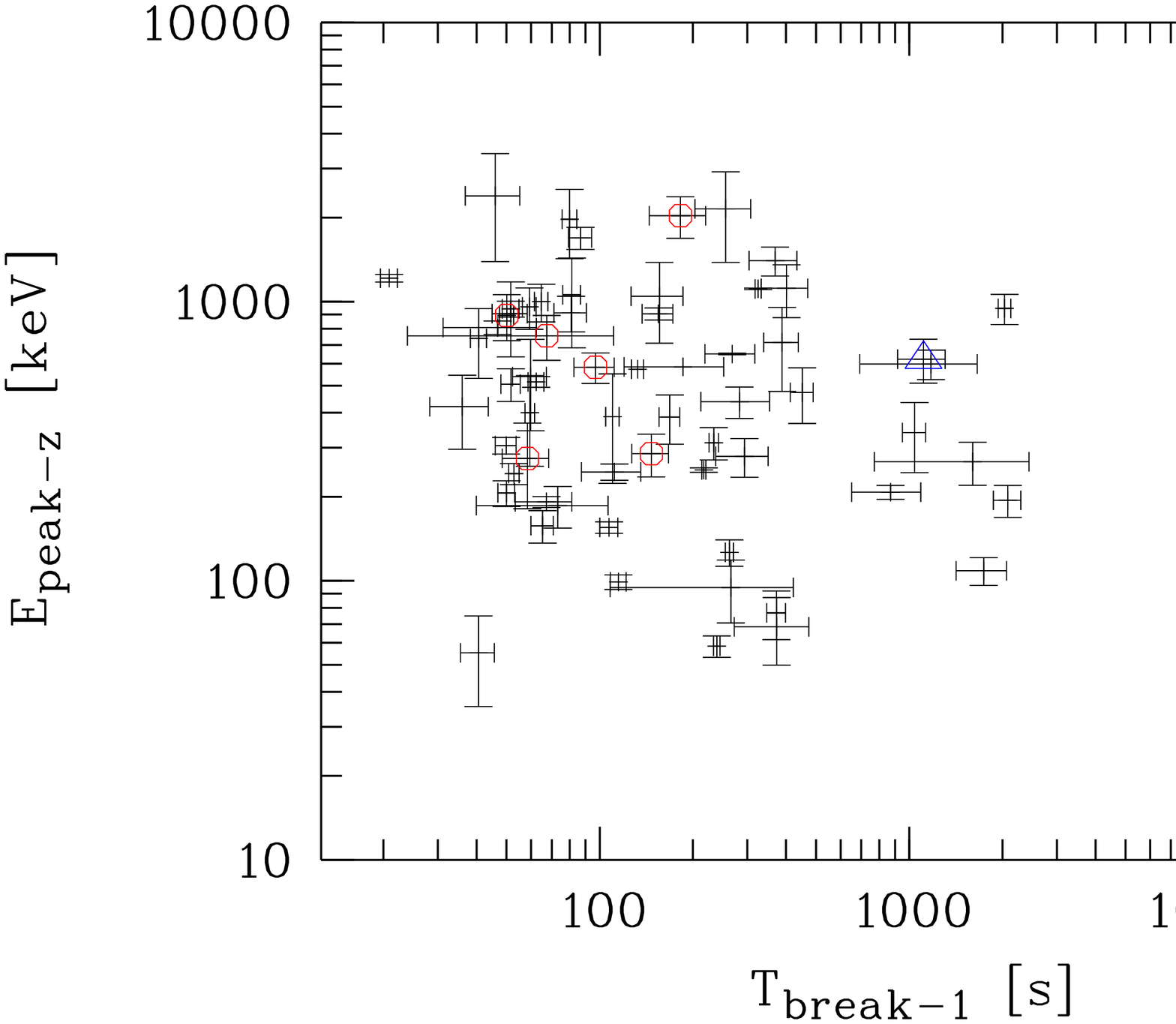}{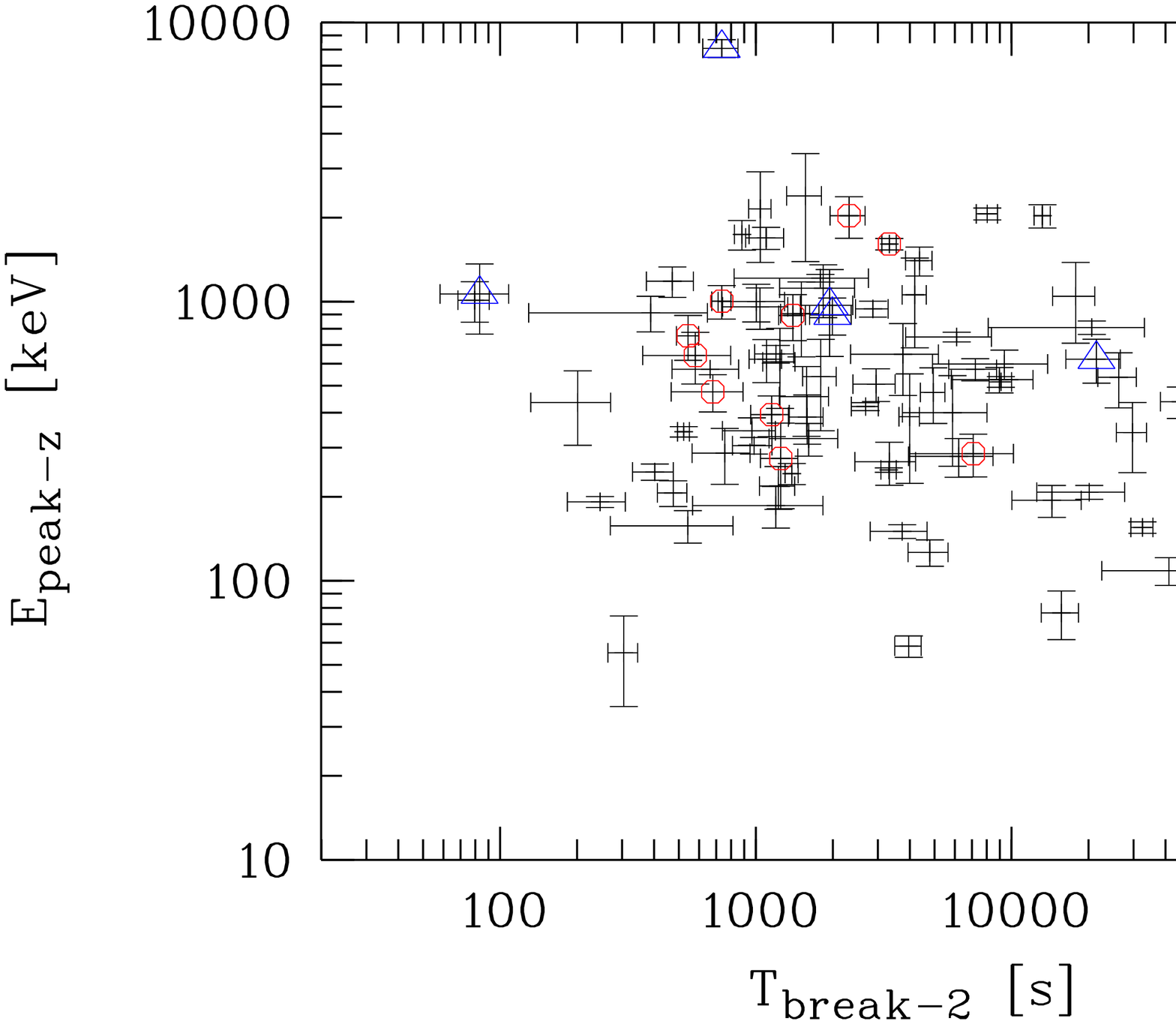}
\caption{\label{plateau_epeak} 
Correlation between the rest frame peak energy $E_{\rm peak, z}$ and the decay
slope during and after the X-ray afterglow plateau phase \axb\ and \axc, and 
the rest-frame break times at the beginning and end of the plateau phase $T_{break 1,z}$  and
$T_{\rm break 2,z}$.
}
\end{figure*}

\clearpage

\begin{figure*}
%\epsscale{0.75}
\epsscale{1.5}
\plottwo{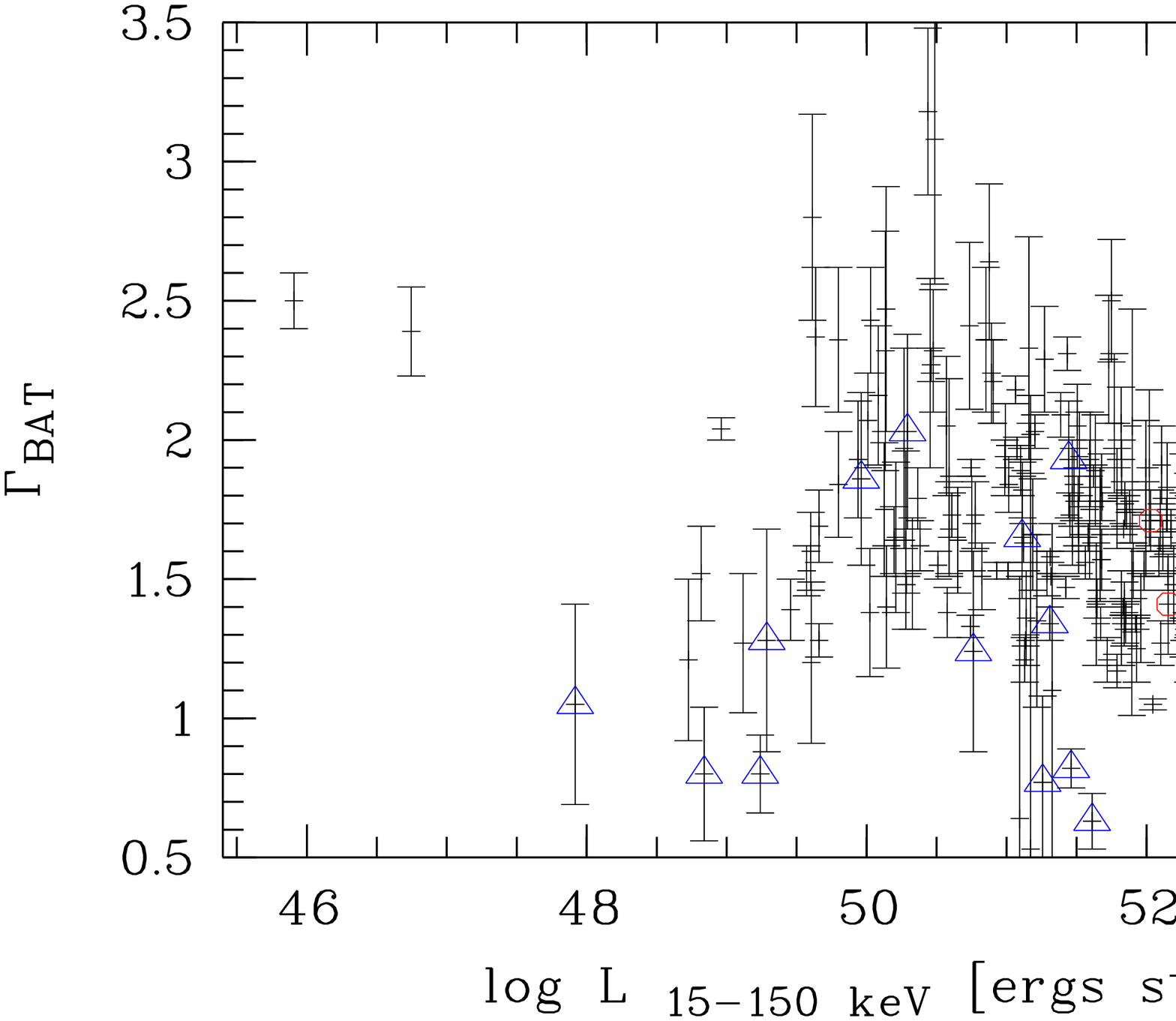}{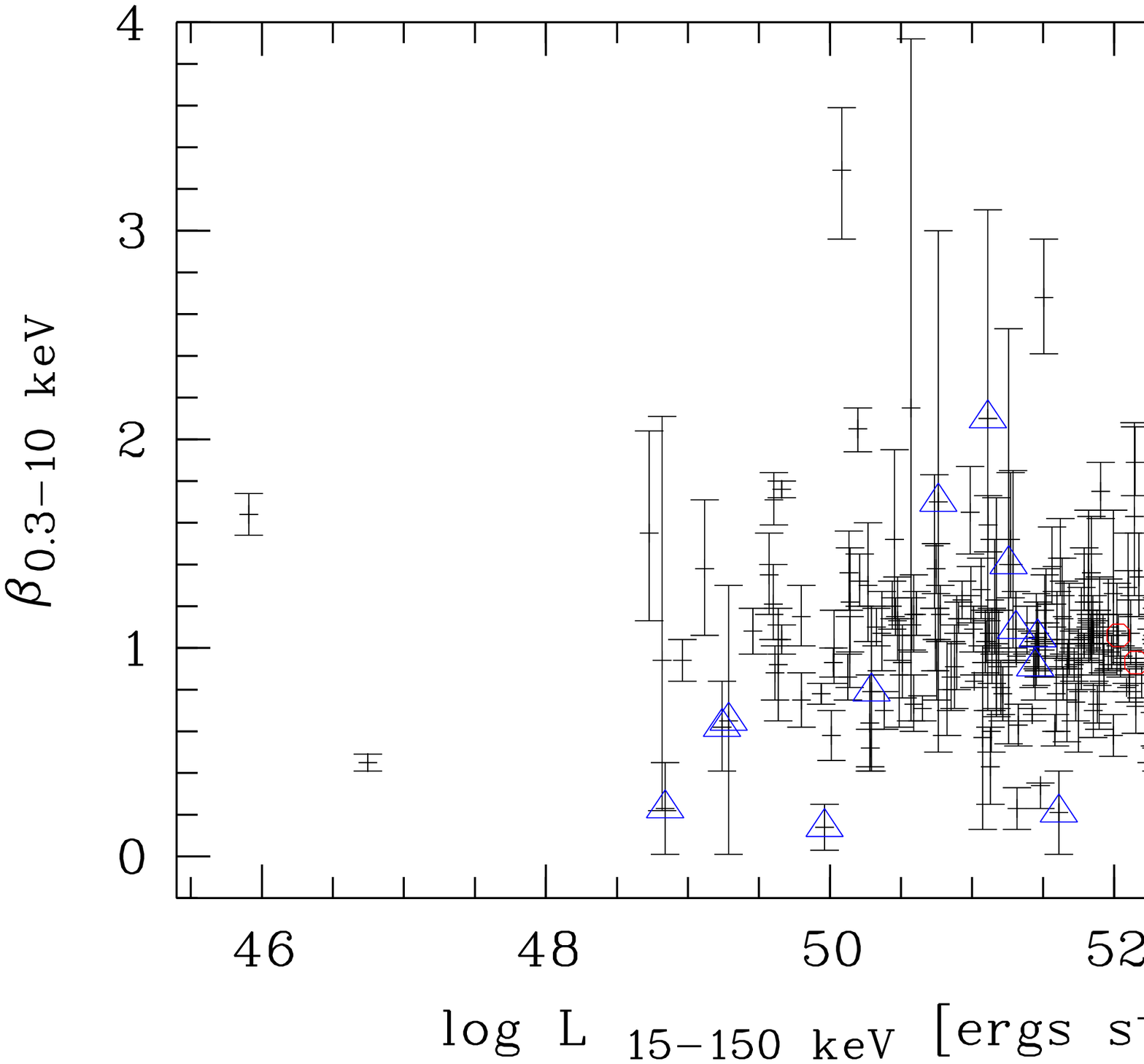}

\plottwo{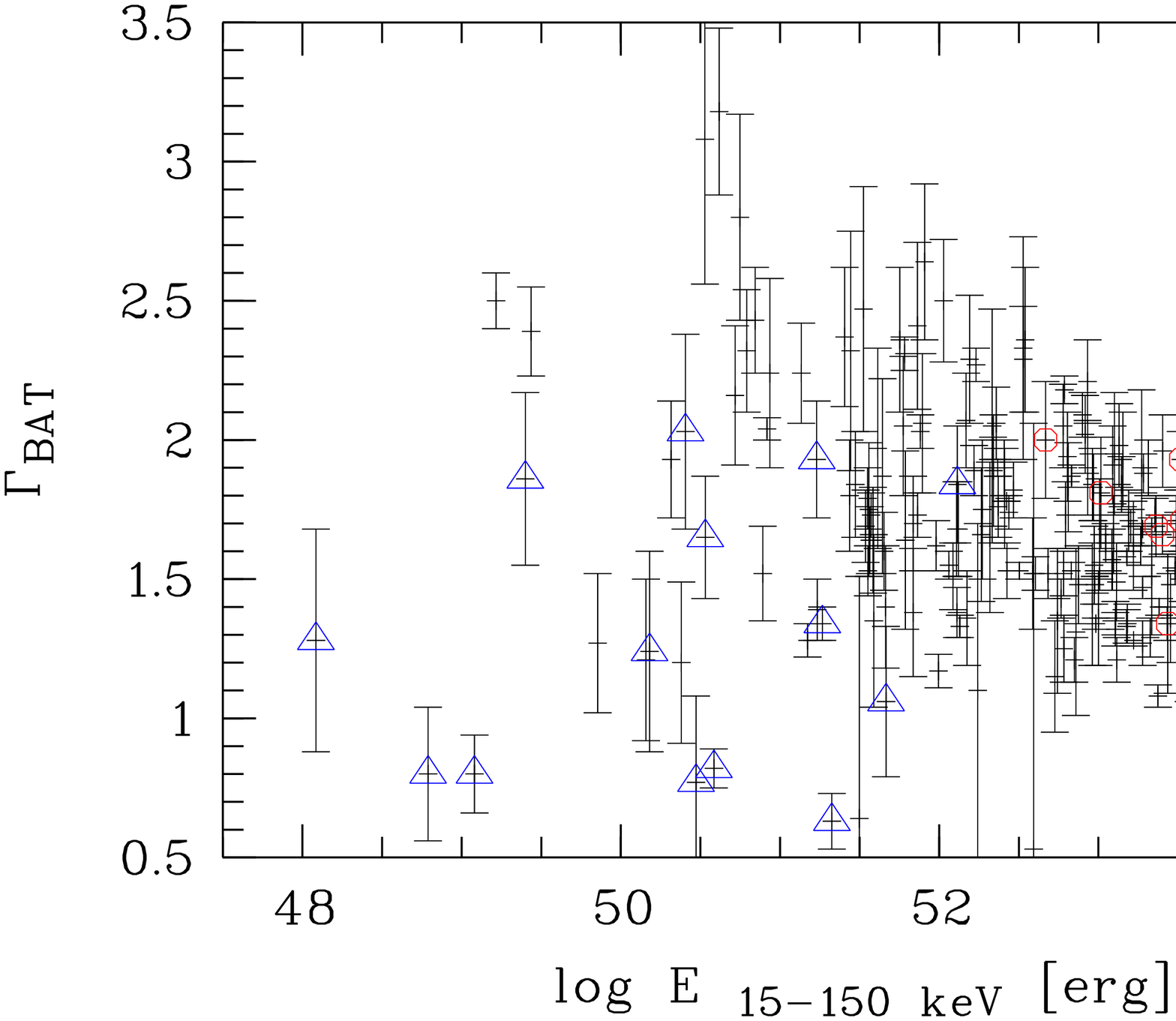}{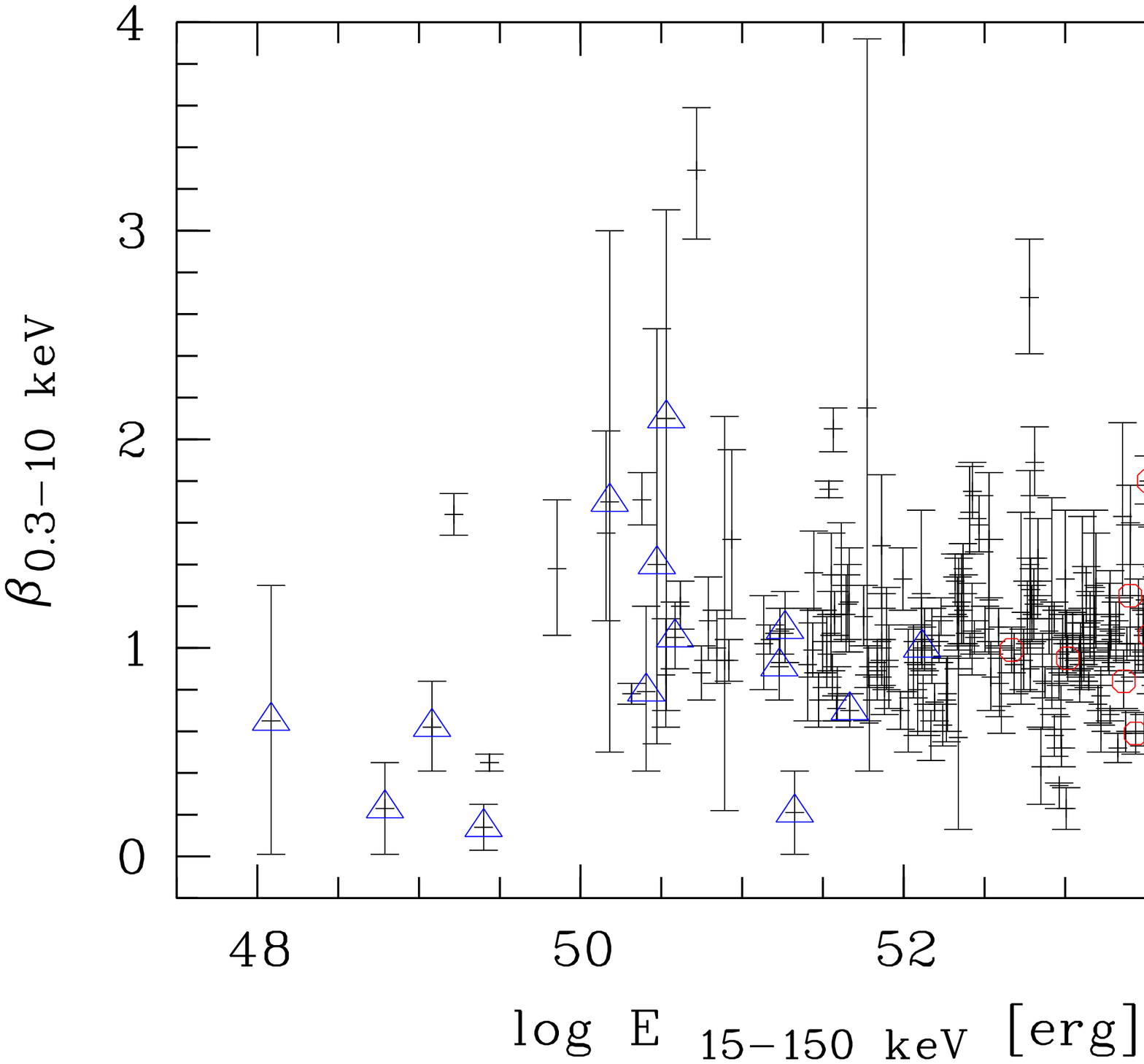}

\plottwo{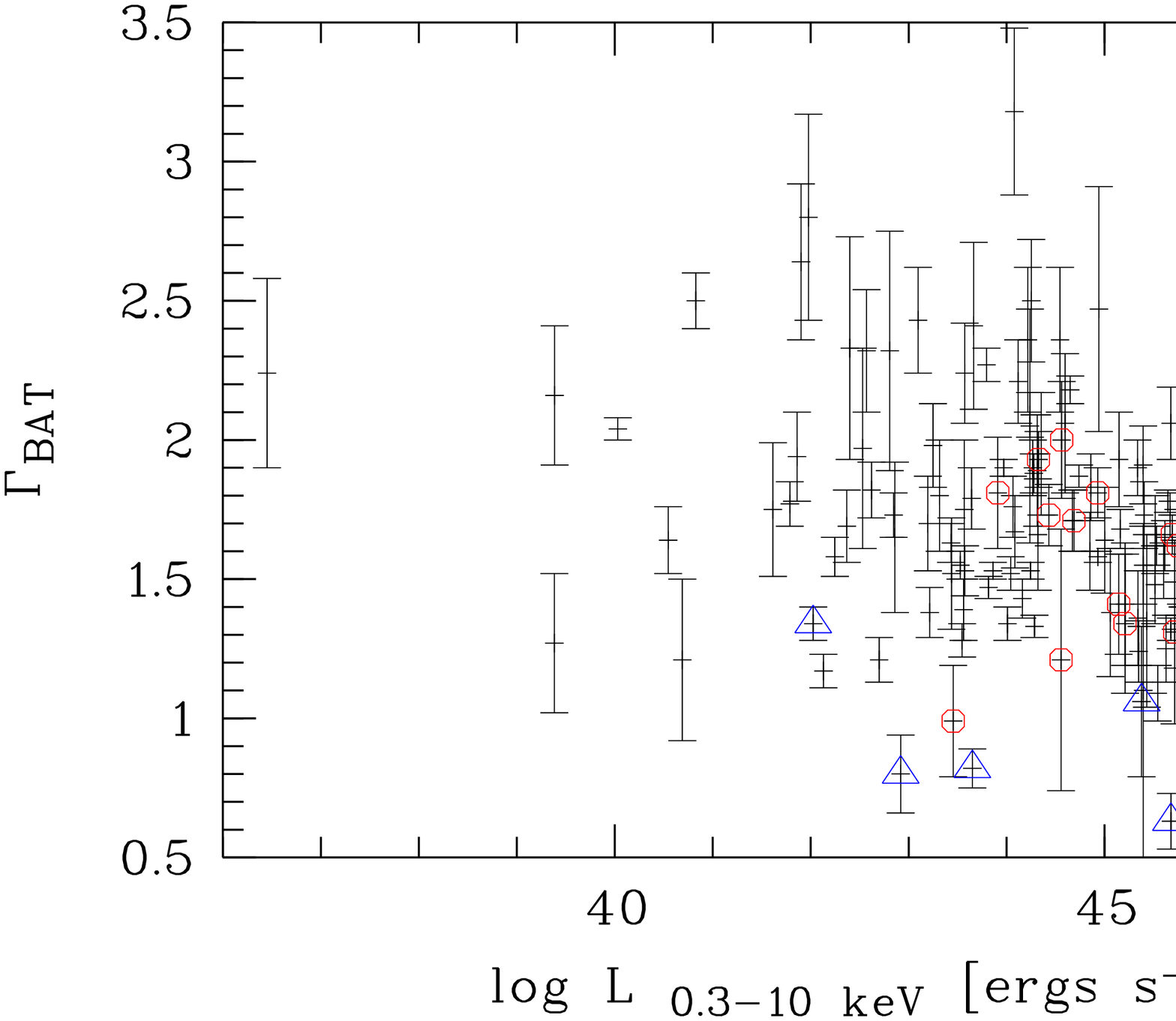}{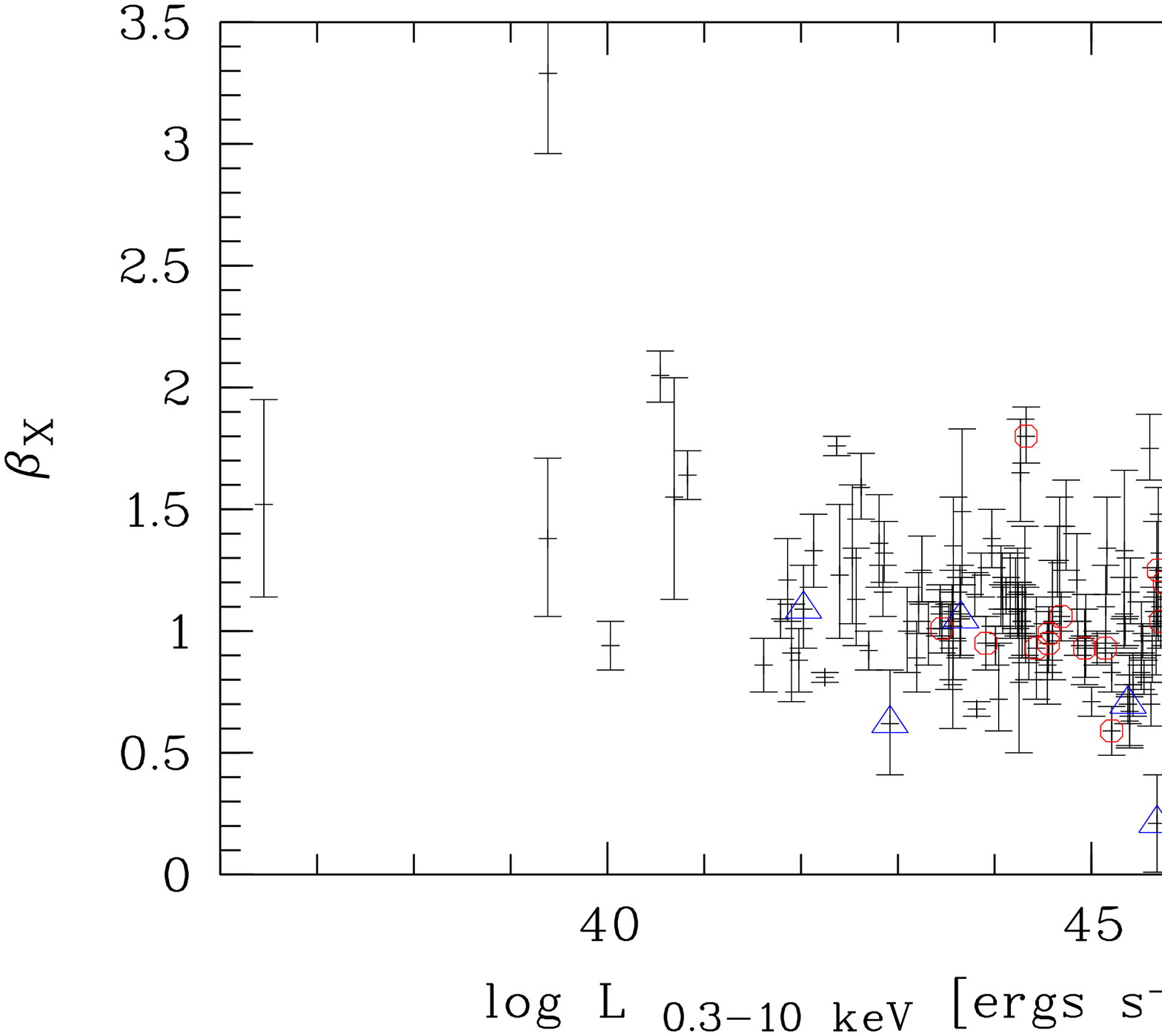}
\caption{\label{lum_gamma_bx} 
Relations of the k-corrected BAT luminosity and isotropic energy, and the luminosity in the 0.3-10 keV 
band
with the BAT photon index $\Gamma$
and the X-ray spectral slope \bx\ (left and right panel, respectively). 
Short
bursts in these plots are marked as triangles.
}
\end{figure*}

\clearpage

\begin{figure*}
%\epsscale{0.75}
\epsscale{1.5}
\plottwo{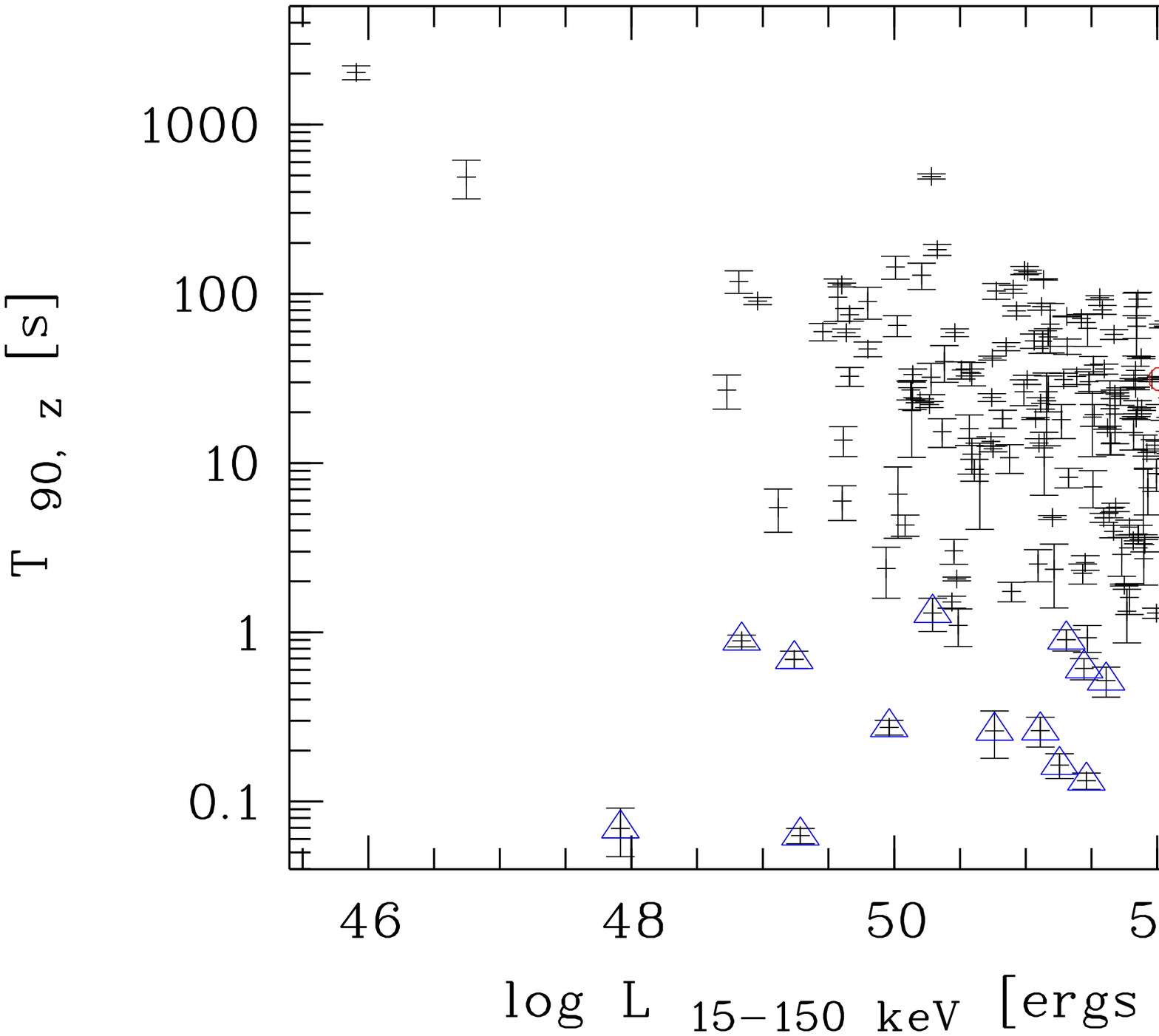}{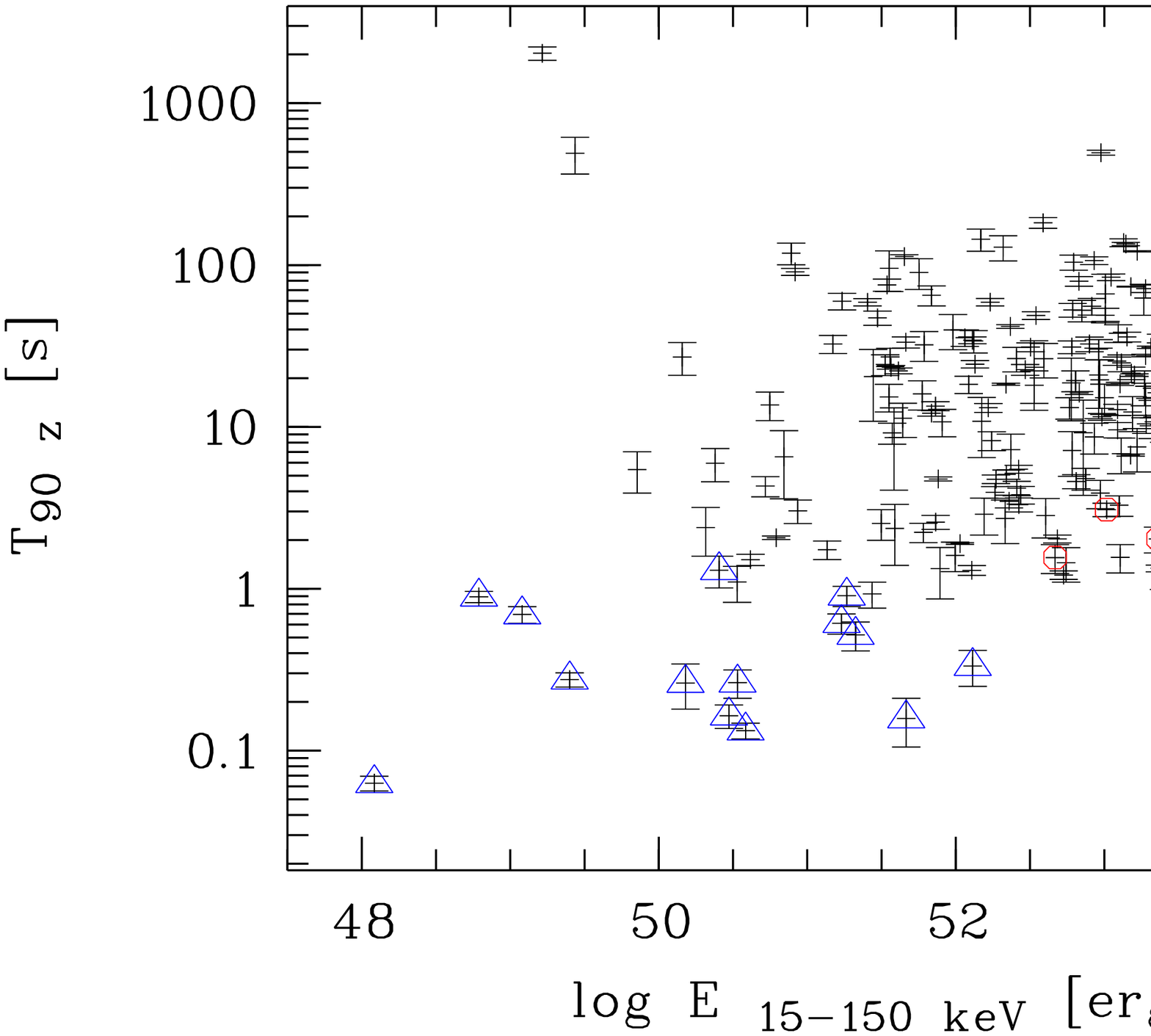}
\caption{\label{lum_t90} 
Anti-correlation between the 15-150 keV luminosity and the rest-frame $T_{90}$
 Short bursts in these plots are marked as triangles.}
\end{figure*}

\clearpage

\begin{appendix}
 
\section{GRB Catalogue \label{grb_catalog}}
Here we present the catalogue of our sample of 754 onboard \swift-BAT-deteced GRBs. 
The ASCII file of the catalogue is available as a machine readable file on the Astrophysical Journal
Supplement website linked to this paper. 
The explanation of the ASCII file is summarized in Table\,\ref{grb_table}.  
The ASCII catalogue is prepared so it can be read directly into MIDAS. In order to read it
into other data manipulations programs, the "*" has to be substituted with the appropriate symbol,
like 'NA' in R for example. A more detained explanation of all parameters used in this paper is 
listed in Appendix\,\ref{parameters}.

\begin{deluxetable}{cllll}
\tabletypesize{\scriptsize}
\tablecaption{GRB Catalogue ASCII file
\label{grb_table}}
\tablewidth{0pt}
\tablehead{
\colhead{Bytes} 
& \colhead{Format}
& \colhead{Units} 
& \colhead{Label}
& \colhead{Explanations}
} 
\startdata
1 - 3     & I3  & ---  &   Num  & Running index number \\
5 - 11    & A7  & ---  &   GRB  & GRB Name  \\
14 - 19   & I6  & ---  & Trigger\# & BAT trigger number \\
22 - 28   & F6.4 & --- & z & Redshift of the GRB \\
30 - 34   & F8.3 & s & $T_{90}$ & observed BAT 15 - 150 keV $T_{90}$ \\
36 - 41   & F7.3 & s & $T_{90}$ err & $T_{90}$ uncertainty \\
43 - 49   & E7.2 & erg cm$^{-2}$ & Fluence & Fluence in the 15-150 keV BAT band \\
51 - 57   & E7.2 & erg cm$^{-2}$ & Fluence err. & Uncertainty in the 15-150 keV BAT fluence \\
59 - 62   & F4.2 & ---  & $\Gamma$ & BAT 15 - 150 keV photon index $\Gamma$ \\
64 - 67   & F4.2 & ---  & $\Gamma_{\rm err}$ & Uncertainty of the BAT 15 - 150 keV photon index $\Gamma$ \\
69 - 73   & F7.2 & keV  & $E_{\rm peak}$ & Peak beak energy in keV at high energies \\
75 - 79   & F7.2 & keV  & $E_{\rm peak, err}$ & Uncertainty in $E_{\rm peak}$ \\
81 - 88   & F7.4 & ---  & log $E_{15-150 keV}$ & k-corrected energy in the 15-150 keV BAT band\tablenotemark{1} given in units of erg \\
90 - 96   & F7.4 & ---  & log $L_{15-150 keV}$ & k-corrected luminosity in the 15-150 keV BAT band\tablenotemark{1} given in units of erg s$^{-1}$. \\
98 - 102  & F6.2 & 10$^{20}$ cm$^{-2}$ & $N_{\rm H, gal}$ & Galactic absorption column density \citep{kalberla05} \\
104 - 109 & F6.2 & 10$^{20}$ cm$^{-2}$ & $N_{\rm H, fit}$ & free fit absorption column density\tablenotemark{2} \\ 
111 - 115 & F6.2 & 10$^{20}$ cm$^{-2}$ & $N_{\rm H, err, -}$ & negative uncertainty in the free fit $N_{\rm H}$  \\
117 - 122 & F6.2 & 10$^{20}$ cm$^{-2}$ & $N_{\rm H, err, -}$ & positive uncertainty in the free fit $N_{\rm H}$  \\
124 - 128 & F7.3 & 10$^{22}$ cm$^{-2}$ & $N_{\rm H, z}$ & absorption column density at the redshift of the burst \\
130 - 134 & F7.3 & 10$^{22}$ cm$^{-2}$ & $N_{\rm H, z, err, -}$ & negative uncertainty in absorption column density at the redshift of the burst. \\
136 - 140 & F7.3 & 10$^{22}$ cm$^{-2}$ & $N_{\rm H, z, err, +}$ & positive uncertainty in absorption column density at the redshift of the burst. \\
142 - 146 & F4.2 & ---  & \bx & 0.3-10 keV energy spectral index \\
148 - 152 & F5.2 & ---  & $\beta_{\rm x, err,-}$ & negative uncertainty in \bx \\
154 - 158 & F5.2 & ---  & $\beta_{\rm x, err,+}$ & positive uncertainty in \bx \\
160 - 164 & I5	 &  s   & $T_{\rm break 1}$ & X-ray light curve break time before the plateau phase \\
166 - 171 & I5	 &  s   & $T_{\rm break 1, err}$ & Uncertainty in $T_{\rm break 1}$ \\
173 - 178 & I6	 &  s   & $T_{\rm break 2}$ & X-ray light curve break time after the plateau phase \\
180 - 185 & I6	 &  s   & $T_{\rm break 2, err}$ & Uncertainty in $T_{\rm break 2}$ \\
187 - 191 & F5.2 & ---  & \axb & X-ray light curve decay slope during the plateau phase \\
193 - 196 & F4.2 & ---  & $\alpha_{\rm x2, err}$ & Uncertainty in \axb \\
198 - 201 & F4.2 & ---  & \axc & X-ray light curve decay slope during the plateau phase \\
203 - 206 & F4.2 & ---  & $\alpha_{\rm x3, err}$ & Uncertainty in \axc \\
\enddata

\tablenotetext{1}{The k-corrected energy and luminosity were based on the k-corrected fluence following the standard k-correction by \citet{oke68}:
$Fluence_{\rm k} = Fluence_{\rm obs}\times {1+z}^{2-\Gamma}$.}
\tablenotetext{2}{The excess absorption above the Galactic value is then given by $\Delta_{\rm NH} = N_{\rm H, fit} - N_{\rm H, gal}$. }
\end{deluxetable}

\section{Explanation of parameters \label{parameters}}

In this section we explain the parameters used in this paper and how they were derived. 
For this purpose we follow the order as listed
in the GRB catalogue, Table\,\ref{grb_table}. 

\begin{itemize}
\item Redshift z: The redshifts prior 2012 were taken primarily from the GRB redshift catalogues by 
 \citet{fynbo09, jakobsson12} and \citet{kruehler12}. All other redshift measurements were taken from GCN circulars.
\item $T_{90}$: This time in which 90\% of the energy is released in the prompt emission. The $T_{90}$ used in this paper
is the time measured in the observed 15-150 keV band in the \swift\ BAT.
\item $T_{90,z}$: rest frame $T_{\rm 90,z} = T_{90}/(1+z)$
\item $Fluence$: This is the fluence in the observed 15-150 keV BAT band. 
\item $Fluence_{\rm k}$: k-corrected fluence with $Fluence_{\rm k} = Fluence_{\rm obs}\times (1+z)^{2-\Gamma}$.
\item $\Gamma$: Photon index of the hard X-ray spectrum in the observed 15-150 keV band.
\item $E_{\rm peak}$: Peak energy of the hard energy spectrum derived from a Band function \citet{band93}. We took the
$E_{\rm peak}$ from GCN circulars. Before the launch of {\it Fermi} $E_{\rm peak}$ was primarily measured by Konus-Wind. After
the {\it Fermi} launch we primarily relied on {\it Fermi} GBM measurements.
\item $E_{\rm peak, z}$: rest frame $E_{\rm peak, z} = E_{\rm peak} \times (1+z)$.
\item $E_{\rm 15-150 keV}$: k-corrected energy in the 15-150 keV BAT band assuming an isotropic energy release.
\item $L_{\rm 15-150 keV}$: k-corrected luminosity in the 15-150 keV BAT band assuming an isotropic energy release, with
$L_{\rm 15-150 keV}$ = $E_{\rm 15-150 keV}$/$T_{90,z}$
\item $N_{\rm H, gal}$: Galactic absorption column density derived from the HI maps by \citet{kalberla05}. Throughout the
paper the Galactic $N_{\rm H}$ is given in units of $10^{20}$ cm$^{-2}$.
\item $N_{\rm H, fit}$: Free-fit absorption column density. This was derived when the 0.3-10 keV X-ray spectrum can not be fit
with just the Galactic value.
\item $\Delta_{\rm NH}$: Excess absorption density as explained in \citet{grupe07} with $\Delta_{\rm NH} = 
N_{\rm H, fit} - N_{\rm H, gal}$
\item $N_{\rm H, z}$: Absorption column density at the redshift of the burst. These column densities were derived by a
power law fit to the 0.3-10 keV X-ray spectrum plus an absorption parameter fixed to the Galactic value.
\item \bx: X-ray energy spectral index in the observed 0.3-10 keV band with ${\nu}F_{\nu} \propto \nu^{- \beta}$
\item $T_{\rm break 1}$: Observed break time in the X-ray afterglow light curve before the plateau phase.
\item $T_{\rm break 2}$: Observed break time in the X-ray afterglow light curve after the plateau phase.
\item $T_{\rm break 1, z}$: Break time in the rest frame in the X-ray afterglow light curve before the plateau phase, with 
$T_{\rm break 1, z} = T_{\rm break 1}/(1+z)$.
\item $T_{\rm break 2, z}$: Break time in the rest frame in the X-ray afterglow light curve after the plateau phase, with 
$T_{\rm break 2, z} = T_{\rm break 2}/(1+z)$.
\item \axb: Decay slope of the plateau phase 
of the X-ray afterglow light curve in the 0.3-10 keV band with ${\nu}F_{\nu} \propto \nu^{- \alpha2}$
\item \axc: Decay slope of the normal decay phase
of the X-ray afterglow light curve in the 0.3-10 keV band with ${\nu}F_{\nu} \propto \nu^{- \alpha3}$
\item $Fluence_{\rm 0.3-10 keV}$: Fluence in the observed 0.3-10 keV band during the plateau phase.
This fluence was determined by integrating over
the afterglow 0.3-10 keV flux light curve starting at the beginning of the plateau phase $T_{\rm break 1}$
to the end of the plateau phase $T_{\rm break 2}$.
\end{itemize}

\section{Additional Distributions \label{distr_append}}
In this section we show additional distributions of GRB parameters observed by \swift\ that have not been shown 
in Section\,\ref{distribution}

As mentioned in Section\,\ref{distribution}, we applied the following statistical visualization tools
to the data sets, which are common in data mining:

\begin{enumerate}
\item The left panel displays
the histogram with the kernel density estimator (solid line) of the distribution. 
On the bottom of the plot the distribution of the real values of this property are shown.
\item The middle plot shows a box diagram for each property for short, long and all GRBs
(top to bottom). The box displays the 1. (mean), 2. (median),  
and 3. quartile and the
'whiskers' which are defined as 
the minimum/maximum values of the distribution or the 1.5 times 
the interquartile range (so between the 1 and 3 quartile, so basically the 95\%
confidence level),  whatever comes first. Values
beyond the 'whiskers' are outliers and are displayed as circles.
\item The right panel displays the quartile-quartile plot (Q-Q-plot). This plot makes it
easy to identify how well (or not) a distribution agrees with a Gaussian distribution
which is shown as the solid line in these plots. The dashed lines mark
 the region of 95\% confidence. 
\end{enumerate}

More information about these tools can be found in \citet[e.g.][]{feigelson12, torgo11, crawley07}.

Figure\,\ref{distr_gamma_bx} displays the
distributions of the observed 15-150 kev photon index in the BAT and the 0.3-10
keV X-ray energy spectral slope \bx. The  distributions of $\Gamma$ is close to a Gaussian
distribution as shown in the Q-Q plot,
with extended tails towards steep spectral slopes. In the
distribution plot of the BAT photon index $\Gamma$, GRB 060202B \citep{aharonian09} is off
this plot with its $\Gamma=4.97$\plm0.49. 
The box plot of the $\Gamma$ distribution makes it apparent that 
 short GRBs tend to have flatter hard X-ray spectra compared with long GRBs, as expected. 
Short GRBs
have a median photon spectral index in the 15-150 keV range of 
$\Gamma$=0.94 while long GRBs have median of $\Gamma$=1.56.
 Note that although $T_{90}$ and 15-150 keV 
fluence measurements exist for all the bursts, three early bursts (041219A, B, and C) did not
have any spectral data available. In addition, GRB 120401 came into the BAT Field of view
during a slew, therefore $T_{90}$ and the fluence could not be measured \citep{palmer12}.
In the 0.3-10 keV X-ray band, however, the spectral 
slopes of the X-ray afterglows of of short and long GRBs  are
 similar with median values of \bx=0.82 and 0.99 for short and long GRBs, respectively. 
  As noticed by \citet{obrien06} from a sample of
40 early \swift\ bursts, the spectral slope of the afterglow data appears to be
softer compared with the 15-150 keV slope during the prompt emission.

Figure\,\ref{distr_fluence} displays the distributions of the observed and k-corrected 
15-150 keV fluence and the objects with the lowest fluence are short duration 
 As we have discussed in Section\,\ref{selection}, 
this is also the result of a selection effect: we can not detect long bursts 
with low fluence because their signal will be dominated by detector background. Bursts with low fluence
therefore need to emit their energy in a short amount of time. 
 Similar to
these distributions is the distribution of the rest-frame 15-150 keV 
k-corrected luminosity and the isotropic energy shown 
in Figure\,\ref{distr_l15_150}.
  
The distributions of the 
peak energies $E_{\rm peak}$ in the observed and rest-frame 
are shown in Figure\,\ref{distr_epeak}. As expected from the 
$E_{\rm peak} - \Gamma$ anti-correlation found by \citet{sakamoto09}, 
short GRBs which have flatter hard X-ray spectra than long-duration GRBs
tend to have very high
peak energies. As a matter of fact the burst with the highest $E_{\rm peak,z}$ in our sample is the short 
GRB 090510, which has also been detected in the {\it Fermi} LAT 
\citep[e.g., ][]{abdo09, depasquale10}. Note, however, that the number of short bursts with redshifts and
$E_{\rm peak}$ measurements is only 6.

\section{Additional Correlation analysis \label{appendix_corr}}

In this Appendix we describe those correlations which have not been discussed in Section\,\ref{correlation}.
We list also Kendall's and Pearson's correlation analysis to the data in addition to the Spearman rank order analysis
discussed in Section\,\ref{correlation}.
The number of
correlation parameter pairs are the same as listed for the
Spearman rank order correlations listed in 
Table\,\ref{correlation_tab_all} for the entire sample and Table\,\ref{correlation_tab_z} for the sample with GRBs with spectroscopic redshift measurements.
Note the the number of parameter pairs for each correlation are listed in thee tables. In the tables below, the Kendall's $\tau$ and the probability are listed
above the diagonal and the Pearson correlation coefficient and probability are listed below. Table\,\ref{correlation_kendall_all} lists the Kendall $\tau$ value and the
Pearson correlation coefficients for all GRBs and Table\,\ref{correlation_tab_z_kendall} for those GRBs with spectroscopic redshift measurements.

\begin{deluxetable}{lccccccccc}
\rotate
\tabletypesize{\tiny}
\tablecaption{Kendall's $\tau$ and Pearson correlation coefficients and probabilities of for the observed parameters of all \swift-detected GRBs\tablenotemark{1}.
 \label{correlation_kendall_all}}
\tablewidth{0pt}
\tablehead{
& \colhead{$\Gamma_{\rm BAT}$} 
& \colhead{\bx} 
& \colhead{\axb} 
& \colhead{\axc} 
& \colhead{$T_{\rm 90}$} 
& \colhead{$T_{\rm break 1}$} 
& \colhead{$T_{\rm break 2}$} 
& \colhead{15-150 keV fluence}
}
\startdata
$\Gamma_{\rm BAT}$   & ---   & 0.123, 4.76$\times 10^{-6}$ & $-$0.176, $<10^{-8}$ & $-$0.215, $<10^{-8}$ & +0.054, 0.0271 & +0.048, 0.1676 & +0.083, 0.0117
& $-$0.1025, 2.89$\times 10^{-5}$ \\
\bx    & $+$0.254, $<10^{-8}$ & --- & $-$0.103, $4.17\times 10^{-4}$ & $-$0.116, $1.88\times 10^{-4}$ & +0.021, 0.4317 & $-$0.049, 0.1504 & +0.1106, $7.95\times 10^{-4}$ 
& +0.006, 0.817\\
\axb   & $-$0.228, $8.31\times 10^{-8}$ & $-$0.161, $1.70 \times 10^{-4}$ & --- & +0.279, $<10^{-8}$ & +0.144, $.6.49\times 10^{-7}$ & $-$0.023, 0.5075 & +0.1725, $1.79\times
10^{-7}$  & +0.1831, $<10^{-8}$ \\
\axc   & $-$0.229, $4.62 \times 10^{-7}$ & $-$0.128, 0.0052 & $+$0.234, $1.48\times 10^{-6}$ & --- & +0.1841, $<10^{-8}$ & +0.086, 0.0339  & +0.163, $8.46\times 10^{-7}$
& +0.1547, 5.18$\times 10^{-7}$ \\
$T_{\rm 90}$       &  $+$0.162, $8.16 \times 10^{-6}$ & $-$0.005, 0.905 & $+$0.212, $7.33\times 10^{-7}$ & $-$0.045, 0.325 & --- & +0.223, $<10^{-8}$ & +0.348, $<10^{-8}$ 
& +0.481, $<10^{-8}$ \\
$T_{\rm break 1}$  &  $+$0.033, 0.5218 & $-$0.059, 0.2451 & $+$0.010, 0.852 & $+$0.075, 0.2143 & $+$0.296, $< 10^{-8}$ & --- & +0.360, $<10^{-8}$ 
& $-$0.042, 0.2169 \\
$T_{\rm break 2}$  &  $+$0.088, 0.0744 & $+$0.166, $6.91\times 10^{-4}$ & $+$0.272, $272\times 10^{-8}$ & $+$0.055, 0.2595 & $+$0.526, $<10^{-8}$ & $+$0.531,
$<10^{-8}$ & ---   & +0.126, 1.25$\times 10^{-3}$ \\
15-150 keV fluence & $-$0.1007, 0.0058 & $-$0.055, 0.165 & $+$0.261, $<10^{-8}$ & $-$0.019, 0.6741 & $+$0.715, $<10^{-8}$ & $-$0.016, 0.7498
& $+$0.235, $1.20\times 10^{-6}$ & --- \\
\enddata

\tablenotetext{1}{Kendall's $\tau$ and Pearson correlation coefficients and probabilities are listed
above and below the diagonal, respectively. The number of parameter pairs are given in Table\,\ref{correlation_tab_all}.
}

\end{deluxetable}

\begin{deluxetable}{lccccccccccc}
\rotate
\tabletypesize{\tiny}
\tablecaption{Kendall's $\tau$ and Pearson correlation coefficients and probabilities of GRBs with spectroscopic redshifts\tablenotemark{1}
\tablenotemark{1}
 \label{correlation_tab_z_kendall}}
\tablewidth{0pt}
\tablehead{
& \colhead{$\Gamma_{\rm BAT}$}
& \colhead{\bx} 
& \colhead{\axb} 
& \colhead{\axc} 
& \colhead{$T_{\rm 90,z}$}
& \colhead{$T_{\rm break 1,z}$}
& \colhead{$T_{\rm break 2,z}$}
& \colhead{$L_{\rm 15-150 keV}$}
& \colhead{$E_{\rm iso}$}
& \colhead{$E_{\rm peak, z}$} 
}
\startdata
$\Gamma_{\rm BAT}$   & ---    & 0.140, $1.70\times 10^{-3}$ & $-$0.186 $7.53\times 10^{-5}$ & $-$0.189, $1.35\times 10^{-4}$ & 0.069, 0.117 & 0.072, 0.2054 & 
  0.107, 0.036 & 0.2675, $<10^{-8}$ &  $-$0.266, $<10^{-8}$ & $-$0.301, $6.51\times 10{-6}$ \\
\bx                  & $+$0.228, 8.82$\times 10^{-4}$ & --- & $-$0.122, 0.0093 & $-1$0.1809, 2.58$\times 10^{-4}$ & 0.050, 0.257 & 0.074, 0.1915 & 0.182, $3.88\times 10^{-4}$ &
  $-$0.090, 0.0443, & $-$0.060, 0.0604 & $-$0.089, 0.1878 \\
\axb                 & $-$0.248, $2.99 \times 10^{-5}$ & $-$0.224, 0.0012  & --- & 0.2189, 2.07$\times 10^{-5}$ & 0.1235, 0.0084 & $-$0.0456, 0.4311 & 
   0.0633, 0.2149, & 0.0887, 0.0592 & 0.1627, 5.31$\times 10^{-4}$ & 0.2267, 1.04$\times 10^{-3}$ \\
\axc                 & $-$0.157, 0.1556 & $-$0.244, $7.34 \times 10^{-4}$ & $+$0.202, 0.0074 & --- & 0.1272, 0.0098 & $-$0.075, 0.2361 & 0.043, 0.4033 &
   $-$0.1100, 0.0256 & 0.2057, 2.97$\times 10^{-5}$ & 0.1904, 0.0093 \\
$T_{\rm 90,z}$       & $+$0.143, 0.0298 & $+$0.029, 0.6643 & $+$0.198, 0.0041 & $+$0.086, 0.2392 & --- & 0.251, 8.46$\times 10^{-6}$ & 
   0.305, $<10^{-8}$ & $-$0.206, 3.09$\times 10^{-6}$ & 0.130, 3.30$\times 10^{-3}$ & $-$0.023, 0.726 \\
$T_{\rm break 1,z}$  & $+$0.050, 0.5524 & $+$0.079, 0.3449 & $-$0.023, 0.7824 & $-$0.101, 0.2836 & $+$0.362, $8.01\times 10^{-6}$ & --- & 
    0.432, $<10^{-8}$ & $-$0.409, $<10^{-8}$ & $-$0.3317, $<10^{-8}$ & $-$0.089, 0.2923 \\
$T_{\rm break 2,z}$  & $+$0.160, 0.03398 & $+$0.335, $5.76\times 10^{-6}$ & $+$0.103, 0.175 & $+$0.070, 0.3535 & $+$0.450, $<10^{-8}$ & $+$0.613, $<10^{-8}$ &
    --- & $-$0.285, $2.05\times 10^{-8}$ & $-$0.131, 0.0100 & $-$0.106, 0.1599 \\
$L_{\rm 15-150 keV}$ & $-$0.343, $9.01 \times 10^{-8}$ & $-$0.129, 0.0512 & $+$0.110, 0.1154 & $+$0.058, 0.4289 & $-$0.238, $2.65\times 10^{-4}$ & 
$-$0.574, $<10^{-8}$ &    $-$0.439, $<10^{-8}$ &  --- & 0.664, $<10^{-8}$ & 0.293, 1.19$\times 10^{-5}$ \\
$E_{\rm iso}$        & $-$0.256, $8.34 \times 10^{-5}$ & $-$0.113, 0.0886 & $+$0.222, $1.3\times 10^{-3}$ & $+$0.110, 0.133 & $+$0.323, $5.33\times 10^{-7}$ & 
$-$0.415, $2.45\times 10^{-7}$ &    $-$0.207, 0.0060 & $+$0.843, $<10^{-8}$ & --- & 0.337, 4.50$\times 10^{-7}$ \\
$E_{\rm peak, z}$    & $-$0.397, $3.06\times 10^{-5}$ & $-$0.131, 0.1889 & $+$0.273, 0.0065 & $-$0.006, 0.959 & $-$0.103, 0.2977 & $-$0.122, 0.3312 &
    $-$0.196, 0.077 & $+$0.384, $6.36\times 10^{-5}$ & $+$0.334, $4.81\times 10^{-4}$ & --- \\
\enddata

\tablenotetext{1}{Kendall's $\tau$ and Pearson correlation coefficients and probabilities are listed
above and below the diagonal, respectively. The number of parameter pairs are given in Table\,\ref{correlation_tab_z}.}

\end{deluxetable}

Somewhat expected are correlations among afterglow parameters. As listed in 
Table\,\ref{correlation_tab_all} we find clear 
correlations between the break times before and after the plateau phase ($P<10^{-8}$), 
and between the decay slopes 
during the plateau and the normal afterglow decay phase \axb\ and \axc\ ($P<10^{-8}$). 
These two relations are displayed 
in Figures\,\ref{tb1_tb2} and \ref{ax2_ax3}, respectively. 
The $T_{\rm break 1}$ - $T_{\rm break, 2}$ relation is shown for the 
observed as well as for the rest-frame times. For the observed break times we 
found $r_s$=0.517 and $T_s$=10.020 (277 GRBs), and $r_s$=0.620, $T_s=$8.463 (117 GRBs)
for the observed and rest-frame times, respectively. In both cases the probability of 
a random result is $P<10^{-8}$. Significant correlations exist between the decay slope during 
the plateau phase \axb\ and the observed
break time after the plateau phase $T_{\rm break, 2}$ with 
$r_s=+0.254$, $T_s=+5.320$ and P=1.75$\times 10^{-7}$ (414 GRBs), and the 
decay slope of the `normal' decay slope \axc\ and the observed break time before the plateau phase with
$r_s=+0.231$, $T_s=+4.825$ and P=2.0$\times 10^{-6}$ (414 GRBs). Note that these correlations disappear
in the rest-frame (Table\,\ref{correlation_tab_z}).

As shown by \citet{sakamoto09} there is a clear anti-correlation between the
photon index in the BAT spectrum and the peak energy
in the spectrum $E_{\rm peak}$ by a
relation log $E_{\rm peak}$ = 3.258-0.829$\times \Gamma$. Especially after the
launch of {\it Fermi} with its all-sky monitor GBM,
the number of GRBs with $E_{\rm peak}$ measurements has significantly
increased. While the pre-{\it Fermi} sample in \citet{sakamoto09} contained 55 bursts
with $E_{\rm peak}$ measurements, the sample in our paper has measurements of 201
bursts with $E_{\rm peak}$ measurements
(including those presented in \citet{sakamoto09}) for which 188 are long GRBs.
This relation is
displayed in the left panel of Figure\,\ref{gamma_epeak}. Of these bursts, 103 had
spectroscopic redshift measurements. As for the long GRBs
 we obtained a regression fit and
found the following relations between the 
observed and rest-frame $E_{\rm peak}$ with $\Gamma$, respectively: 
$log~E_{\rm peak} = (2.91\pm0.12)-(0.49\pm0.08)\times\Gamma$ and 
$log~E_{\rm peak, z} = (3.32\pm0.16)-(0.43\pm0.11)\times\Gamma$. 
The relation of $\Gamma$ with $E_{\rm peak, z}$ in
the rest-frame of the GRB is shown in the right panel of
Figure\,\ref{gamma_epeak}. In both cases there is a clear anti-correlation between
the two properties with $r_s=-0.462$, $T_s=-7.455$, $P<10^{-8}$, and --0.416,
--4.625, $1.10\times10^{-5}$ for the observed and rest-frame $E_{\rm peak}$
respectively (206 and 104 GRBs respectively).
The solid line in these plots display the relations between $E_{\rm peak}$ and $\Gamma$ 
that we found for our \swift\ GRB sample and the dotted line shows the relation found by \citet{sakamoto09}.

The best-known relationships of the peak energy $E_{\rm peak}$,  are those
between the isotropically radiated energy $E_{\rm iso}$ and the collimation-corrected
energy $E_{\gamma}$, the Amati and Ghirlanda relations, respectively
\citep{amati02, ghirlanda04}. Similar to these relations is the $E_{\rm peak} -
L_{\rm peak}$ relation found by \citet{yonetoku04} from BATSE detected bursts. 
Previously \citet{schaefer07} found this relation for pre-\swift\ bursts, however,  
it has also been found in \swift-detected 
bursts as reported by \citep{nava12} who used 58 bursts with photon peak 
fluxes in the 15-150 BAT 
energy window with $P>2.6$ photons cm$^{-2}$ s${-1}$. 
We
found a similar correlation between the rest-frame $E_{\rm peak_z}$ and the prompt
emission  luminosity and isotropic energy $E_{\rm iso}$
as displayed in Figure\,\ref{epeak_lum}.
These correlations (103 long GRBs) are
quite tight with $r_s$=0.426, $T_s$=4.735 and $P=7.0\times 10^{-6}$ for the luminosity
and $r_s$=0.459, $T_s$=5.187 with $P=1.07\times 10^{-6}$ for $E_{\rm iso}$.

The question is: does $E_{\rm peak, z}$ also correlate with other burst
properties? What we found was that among the long bursts 
there is a correlation between $E_{\rm peak, z}$ and the 
decay slope during the plateau phase \axb, with $r_s$=0.344,
$T_s$=3.470, and $P=7.94\times 10^{-4}$ (92 GRBs), as displayed in
Figure\,\ref{plateau_epeak}.
This again demonstrates 
a clear connection between prompt and
afterglow emission properties: bursts with steeper decay slopes during the plateau phase show peak
energies in the rest-frame  $E_{\rm peak_z}$ at high energies than bursts with flatter decay slopes.  
All other properties, such as \bx, or \axc\ do not
show any significant correlation with $E_{\rm peak_z}$.

The relations of the 
K-corrected 15-150 keV luminosity and the 
spectral slope in the BAT band $\Gamma_{\rm BAT}$ and the X-ray spectral slope of the 
afterglow emission are displayed in Figure\,\ref{lum_gamma_bx}. While there is no correlation
with \bx\  (right panel in Figure\,\ref{lum_gamma_bx}), there is a strong correlation with the 
BAT photon index $\Gamma_{\rm BAT}$. Here we found an anti-correlation for the long bursts
with $r_s=-0.457$, $T_s=-7.519$ and a probability $P<10^{-8}$ (216 GRBs).
In other words, bursts with steep 
BAT photon indices tend to be less luminous than bursts with flatter $\Gamma_{\rm BAT}$.
Note that short duration GRBs do not follow this anti-correlation. 
Table\,\ref{correlation_tab_z} lists the
correlations of all bursts, including short bursts. The anti-correlation between $\Gamma$ and $L_{\rm 15-150 keV}$ becomes sightly
weaker for all GRBs. 
 Although this anti-correlation is a relation between a redshift dependent
parameter ($\Gamma$) and a redshift independent parameter ($L_{\rm 15-150 keV}$), GRBs still can not be used as
standard candles. Not only is the scatter in the relation quite large, but $L_{\rm 15-150 keV}$ is not independent
of $\Gamma$ because we used $\Gamma$ as an input parameter in the k-correction. 
Now lets see how $\Gamma$ and \bx\ depend on the 15-150 keV energy release. These relations are displayed in the 
lower panels in Figure\,\ref{lum_gamma_bx} where  $\Gamma$ anti-correlates strongly with the isotropic 
energy in the 15-150 keV band $E_{\rm 15-150 keV}$ ($r_{\rm s}$=--0.492, $T_{\rm s}$=--8.268, $P<10^{-8}$ for the 216 long 
GRBs. Again, 
as listed in Table\,\ref{correlation_tab_all} the relation becomes weaker for all GRBs. 
As for the 0.3-10 keV spectral slope \bx, this is only a 
weak trend with $r_{\rm s}$=-0.177, $T_{\rm s}=-2.634$, and $P=0.0090$ for the long GRBs (215).

Figure\,\ref{lum_t90} displays the relation between the rest-frame $T_{\rm 90, z}$ and the
luminosity in the 15-150 keV band. For the whole GRB sample we found a Spearman rank order
correlation coefficient of $r_{\rm s}$=--0.290 and a Student's T-test of $T_{\rm s}$=--4.580
with a probability of $P=8.41\times 10^{-6}$ (231 GRBs)
However, if we only look at the long GRBs in the
sample with spectroscopic redshifts (216 GRBs), this correlation becomes even stronger with
$r_{\rm s}$=--0.420, $T_{\rm s}$=--6.777, and $P<10^{-8}$.
This relation suggests a strong
anti-correlation with high luminosity bursts having shorter $T_{90}$. However, keep in mind
that the luminosity and $T_{90}$ are not independent parameters. The luminosity was calculated
by $L=E/T_{90}$ so bursts with longer $T_{90}$ will have lower luminosities if we assume that
their isotropic energies are somewhat similar. Although the $T_{90}$ - luminosity relation
looks like a significant correlation it may not be a physical relation. If more energetic
bursts have really shorter $T_{90}$ then we should expect such a relation between the
isotropic energy in the 15-150 keV band and $T_{90}$. This is, however, a weak correlation
with $r_{\rm s}$=+0.197 and $T_{\rm s}$=3.044 with $P=2.60\times 10^{-3}$ of a random
distribution. This relation, however, is driven by the short bursts. When applying this relation
to only long bursts, the correlation disappears completely.

\section{Redshift estimate based in BAT parameters \label{z_predict}}

In Section \,\ref{redshift} we showed that it may be possible to determine the
15-150 keV $E_{\rm iso}$ based on the observed \swift-BAT parameter. 
The relations from
the PCA as described in Section\,\ref{redshift} are shown in the 
the eigenvector 1  - $E_{\rm iso}$ diagram,  
Figure\,\ref{ev1_eiso}. In order to place a new GRB onto this digram we give the
equations of the normalized BAT parameters 
$T_{90}$, $\Gamma$, and fluence below:

\begin{equation}
log T_{\rm 90, norm} = (log T_{90}-1.636871)/0.557497
\end{equation}

\begin{equation}
  \Gamma_{\rm norm} = (\Gamma-1.643410)/0.4071438
\end{equation}

\begin{equation}
log Fluence_{\rm norm} = (log Fluence+5.650423)/0.5962325
\end{equation}

Now we can calculate eigenvector 1 based on the PCA performed on the long GRBs with
spectroscopic redshift measurements:

\begin{equation}
eigenvector 1 = -0.558370*T_{\rm 90, norm}+0.455017*\Gamma_{\rm norm} -0.693674*Fluence_{\rm norm}
\end{equation}

\section{Separating Short and Long Duration GRBs \label{long_short}}

In this section we give the relations that will allow the reader to place a 
new GRB in the eigenvector 1 -  eigenvector 2 diagram, 
Figure\,\ref{cluster_ev1_ev2} as described in Section\,\ref{long-short}.
We first list the equation for the normalized $T_{90}$, $\Gamma$, and fluence:

\begin{equation}
log T_{\rm 90, norm} = (log T_{90}-1.43508)/0.798325
\end{equation}

\begin{equation}
  \Gamma_{\rm norm} = (\Gamma-1.614035)/0.4201068
\end{equation}

\begin{equation}
log Fluence_{\rm norm} = (log Fluence+5.919272)/0.6762907
\end{equation}

Now we can calculate the eigenvectors:

\begin{equation}
eigenvector 1 = 0.714278*T_{\rm 90, norm}+0.06593*\Gamma_{\rm norm} +0.696829*Fluence_{\rm norm}
\end{equation}

\begin{equation}
eigenvector 2 = 0.120528*T_{\rm 90, norm}+0.969348*\Gamma_{\rm norm} -0.214095Fluence_{\rm norm}
\end{equation}

\end{appendix}


\begin{thebibliography}{}
\bibitem[Abdo et al.(2009)]{abdo09} Abdo, A.A., et al., Nature, 462, 331
\bibitem[Aharonian et al.(2009)]{aharonian09} Aharonian, F., et al., 2009, \apj,
690, 1068
\bibitem[Amati et al.(2002)]{amati02} Amati, L., et al., 
2002, \aap, 390, 81 
\bibitem[Aptekar et al. (1995)]{aptekar95} Aptekar, R.L., et al., 1995, \ssr, 71, 1, 265
\bibitem[Arnaud (1996)]{arnaud96} Arnaud, K.~A., 1996, ASP
Conf.~Ser.~101: Astronomical Data Analysis Software and Systems V, 101, 17
\bibitem[Atwood et al.(2009)]{atwood09} Atwood, W.B., et al., 2009, \apj, 697, 1071
\bibitem[{{Bal{\'a}zs} \& {Veres}(2011)}]{Balazs+11multi}
{Bal{\'a}zs}, L.~G., \& {Veres}, P. 2011, Advances in Space Research, 47, 1404
\bibitem[Band et al.(1993)]{band93} Band, D., et al., 1993, \apj, 413, 281
\bibitem[Barthelmy et al.(1995)]{barthelmy95} Barthelmy, S.D., Butterworth, P.,
Cline, T., Gehrels, N., Fishman, G.J., Kouveliotou, C., \& Meegan, C.A., 1995,
Ap\&SS, 231, 235
\bibitem[Barthelmy (2005)]{barthelmy05} Barthelmy, S.D., 2005, Space Science
Reviews, 120, 143
\bibitem[{{Beloborodov}(2010)}]{Beloborodov10phot}
{Beloborodov}, A.~M. 2010, \mnras, 407, 1033, 0907.0732
\bibitem[Berger et al.(2005a)]{berger05b} Berger, E., Kulkarni, S.R., Fox,
D.B., et al., 2005a, \apj, 629, 328
\bibitem[Berger et al.(2005b)]{berger05c} Berger, E., et al., 2005b, GCN 3368 
\bibitem[Berger et al. (2006a)]{berger06a} Berger, E., Price, P.A., \& Fox,
D.B., 2006a, GCN 4622
\bibitem[Berger et al. (2006b)]{berger06b} Berger, E., Kulkarni, S.R., Rau, A.,
\& Fox, D.B., 2006b, GCN 4815
\bibitem[Berger et al. (2007)]{berger06d} Berger, E., et al., 2007, \apj, 664, 1000
\bibitem[Berger \& Gladders (2006)]{berger06c} Berger, E., \& Gladders, M.,
2006, GCN 5170
\bibitem[Berger (2006)]{berger06d} Berger, E., 2006, GCN 5962
\bibitem[{{Bernardini} {et~al.}(2012){Bernardini}, {Margutti}, {Mao},
  {Zaninoni}, \& {Chincarini}}]{Bernardini+12xrayag}
{Bernardini}, M.~G., {Margutti}, R., {Mao}, J., {Zaninoni}, E., \&
  {Chincarini}, G. 2012, \aap, 539, A3
  \bibitem[Bernardini et al. (2012b)]{bernardini12} Bernardini, M.G.,
 Margutti, R., Zaninoni, E., \& Chincarini, G., 2012b, \mnras, 425, 1199
\bibitem[{{Blandford} \& {Znajek}(1977)}]{Blandford+77Znajek}
{Blandford}, R.~D., \& {Znajek}, R.~L. 1977, \mnras, 179, 433
\bibitem[Bloom et al.(2005)]{bloom05} Bloom, J.S., Perley, D., Foley, R.,
Prochaska, J.X., Chen, H.W., \& Starr, D., 2005, GCN 3758
\bibitem[Bloom et al. (2006a)]{bloom06} Bloom, J.S., Foley, R.J.,
Koceveki, D., \& Perley, D., 2006a, GCN 5217
\bibitem[Bloom et al. (2006b)]{bloom06b} Bloom, J.S., Perley, D., \& Chen, H.W.,
2006b, GCN 5826
\bibitem[Boroson (2002)]{boroson02} Boroson, T.A., 2002, \apj, 565, 78 
\bibitem[Boroson \& Green (1992)]{boroson92} Boroson, T.A., \& Green, R.F., 1992,
\apjs, 80, 109
\bibitem[Bromberg et al.(2013)]{bromberg12} Bromberg, O., Nakar, E., Piran, T.,
 \& Sari, R, 2013, \apj, 769, 179
\bibitem[Burrows et al.(2005)]{burrows05} Burrows, D.N., et al., 2005, 
Space Science Reviews, 120, 165
\bibitem[Burrows et al.(2006)]{burrows06} Burrows, D.N., et al., \apj, 653, 458
\bibitem[Campana et al.(2006a)]{campana06} Campana, S., et al., 2006a, 
\aap, 449, 61
\bibitem[Campana et al.(2006b)]{campana06b} Campana, S., et al., 2006b,
Nature, 442, 1008
\bibitem[Campisi \& Li(2008)]{campisi08} Campisi, M.A., \& Li, L-X., 2008, \mnras,
391, 935
\bibitem[Cao et al.(2011)]{cao11} Cao, X.-F., Yu, Y.-W., Cheng, K.S., \& Zheng, X.-P, 2011, \mnras, 416, 2174
\bibitem[Castro-Tirado et al. (2006)]{castro06} Castro-Tirado, A.J., 
Amado, P., Negueruela, I., Gorosabel, J., Jel\'inek, M.,
\& A. de Ugarte Postigo, 2006, GCN 5218
\bibitem[Cenko et al.(2005)]{cenko05} Cenko, S.B., et al., 2005, GCN 3542
\bibitem[Cenko et al.(2006)]{cenko06} Cenko, S.B., Berger, E., Djorgovski, S.G.,
Mahabal, A.A., \& Fox, D.B., 2006, GCN 5155
\bibitem[Chen et al.(2005)]{chen05} Chen, H.-W., et al., 2005, GCN 3706
\bibitem[{{Corsi} \& {M{\'e}sz{\'a}ros}(2009)}]{Corsi+09mag}
{Corsi}, A., \& {M{\'e}sz{\'a}ros}, P. 2009, \apj, 702, 1171
\bibitem[Costa et al.(1999)]{costa99} Costa, E., et al., 1999, A\&AS, 138, 425
\bibitem[Coward et a.(2013)]{coward12} Coward, D.M., Howell, E.J., Branchesi, M., Stratta, G, Guetta, D.,
Gendre, B., \& Macpherson, D., 2013, MNRAS, 432, 2141
\bibitem[Crawley (2007)]{crawley07} Crawley, M.J., 2007, The R Book, Wiley \& Sons
\bibitem[Cucchiara et al. (2006a)]{cucchiara06a} Cucchiara, A., Fox, D.B., \&
Berger, E., 2006a, GCN 4729
\bibitem[Cucchiara et al. (2006b)]{cucchiara06b} Cucchiara, A., Price, P.A.,
Fox, D.B., Cenko, S.B., \& Schmidt, B.P.,  2006b, GCN 5052
\bibitem[{{Dai} \& {Lu}(1998)}]{dailu98b}
{Dai}, Z.~G., \& {Lu}, T. 1998, \aap, 333, L87
\bibitem[Dainotti et al.(2008)]{dainotti08} Dainotti, M.G., Cardone, V.F., 
\& Capozziello, S, 2008, \mnras, 391, 79
%\bibitem[Dainotti et al.(2011)]{dainotti11} Dainotti, M.G., Ostrowski, M., \&
%Willingale, R., 2011, \mnras, 418, 2202
\bibitem[{Dall'Osso} {et~al.}(2011)]{DallOsso+11injection}
{Dall'Osso}, S., {Stratta}, G., {Guetta}, D., {Covino}, S., {De Cesare}, G., \&
  {Stella}, L. 2011, \aap, 526, A121
\bibitem[D'Avanzo et al. (2012)]{davanzo12} D'Avanzo, P., et al., 2012, \mnras, 
425, 506
\bibitem[De Pasquale et al.(2006)]{depasquale06} De Pasquale, M., et al., 2006, \aap, 455, 813
\bibitem[De Pasquale et al.(2010)]{depasquale10} De Pasquale, M., et al., 2010, \apj, 709, L146
\bibitem[Dickey \& Lockman(1990)]{dic90} Dickey, J.M., \& Lockman, F.J., 1990,
\araa, 28, 215
\bibitem[{{Drenkhahn} \& {Spruit}(2002)}]{Drenkhahn+02}
{Drenkhahn}, G., \& {Spruit}, H.~C. 2002, \aap, 391, 1141
\bibitem[{{Eichler} \& {Granot}(2006)}]{eichler06}
{Eichler}, D., \& {Granot}, J. 2006, \apjl, 641, L5
\bibitem[Evans et al.(2007)]{evans07} Evans, P.A., et al., 2007, \aap, 469, 379
\bibitem[Evans et al.(2009)]{evans09} Evans, P.A., et al., 2009, \mnras, 397,
1177
\bibitem[Everitt \& Hothorn (2010)]{everitt10} Everitt, B.S., \& Hothorn, T., 2010, ``A
Handbook of Statistical Analysis using R'', Chapman \& Hall, Boca Raton, FL
\bibitem[Everitt et al. (2011)]{everitt11} Everitt, B., Landau, S., Leese, M., \& Stahl, D., 2011, 
``Cluster Analysis, 5th edition'', John Wiley \& Sons, Ltd, Chichester, UK. 
\bibitem[Falcone et al.(2007)]{falcone07} Falcone, A.D., et al., 2007, \apj,
671, 1921
\bibitem[Feigelson \& Babu (2012)]{feigelson12} Feigelson, E.D., \& Babu, G.j., 2012,
``Modern Statistical Methods for Astronomy'', Cambridge University Press
\bibitem[Fong et al.(2013)]{fong13} Fong W., et al., 2013, \apj, 769, 56
\bibitem[Frail et al.(2001)]{frail01} Frail, D.A., et al., 2001, \apj, 562, L55
\bibitem[Francis \& Wills(1999)]{francis99} Francis, P.J., \& Wills, B.J., 1999, ASP Conf. Series, 162, 363
\bibitem[Fynbo et al.(2009)]{fynbo09} Fynbo, J.P.U., et al., 2009, \apjs, 185,
526
\bibitem[Gehrels et al.(2004)]{gehrels04} Gehrels, N., et al., 2004, ApJ, 611,
1005
\bibitem[Gehrels et al. (2006)]{gehrels06} Gehrels, N., et al., Nature, 444, 1044
\bibitem[Gehrels et al.(2008)]{gehrels08} Gehrels, N., et al., \apj, 689, 1161
\bibitem[Gehrels \& Cannizzo (2012)]{gehrels12} Gehrels, N., \& Cannizzo, J.K., 2012, proceedings of the 
"New windows on transients across the Universe issue" Discussion Meeting issue 
of Philosophical Transactions A, ed. P. O'Brien, S. Smartt, R. Wijers \&; K. Pounds 
\bibitem[{{Genet} {et~al.}(2007){Genet}, {Daigne}, \&
  {Mochkovitch}}]{Genet+07shallow}
{Genet}, F., {Daigne}, F., \& {Mochkovitch}, R. 2007, \mnras, 381, 732
\bibitem[Ghirlanda et al.(2004)]{ghirlanda04} Ghirlanda, G., Ghisellini, G., \&
Lazzati, D., 2004, \apj, 616, 331
\bibitem[{{Ghisellini} {et~al.}(2007){Ghisellini}, {Ghirlanda}, {Nava}, \&
  {Firmani}}]{Ghisellini:07a-late}
{Ghisellini}, G., {Ghirlanda}, G., {Nava}, L., \& {Firmani}, C. 2007, \apjl,
  658, L75
\bibitem[Godet et al.,(2009)]{godet09} Godet, O., et al., 2009, \aap, 494, 775
\bibitem[{{Granot} \& {Kumar}(2006)}]{granot06}
{Granot}, J., \& {Kumar}, P. 2006, \mnras, 366, L13, arXiv:astro-ph/0511049
\bibitem[Greiner et al. (2008)]{greiner08} Greiner, J., et al., \pasp, 120, 405
\bibitem[Grieco et al.(2012)]{grieco12} Grieco, V., Matteucci, F., Meynet, G., Longo, F., Della Valle, M., \&
Salvaterra, R., 2012, \mnras, 432, 2141
\bibitem[Grupe (2004)]{grupe04} Grupe, D., 2004, \aj, 127, 1799
\bibitem[Grupe (2012)]{grupe12} Grupe, D., 2012, Proceedings of the Gamma-Ray 
Bursts 2012 Conference, Munich/Germany, PoS(GRB 2012)065
\bibitem[Grupe et al. (2006)]{grupe06} Grupe, D., et al., 2006, \apj, 645, 464
\bibitem[Grupe et al. (2007)]{grupe07} Grupe, D., et al., 2007, \aj, 133, 2216
\bibitem[Halpern \& Mirabal (2006)]{halpern06} Halpern, J.P., \& Mirabal, N, 2006, GCN 5982
\bibitem[Hill et al. (2004)]{hill04} Hill, J.E., et al., 2004, SPIE, 5165, 217
\bibitem[Hopkins \& Beacom (2006)]{hopkins06} Hopkins, A.M., \& Beacom, J.F., 2006, \apj, 651, 142
\bibitem[Hullinger et al.(2005)]{hullinger05} Hullinger, D., et al., 2005, GCN circ. 4237
\bibitem[Humason et al.(1956)]{humason56} Humason, M.L., Mayall, N.U., \& Sandage,
A., 1956, \aj, 61, 97
\bibitem[{{Ioka} {et~al.}(2006){Ioka}, {Toma}, {Yamazaki}, \&
  {Nakamura}}]{ioka06}
{Ioka}, K., {Toma}, K., {Yamazaki}, R., \& {Nakamura}, T. 2006, \aap, 458, 7
\bibitem[Jakobsson et al.(2012)]{jakobsson12} Jakobsson, P., et al., 2012, \apj,
752, 62
\bibitem[Kalberla et al.(2005)]{kalberla05} Kalberla, P.M.W., et al., \aap, 440,
775
\bibitem[Kovecski (2012)]{kovecski12} Kovecski, D., 2012, \apj, 747, 146
\bibitem[Kouveliotou et al.(1993)]{kouveliotou93} Kouveliotou, C., Meegan, C.A., Fishman, G.J., Bhat, N.P., Briggs, M.S., 
Koshut, T.M., Paciesas, W.S., \& Pendleton, G.N., 1993, \apj, 413, L101
\bibitem[Kr\"uhler et al.(2011)]{kruehler11} Kr\"uhler, T., et al., 2011, \aap, 526, 153
\bibitem[Kr\"uhler et al.(2012)]{kruehler12} Kr\"uhler, T., et al., 2012, \apj,
758, 46
\bibitem[Kumar \& Piran (2000)]{kumar00} Kumar, P., \& Piran, T., 2000, \apj, 535,
152
\bibitem[{{Kumar} {et~al.}(2008){Kumar}, {Narayan}, \& {Johnson}}]{Kumar+08acc}
{Kumar}, P., {Narayan}, R., \& {Johnson}, J.~L. 2008, Science, 321, 376
\bibitem[{{Liang} {et~al.}(2007){Liang}, {Zhang}, \& {Zhang}}]{liang07b}
{Liang}, E.-W., {Zhang}, B.-B., \& {Zhang}, B. 2007, \apj, 670, 565
\bibitem[Littlejohns et al. (2013)]{littlejohns13} Littlejohns,  O., Tanvir, N., Willingale, R., O'Brien, P., Evans, P., \& Levan, A.,
2013, 7th Huntsville Gamma-Ray Burst Symposium, GRB 2013: paper 36 in eConf Proceedings C1304143
\bibitem[L\"u et al. (2010)]{lu10} L\"u, H.,  Liang. E.-W., Zhang, B.B., \& Zhang, B.,
2010, \apj, 725, 1965 
\bibitem[{{Lyons} {et~al.}(2010){Lyons}, {O'Brien}, {Zhang}, {Willingale},
  {Troja}, \& {Starling}}]{Lyons+10magnetar}
{Lyons}, N., {O'Brien}, P.~T., {Zhang}, B., {Willingale}, R., {Troja}, E., \&
  {Starling}, R.~L.~C. 2010, \mnras, 402, 705
  \bibitem[Margutti et al. (2011a)]{margutti11a} Margutti, R., et al., 2011a,
\mnras, 410, 1064
\bibitem[Margutti et al. (2011b)]{margutti11b} Margutti, R., et al., 2011b,
\mnras, 417, 2144
\bibitem[Margutti et al.(2013)]{margutti12} Margutti, R., et al., 2013, 
\mnras, 428, 729
\bibitem[{{McKinney} \& {Uzdensky}(2012)}]{McKinney+11switch}
{McKinney}, J.~C., \& {Uzdensky}, D.~A. 2012, \mnras, 419, 573
\bibitem[Meegan et al. (2009)]{meegan09} Meegan, C.A., et al., 2009, \apj, 702, 791
\bibitem[\meszaros (2006)]{meszaros06} \meszaros, P., 2006, Rep. Prog. Phys. 
69, 2259
\bibitem[{{M\'esz\'aros} \& {Rees}(1997)}]{Meszaros+97ag}
{M\'esz\'aros}, P., \& {Rees}, M.~J. 1997, \apj, 476, 232,
\bibitem[{{Metzger} {et~al.}(2011){Metzger}, {Giannios}, {Thompson},
  {Bucciantini}, \& {Quataert}}]{Metzger+11grbmag}
{Metzger}, B.~D., {Giannios}, D., {Thompson}, T.~A., {Bucciantini}, N., \&
  {Quataert}, E. 2011, \mnras, 413, 2031
\bibitem[Morgan et al.(2012)]{morgan11} Morgan, A.N., Long, J., Richanrds, J.W.,
Broderick, T., Butler, N.R., \& Bloom, J.S., 2012, \apj, 746, 170 
\bibitem[Mukherjee et al., (1998)]{mukherjee98} Mukherjee, S., et al., 1998, \apj, 508, 314 
\bibitem[Nava et al. (2012)]{nava12} Nava, L., et al., 2012, \mnras, 421, 1256
\bibitem[Norris et al. (2000)]{norris00} Norris, J.P/. Marani, G.F., \& Bonnell, J.T., 2000, 
\apj, 534, 248
\bibitem[Nousek et al.(2006)]{nousek06} Nousek, J.A., et al., 2006, \apj, 642, 389
\bibitem[O'Brien et al. (2006)]{obrien06} O'Brien, P.T., et al., 2006, \apj,
647, 1213
\bibitem[Oke \& Sandage (1968)]{oke68} Oke, J.B., \& Sandage, A., 1968, \apj, 154, 21
\bibitem[Palmer et al.(2012)]{palmer12} Palmer, D.M., et al., 2012, GCN circ. 13186
\bibitem[{{Panaitescu} \& {Kumar}(2000)}]{panaitescu00}
{Panaitescu}, A., \& {Kumar}, P. 2000, \apj, 543, 66
\bibitem[{{Pe'er} {et~al.}(2006){Pe'er}, {M{\'e}sz{\'a}ros}, \&
  {Rees}}]{Peer+06phot}
{Pe'er}, A., {M{\'e}sz{\'a}ros}, P., \& {Rees}, M.~J. 2006, Astrophys.J., 642,
  995
\bibitem[Pearson (1901)]{pearson1901} Pearson, K., 1901, Philosophical Magazine 2 (6), 559
\bibitem[Perley et al.(2005)]{perley05} Perley, D.A., Foley, R.J., Bloom, J.S., \& Butler, N.R.,
2005, GCN circ. 5387 
\bibitem[Qin et al.(2013)]{qin13} Qin, Y., et al., 2013, \apj, 763, 15 
\bibitem[Racusin et al.(2009)]{racusin09} Racusin, J.L., et al., 2009, \apj,
698, 43
\bibitem[Racusin et al.(2011)]{racusin11} Racusin, J.L., et al., 2011, \apj, 738, 138
\bibitem[{{Rees} \& {M\'esz\'aros}(1998)}]{Rees+98refresh}
{Rees}, M.~J., \& {M\'esz\'aros}, P. 1998, \apjl, 496, L1
\bibitem[Richards et al(2006)]{richards06} Richards, G.T., et al., 2006, \aj, 131, 2766
\bibitem[{{Rhoads}(1999)}]{Rhoads99jet}
{Rhoads}, J.~E. 1999, \apj, 525, 737
\bibitem[Roming et al.(2005a)]{roming05} Roming, P.W.A., et al., 2005a, Space
Science Reviews, 120, 95
\bibitem[Roming et al. (2006)]{roming06} Roming, P.W.A., et al., 2006, \apj,
652, 1416
\bibitem[Ross et al(2012)]{ross12} Ross, N.P., et al., 2012, \apj, 773, 14
\bibitem[{{Rowlinson} {et~al.}(2013){Rowlinson}, {O'Brien}, {Metzger},
  {Tanvir}, \& {Levan}}]{Rowlinson+13magnetar}
{Rowlinson}, A., {O'Brien}, P.~T., {Metzger}, B.~D., {Tanvir}, N.~R., \&
  {Levan}, A.~J. 2013, \mnras, 430, 1061
 \bibitem[The R-Team (2009)]{rteam09} R Development Core Team (2009). R: A language and environment
  for statistical computing. R Foundation for Statistical
  Computing, Vienna, Austria. ISBN 3-900051-07-0, URL
  http://www.R-project.org. 
\bibitem[Sakamoto et al.(2008)]{sakamoto08} Sakamoto, T., et al., 2008, \apj, 679, 570
\bibitem[Sakamoto et al.(2009)]{sakamoto09} Sakamoto, T., et al., 2009, \apj, 693,
922
\bibitem[Salvaterra et al.(2008)]{salvaterra08} Salvaterra, R., Cerutti, A.,
Chincarini, G., Colpi, M., Guidorzi, C., \& Romano, P., 2008, \mnras, 388, L6
\bibitem[Salvaterra et al.(2012)]{salvaterra12} Salvaterra, R., et al., 2012, \apj, 749, 68
\bibitem[Savaglio, S., et al.(2007)]{savaglio07} Savaglio, S., Palazzi, E., Ferrero, P., \& Klose, S., 2007, GCN 6166
\bibitem[Schaefer (2007)]{schaefer07} Schaefer, B.E., 2007, \apj, 660, 16
\bibitem[Schmidt (2009)]{schmidt09} Schmidt, M., 2009, \apj, 700, 633
\bibitem[Soderberg et al.(2005)]{soderberg05} Soderberg, A.M., Berger, E., \& Ofek, E., 2005, GCN
4186
\bibitem[Soderberg et al. (2006)]{soderberg06} Soderberg, A.M., et al., 2006, Nature, 442, 1024.
\bibitem[Still et al. (2006)]{still06} Still, M., et al., 2006, GCN 5226
\bibitem[Stratta et al. (2004)]{stratta04} Stratta, G., Fiore, F., Antonelli,
L.A., Piro, L., \& De Pasquale, M., 2004, \apj, 608, 846
\bibitem[Stratta et al.(2009)]{stratta09} Stratta, G., Guetta, D., D'Elia, V.D.,
Perri, M., Corvino, S., \& Stella, L., 2009, \aap, 494, L9
\bibitem[Tagliaferri et al.(2005)]{tagliaferri05} Tagliaferri, G., et al., 2005, GCN Circ. 4222
\bibitem[Tanvir et al.(2012)]{tanvir12} Tanvir, N.R., et al., 2012, \apj, 754, 46
\bibitem[Thoene et al.(2006a)]{thoene06} Thoene, C.C., et al., 2006a, GCN 5373
\bibitem[Thoene et al.(2006b)]{thoene06b} Thoene, C.C., Fynbo, J.P.U., Jakobsson, P., Vreeswijk, P.M., \& Hjorth, J., 
2006b, GCN 5812
\bibitem[{{Thompson}(1994)}]{Thompson94}
{Thompson}, C. 1994, \mnras, 270, 480
\bibitem[{{Toma} {et~al.}(2006){Toma}, {Ioka}, {Yamazaki}, \&
  {Nakamura}}]{toma06}
{Toma}, K., {Ioka}, K., {Yamazaki}, R., \& {Nakamura}, T. 2006, \apjl, 640,
  L139
\bibitem[Torgo (2011)]{torgo11} Torgo, L., 2011, ``Data Mining in R'', Chapman \&
Hall/CRC
\bibitem[{{Uhm} \& {Beloborodov}(2007)}]{Uhm+07shallow}
{Uhm}, Z.~L., \& {Beloborodov}, A.~M. 2007, \apjl, 665, L93
\bibitem[Ukwatta et al.(2010)]{ukwatta10} Ukwtta, T., et al., 2010, \apj, 711, 1073
\bibitem[Ukwatta et al.(2012)]{ukwatta12} Ukwtta, T., et al., 2012, \mnras, 419, 614
\bibitem[Veres et al.(2010)]{veres10} Veres, P., Bagoly, Z., Horvath, I., \meszaros, A., \& Balazs, L.G., 2010, \apj, 725, 1955
\bibitem[Wang et al.(2004)]{wang04} Wang, J.X., Malhatra, S., Rhoads, J.E., \&
Norman, C.A., 2004, \apj, 612, L109
\bibitem[Wanderman \& Piran (2010)]{wanderman10} Wanderman, D., \& Piran, T.,
2010, \mnras, 406, 1944
\bibitem[Watson et al.(2006)]{watson06} Watson, D., et al., \apj, 652, 1011
\bibitem[Willingale et al.(2007)]{willingale07}
 Willingale, R., et al., 2007, \apj, 662, 1193
\bibitem[Wright (2006)]{wright06} Wright, E.L., 2006, \pasp, 118, 1711
\bibitem[Yonetoku et al.(2004)]{yonetoku04} Yonetoku, D., Murakami, T., Nakamura,
T, Yamazaki, R., Inoue, A.K., \& Ioka, K., 2004, \apj, 609, 935
\bibitem[{{Zalamea} \& {Beloborodov}(2011)}]{Zalamea+11nu}
{Zalamea}, I., \& {Beloborodov}, A.~M. 2011, \mnras, 410, 2302
\bibitem[{{Zhang} \& {M{\'e}sz{\'a}ros}(2001)}]{zhangmeszaros01a}
{Zhang}, B., \& {M{\'e}sz{\'a}ros}, P. 2001, \apjl, 552, L35
\bibitem[Zhang \& Meszaros (2004)]{zhang04} Zhang, B., \& Meszaros, P., 2004, Int. Jour. Mod. Phys. A, 19, 2385
\bibitem[Zhang et al.(2006)]{zhang06} Zhang, B., et al., 2006, \apj, 642, 354
\bibitem[{{Zhang}(2007)}]{zhangcjaa07}
{Zhang}, B. 2007, Chinese Journal of Astronomy and Astrophysics, 7, 1
\bibitem[Zhang (2009)]{zhang09} Zhang, B., et al., 2009, \apj, 703, 1696
\bibitem[Zhang (2012)]{zhang12} Zhang, B., 2012, Proceedings of the IAU 
Symposium 279, arXiv:1204.4919 
\bibitem[Zhang et al.(2007)]{zhangbb07} Zhang, B.-B., Liang, E.-W., \& Zhang, B.,
2007, \apj, 666, 1002 
\bibitem[{{Zhang} \& {Yan}(2011)}]{Zhang+11icmart}
{Zhang}, B., \& {Yan}, H. 2011, \apj, 726, 90
\bibitem[Zhang et al. (2009)]{zhangbb09} Zhang, B.-B., Zhang, B., Liang, E.-W., 
\& Wang, X.-Y., 2009, \apjl, 690, L10 
\end{thebibliography}
\end{document}